\title{\bf Neutrino Mass and Mixing: from Theory to Experiment}

\author{ 
  {Stephen F.~King}$^a$,
  {Alexander Merle}$^a$,
  {Stefano Morisi}$^b$, 
  {Yusuke Shimizu}$^c$,
  {Morimitsu\,Tanimoto}$^d$ \\
\\
  $^a${\small \it Physics and Astronomy, University of Southampton, Southampton, SO17 1BJ, U.K.}\\
  $^b${\small \it Institut f{\"u}r Theoretische Physik und Astrophysik,}\\
  {\small \it Universit{\"a}t W{\"u}rzburg, 97074 W{\"u}rzburg, Germany}\\
  $^c${\small \it Max-Planck-Institut f\"ur Kernphysik, Saupfercheckweg 1, D-69117 Heidelberg, Germany}\\
  $^d${\small \it Department of Physics, Niigata University, Niigata 950-2181, Japan}\\
       }

\date{\today}
\documentclass[11pt]{article}

\usepackage[margin=2cm]{geometry}
\usepackage{color}
\usepackage{graphicx}
\usepackage{amssymb}
\usepackage{amsmath} 
\usepackage{pifont} 
\usepackage{hyperref}
\usepackage[noadjust]{cite}
\usepackage[center,footnotesize,hang]{subfigure}
\usepackage{multirow}

\def\2tvec#1#2{
\left(
\begin{array}{c}
#1  \\
#2  \\   
\end{array}
\right)}

\def\mat2#1#2#3#4{
\left(
\begin{array}{cc}
#1 & #2 \\
#3 & #4 \\
\end{array}
\right)
}

\def\Mat3#1#2#3#4#5#6#7#8#9{
\left(
\begin{array}{ccc}
#1 & #2 & #3 \\
#4 & #5 & #6 \\
#7 & #8 & #9 \\
\end{array}
\right)
}

\def\3tvec#1#2#3{
\left(
\begin{array}{c}
#1  \\
#2  \\   
#3  \\
\end{array}
\right)}

\def\4tvec#1#2#3#4{
\left(
\begin{array}{c}
#1  \\
#2  \\   
#3  \\
#4  \\
\end{array}
\right)}

\def\5tvec#1#2#3#4#5{
\left(
\begin{array}{c}
#1  \\
#2  \\
#3  \\
#4  \\
#5  \\
\end{array}
\right)}

\begin{document}

\maketitle

\begin{abstract}

The origin of fermion mass hierarchies and mixings is one of the unresolved and most difficult problem in high-energy physics. One possibility to address the flavour problem is by extending the Standard Model to include a family symmetry. In the recent years it has become very popular to use non-Abelian discrete flavour symmetries because of their power in the prediction of the large leptonic mixing angles relevant for neutrino oscillation experiments. Here we give an introduction to the flavour problem and to discrete groups which have been used to attempt a solution for it. We review the current status of models in the light of the recent measurement of the reactor angle and we consider different model building directions taken. The use of the flavons or multi Higgs scalars in model building is discussed as well as the direct vs.\ indirect approaches. We also focus on the possibility to distinguish experimentally  flavour symmetry models by means of  mixing sum rules and mass sum rules. 
In fact, we illustrate in this review the complete path from mathematics, via model building, to
experiments, so that any reader interested to start working in the field could use this text as a starting point in order to get a broad overview of the different subject areas.

\end{abstract}

{\it Pacs}: 14.60.Pq, 12.60.Jv, 14.80.Cp.

\vskip5.mm

\tableofcontents
\newpage




\section{Introduction}

In the Standard Model (SM) we have three families of fermions. In each family we have two kinds of quarks, the up-quark with electric charge $Q=2/3$ and the down-quark with $Q=-1/3$, as well as two kinds of leptons, the charged leptons with $Q=-1$ and the neutrino with $Q=0$. We have three copies of such families: all the quantum numbers\footnote{In the SM the quantum numbers are the hypercharge $Y$, the weak isospin $T_3$, and the colour charge.} of each particle in the first family are identical to the quantum numbers of the corresponding particles in the second and third families.
\begin{table}[h!]
\begin{center}
\begin{tabular}{|c|c|c|c|}
    \hline
     Families & 1st & 2nd & 3rd\\
    \hline
    \multirow{2}{*}{Quarks} & $u$ & $c$ & $t$ \\
                                           & $d$ & $s$ & $b$ \\
    \hline
    \multirow{2}{*}{Leptons} & $e$ & $\mu$ & $\tau$ \\
                                           &$\nu_e$ & $\nu_\mu$ & $\nu_\tau$ \\
    \hline
\end{tabular}\caption{SM fermions.}\label{tabfl}
\end{center}
\end{table}
But particles with equal quantum numbers that belong to different families have different masses~\cite{Beringer:1900zz}, for instance the up-type quarks $u,\,c,\,t$ have equal quantum numbers but their masses are, respectively, about $0.0023$, $1.28$, $173\,$GeV. Instead, the down-quarks and charged leptons have similar masses: about $0.00055$, $0.11$, $1.8\,$GeV for the charged leptons and $0.0048$, $0.95$, $4.2\,$GeV for the down-type quarks. For many years neutrinos have been considered to be massless, but quite recently the experimental discovery of neutrino oscillations has confirmed that neutrinos are in fact massive, with a mass of the order of $10^{-10}$GeV.

One of the first problems that emerge from this picture is why we have exactly three copies of fermions and not another number. What is the origin of such a replication? The second question that shows up is why fermion masses are so hierarchical and are not of the same order. A third question is why neutrinos have masses so small compared to the charged fermions, which suggests to investigate the origin of neutrino masses. These questions are all part of the so-called {\it flavour problem} which will be the topic of this review.
 
In order to study the flavour problem we first have to understand the origin of fermion masses in the SM. In the electrically charged sector this is straightforward, as the masses arise from the Yukawa interactions,
\begin{equation}
\mathcal{L_Y}= Y_{u_{ij}} \overline{Q}_{L_{i}} H u_{R_j}+Y_{d_{ij}} \overline{Q}_{L_{i}} \tilde{H} d_{R_j}+
Y_{\ell_{ij}} \overline{L}_{L_{i}} H l_{R_j} + H.c.,
\end{equation}
where $\tilde{H}=\sigma_2 H^*$, $Q_{L_i}=(u_{L_i},d_{L_i})^T$, $L_{L_i}=(e_{L_i},\nu_{L_i})^T$, and where $L$($R$) means left-handed (right-handed) chirality. The Yukawa couplings $Y_{u,d,\ell}$ are arbitrary (but ideally perturbative) complex $3\times 3 $ matrices leading to a large number of free parameters, but not all of them are physical (some parameters can be reabsorbed). In particular in the quark sector it is possible to show that we can go in the basis where the up- and down-type quark mass matrices, $M_{u,d}=Y_{u,d} \, v_H$\,,\footnote{Where $v_H = 174$~GeV is the vacuum expectation value (VEV) of the Higgs doublet.} are diagonal. In this basis the charged current interactions are \emph{not} diagonal, 
\begin{equation}
V_{ij}\, \overline{u}_{L_i}\gamma^\mu d_{L_j}\,W_\mu + H.c.,
\end{equation}
where $V$ is a $3\times 3$ unitary matrix called the CKM (from Cabibbo-Kobayashi-Maskawa) matrix. This matrix represents the mismatch between the flavour and mass bases and it is parametrised by three mixing angles and one complex phase. The corresponding observables are well measured and give an almost diagonal CKM matrix with one angle of the order of $\lambda_C\equiv \sin\theta_C\approx 0.2$ (Cabibbo angle), two small angles (of the orders of $\lambda_C^2$ and $\lambda_C^3$, respectively), and a large $CP$ violating phase $\delta \approx 69^\circ$. Thus a fourth question naturally emerges: why are quarks of different flavours mixed?

As for quarks, the lepton charged currents are not diagonal in the basis where charged lepton and neutrino mass matrices are diagonal. In principle, we could repeat here the same argument give above for quarks. However, the origin of neutrino mass it is not clear, in contrast to the charged lepton masses, since we have theoretical uncertainties: in fact, being electrically neutral fermions, left and right-handed neutrinos could have a Dirac or a Majorana mass term\footnote{To be more specific, it is also possible to have intermediate cases like pseudo-Dirac~\cite{Wolfenstein:1981kw}, quasi-Dirac~\cite{Valle:1982yw}, schizophrenic~\cite{Allahverdi:2010us}, and so on, but in this review we will consider only the Dirac and Majorana cases.}
\begin{equation}
\nu_L^\dagger \nu_R + H.c.\quad\mbox{(Dirac)}\,,\qquad 
\nu_L^T \sigma_2 \nu_L + H.c.\quad\mbox{(Majorana)}\,.
\end{equation}
These two mass terms are phenomenologically very different, because in the Dirac case lepton number is conserved while in the Majorana case it is broken by 2 units. Moreover we observe that, in order to write a mass term for Dirac neutrinos, we must extend the SM to include at least 2 right-handed neutrinos $\nu_{R_i}$ ($i\ge 2$). If only two right-handed neutrino are assumed, the lightest neutrino mass is zero~\cite{Schechter:1980gr,Xing:2007uq}. For simplicity we will assume that right-handed neutrinos come sequentially in three generations, mimicking the electrically charged sector.\footnote{This assumption is quite reasonable having in mind $SO(10)$ grand unified frameworks, where all SM fermions and the right-handed neutrino belong to a ${\bf 16}$ multiplet.} In that case we can indeed repeat the same argument given for quarks: the resulting charged lepton current is not diagonal in the basis where the charged leptons and neutrino mass matrices are diagonal, but it must be multiplied by a unitary mixing matrix, the so-called 
lepton mixing matrix matrix. The lepton mixing matrix matrix is parametrised in the Dirac case by three mixing angles and one $CP$ phase, just like the CKM in the quark sector. 

Within the SM, it is also possible to write a Majorana mass term (without the introduction of any new field) using a non-renormalisable operator of dimension five (introduced by Weinberg)~\cite{Weinberg:1979sa},
\begin{equation}
\frac{\lambda_{ij}}{\Lambda}L_i\tilde{H}\,L_j\tilde{H}  + H.c.,
\end{equation}
where $\Lambda$ is an effective energy scale. This operator hides new physics at the scale $\Lambda$ that must be close to the grand unified scale if $\mathcal{O}(1)$ parameters are desired for $\lambda$. The Weinberg operator can arise from different seesaw mechanisms (type~I~\cite{Minkowski:1977sc,Yanagida:1979as,GellMann:1980vs,Mohapatra:1979ia}, II~\cite{Magg:1980ut,Schechter:1980gr,Wetterich:1981bx,Mohapatra:1980yp,Cheng:1980qt} or III~\cite{Foot:1988aq}\,\footnote{
The terminology I, II and III has been introduced in \cite{Ma:1998dn}.}) at high scales or from low energy seesaw-type mechanisms (inverse~\cite{Mohapatra:1986bd}, linear~\cite{Akhmedov:1995vm,Akhmedov:1995ip,Malinsky:2005bi}, or scotogenic~\cite{Ma:2006km}). If neutrinos are Majorana particles, the lepton mixing matrix is parametrized by three mixing angles  like in the Dirac case described above, but the phases are three instead of one, one called Dirac phase and the other two Majorana phases.

Experimentally we do not know whether neutrinos are Dirac or Majorana particles, and this ambiguity leads to another important question related to the flavour problem. The observation of processes with lepton number violation by 2 units, like neutrinoless double beta decay~\cite{Barabash:2011mf,Rodejohann:2011mu,Vergados:2012xy}, would prove that neutrino are Majorana particles~\cite{Schechter:1981bd,Duerr:2011zd}. These three phases have not been measured so far, and we do not know if $CP$ is at all violated in the lepton sector as for quarks, posing a further question around the flavour problem. What we know to a good precision in the neutrino sector are the mixing parameters. Two angles are large, in particular one (the atmospheric angle) is compatible with the maximal value, $\sin\theta_{23}=1/\sqrt{2}$, and the other (the solar angle) is almost trimaximal, $\sin\theta_{12}=1/\sqrt{3}$. The reactor angle is smaller (of the order of $\sin\theta_{13}\sim \lambda_C$).

We observe that the mixing parameters in the lepton mixing matrix are very different from the corresponding one in the CKM, which leads us to one more question connected to the flavour problem: why are the quark and lepton mixings so different?

Neutrino oscillation probabilities are not only functions of the mixing parameters but also of the neutrino masses, but from such experiments it is only possible to obtain the following square mass differences $|\Delta m^2_{\rm atm}|= |m_3^2-m_1^2|$ and $\Delta m^2_{\rm sol} = m_2^2-m_1^2$ and we know that $\Delta m^2_{\rm atm}\gg \Delta m^2_{\rm sol}$. The absolute scale of the neutrino mass is unknown as well as the sign of $\Delta m^2_{\rm atm}$ and therefore we do not know whether $m_1^2 < m_2^2 < m_3^2$ or $m_3^2 < m_1^2 < m_2^2$. These two possibilities are referred to, respectively, as normal and inverted neutrino mass ordering (NO and IO), and its determination is one important experimental task and is part of the flavour problem as well.\\

In short the {\em theoretical} questions of the flavour problem are summarised as follows:

\begin{itemize}
\item Why are there three families of quarks and leptons?
\item Why are all charged fermion masses so hierarchical with down-type quark masses being of the same order as charged lepton masses, and up-type quark masses are much more hierarchical?
\item Why are at least two neutrino masses not very hierarchical?
\item What is the origin of the neutrino mass?
\item Why are neutrino masses so tiny compared to charged fermion masses? 
\item What is the origin of fermion mixing (CKM and lepton mixing matrix)?
\item Why are CKM mixing angles smaller then lepton mixing matrix mixing angles apart from the Cabibbo angle
which is of the same order as the reactor angle?
\end{itemize}

Note that many of these questions are related to neutrinos. In the neutrino sector the current open {\em experimental} questions can be summarised as below:

\begin{itemize}
\item Is the atmospheric neutrino angle in the first or second octant?
\item Do neutrino mass eigenvalues have a normal or inverted ordering?
\item What is the value of the lightest neutrino mass?
\item Are neutrinos Dirac or Majorana?
\item Is $CP$ violated in the leptonic sector?
\item What are the values of the $CP$ violating phase(s)?
\end{itemize}

In order to study the flavour problem, one possibility is to introduce a new symmetry $G$ acting on the three families, also called {\it family symmetry}. The family symmetry is  not  a gauge symmetry or vertical symmetry with respect to the convention of Tab.~\ref{tabfl}, and for this reason such a symmetry is also called \emph{horizontal}. The SM is extended as $SU(3)_C\times SU(2)_L\times U(1)_Y\times G$. To give a simple example of a flavour symmetry, we can consider the group $G=SU(3)$ mimicking the colour in the quark sector. We know that the quarks come in three colours that belong to a triplet irreducible representation of $SU(3)_C$. In a similar way, we can assume that the three flavours could come in a triplet irreducible representation of $SU(3)$, thereby providing a clue on the origin of the three families. 

\begin{figure*}[h!]
\centering
\includegraphics[height=4.2cm,width=5cm]{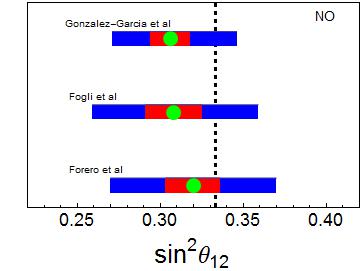}
\hskip3.mm
\includegraphics[height=4.2cm,width=5cm]{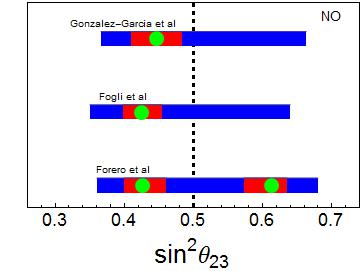}
\hskip3.mm
\includegraphics[height=4.2cm,width=5cm]{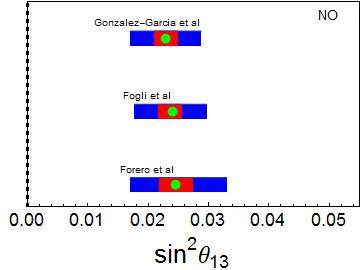}\\
\vskip8.mm
\includegraphics[height=4.2cm,width=5cm]{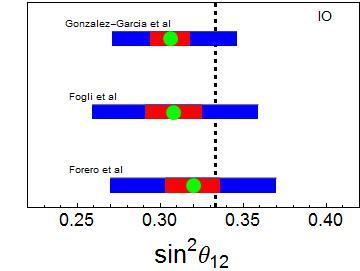}
\hskip3.mm
\includegraphics[height=4.2cm,width=5cm]{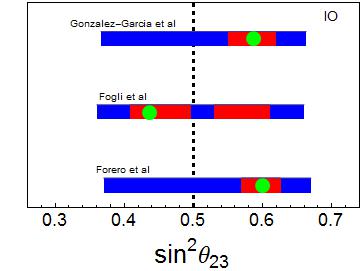}
\hskip3.mm
\includegraphics[height=4.2cm,width=5cm]{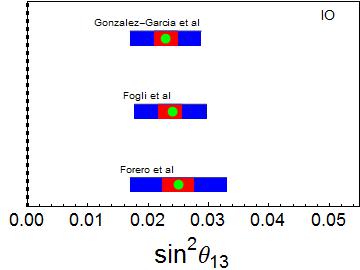}
\caption{ 
In each plot the three bands refer to three different global neutrino data analysis, namely~\cite{GonzalezGarcia:2012sz},~\cite{Capozzi:2013csa}, and~\cite{Tortola:2012te}, each for normal (top) and inverted (bottom) mass ordering. The vertical indicate give the TBM values.}\label{tbmvsdata}
\end{figure*}

In the recent years, the use of non-Abelian discrete subgroups of $SU(3)$ as flavour symmetries has become more and more popular (see Sec.~\ref{sec:groups} for a mathematical introduction and Refs.~\cite{Luhn:2011ip,Merle:2011vy,Merle:2012xr} for a discussion on how symmetry breaking from $SU(3)$ to one of those subgroups could occur). The origin of this popularity partially arose from the fact that, before 2012, the neutrino data was in very good agreement with the so called {\it tri-bi-maximal} neutrino mixing ansatz (TBM), proposed in 2002 by Harrison, Perkins, and Scott~\cite{Harrison:2002er}:
\begin{equation}
U_{\rm TBM} =
\left(
\begin{array}{ccc}
2/\sqrt{6} & 1/\sqrt{3}& 0\\
-1/\sqrt{6} & 1/\sqrt{3}&1/\sqrt{2}\\
1/\sqrt{6} & -1/\sqrt{3}&1/\sqrt{2}
\end{array}
\right),
\label{eq:TBM}
\end{equation}
where the third column corresponds to maximal mixing of the $| \nu_\mu \rangle$ and $| \nu_\tau \rangle$ states (bimaximal), while the second column encodes the equal mixing of the states $| \nu_e \rangle$, $| \nu_\mu \rangle$, and $| \nu_\tau \rangle$ (trimaximal). We observe that the TBM ansatz yields a zero reactor angle. In Fig.~\ref{tbmvsdata}, we confront the TBM ansatz with the newest global fit.

After the proposal of TBM there was a strong reaction in the model building community in order to explain its origin. Indeed, the entries of the matrix in Eq.~\eqref{eq:TBM} look like the Clebsch-Gordan coefficients of some symmetry. Thus, a lot of effort has been dedicated to its identification. The first model where TBM has been successfully obtain with an {\it ad hoc} relation between some parameters of the model  was in 2004\,\cite{Ma:2004zv}  by means of the $A_4$ group.
 A solution was proposed  in 2005~\cite{Altarelli:2005yp,Altarelli:2005yx,Babu:2005se,deMedeirosVarzielas:2005qg}. Such a group was previously proposed in~\cite{Ma:2001dn,Babu:2002dz} but the solar angle was not predicted in these models. Then other group as been considered during the last ten years like for instance $U(1)^3\times Z_2^3 \rtimes S_3$~\cite{Grimus:2008vg}, $Z_2^3 \rtimes S_3$~\cite{Mohapatra:2006pu}, $S_4$~\cite{Lam:2008rs,Bazzocchi:2008ej}, $T'$~\cite{Feruglio:2007uu,Carr:2007qw}, $\Delta(27)$~\cite{deMedeirosVarzielas:2006fc}, $\Delta(54)$~\cite{Ishimori:2008uc}.\\

In 2012 with the measure of the reactor angle with great precision from T2K~\cite{Abe:2013nka}, Daya Bay~\cite{An:2013zwz}, MINOS~\cite{Adamson:2013whj}, and RENO~\cite{Ahn:2012nd} experiments, TBM has been ruled out. The observed value is 
\begin{equation}
\sin^2 2\theta_{13}=
\begin{array}{c}
0.140^{+0.038}_{-0.032}\\
0.170^{+0.045}_{-0.037}
\end{array} 
\,{\rm (T2K)}\,,\,
0.090^{+0.008}_{-0.009}
\,{\rm (DayBay)}\,,
0.095^{+0.035}_{-0.036}
\,{\rm (MINOS)}\,,
0.100^{+0.025}_{-0.025}
\,{\rm (RENO)}\,,
\end{equation}
where in the case of T2K the upper and lower values are for NO and IO, respectively.

After the measurement of the reactor angle, the model building community took different ways to explain such a large angle, which we can generically classify as:
\begin{itemize}
\item deviations from TBM, 
\item different ansatz, 
\item fit of the lepton mixing matrix,
\item anarchy.  
\end{itemize}

Deviations from the TBM can arise from the neutrino sector (see, e.g., Ref.~\cite{Ma:2011yi}) or the charged lepton sector (see, e.g., Refs.~\cite{Antusch:2011qg,Acosta:2012qf}). Many different lepton mixing patterns have been proposed in order to obtain the reactor angle, some of them are listed below. We observe that most of the new approaches that have been used  are specific cases of the general parametrisation of the lepton mixing matrix in terms of deviations from TBM~\cite{King:2007pr},
\begin{equation}
\sin\theta_{12} = \frac{1}{\sqrt{3}} (1+s)\,,\qquad
\sin\theta_{23} = \frac{1}{\sqrt{2}} (1+a)\,,\qquad
\sin\theta_{12} = \frac{r}{\sqrt{2}}\,,\qquad
\end{equation}
where $s,\,a$, and $r$ are free parameters. Using this parametrisation, the lepton mixing matrix matrix gets the structure
\begin{equation}\label{TBMd}
U_{\rm } =
\left( \begin{array}{ccc}
\sqrt{\frac{2}{3}}(1-\frac{1}{2}s)  & \frac{1}{\sqrt{3}}(1+s) & \frac{1}{\sqrt{2}}re^{-i\delta } \\
-\frac{1}{\sqrt{6}}(1+s-a + re^{i\delta })  & \frac{1}{\sqrt{3}}(1-\frac{1}{2}s-a- \frac{1}{2}re^{i\delta })
& \frac{1}{\sqrt{2}}(1+a) \\
\frac{1}{\sqrt{6}}(1+s+a- re^{i\delta })  & -\frac{1}{\sqrt{3}}(1-\frac{1}{2}s+a+ \frac{1}{2}re^{i\delta })
 & \frac{1}{\sqrt{2}}(1-a)
\end{array}
\right)\,P ,
\end{equation}
where $\delta$ is the Dirac phase and $P$ is the diagonal matrix containing the two Majorana phases. It is clear that in the limit $a=s=r=0$ we obtain the TBM.\\

{\bf Trimaximal (TM):} This is a specific limit of the the matrix given in Eq.~\eqref{TBMd} where the solar deviation $s$ is set to be zero $s=0$ and we have~\cite{Grimus:2008tt}
\begin{equation}\label{TM}
U_{\rm TM} =
\left(
\begin{array}{ccc}
\frac{2}{\sqrt{6}}         & \frac{1}{\sqrt{3}} & \frac{r}{\sqrt{2}} e^{-i\delta}\\
\frac{-1}{\sqrt{6}}(1-a+r e^{-i\delta}) & \frac{1}{\sqrt{3}} (1-a- \frac{1}{2}r e^{-i\delta})  & \frac{1}{\sqrt{2}}(1+a)\\
\frac{1}{\sqrt{6}} (1+a-r e^{-i\delta}) & \frac{-1}{\sqrt{3}} (1+a+ \frac{1}{2}r e^{-i\delta})  & \frac{r}{\sqrt{2}}(1-a)
\end{array}
\right)\,P,
\end{equation}
where clearly the solar angle is $\sin\theta_{12}=1/\sqrt{3}$. From Eq.~\eqref{TM}, we can obtain two interesting limits. The first one is for
\begin{equation}
a=r\cos\delta\,,\qquad {\rm TM}_1\ .
\end{equation}
In this case it easy to check that  the first column of the matrix in Eq.~\eqref{TM} corresponds to the first column of the TBM in  Eq.~\eqref{eq:TBM} and is called ${\rm TM}_1$.

The second case is when 
\begin{equation}
a=- \frac{1}{2}r\cos\delta\,,\qquad {\rm TM}_2\ ,
\end{equation}
where the second column of the matrix in Eq.~\eqref{TM} is equal to the second column of the TBM in Eq.~\eqref{eq:TBM} and is called ${\rm TM}_2$.

{\bf Bimaximal (BM):} This lepton mixing matrix has been proposed in~\cite{Vissani:1997pa},
\begin{equation}\label{BM}
U_{\rm BM} =
\left(
\begin{array}{ccc}
\frac{1}{\sqrt{2}} & -\frac{1}{\sqrt{2}} & 0\\
\frac{1}{2} & \frac{1}{2} & -\frac{1}{\sqrt{2}}\\
\frac{1}{2} & \frac{1}{2} & \frac{1}{\sqrt{2}}
\end{array}
\right)\,P\ .
\end{equation}
In this case a large deviation of the order of the Cabibbo angle is necessary to correct both the solar and the reactor angles as has been shown in~\cite{Altarelli:2009gn}.


{\bf Tri-bimaximal Cabibbo (TBC):}
Recent data are in very good agreement with the ansatz~\cite{King:2012vj} 
\begin{equation}
\sin\theta_{12} = \frac{1}{\sqrt{3}} \,,\qquad
\sin\theta_{23} = \frac{1}{\sqrt{2}} \,,\qquad
\sin\theta_{13} = \frac{\lambda_C}{\sqrt{2}}\,,\qquad
\end{equation}
because after last T2K we had strong hints in favour of a maximal atmospheric angle. Therefore such a matrix could be a good starting 
point for model builders. Examples of models with such an ansatz are~\cite{King:2011ab,Morisi:2011pm}.


{\bf Bi-large (BL):} In this case the reactor angle is taken to be the seed for the solar and the atmospheric angles~\cite{Boucenna:2012xb}
\begin{equation}
\sin\theta_{13} = \lambda \,,\qquad
\sin\theta_{23} = a'\lambda \,,\qquad
\sin\theta_{12} = s' \lambda \,,\qquad
\end{equation}
where $a'$ and $s'$. This pattern was proposed before the last T2K result and its interest is that without considering this result, the global fits give indications for $s'\simeq a'$. In particular in the BL case it is assumed that $s'=a'\simeq 3$ and $\lambda\simeq \lambda_C$:
\begin{equation}
\sin\theta_{23} = 
\sin\theta_{12} \approx  0.63\,,\qquad
\sin\theta_{13} \approx 0.21 \,,
\end{equation}
as approximate starting point. It is interesting to notice that in this case the three mixing angles have almost a similar value revitalising the anarchy idea. This ansatz has been implemented in~\cite{Ding:2012wh}. 


{\bf Bi-trimaximal (BT):} 
In this case the mixing is due to the distinctive St George's cross feature of the middle row and column being of the tri-maximal form~\cite{King:2012in},
\begin{equation}\label{BT}
U_{\rm BT} =
\left(
\begin{array}{ccc}
a_+        & \frac{1}{\sqrt{3}} & a_-\\
\frac{-1}{\sqrt{3}} & \frac{1}{\sqrt{3}}  & \frac{1}{\sqrt{3}}\\
a_- & \frac{-1}{\sqrt{3}} &a_+
\end{array}
\right)\,P\ ,
\end{equation}
where $a_\pm = (1\pm 1/\sqrt{3})/2$, and leads to 
\begin{equation}
\sin\theta_{12}= \sin\theta_{23}= \sqrt{\frac{8-2\sqrt{2}}{13}}\approx0.591\,,\qquad \sin\theta_{13}=a_-\approx0.211 \ .
\end{equation}

Another possibility is to directly fit the lepton mixing observable without introducing any particular ansatz but using some non-Abelian discrete flavour symmetry to predict the mixing angles, like for instance in~\cite{Toorop:2011jn}. 

{\bf Golden Ratio (GR):}
For golden ratio (GR) mixing~\cite{Datta:2003qg,Everett:2008et,Feruglio:2011qq}
the solar angle is given by
$\tan \theta_{12}=1/\phi$, where $\phi = (1+\sqrt{5})/2$ is the golden ratio
which implies $\theta_{12}=31.7^\circ$. 
There is an alternative version where 
$\cos \theta_{12} =\phi/2$ and $\theta_{12}=36^\circ$~\cite{Rodejohann:2008ir,Adulpravitchai:2009bg} 
which we refer to as GR$'$ mixing. 

If experiments give us an indication for a non-maximal atmospheric angle, then the three mixing angles could be considered in first approximation about of similar order and may be difficult to see any underlying mixing pattern. In this case the neutrino mass matrix can be anarchical~\cite{Hall:1999sn}. This possibility has been recently revived in a series of works, e.g.~\cite{deGouvea:2012ac,Altarelli:2012ia}.

We should mention that there are already several excellent and up-to-date reviews in the literature~\cite{Altarelli:2010gt,Ishimori:2010au,King:2013eh}. This review attempts to span a very broad range of topics from the abstract group theory of finite groups, through model building applications, to the experimental tests of such theories, \emph{providing all the links in the chain between mathematics and experiments} that characterise this branch of neutrino physics. We also go beyond the three neutrino paradigm into the realms of sterile neutrinos and Dark Matter. In addition, several of the topics and models are reviewed here for the first time.

This review is organised into three main sections which may be roughly described as {\em mathematics}, 
{\em model building} and {\em experiment}. In Sec.~\ref{sec:groups}
we give an introduction to the discrete non-Abelian groups that will be used. In Sec.~\ref{sec:model-building} we will show with some example the use of such a groups as flavour symmetries. In Sec.~\ref{sec:experiments} we discuss the possibilities to experimentally distinguish different models. In Sec.~\ref{sec:outlook} we give a short outlook before concluding in Sec.~\ref{sec:conc}.

\section{\label{sec:groups}{\em Mathematics}: Introduction to finite groups}

\subsection{Basic of finite groups}

Non-Abelian discrete symmetries are not familiar to all particle physicists, unlike non-Abelian continuous symmetries. Therefore, at first, we introduce some group-theoretical aspects for many concrete groups explicitly, such as representations and their tensor products. 

Let us begin with introducing the basic aspects of  finite groups, which are presented in the references~\cite{Ishimori:2010au,Ishimori:2012zz,Ishimori:2013woa,Georgi:1982jb,Frampton:1994rk,Ludl:2009ft,Grimus:2005mu,ramond,miller,Hamermesh,Fairbairn}. A group, $G$,  is a set, where a multiplication is defined such that the following four conditions are satisfied:
\begin{enumerate}

\item {\bf Closure} If $a$ and $b$ are elements of the group $G$, $c=ab$ is also its element.
\item{\bf Associativity} $(ab)c=a(bc)$ for $a,b,c \in G$.
\item{\bf Identity} The group $G$ includes an identity element $e$, which satisfies $ae = ea =a$ for any element $a \in G$.
\item{\bf Inverse} The group $G$ includes an inverse element $a^{-1}$ for any element $a \in G$ such that $aa^{-1}=a^{-1}a=e$.

\end{enumerate}

The {\bf order} $N_G$ of a group $G$ is defined as the number of elements in $G$. Trivially, the order is finite for a finite group. For example, the order of the $Z_N$ group is $N$, while the order of the $S_N$ group is $N!$.

The group $G$ is {\bf Abelian} if all of their elements are commutable each other, i.e.\ $ab = ba$ for any elements $a$ and $b$ in $G$. If any two elements do not satisfy the commutativity, the group is called {\bf non-Abelian}.

If a subset $H$ of the group $G$ is a group by itself, $H$ is called the {\bf subgroup} of $G$. The order of the subgroup $H$ must be a divisor of the order of $G$. That is {\bf Lagrange's theorem}. If a subgroup $N$ of $G$ satisfies $g^{-1}Ng=N$ for any element $g \in G$, the subgroup $N$ is called a {\bf normal subgroup} or an {\bf invariant subgroup}. Any subgroup $H$ and normal subgroup $N$ of $G$ satisfy $HN=NH$, where $HN$  denotes
\begin{equation}
\{h_i n_j | h_i \in H, n_j \in N    \},
\end{equation}
and $NH$ denotes a similar meaning.

If $a^h=e$ for an element $a \in G$, the number $h$ is called the {\bf order} of $a$. The elements $\{ e, a, a^2, \cdots, a^{h-1} \}$ always form a subgroup of $G$, which is the Abelian group $Z_h$ with the order $h$.

The elements $g^{-1}ag$ for $g \in G$ are called {\bf conjugate} to the element $a$. The set of all elements conjugate to an element $a$ of $G$, $\{ g^{-1}ag, \ \bigm | g \in G \}$, is called a {\bf conjugacy class}. All elements in a conjugacy class have the same order, since 
\begin{eqnarray}
(gag^{-1})^h=ga(g^{-1}g)a(g^{-1}g)\cdot\cdot\cdot ag^{-1}=ga^{h}g^{-1}
=geg^{-1}=e .
\end{eqnarray}
The conjugacy class including the identity $e$ consists of the single element $e$.

We now consider two groups, $G$ and $G'$, and a map $f$ of  $G$ on  $G'$. This map is {\bf homomorphic} if and only if the map preserves the multiplication structure, that is,
\begin{equation}
f(a)f(b) = f(ab),
\end{equation}
for any $a,b \in G$. Furthermore, the map is called {\bf isomorphic} if it furnishes a one-to-one correspondence.

A {\bf representation} of $G$ is a homomorphic map of elements of $G$ onto matrices, $D(g)$ for $g \in G$. The representation matrices must satisfy $D(a)D(b)=D(c)$ if $ab=c$ for $a,b,c \in G$. The vector space $v_j$, on which the representation matrices act, is called the {\bf representation space}, such as $D(g)_{ij} v_j$ $(j = 1,\cdots,n)$. The dimension $n$ of the vector space $v_j$ $(j = 1,\cdots,n)$ is called the {\bf dimension} of the representation. A subspace in the representation space is called {\bf invariant subspace} if, for any vector $v_j$ in the subspace and any element $g \in G$,  $D(g)_{ij}v_j$ also corresponds to a vector in the same subspace. If a representation has an invariant subspace, such a representation is called {\bf reducible}. A representation is {\bf irreducible} if it has no invariant subspace. In particular, a representation is called {\bf completely reducible} if $D(g)$ for $g \in G$ are written as the following block diagonal form,
\begin{eqnarray}
\left(
\begin{array}{cccc}
D_1(g) & 0 &  &  \\
0 & D_2(g) &  &  \\
  &    & \ddots &   \\
  &    &    & D_r(g)  \\
\end{array}
\right),
\end{eqnarray}
where each $D_\alpha(g)$ for $\alpha=1,\cdots, r$ is irreducible. This implies that a reducible representation $D(g)$ is the direct sum of $D_\alpha(g)$,
\begin{eqnarray}
 \bigoplus_{\alpha =1}^r D_\alpha(g).
\end{eqnarray}
Every (reducible) representation of a finite group is completely reducible. Furthermore, every representation of a finite group is equivalent to a unitary representation. The simplest (irreducible) representation is found if $D(g)=1$ for all elements $g$, that is, a trivial singlet. Matrix representations satisfy the following orthogonality relation,
\begin{eqnarray}
\sum_{g \in G}  D_\alpha(g)_{i \ell} D_\beta(g^{-1})_{mj} = 
\frac{N_G}{d_\alpha} \delta_{\alpha \beta}\delta_{ij} \delta_{\ell m},
\end{eqnarray}
where $N_G$ is the order of $G$ and $d_\alpha$ is the dimension of the $D_\alpha(g)$.

The {\bf character} $\chi_D(g)$ of a representation matrix $D(g)$ is its trace,
\begin{eqnarray}
\chi_D(g) = {\rm tr}~D(g) = \sum_{i=1}^{d_\alpha} D(g)_{ii}.
\end{eqnarray}
Any element conjugate to $a$ has the same character as the element $a$ itself, because of the cyclic property of the trace:
\begin{eqnarray}
{\rm tr}~D(g^{-1}a g) = {\rm tr}~\left( D(g^{-1})D(a) D(g) \right)
= {\rm tr}~D(a),
\end{eqnarray}
that is, the characters are constant within a conjugacy class. The characters satisfy the following orthogonality relation,
\begin{eqnarray}\label{eq:character-1}
\sum_{g \in G} \chi_{D_\alpha}(g)^* \chi_{D_\beta}(g) 
= N_G \delta_{\alpha \beta}.
\end{eqnarray}
That is, the characters of different irreducible representations are orthogonal and generally different from each other. {\it Furthermore it is found that the number of irreducible representations must be equal to the number of conjugacy classes.} In addition, the characters satisfy a second orthogonality relation,
\begin{eqnarray}\label{eq:character-2}
\sum_{\alpha} \chi_{D_\alpha}(g_i)^* \chi_{D_\alpha}(g_j) 
= \frac{N_G}{n_i} \delta_{C_i C_j},
\end{eqnarray}
where $C_i$ denotes the conjugacy class of $g_i$ and $n_i$ denotes the number of elements in the conjugacy class $C_i$. The above equation means that the right hand side is equal to $\frac{N_G}{n_i}$ if $g_i$ and $g_j$ belong to the same conjugacy class, while otherwise the sum must vanish. A trivial singlet, $D(g)=1$ for any $g\in G$, must always be included. Thus, the corresponding character satisfies $\chi_1(g)=1$ for any $g\in G$.

Suppose that there are $m_n$ $n$-dimensional irreducible representations, that is, $D(g)$ are represented by $(n \times n)$ matrices. The identity $e$ is always represented by the $(n \times n)$ identity matrix. Obviously, the character $\chi_{D_\alpha}(C_1)$ for the conjugacy class $C_1=\{ e\}$ is found that $\chi_{D_\alpha}(C_1) = n$ for an $n$-dimensional representation. Then, the orthogonality relation~\eqref{eq:character-2} requires 
\begin{eqnarray}\label{eq:character-2-e}
&&\sum_\alpha[\chi_\alpha(C_1)]^2=\sum_nm_nn^2=m_1+4m_2+9m_3+\cdots=N_G,
\end{eqnarray}
where $m_n \geq 0$. Furthermore, $m_n$ must satisfy 
\begin{eqnarray}\label{eq:sum-dim}
&&\sum_nm_n = 
{\rm the~number~of~conjugacy~classes},
\end{eqnarray}
because the number of irreducible representations is equal to the number of conjugacy classes. Eqs.~\eqref{eq:character-2-e} and~\eqref{eq:sum-dim} as well as Eqs.~\eqref{eq:character-1} and~\eqref{eq:character-2} are often used in the following sections to determine characters.

We can construct a larger group from more than two groups, $G_i$, by certain products. A rather simple one is the {\bf direct product}. Let us consider two groups $G_1$ and $G_2$. Their direct product is denoted as $G_1 \times G_2$, and its multiplication rule is defined as 
\begin{equation}
(a_1,a_2)(b_1,b_2)=(a_1b_1,a_2b_2),
\end{equation}
for $a_1,b_1 \in G_1$ and $a_2,b_2 \in G_2$.

The {\bf semi-direct product} is a more non-trivial product between two groups $G_1$ and $G_2$, and it is defined as 
\begin{equation}
(a_1,a_2)(b_1,b_2)=(a_1f_{a_2}(b_1),a_2b_2),
\end{equation}
for $a_1,b_1 \in G_1$ and $a_2,b_2 \in G_2$, where $f_{a_2}(b_1)$ denotes a homomorphic map from $G_2$ to the automorphism of $G_1$. This semi-direct product is denoted as $G_1 \rtimes  G_2$. Let us now consider the group G, its subgroup $H$, and normal subgroup $N$, whose elements are $h_i$ and $n_j$, respectively. When $G=NH=HN$ and $N \cap H = \{ e \}$, the semi-direct product $N \rtimes H$ is isomorphic to $G$, $G \simeq  N \rtimes H$, where we use the map $f$ as 
\begin{equation}
f_{h_i}(n_j) = h_i n_j (h_i)^{-1}.
\end{equation}

In the following subsections, we explicitly present group-theoretical aspects in detail for several concrete groups.

\subsection{$\boldsymbol{S}_3$ group}

Let us present a simple example of a non-Abelian finite group, $S_3$. All possible permutations among $N$ objects $x_i$ with $i=1,\cdots, N$, form a group,
\begin{eqnarray}
(x_1, \cdots, x_N) \rightarrow (x_{i_1},\cdots,x_{i_N}).
\end{eqnarray} 
This group is denoted by $S_N$, it has the order $N !$, and it is called the symmetric group.

For the case of $N=3$, that is $S_3$, the order is $3! = 6$. These six elements correspond to the following transformations,
\begin{eqnarray}\label{eq:s3-permutation}
e   &:& (x_1,x_2,x_3)\to (x_1,x_2,x_3), \quad a_1 ~:~ (x_1,x_2,x_3)\to (x_2,x_1,x_3),\nonumber \\
a_2 &:& (x_1,x_2,x_3)\to (x_3,x_2,x_1),\quad a_3 ~:~ (x_1,x_2,x_3)\to (x_1,x_3,x_2),\\
a_4 &:& (x_1,x_2,x_3)\to (x_3,x_1,x_2),\quad a_5 ~:~ (x_1,x_2,x_3)\to (x_2,x_3,x_1).\nonumber 
\end{eqnarray}
Their multiplication forms a closed algebra. By defining $a_1=a,a_2=b$, all of elements are written as 
\begin{eqnarray}
\{e,a,b,ab,ba,bab\}.
\end{eqnarray}
Note that $aba=bab$. $S_3$ is isomorphic to the symmetry group of an equilateral triangle. For example, the elements $a$ and $ba$ correspond to a reflection and the $2\pi/3$ rotation, respectively.

\vskip .5cm
{$\bullet$ \bf Conjugacy classes}

The elements of $S_3$ are classified into three conjugacy classes,
\begin{eqnarray}
C_1:\{e\},\quad C_2:\{ab,ba\},\quad C_3:\{a,b,bab\}.
\end{eqnarray}
Here, the subscript in $C_n$ denotes the number of elements in that conjugacy class. Their orders are found as 
\begin{eqnarray}
(ab)^3=(ba)^3=e, \qquad a^2=b^2=(bab)^2=e.
\end{eqnarray}
The elements $\{e,ab,ba\}$ correspond to even permutations, while the elements $\{a, b, bab\}$ are odd permutations.

\vskip .5cm
{$\bullet$ \bf Characters and representations}

Let us study irreducible representations of $S_3$. The number of irreducible representations must be equal to three, because there are three conjugacy classes. We assume that there are $m_n$ $n$-dimensional representations, that is, $D(g)$ are represented by $(n \times n)$ matrices. Here, $m_n$ must satisfy $\sum_n m_n =3$. Furthermore, the orthogonality relation~\eqref{eq:character-2-e} requires 
\begin{eqnarray}\label{eq:character-2-S3}
&&\sum_\alpha[\chi_\alpha(C_1)]^2=\sum_nm_nn^2=m_1+4m_2+9m_3+\cdots=6,
\end{eqnarray}
where $m_n \geq 0$. This equation has only two possible solutions, $(m_1,m_2)=(2,1)$ and $(6,0)$, but only the former, $(m_1,m_2)=(2,1)$, satisfies $m_1+m_2=3$. Thus, the irreducible representations of $S_3$ include two singlets ${\bf 1}$ and ${\bf 1'}$, and one doublet ${\bf 2}$. We denote their characters by $\chi_1(g), \chi_{1'}(g)$, and $\chi_2(g)$, respectively. Obviously, it is found that $\chi_1(C_1)=\chi_{1'}(C_1)=1$ and $\chi_2(C_1)=2$. Furthermore, one of singlet representations corresponds to a trivial singlet, that is, $\chi_1(C_2)=\chi_1(C_3)=1$. The characters, which are not fixed at this stage, are $\chi_{1'}(C_2)$, $\chi_{1'}(C_3)$, $\chi_{2}(C_2)$, and $\chi_{2}(C_3)$. Let us determine them. For the non-trivial singlet ${\bf 1'}$, the representation matrices are nothing but characters, $\chi_{1'}(C_2)$ and $\chi_{1'}(C_3)$. They must satisfy 
\begin{eqnarray}
\left( \chi_{1'}(C_2)\right)^3 =1, \qquad \left( \chi_{1'}(C_3) 
\right)^2=1.
\end{eqnarray}
Thus, $\chi_{1'}(C_2)$ is equal to $1$, $\omega$, or $\omega^2$, where $\omega \equiv \exp[2\pi i/3]$, and $\chi_{1'}(C_3) $ is $1$ or $-1$. On top of that, the orthogonality relation~\eqref{eq:character-1} requires 
\begin{eqnarray}
\sum_g \chi_{1}(g) \chi_{1'}(g)= 1 + 2\chi_{1'}(C_2)
+3\chi_{1'}(C_3) =0.
\end{eqnarray}
The unique solution of this equation is $\chi_{1'}(C_2)=1$ and $\chi_{1'}(C_3)=-1$. Furthermore, the orthogonality relations~\eqref{eq:character-1} and~\eqref{eq:character-2} require 
\begin{eqnarray}
\sum_g \chi_{1}(g) \chi_{2}(g) &=& 2 + 2\chi_{2}(C_2)
+3\chi_{2}(C_3) =0, \\
\sum_\alpha \chi_{\alpha}(C_1)^* \chi_{\alpha}(C_2) 
&=& 1 + \chi_{1'}(C_2)+2\chi_{2}(C_2) =0.
\end{eqnarray}
Their solution is given by $\chi_2(C_2)=-1$ and $\chi_2(C_3)=0$. These results are summarised in Table~\ref{tab:S3-character}.

\begin{table}[t]
\begin{center}
\begin{tabular}{|c|c|c|c|c|}
\hline
     &$h$&$\chi_1$&$\chi_{1'}$&$\chi_2$ \\ \hline
$C_1$&$1$&  $1$   &   $1$     &   $2$   \\ \hline
$C_2$&$3$&  $1$   &   $1$     &   $-1$  \\ \hline
$C_3$&$2$&  $1$   &   $-1$    &   $0$   \\ 
\hline
\end{tabular}
\end{center}
\caption{Characters of $S_3$ representations.}
\label{tab:S3-character}
\end{table}

Next, let us figure out representation matrices $D(g)$ of $S_3$ by using the characters in Table~\ref{tab:S3-character}. For singlets, the characters are identical to the representation matrices. Thus, let us consider representation matrices $D(g)$ for the doublet, where $D(g)$ are $(2 \times 2)$ unitary matrices. Obviously, $D_2(e)$ is the $(2 \times 2)$ identity matrix. Because of $\chi_2(C_3)=0$, one can diagonalize one element of the conjugacy class $C_3$. Here we choose e.g. $a$ in $C_3$ as the diagonal element,
\begin{eqnarray}
a = \left(
\begin{array}{cc}  
1 & 0 \\
0 & -1 \\
\end{array}
\right).
\end{eqnarray}
The other elements in $C_3$  as well as $C_2$ are non-diagonal matrices. Recalling that $b^2=e$, we can write 
\begin{eqnarray}
b=\mat2{\cos\theta}{\sin\theta}{\sin\theta}{-\cos\theta}, \quad
bab=\mat2{\cos (2\theta)}{\sin (2\theta)}{\sin (2\theta)}{-\cos (2\theta)} .
\end{eqnarray}
This allows to write the elements in $C_2$ as 
\begin{eqnarray}
ab=\mat2{\cos\theta}{\sin\theta}{-\sin\theta}{\cos\theta}, \quad
ba=\mat2{\cos\theta}{-\sin\theta}{\sin\theta}{\cos\theta} .
\end{eqnarray}
Recall that the trace of the elements in $C_2$ must be equal to $-1$, which implies that $\cos\theta=-1/2 $, that is, $\theta={2\pi}/{3},{4\pi}/{3}$. When we choose $\theta={4\pi}/{3}$, we obtain the $(2 \times 2)$ matrix representation of $S_3$ as 
\begin{eqnarray}\label{eq:s3-2-rep}
&&e=\mat2{1}{0}{0}{1} ,\quad a=\mat2{1}{0}{0}{-1}, \quad
 b=\mat2{-\frac12}{-\frac{\sqrt{3}}{2}}{-\frac{\sqrt{3}}{2}}{\frac12} ,
 \nonumber \\
&&ab=\mat2{-\frac12}{-\frac{\sqrt{3}}{2}}{\frac{\sqrt{3}}{2}}{-\frac12},\quad
ba=\mat2{-\frac12}{\frac{\sqrt{3}}{2}}{-\frac{\sqrt{3}}{2}}{-\frac12},\quad
bab=\mat2{-\frac12}{\frac{\sqrt{3}}{2}}{\frac{\sqrt{3}}{2}}{\frac12} .
\end{eqnarray}

\vskip .5cm
{$\bullet$ \bf Tensor products}

Finally, we consider tensor products of irreducible representations. Let us start with the tensor products of two doublets, $(x_1,x_2)$ and $(y_1,y_2)$. For example, each element $x_i y_j$ transforms under $b$ as 
\begin{eqnarray}
x_1y_1&\to &\frac{x_1y_1+3x_2y_2+\sqrt{3}(x_1y_2+x_2y_1)}{4}, 
\quad \
x_1y_2\to \frac{\sqrt{3}x_1y_1-\sqrt{3}x_2y_2-x_1y_2+3x_2y_1}{4},
\nonumber \\
x_2y_1&\to &\frac{\sqrt{3}x_1y_1-\sqrt{3}x_2y_2-x_2y_1+3x_1y_2}{4}, 
\quad 
x_2y_2\to \frac{3x_1y_1+x_2y_2-\sqrt{3}(x_1y_2+x_2y_1)}{4}. 
\end{eqnarray}
Thus, it is found that
\begin{eqnarray}
b(x_1y_1+x_2y_2) = (x_1y_1+x_2y_2),\quad
b(x_1y_2-x_2y_1) = - (x_1y_2-x_2y_1) .
\end{eqnarray}
This implies that these linear combinations correspond to the singlets,
\begin{eqnarray}
{\bf 1}:x_1y_1+x_2y_2,\quad {\bf 1'}:x_1y_2-x_2y_1  .
\end{eqnarray}
Furthermore, it is found that 
\begin{eqnarray}
b \2tvec{x_2y_2-x_1y_1}{x_1y_2+x_2y_1} = 
\mat2{-\frac12}{-\frac{\sqrt{3}}{2}}{-\frac{\sqrt{3}}{2}}{\frac12}
\2tvec{x_2y_2-x_1y_1}{x_1y_2+x_2y_1} .
\end{eqnarray}
Hence, $(x_2y_2-x_2y_2,x_1y_2+x_2y_1)$ corresponds to the doublet, i.e.,
\begin{eqnarray}
{\bf 2}=\2tvec{x_2y_2-x_1y_1}{x_1y_2+x_2y_1} .
\end{eqnarray}

Similarly, we can study the tensor product of the doublet $(x_1,x_2)$ and the ${\bf 1'}$ singlet $y'$. Their products $x_iy'$ transform under $b$ as 
\begin{equation}
x_1y' \to \frac12x_1y'+\frac{\sqrt{3}}{2}x_2y' , 
\quad 
x_2y' \to \frac{\sqrt{3}}{2}x_1y'-\frac12x_2y' .
\end{equation}
Thus they form a doublet, 
\begin{eqnarray}
{\bf 2}:\2tvec{-x_2y'}{x_1y'} .
\end{eqnarray}

These results are summarised as follows,
\begin{eqnarray}
&&\2tvec{x_1}{x_2}_{\bf 2}\otimes\2tvec{y_1}{y_2}_{\bf 2}
=(x_1y_1+x_2y_2)_{\bf 1} \oplus (x_1y_2-x_2y_1)_{{\bf 1}'}
 \oplus \2tvec{x_1y_2+x_2y_1}{x_1y_1-x_2y_2}_{\bf 2}, \nonumber \\
&&\2tvec{x_1}{x_2}_{\bf 2}\otimes (y')_{{\bf
    1}'}=\2tvec{-x_2y'}{x_1y'}_{\bf 2}, \quad 
(x')_{{\bf 1}'}\otimes(y')_{{\bf 1}'}=(x'y')_{\bf 1}  .\nonumber
\end{eqnarray}
In addition, obviously the tensor product of two trivial singlets corresponds to a trivial singlet. One can also find representations in a different basis, see~\cite{Ishimori:2010au}.

Tensor products are important to applications for particle phenomenology. Matter and Higgs fields may be assigned to certain representations of a discrete symmetry. The Lagrangian must then be invariant under the discrete symmetry. This implies that only $n$-point couplings corresponding to trivial singlets can appear in the Lagrangian.

\subsection{A flavour model with $\boldsymbol{S}_3$}

The $S_3$ symmetry has been often discussed for flavour symmetry because it tends to yield large leptonic mixing angles. Here we illustrate a typical example to realize the so-called $\mu-\tau$ flavour symmetry in the neutrino mass matrix~\cite{Grimus:2005mu}.

Let us that the three left-handed lepton $SU(2)_L$ doublets  by $D_{e, \mu, \tau}$, the three right-handed charged-leptons by $(e_R, \mu_R, \tau_R)$ and the three $SU(2)_L$ singlet right-handed neutrinos by $(\nu_{eR}, \nu_{\mu R}, \nu_{\tau R})$. In addition,  three Higgs $SU(2)_L$ doublets $\phi_{1,2,3}$ and  one complex neutral scalar $SU(2)_L$ singlet $\chi$ are introduced. Here, we introduce  the auxiliary symmetry $Z_2^{\rm (aux)}$, which serves the purpose of allowing for $m_\mu \not = m_\tau$ while preserving the appropriate form of the neutrino mass matrix as we shall see soon. The $Z_2^{\rm (tr)}$ symmetry is also introduced to make a transposition of the multiplets of the $\mu$ and $\tau$ families. The fields transform under the $Z_2^{\rm (aux)}$ and  $Z_2^{\rm (tr)}$ as follows: 
\begin{equation}
{Z}_2^{\rm (aux)}:
\quad
\nu_{eR},\ \nu_{\mu R},\ \nu_{\tau R},\ \phi_1,\ e_R\ \, \rightarrow
-\nu_{eR},\ -\nu_{\mu R},\ -\nu_{\tau R},\ -\phi_1,\ -e_R\ \,
\end{equation}
and
\begin{equation}
Z_2^{\rm (tr)}:
\quad
D_\mu \leftrightarrow D_\tau,\
\mu_R \leftrightarrow \tau_R,\
\nu_{\mu R} \leftrightarrow \nu_{\tau R},\
\chi \to \chi^\ast,\
\phi_3 \to - \phi_3.
\label{Z2tr}
\end{equation}
$Z_2^{\rm (tr)}$ is the $\mu-\tau$ interchange symmetry. We introduce a symmetry $Z_3$ which, together with $Z_2^{\rm (tr)}$, generates a group $S_3$. With $\omega \equiv \exp{\left( 2 i \pi / 3 \right)}$, it is imposed that
\begin{equation}
Z_3:
\quad \left\{
\begin{array}{ll}
D_\mu \to \omega D_\mu, & D_\tau \to \omega^2 D_\tau,
\\*[1mm]
\mu_R \to \omega \mu_R, & \tau_R \to \omega^2 \tau_R,
\\*[1mm]
\nu_{\mu R} \to \omega \nu_{\mu R}, & \nu_{\tau R} \to \omega^2 \nu_{\tau R},
\\*[1mm]
\chi \to \omega \chi, & \chi^\ast \to \omega^2 \chi^\ast.
\end{array} \right.
\label{z3}
\end{equation}
Thus, $\left( D_\mu, D_\tau \right)$, $\left( \mu_R, \tau_R \right)$, $\left( \nu_{\mu R}, \nu_{\tau R} \right)$, and $\left( \chi, \chi^\ast \right)$ are doublets of $S_3$. The Higgs $SU(2)_L$ doublet $\phi_3$ changes sign under the odd permutations of $S_3$, but stays invariant under the cyclic permutations.

The Yukawa Lagrangian symmetric under $S_3 \times Z_2^{(\rm aux)}$ is
\begin{eqnarray}
{L}_{\rm Y} &=&
- \left[ y_1 \bar D_e \nu_{eR}
+ y_2 \left( \bar D_\mu \nu_{\mu R} + \bar D_\tau \nu_{\tau R} \right)
\right] \tilde \phi_1
\nonumber \\ &&
- y_3 \bar D_e e_R \phi_1
- y_4 \left( \bar D_\mu \mu_R + \bar D_\tau \tau_R \right) \phi_2
- y_5 \left( \bar D_\mu \mu_R - \bar D_\tau \tau_R \right) \phi_3
\nonumber  \\ &&
+ y_\chi^\ast\, \nu_{eR}^T C^{-1}
\left( \nu_{\mu R} \chi^\ast + \nu_{\tau R} \chi \right)
\nonumber  \\ &&
+ \frac{z_\chi^\ast}{2} \left( \nu_{\mu R}^T C^{-1} \nu_{\mu R} \chi
+ \nu_{\tau R}^T C^{-1} \nu_{\tau R} \chi^\ast
\right) + H.c.,
\label{yukawas}
\end{eqnarray}
where $\tilde \phi_1 \equiv i \tau_2 \phi_1^\ast$. There is also an $S_3 \times Z_2^{(\rm aux)}$-invariant Majorana mass term 
\begin{equation}
\mathcal{L}_{\rm M} = \frac{m^\ast}{2}\,
\nu_{eR}^T C^{-1} \nu_{eR}
+\frac{ {m^\prime}^\ast}{2} \nu_{\mu R}^T C^{-1} \nu_{\tau R} + \frac{ {m^\prime}^\ast}{2} \nu_{\tau R}^T C^{-1} \nu_{\mu R}
\ .
\label{LM}
\end{equation}

With VEVs $\left\langle 0 \left| \phi_j^0 \right| 0 \right\rangle = v_j$ for $j = 1, 2, 3$, one obtains
\begin{equation}
m_e = \left| y_3 v_1 \right|, \ \ 
m_\mu = \left| y_4 v_2 + y_5 v_3 \right|,
\ \
m_\tau = \left| y_4 v_2 - y_5 v_3 \right|.
\end{equation}
The $\mu-\tau$ interchange symmetry $Z_2^\mathrm{(tr)}$ is spontaneously broken by the VEV of $\phi_3^0$, so that the $\mu$ and $\tau$ charged leptons acquire different masses.

The neutrino Dirac mass matrix is given by
\begin{equation}
m^D = {\rm diag} \left( a, b, b \right),\ {\rm with}\
a \equiv y_1^\ast v_1,\ b \equiv y_2^\ast v_1.
\label{MD}
\end{equation}
When the singlet $\chi$ acquires a VEV $\left\langle 0 \left| \chi \right| 0 \right\rangle = W$, one obtains Majorana mass terms for the right-handed neutrinos:
\begin{equation}
\mathcal{L}_{M_R} = - \frac{1}{2} \left(
\bar \nu_{eR},\ \bar \nu_{\mu R},\ \bar \nu_{\tau R}
\right) M_R\, C \left( \begin{array}{c}
\bar \nu_{eR}^T \\ \bar \nu_{\mu R}^T \\ \bar \nu_{\tau R}^T
\end{array} \right) + \mathrm{H.c.},
\label{eq:S3-ex_Mat}
\end{equation}
with
\begin{equation}
M_R = \left( \begin{array}{ccc}
m & y_\chi W & y_\chi W^\ast \\
y_\chi W & z_\chi W^\ast & m^\prime \\
y_\chi W^\ast & m^\prime & z_\chi W
\end{array} \right).
\end{equation}
After rephasing  the fields, the matrix $M_R$ has become $\mu-\tau$ symmetric. Since $m^D$ has  the $\mu-\tau$ interchange symmetry, it follows by applying the seesaw mechanism that
\begin{equation}
m^\nu_{LL} = - {m^D}^T M_R^{-1} m^D
\end{equation}
is $\mu-\tau$ symmetric. We thus find that it is possible to produce a neutrino mass matrix, which leads to $\theta_{13} = 0$ and $\theta_{23} = \pi / 4$ in the  $S_3 \times{Z}_2^{(\rm aux)}$ flavour model. Since the experimental data indicates the non-zero $\theta_{13} = 0$, the model should be modified. For example, the partial $\mu-\tau$ interchange symmetry~\cite{Grimus:2005mu} is required instead of the  full $\mu-\tau$ symmetry.

\subsection{\label{subsec:S4}$\boldsymbol{S}_4$ group}

Next, we present the $S_4$ group, which is often used in flavour models. It consists of all permutations of four objects, $(x_1,x_2,x_3,x_4)$,
\begin{eqnarray}
 (x_1,x_2,x_3,x_4) \quad \to\quad (x_i,x_j,x_k,x_l).
\end{eqnarray}
The order of $S_4$ is equal to $4 ! = 24$. We can write down all $S_4$ elements explicitly, 
\begin{eqnarray}
&&a_1:(x_1,x_2,x_3,x_4),~a_2:(x_2,x_1,x_4,x_3),
~a_3:(x_3,x_4,x_1,x_2),~a_4:(x_4,x_3,x_2,x_1), 
\nonumber \\
&&b_1:(x_1,x_4,x_2,x_3),~b_2:(x_4,x_1,x_3,x_2),
~b_3:(x_2,x_3,x_1,x_4),~b_4:(x_3,x_2,x_4,x_1), 
\nonumber \\
&&c_1:(x_1,x_3,x_4,x_2),~c_2:(x_3,x_1,x_2,x_4),
~c_3:(x_4,x_2,x_1,x_3),~c_4:(x_2,x_4,x_3,x_1), 
\nonumber \\
&&d_1:(x_1,x_2,x_4,x_3),~d_2:(x_2,x_1,x_3,x_4),
~d_3:(x_4,x_3,x_1,x_2),~d_4:(x_3,x_4,x_2,x_1),\\
&&e_1:(x_1,x_3,x_2,x_4),~e_2:(x_3,x_1,x_4,x_2),
~e_3:(x_2,x_4,x_1,x_3),~e_4:(x_4,x_2,x_3,x_1), \nonumber \\
&&f_1:(x_1,x_4,x_3,x_2),~f_2:(x_4,x_1,x_2,x_3),
~f_3:(x_3,x_2,x_1,x_4),~f_4:(x_2,x_3,x_4,x_1), \nonumber 
\end{eqnarray}
where we have shown the ordering of four objects $(x_1,x_2,x_3,x_4)$ after permutations. Note that $S_4$ is isomorphic to the symmetry group $O$ of a cube.

It is obvious that $x_1+x_2+x_3+x_4$ is invariant under any permutation of $S_4$, that is, a trivial singlet. Thus, we can make use of the vector space which is orthogonal to this singlet direction, 
\begin{eqnarray}\label{eq:S4-3}
{\bf 3}:\3tvec{A_x}{A_y}{A_z}=
\3tvec{x_1+x_2-x_3-x_4}{x_1-x_2+x_3-x_4}{x_1-x_2-x_3+x_4} ,
\end{eqnarray}
in order to construct a matrix representation of $S_4$, that is, namely a triplet representation. In this triplet vector space, the $S_4$ elements are represented by the following matrices,
\begin{eqnarray}\label{eq:S4-3-2}
&&
a_1=\Mat3{1}{0}{0} {0}{1}{0} {0}{0}{1},~
a_2=\Mat3{1}{0}{0} {0}{-1}{0}{0}{0}{-1},~
a_3=\Mat3{-1}{0}{0} {0}{1}{0} {0}{0}{-1},~
a_4=\Mat3{-1}{0}{0} {0}{-1}{0} {0}{0}{1}, \nonumber \\
&&
b_1=\Mat3{0}{0}{1} {1}{0}{0} {0}{1}{0},~
b_2=\Mat3{0}{0}{1} {-1}{0}{0} {0}{-1}{0},~
b_3=\Mat3{0}{0}{-1} {1}{0}{0} {0}{-1}{0},~
b_4=\Mat3{0}{0}{-1} {-1}{0}{0} {0}{1}{0},\nonumber \\
&&
c_1=\Mat3{0}{1}{0} {0}{0}{1} {1}{0}{0},~
c_2=\Mat3{0}{1}{0} {0}{0}{-1} {-1}{0}{0},~
c_3=\Mat3{0}{-1}{0} {0}{0}{1} {-1}{0}{0},~
c_4=\Mat3{0}{-1}{0} {0}{0}{-1} {1}{0}{0}, \\
&&
d_1=\Mat3{1}{0}{0} {0}{0}{1} {0}{1}{0},~
d_2=\Mat3{1}{0}{0} {0}{0}{-1} {0}{-1}{0},~
d_3=\Mat3{-1}{0}{0} {0}{0}{1} {0}{-1}{0},~
d_4=\Mat3{-1}{0}{0} {0}{0}{-1} {0}{1}{0},  \nonumber \\
&&
e_1=\Mat3{0}{1}{0} {1}{0}{0} {0}{0}{1},~
e_2\Mat3{0}{1}{0} {-1}{0}{0} {0}{0}{-1},~
e_3=\Mat3{0}{-1}{0} {1}{0}{0} {0}{0}{-1},~
e_4=\Mat3{0}{-1}{0} {-1}{0}{0} {0}{0}{1},  \nonumber \\
&&
f_1=\Mat3{0}{0}{1} {0}{1}{0} {1}{0}{0},~
f_2=\Mat3{0}{0}{1} {0}{-1}{0} {-1}{0}{0},~
f_3=\Mat3{0}{0}{-1} {0}{1}{0} {-1}{0}{0},~
f_4=\Mat3{0}{0}{-1} {0}{-1}{0} {1}{0}{0}. \nonumber 
\end{eqnarray}

\vskip .5cm
{$\bullet$ \bf Conjugacy classes}

The $S_4$ elements can be classified by the order $h$ of each element, where $a^h=e$, as 
\begin{eqnarray}
\begin{array}{ccc}
h=1\quad:\quad & \{a_1\}, \\
h=2\quad:\quad &\{a_2,a_3,a_4,d_1,d_2,e_1,e_4,f_1,f_3 \},  \\
h=3\quad:\quad &\{b_1,b_2,b_3,b_4,c_1,c_2,c_3,c_4\}, \\
h=4\quad:\quad &\{d_3,d_4,e_2,e_3,f_2,f_4\} . 
\end{array}
\end{eqnarray}

Moreover, they are classified by the conjugacy classes as 
\begin{eqnarray}
\begin{array}{ccc}
 C_1~: &\{a_1\} , & h=1,  \\
 C_3~: &~\{a_2,a_3,a_4\} , & h=2, \\
 C_6~: &~\{d_1,d_2,e_1,e_4,f_1,f_3\}, & h=2, \\
 C_8~: &~\{b_1,b_2,b_3,b_4,c_1,c_2,c_3,c_4\}, & h=3,  \\
 C_{6'}~: &~\{d_3,d_4,e_2,e_3,f_2,f_4\}, & h=4. 
\end{array}
\end{eqnarray}

\vskip .5cm
{$\bullet$ \bf Characters and representations}

$S_4$ has five conjugacy classes, which implies that there are five irreducible representations. For example, all elements can be  written as products of $b_1$ in $C_8$ and $d_4$ in $C_{6'}$, which satisfy 
\begin{eqnarray}
& & (b_1)^3=e, \quad (d_4)^4=e, \quad 
 d_4 (b_1)^2d_4 = b_1, \quad d_4 b_1 d_4 = b_1 (d_4)^2b_1.
\end{eqnarray} 
The orthogonality relation~\eqref{eq:character-2-e} requires 
\begin{eqnarray}\label{eq:chracter-2-S4}
&&\sum_\alpha[\chi_\alpha(C_1)]^2=\sum_nm_nn^2=m_1+4m_2+9m_3+\cdots=24,
\end{eqnarray}
like Eq.~\eqref{eq:character-2-S3}, and the $m_n$ also satisfy $m_1+m_2+m_3+\cdots= 5$, because there are five irreducible representations. The unique solution is obtained as $(m_1,m_2,m_3)=(2,1,2)$. That is, irreducible representations of $S_4$ include two singlets ${\bf 1}$ and ${\bf 1}'$, one doublet ${\bf 2}$, and two triplets ${\bf 3}$ and ${\bf 3}'$, where ${\bf 1}$ corresponds to a trivial singlet and ${\bf 3}$ corresponds to~\eqref{eq:S4-3} and~\eqref{eq:S4-3-2}. We can compute the characters for each representation by an analysis similar to the $S_3$ case presented in the previous section. The results are shown in Table~\ref{tab:S4-character}.

\begin{table}[t]
\begin{center}
\begin{tabular}{|c|c|c|c|c|c|c|}
\hline
        &$h$&$\chi_1$&$\chi_{1'}$&$\chi_2$&$\chi_3$&$\chi_{3'}$ \\ \hline
$C_1$   &$1$&  $1$   &   $1$     &  $2$   &   $3$  &   $3$    \\ \hline
$C_3$   &$2$&  $1$   &   $1$     &  $2$   &   $-1$ &  $-1$    \\ \hline
$C_6$   &$2$&  $1$   &   $-1$    &  $0$   &   $1$  &  $-1$    \\ \hline 
$C_{6'}$&$4$&  $1$   &   $-1$    &  $0$   &   $-1$ &  $1$     \\ \hline
$C_8$   &$3$&  $1$   &   $1$     & $-1$   &   $0$  &   $0$    \\
\hline
\end{tabular}
\end{center}
\caption{Characters of $S_4$ representations.}
\label{tab:S4-character}
\end{table}

For ${\bf 2}$, the representation matrices are written as, e.g.,
\begin{eqnarray}\label{eq:S4-2}
&&
a_2({\bf2})=\mat2{1}{0}{0}{1},~b_1({\bf2})=\mat2{\omega}{0}{0}{\omega^2},
\nonumber \\
&&
d_1({\bf2})=d_3({\bf2})=d_4({\bf2})=\mat2{0}{1}{1}{0}.
\end{eqnarray}
For ${\bf 3}'$, the representation matrices are written as 
e.g. 
\begin{eqnarray}
&& 
a_2({\bf3'})=\Mat3{1}{0}{0} {0}{-1}{0} {0}{0}{-1},~
b_1({\bf3'})=\Mat3{0}{0}{1} {1}{0}{0} {0}{1}{0}, \\
&&
d_1({\bf3'})=\Mat3{-1}{0}{0} {0}{0}{-1} {0}{-1}{0},~ 
d_3({\bf3'})=\Mat3{1}{0}{0} {0}{0}{-1} {0}{1}{0},~
d_4({\bf3'})=\Mat3{1}{0}{0} {0}{0}{1} {0}{-1}{0}.\nonumber 
\end{eqnarray}
Note that $a_2({\bf 3'}) = a_2({\bf 3})$ and 
$b_1({\bf 3'}) = b_1({\bf 3})$, but $d_1({\bf3'}) = - d_1({\bf3}) $, 
$d_3({\bf3'})= - d_3({\bf3})$ and $d_4({\bf3'})= - d_4({\bf3})$.
This aspect would be obvious from the above character table.

\vskip .5cm
{$\bullet$ \bf Tensor products}

Finally, we show the tensor products. The tensor products of ${\bf 3} \times {\bf 3}$ can be decomposed as
\begin{eqnarray}
({\bf A})_{\bf3}\otimes({\bf B})_{\bf3}=({\bf A}\cdot{\bf B})_{\bf1}
\oplus \2tvec{ {\bf A}\cdot\Sigma\cdot{\bf B} } { {\bf A}\cdot\Sigma^*\cdot{\bf B} }_{\bf2}
\oplus \3tvec{\{A_yB_z\}} {\{A_zB_x\}} {\{A_xB_y\}}_{\bf3}
\oplus \3tvec{\left [A_yB_z\right ]} {\left [A_zB_x\right ]} {\left [A_xB_y\right ]}_{\bf3'},
\end{eqnarray}
where 
\begin{eqnarray}
{\bf A}\cdot{\bf B}&=&A_xB_x+A_yB_y+A_zB_z, \nonumber  \\
\{A_iB_j\}&=&A_iB_j+A_jB_i, \nonumber \\
\left [A_yB_z\right ]&=&A_iB_j-A_jB_i, \\
{\bf A}\cdot\Sigma\cdot{\bf B} &=&A_xB_x+\omega A_yB_y+\omega^2A_zB_z,
\nonumber \\
{\bf A}\cdot\Sigma^*\cdot{\bf B} &=&A_xB_x+\omega^2 A_yB_y+\omega
A_zB_z . \nonumber 
\end{eqnarray}

The tensor products of other representations are also decomposed as, e.g.,
\begin{eqnarray}
({\bf A})_{\bf3'}\otimes({\bf B})_{\bf3'}=({\bf A}\cdot{\bf B})_{\bf1}
\oplus \2tvec{ {\bf A}\cdot\Sigma\cdot{\bf B} } { {\bf A}\cdot\Sigma^*\cdot{\bf B} }_{\bf2}
\oplus \3tvec{\{A_yB_z\}} {\{A_zB_x\}} {\{A_xB_y\}}_{\bf3}
\oplus \3tvec{\left [A_yB_z\right ]} {\left [A_zB_x\right ]} {\left [A_xB_y\right]}_{\bf3'}, \\
({\bf A})_{\bf3}\otimes({\bf B})_{\bf3'}=({\bf A}\cdot{\bf B})_{\bf1'}
\oplus \2tvec{ {\bf A}\cdot\Sigma\cdot{\bf B} } {-{\bf A}\cdot\Sigma^*\cdot{\bf B} }_{\bf2}
\oplus \3tvec{\{A_yB_z\}} {\{A_zB_x\}} {\{A_xB_y\}}_{\bf3'}
\oplus \3tvec{\left [A_yB_z\right ]} {\left [A_zB_x\right ]} {\left [A_xB_y\right ]}_{\bf3},
\end{eqnarray}
and 
\begin{eqnarray}
({\bf A})_{\bf2}\otimes({\bf B})_{\bf2}= \{A_xB_y\}_{\bf1} \oplus \left [A_xB_y\right ]_{\bf1'} \oplus \2tvec{A_yB_y} {A_xB_x}_{\bf2} ,
\end{eqnarray}

\begin{eqnarray}
\2tvec{A_x}{A_y}_{\bf 2}\otimes \3tvec{B_x}{B_y}{B_z}_{\bf 3}
&=&\3tvec{(A_x+A_y)B_x}{(\omega^2A_x+\omega A_y)B_y}{(\omega
  A_x+\omega^2 A_y)B_z}_{\bf 3} \oplus 
\3tvec{(A_x-A_y)B_x}{(\omega^2A_x-\omega A_y)B_y}{(\omega A_x-\omega^2
  A_y)B_z}_{{\bf 3}'}, \\
\2tvec{A_x}{A_y}_{\bf 2}\otimes \3tvec{B_x}{B_y}{B_z}_{{\bf 3}'}
&=&\3tvec{(A_x+A_y)B_x}{(\omega^2A_x+\omega A_y)B_y}{(\omega
  A_x+\omega^2 A_y)B_z}_{{\bf 3}'} \oplus 
\3tvec{(A_x-A_y)B_x}{(\omega^2A_x-\omega A_y)B_y}{(\omega A_x-\omega^2
  A_y)B_z}_{\bf 3} .
\end{eqnarray}
Furthermore, we have ${\bf 3}\otimes {\bf 1'}={\bf 3}'$ and ${\bf 3}'\otimes {\bf 1'}={\bf 3}$ and  ${\bf 2}\otimes {\bf 1'}={\bf 2}$.

In the literature, several bases are used for $S_4$. The decomposition of tensor products, ${\bf r} \otimes {\bf r}' = \bigoplus_m {\bf r}_m$, does not depend on the basis. For example, we obtain ${\bf 3} \otimes {\bf  3}' = {\bf 1}' \oplus {\bf 2} \oplus {\bf 3} \oplus {\bf  3}'$ in any basis. However, the multiplication rules written by components depend on the basis, which we use. We have used the basis~\eqref{eq:S4-2}. One can write down relations between several bases and give explicitly the multiplication rules in terms of components, see Ref.~\cite{Ishimori:2010au}.

As well as  the $S_3$ symmetry, there are many interesting flavour models of neutrinos using the $S_4$ symmetry, many of which are listed in Ref.~\cite{Ishimori:2010au}.

\subsection{\label{sec:A4}$\boldsymbol{A}_4$ group}

All even permutations among $S_N$ form a group, which is $A_N$. It is called the alternating group. Therefore,  the order of this group is  $N!/2$. 
Let us consider a simple example.
 In $S_3$ 
the even permutations include
\begin{eqnarray}
e   &:& (x_1,x_2,x_3)\to (x_1,x_2,x_3), \nonumber\\
a_4 &:& (x_1,x_2,x_3)\to (x_3,x_1,x_2),\\
a_5 &:& (x_1,x_2,x_3)\to (x_2,x_3,x_1), \nonumber
\end{eqnarray}
while the odd permutations include
\begin{eqnarray}
a_1 &:& (x_1,x_2,x_3)\to (x_2,x_1,x_3), \nonumber\\
a_2 &:& (x_1,x_2,x_3)\to (x_3,x_2,x_1),\\
a_3 &:& (x_1,x_2,x_3)\to (x_1,x_3,x_2). \nonumber
\end{eqnarray}
The three elements of even permutations, $\{e, a_4, a_5 \}$ form the 
group, which is  $A_3$.
Since $(a_4)^2=a_5$ and $(a_4)^3=e$, 
the group $A_3$ is nothing but $Z_3$.
$A_4$ is the smallest non-Abelian group and it is frequently used for flavour models of leptons.

The $A_4$ group is formed by all even permutations of $S_4$. Thus, its order is equal to $4!/2=12$. $A_4$ group is isomorphic to the symmetry group $T$ of a tetrahedron. Using the notation in Eq.~\eqref{eq:S4-3-2}, all 12 elements can be written as 
\begin{eqnarray}\label{eq:A4-ABC}
&&
a_1=\Mat3{1}{0}{0} {0}{1}{0} {0}{0}{1},~a_2=\Mat3{1}{0}{0} {0}{-1}{0}{0}{0}{-1},~
a_3=\Mat3{-1}{0}{0} {0}{1}{0} {0}{0}{-1},~a_4=\Mat3{-1}{0}{0} {0}{-1}{0} {0}{0}{1}, \nonumber \\
&&
b_1=\Mat3{0}{0}{1} {1}{0}{0} {0}{1}{0},~b_2=\Mat3{0}{0}{1} {-1}{0}{0} {0}{-1}{0},~
b_3=\Mat3{0}{0}{-1} {1}{0}{0} {0}{-1}{0},~b_4=\Mat3{0}{0}{-1} {-1}{0}{0} {0}{1}{0},\\
&&
c_1=\Mat3{0}{1}{0} {0}{0}{1} {1}{0}{0},~c_2=\Mat3{0}{1}{0} {0}{0}{-1} {-1}{0}{0},~
c_3=\Mat3{0}{-1}{0} {0}{0}{1} {-1}{0}{0},~c_4=\Mat3{0}{-1}{0} {0}{0}{-1} {1}{0}{0}.\nonumber
\end{eqnarray}

They are classified by the conjugacy classes as 
\begin{eqnarray}
\begin{array}{ccc}
C_1~: & \{a_1\}, & h=1,  \\
C_3~: & \{a_2,a_3,a_4\}, & h=2,  \\
 C_4~:& \{b_1,b_2,b_3,b_4,\}, & h=3, \\
 C_{4'}~:& \{c_1,c_2,c_3,c_4,\}, & h=3,
\end{array}
\end{eqnarray}
where we have also shown the orders of each element in the conjugacy class by $h$. There are four conjugacy classes and there must be four irreducible representations, i.e.\ $m_1+m_2+m_3 + \cdots = 4$.

The orthogonality relation~\eqref{eq:character-2} requires 
\begin{eqnarray}\label{eq:chracter-2-A4}
&&\sum_\alpha[\chi_\alpha(C_1)]^2=\sum_nm_nn^2=m_1+4m_2+9m_3+\cdots=12,
\end{eqnarray}
for $m_i$, which satisfy $m_1+m_2+m_3 + \cdots = 4$. There exists one solution, $(m_1,m_2,m_3)=(3,0,1)$. That is, the $A_4$ group has three singlets, ${\bf 1}$, ${\bf 1}'$, and ${\bf 1}''$, and one triplet ${\bf 3}$, where the triplet representation corresponds to Eq.~\eqref{eq:A4-ABC}.

Another algebraic definition of $A_4$ is often used in the literature. We denote $a_1 = e$, $a_2=s$ and $b_1=t$. They satisfy the following algebraic relations,
\begin{eqnarray}\label{eq:T-st}
s^2=t^3=(st)^3=e.
\end{eqnarray}
The closed algebra of these elements, $s$ and $t$, is defined as the $A_4$ group. That is, $s$ and $t$ are generators of $A_4$. It is straightforward to write all elements $a_i, b_i$, and $c_i$ in terms of $s$ and $t$. Then, the conjugacy classes are rewritten as 
\begin{eqnarray}
\begin{array}{ccc}
C_1:&\{e\}, & h=1, \\
C_3:&\{s,tst^2,t^2st\},& h=2, \\
C_4:&\{t,ts,st,sts\}, & h=3,\\
C_{4'}:&\{t^2,st^2,t^2s,tst\}, & h=3. 
\end{array}
\end{eqnarray}
Using them, we can study characters. First, we consider characters of the three singlets. Because of $s^2=e$, there are two possibilities for the character of $C_3$, $\chi_\alpha(C_3) = \pm 1$. However, the two elements $t$ and $ts$ belong to the same conjugacy class $C_4$. This implies that $\chi_\alpha(C_3)$ should have the unique value $\chi_\alpha(C_3) =  1$. Similarly, because of $t^3=e$, the character $\chi_\alpha(t)$ could correspond to three values, i.e.\ $\chi_\alpha(t)=\omega^n$, $n=0,1,2$, where all three values are consistent with the above structure of conjugacy classes. Thus, the three singlets, ${\bf 1}$, ${\bf 1'}$, and ${\bf 1''}$ are classified by these three values, $\chi_\alpha(t)=1, \omega$, and $\omega^2$, respectively. Obviously, it is true that $\chi_\alpha(C_{4'}) =(\chi_\alpha(C_{4}))^2$. Thus, the generators such as $s=a_2,t=b_1,t^2=c_1$ are represented on the non-trivial singlets ${\bf 1'}$ and ${\bf 1''}$ as 
\begin{eqnarray}
&&
s({\bf 1'})={a_2}({\bf 1'})=1,
\quad t({\bf 1'})={b_1}({\bf  1'})=\omega,
\quad t^2({\bf 1'})={c_1}({\bf 1'})=\omega^2,
\nonumber \\ &&
s({\bf 1''})={a_2}({\bf 1''})=1,
\quad t({\bf 1''})={b_1}({\bf  1''})=\omega^2,
\quad t^2({\bf 1''})={c_1}({\bf 1''})=\omega .
\end{eqnarray}
These characters are shown in Table~\ref{tab:A4-character}. Next, we consider the characters of the triplet representation. Obviously, the matrices in Eq.~\eqref{eq:A4-ABC} correspond to the triplet representation. Thus, we can directly read off their characters. The result is shown in Table~\ref{tab:A4-character}.

The tensor product of ${\bf 3} \otimes {\bf 3}$ can be decomposed as 
\begin{eqnarray}
({\bf A})_{\bf3}\otimes({\bf B})_{\bf3}&=&({\bf A}\cdot{\bf B})_{\bf1}
\oplus ({\bf A}\cdot\Sigma\cdot{\bf B})_{\bf1'}
\oplus ({\bf A}\cdot\Sigma^*\cdot{\bf B})_{\bf1''}  \nonumber \\
& & \oplus \3tvec{\{A_yB_z\}} {\{A_zB_x\}} {\{A_xB_y\}}_{\bf3}
\oplus \3tvec{\left [A_yB_z\right ]} {\left [A_zB_x\right ]} {\left [A_xB_y\right ]}_{\bf3}.
\end{eqnarray}
The same representation in another basis is shown in Ref.~\cite{Ishimori:2010au}.

\begin{table}[t]
\begin{center}
\begin{tabular}{|c|c|c|c|c|c|}
\hline
        &$h$&$\chi_1$&$\chi_{1'}$&$\chi_{1''}$&$\chi_3$ \\ \hline
$C_1$   &$1$&   $1$  &    $1$    &    $1$     &$3$   \\ \hline
$C_3$   &$2$&   $1$  &    $1$    &    $1$     &$-1$  \\ \hline
$C_4$   &$3$&   $1$  & $\omega$  & $\omega^2$ &$0$   \\ \hline
$C_{4'}$&$3$&   $1$  & $\omega^2$&  $\omega$  &$0$   \\
\hline
\end{tabular}
\end{center}
\caption{Characters of $A_4$ representations.}
\label{tab:A4-character}
\end{table}

It is well-known that flavour models with $A_4$ gives the tri-bimaximal mixing of leptons by adopting the $A_4$ triplet for the left-handed leptons.

\subsection{Other finite groups}

We summarise some other finite groups which are applied to the flavour models.

\subsubsection{$\boldsymbol{D_N}$}

The first one is the  dihedral group $D_N$. It is isomorphic to the symmetry group of a regular polygon with $N$ sides. It is furthermore isomorphic to $Z_N \rtimes Z_2$ and is also denoted by $\Delta(2N)$. It consists of cyclic rotation, $Z_N$ and  reflection. That is, it is generated by two generators $a$ and $b$, which act on $N$ edges $x_i$ ($i=1,\cdots,N$) of $N$-polygon as 
\begin{eqnarray}\label{eq:Dn-n}
a&:&(x_1,x_2 \cdots, x_N) \rightarrow (x_{N},x_{1}\cdots,x_{N-1}),\\
b&:&(x_1,x_2 \cdots, x_N) \rightarrow (x_1,x_N \cdots, x_2).
\end{eqnarray} 
These two generators satisfy 
\begin{eqnarray}\label{eq:Dn-AB}
a^N=e, \qquad b^2=e, \qquad bab=a^{-1} ,
\end{eqnarray}
where the third equation is equivalent to $aba=b$. The order of $D_N$ is equal to $2N$, and all the $2N$ elements can be written as $a^mb^k$ with $m=0,\cdots, N-1$ and $k=0,1$. The third equation in~\eqref{eq:Dn-AB} implies that the $Z_N$ subgroup including $a^m$ is a normal subgroup of $D_N$. Thus, $D_N$ corresponds to a semi-direct product between $Z_N$ including $a^m$ and $Z_2$ including $b^k$, i.e., $Z_N \rtimes Z_2$. Eq.~\eqref{eq:Dn-n} corresponds to the (reducible) $N$-dimensional representation. The simple doublet representation is written as
\begin{eqnarray}\label{eq:Dn-2}
a=\left(
\begin{array}{cc}
\cos (2\pi /N) & -\sin (2 \pi /N) \\
 \sin (2 \pi /N) & \cos (2\pi /N) 
\end{array}
\right), \qquad 
b = \left(
 \begin{array}{cc}
 1 & 0 \\
0 & -1
\end{array}
\right).
\end{eqnarray} 

\subsubsection{$\boldsymbol{Q_N}$}

The binary dihedral group is called  as $Q_N$, where $N$ is  even. It  consists of the elements, $a^mb^k$ with $m=0,\cdots, N-1$ and $k=0,1$, where the generators $a$ and $b$ satisfy
\begin{eqnarray}\label{eq:Qn-AB}
a^N=e, \qquad b^2=(a^{N/2}), \qquad b^{-1}ab=a^{-1} .
\end{eqnarray}
The order of $Q_N$ is equal to $2N$. The generator $a$ can be represented by the same $( 2 \times 2)$ matrices as $D_N$, i.e.,
\begin{eqnarray}
a=\left(
\begin{array}{cc}
\exp [2\pi ik/N] & 0 \\
 0 & \exp [-2\pi ik/N]
\end{array}
\right).
\end{eqnarray}
It is noted  that $a^{N/2}=e$ for $k=$ even and $a^{N/2}=-e$ for $k=$ odd. That leads to that $b^2=e$ for $k=$ even and $b^2 =-e$ for $k=$ odd. Thus, the generators  $a$ and $b$ are represented by $(2 \times 2)$ matrices, e.g., as  
\begin{eqnarray}\label{eq:Qn-2k-2}
a=\left(
\begin{array}{cc}
\exp [2\pi ik/N] & 0 \\
 0 & \exp [-2\pi ik/N]
\end{array}
\right), \qquad 
b = \left(
 \begin{array}{cc}
 0 & i \\
 i & 0
\end{array}
\right),
\end{eqnarray}
for $k=$ odd, and
\begin{eqnarray}\label{eq:Qn-2k-2-2}
a=\left(
\begin{array}{cc}
\exp [2\pi ik/N] & 0 \\
 0 & \exp [-2\pi ik/N]
\end{array}
\right), \qquad 
b = \left(
 \begin{array}{cc}
 0 & 1 \\
 1 & 0
\end{array}
\right),
\end{eqnarray}
for $k=$ even.
 
\subsubsection{$\bf \Sigma(2 \boldsymbol{N}^2)$}

The discrete group $\Sigma (2N^2)$ is isomorphic to $(Z_N \times Z_N')\rtimes Z_2$. $\Sigma(2)$ is nothing but $Z_2$, and $\Sigma(8)$ is isomorphic to  $D_4$. The simplest non-trivial example is $\Sigma(18)$. The next ones are  $\Sigma(32)$ and $\Sigma(50)$.

Let us denote the generators of $Z_N$ and $Z_N'$ by $a$ and $a'$, respectively, and the $Z_2$ generator  by $b$. These generators satisfy
\begin{eqnarray}
& & a^N = {a'}^N = b^2 = e, \nonumber \\
& &  aa' = a'a, \qquad bab=a'.\label{daisu-kankei}
\end{eqnarray}
Therefore, all $\Sigma (2N^2)$ elements can be written as 
\begin{eqnarray}
 & & g=b^{k}a^{m}a'^{n} ,
\end{eqnarray}
for $k=0,1$ and $m,n=0,1,...,N-1$.

Since these generators, $a$, $a'$ and $b$, are represented, e.g. as 
\begin{eqnarray}\label{eq:sigma-2-1}
a=\mat2{1}{0} {0}{\rho},
\qquad a'=\mat2{\rho}{0} {0}{1}, \qquad b=\mat2{0}{1}{1}{0},
\end{eqnarray}
where $\rho=e^{2\pi i/N}$, all of $\Sigma (2N^2)$ elements are expressed by the $2\times 2$ matrices as 
\begin{eqnarray}
\mat2{\rho^m}{0}{0}{\rho^n}, \qquad \mat2{0}{\rho^m}{\rho^n}{0}.
\end{eqnarray}

\subsubsection{$\bf \Delta(3\boldsymbol{N}^2)$}

The discrete group $\Delta (3N^2)$ is isomorphic to $(Z_N \times Z_N')\rtimes Z_3$. $\Delta (3)$ is nothing but $Z_3$, and $\Delta (12)$ is isomorphic to $A_4$. The first non-trivial example is $\Delta (27)$.

Let us denote the generators of $Z_N$ and $Z_N'$ by $a$ and $a'$, respectively, and the $Z_3$ generator by $b$, which  satisfy
\begin{eqnarray}
& & a^N = {a'}^N = b^3 = e, \qquad aa' = a'a, \nonumber \\
& &  bab^{-1}=a^{-1}(a')^{-1}, \qquad ba'b^{-1}=a.
\end{eqnarray}
Therefore, all elements of $\Delta (3N^2)$ can be written as 
\begin{eqnarray}
 & & g=b^{k}a^{m}a'^{n},
\end{eqnarray}
for $k=0,1,2$ and $m,n=0,1,2,\cdots ,N-1$.

Since the generators, $a$, $a'$, and $b$, are represented as, e.g., 
\begin{eqnarray}\label{eq:delta-3n-3}
b=\Mat3{0}{1}{0}{0}{0}{1}{1}{0}{0},\quad a=
\Mat3{\rho}{0}{0} {0}{1}{0} {0}{0}{\rho^{-1}},\quad 
a'=\Mat3{\rho^{-1}}{0}{0} {0}{\rho}{0} {0}{0}{1},
\end{eqnarray}
where $\rho=e^{2\pi i/N}$, all elements of $\Delta(3N^2)$ can be displayed as
\begin{eqnarray}
\Mat3{\rho^m}{0}{0}{0}{p^n}{0}{0}{0}{\rho^{-m-n}},\quad
\Mat3{0}{\rho^m}{0}{0}{0}{\rho^n}{\rho^{-m-n}}{0}{0},\quad
\Mat3{0}{0}{\rho^m}{\rho^n}{0}{0}{0}{\rho^{-m-n}}{0}, 
\end{eqnarray}
for $m,n=0,1,2,\cdots ,N-1$.

\subsubsection{$\bf \Sigma(3 \boldsymbol{N}^3)$}

The discrete group $\Sigma (3N^3)$ is defined as a closed algebra of three Abelian symmetries, $Z_N$, $Z_N'$, and $Z_N''$, which commute with each other, and their $Z_3$ permutations. Let us denote the generators of $Z_N$, $Z_N'$, and $Z_N''$ by $a$, $a'$, and $a''$, respectively, and the $Z_3$ generator  by $b$. All $\Sigma(3N^3)$ elements can be written as 
\begin{equation}
g=b^ka^ma'^na''^{\ell} ,
\end{equation}
with $k=0,1,2$, and $m,n,\ell=0,...,N-1$, 
where $a$, $a'$, $a''$, and $b$ satisfy
\begin{eqnarray}
&&a^N=a'^N=a''^N=1,
\quad
aa'=a'a, ~~aa''=a''a, ~~a''a'=a'a'',
\quad
b^3=1 ,
\nonumber\\
&&b^2ab=a'',\quad b^2a'b=a, \quad  b^2a''b=a'.
\end{eqnarray}
The generators, $a$, $a'$, $a''$, and $b$, are represented as, e.g,
\begin{eqnarray}
b=\Mat3{0}{1}{0} {0}{0}{1} {1}{0}{0},~~ 
a=\Mat3{1}{0}{0} {0}{1}{0} {0}{0}{\rho} ,~~
a'=\Mat3{1}{0}{0} {0}{\rho}{0} {0}{0}{1} ,~~ a''=\Mat3{\rho}{0}{0} 
{0}{1}{0} {0}{0}{1} ,
\end{eqnarray}
where $\rho=e^{2i\pi/N}$. Then, all elements of $\Sigma(3N^3)$ can be written as
\begin{eqnarray}
\Mat3{0}{\rho^n}{0} {0}{0}{\rho^m} {\rho^{\ell}}{0}{0},\quad 
\Mat3{\rho^{\ell}}{0}{0} {0}{\rho^m}{0} {0}{0}{\rho^n},
\quad
\Mat3{0}{0}{\rho^m} {\rho^{\ell}}{0}{0} {0}{\rho^n}{0}.
\label{sigma3n-rep}
\end{eqnarray}

For the case of  $N=2$, the element $aa'a''$ commutes with all other elements. In addition, if we define $\tilde a=aa''$ and $\tilde a'=a'a''$, we find a closed algebra among $\tilde a$, $\tilde a'$, and $b$, which corresponds to a $\Delta(12)$ subgroup. Since the element $aa'a''$ is not included in this closed algebra, the resulting group is isomorphic to $Z_2 \times \Delta(12)$.

The situation for $N=3$ is a little different. Similarly, the element $aa'a''$ commutes with all other elements. Furthermore, if we define $\tilde a=a^2a''$ and $\tilde a'=a'a''^{2}$, the closed algebra among $\tilde a$, $\tilde a'$, and $b$ corresponds to $\Delta(27)$. However, the element $aa'a''$ can be rewritten as $aa'a'' = \tilde a^2 \tilde a' $ in this case. Thus, $aa'a''$  is an element of $\Delta(27)$. This implies that the group $\Sigma(81)$ is \emph{not} isomorphic to $Z_3 \times \Delta(27)$, but instead to $(Z_3 \times Z_3' \times Z_3'') \rtimes Z_3$.

Similarly, for generic values of $N$, the element $aa'a''$ generally commutes with all other elements. If we define $\tilde a=a^{N-1}a''$ and $\tilde a'=a'a''^{N-1}$, the closed algebra among $\tilde a$, $\tilde a'$, and $b$ corresponds to $\Delta(3N^2)$. For the case of $N/3 \neq $ integer, the element $aa'a''$ is not included in $\Delta(3N^2)$. Thus, we find that this group is isomorphic to $Z_N\times \Delta(3N^2)$. On the other hand, if $N/3 = $ integer, the element  $(aa'a'')^{Nk/3}$ with $k=0, 1, 2$ is included in $\Delta(3N^2)$. Therefore, the group $\Sigma (3N^3)$ cannot be isomoprhic to $Z_N\times \Delta(3N^2)$.

\subsubsection{$\bf \Delta(6 \boldsymbol{N}^2)$}

The discrete group $\Delta (6N^2)$ is isomorphic to $(Z_N \times Z_N')\rtimes S_3$. $\Delta (6)$ is nothing but $S_3$, and $\Delta (24)$ is isomorphic to $S_4$. The simplest non-trivial examples are $\Delta (54)$ and $\Delta (96)$.

Let us denote the generators of $Z_N$ and $Z_N'$ by $a$ and $a'$, respectively. We furthermore denote the $S_3$ generators by $b$ and $c$, where $b$ and $c$ are the $Z_3$ and $Z_2$ subgroup generators of $S_3$, respectively. These generators  satisfy
\begin{eqnarray}
& & a^N = {a'}^N = b^3=c^2=(bc)^2 = e, \quad aa' = a'a, 
\nonumber \\
& &  bab^{-1}=a^{-1}(a')^{-1}, \quad ba'b^{-1}=a,
\nonumber \\
& &  cac^{-1}=(a')^{-1}, \quad ca'c^{-1}=a^{-1}.
\end{eqnarray}
Using them, all $\Delta (6N^2)$ elements can be written as 
\begin{eqnarray}
 & & g=b^{k}c^{\ell}a^{m}a'^{n},
\end{eqnarray}
for $k=0,1,2$, $\ell=0,1$ and $m,n=0,1,2,\cdots ,N-1$.

Note that the $\Delta(6N^2)$ group includes the subgroup $\Delta(3N^2)$, whose elements can be written as $b^{k}a^{m}a'^{n}$. Thus, some group-theoretical aspects of $\Delta(6N^2)$ can be derived from those of $\Delta(3N^2)$.

\section{\label{sec:model-building}{\em Model Building}: Discrete non-Abelian family symmetries and GUTs}

\subsection{Flavons vs.\ Multi-Higgs }

Typically in flavour models it is required to introduce new scalar fields. A simple argument to understand why this extension of the SM is needed,  is the following. Let consider for simplicity a flavour group that has a (not-trivial) real irreducible representation $\mathbf{r}=\overline{\mathbf{r}}$ and that $\mathbf{r} \otimes \mathbf{r} \supset \mathbf{1}$ and $\mathbf{r} \otimes \mathbf{r} \otimes \mathbf{r} \supset \mathbf{1}$ where  $\mathbf{1}$ is a singlet of $G$. 

Suppose that the left and right-handed leptons $L$ and $l_R$ transform as $\mathbf{r}$ under $G$. The Yukawa interaction as well as the dimension-5 Weinberg operator are invariant under $G$,
\begin{equation}\label{FM1}
Y^\ell_{ij} L_i l_{j}^c\,H +\frac{Y^\nu_{ij}}{\Lambda} L_iL_j\,\tilde{H}\tilde{H} + H.c.,
\end{equation}
where $H$ is the Higgs $SU(2)_L$ doublet, $\tilde{H}=i\sigma_2 H^*$ and the structure of $Y^\ell$ and $Y^\nu$ are given by the contraction rules of $G$. It is clear that in this simple case, $Y_\ell$ and $Y_\nu$ are similar matrices (because they originate from the same $G$-product), giving rise to a diagonal lepton mixing matrix. In order to obtain the observed large lepton mixing angles we have to add new scalar fields to allow not trivial contractions of $ L l^c$ and $LL$.

Here we have in general two possibilities, depending on whether the matter content of the model is made of multi-Higgs $SU(2)_L$ doublets or extra flavon scalar fields that are $SU(2)_L$-singlets transforming under $G$. In the former case we have
\begin{equation}\label{FM2}
y^\ell_{ijk} L_i l_{J}^c\,H_k +\frac{y^\nu_{ijkm}}{\Lambda}L_i L_j \tilde{H}_k\tilde{H}_m + H.c.,
\end{equation}
where  $H_i$ represent a set of multi-Higgs doublets that transforms as $\mathbf{r}$ under $G$, while in the latter case we have, in addition to the terms in Eq.~\eqref{FM1},
\begin{equation}\label{FM3}
\frac{y^\ell_{ijk}}{\Lambda} L_i l_{J}^c\,H \phi_k +\frac{y^\nu_{ijk}}{\Lambda^2} L_i L_j \tilde{H}\tilde{H}\phi_k
+\frac{y^\nu_{ijkm}}{\Lambda^3} L_i L_j \tilde{H}\tilde{H}\phi_k\phi_m + H.c.,
\end{equation}
where $\phi$ is a scalar flavon that transforms as $\mathbf{r}$ under $G$ and $y^\ell_{ijk}$, $y^\nu_{ijk}$, and $y^\nu_{ijkm}$ are tensors of $G$.

In both cases given, respectively, by Eqs.~\eqref{FM2} and~\eqref{FM3}, we can have a mismatch between the charged and neutral lepton sectors, differently from the case of Eq.~\eqref{FM1}, and therefore the lepton mixing matrix can be different from the identity. Notice that the first terms in Eqs.~\eqref{FM2} and~\eqref{FM3} are equivalent (i.e., they give the same pattern for the charged lepton mass matrix) and this is because formally we have replaced the $\mathbf{r}$-multiplet $H_i$ with the $\mathbf{r}$-multiplet $H\,\phi_i$. But the neutral sectors in Eqs.~\eqref{FM2} and~\eqref{FM3} can be different. Therefore, the replacement $H_i\to H\,\phi_i$ is not only a simple academic exercise because it can lead to different models.

Moreover, having $\mathbf{r}$-Higgs $SU(2)_L$ doublets that live around the weak scale gives a very different phenomenology (for instance at the Large Hadron Collider (LHC), in what regards flavour changing neutral currents (FCNCs), and so on) with respect to the case of $\mathbf{r}$-scalar $SU(2)_L$ singlets (typically at scales between TeV and the inflation scale). Some examples of models with flavon scalar fields will be presented in the next sections. Here we focus on the multi-Higgs case.\\

The first model that we are going to present is based on $D_4$~\cite{Grimus:2003kq}\footnote{The dihedral group $D_4$ is isomorphic to the group of permutation of three objects $S_3$. For models with multi-Higgs and based on $S_3$, see for instance~\cite{Kubo:2003iw,Mondragon:2007af}.} that has five irreducible representations, four singlets $\mathbf{1}_{++}$, $\mathbf{1}_{+-}$, $\mathbf{1}_{-+}$, $\mathbf{1}_{--}$, and one doublet $\mathbf{2}$. The product of two doublets is $\mathbf{2} \otimes \mathbf{2} = \mathbf{1}_{++} \oplus \mathbf{1}_{+-} \oplus \mathbf{1}_{-+} \oplus \mathbf{1}_{--}$ and the product of singlets is trivial (for example $\mathbf{1}_{+-} \otimes \mathbf{1}_{-+} = \mathbf{1}_{--}$). Here, the SM is extended by adding three right-handed neutrinos $\nu^c_{1,2,3}$, three Higgs doublets $H_{1,2,3}$, and two neutral singlets $\chi_{1,2}$. The matter content of the model is given in Table\,\ref{tab:D4}.

\begin{table}
\begin{center}
\begin{tabular}{ccccccccccc}
\hline
 & $L_e$ & $e^c$ & $L_{\mu,\tau} $ & $\mu^c,\tau^c$ & $\nu^c_1$ & $\nu^c_{2,3}$ & $H_1$ & $H_{2}$ & $H_{3}$ & $\chi_{1,2}$  \\
\hline
$D_4$ & $\mathbf{1}_{++}$ & $\mathbf{1}_{++}$ & $\mathbf{2}$ & $\mathbf{2}$ & $\mathbf{1}_{++}$ & $\mathbf{2}$ & $\mathbf{1}_{++}$ & $\mathbf{1}_{++}$ & $\mathbf{1}_{+-}$ & $\mathbf{2}$\\
$Z_2^{\rm aux}$ & $+$ & $-$ & $+$ & $+$ & $-$ & $-$ & $-$ & $+$ & $-$ & $+$ \\
\hline
\end{tabular}
\caption{\label{tab:D4}Matter content of the model from Ref.~\cite{Grimus:2003kq}.}
\end{center}
\end{table}

The Lagrangian is given by 
\begin{equation}
\begin{array}{ccl}
\mathcal{L}&=&[y_1 L_e\nu^c_1 +y_2(L_\mu \nu^c_2+L_\tau \nu^c_3)] \tilde{H}_1+\\
&& +y_3  L_e e^c_1 H_1 +y_4(L_\mu \mu^c+L_\tau \tau^c) H_2 +y_5(L_\mu \mu^c-L_\tau \tau^c) H_3+\\
&& + y_\chi{\nu_1^c}^T (\nu_2^c \chi_1+\nu_2^c \chi_2)+ M {\nu_1^c}^T \nu_1^c  + 
M' ({\nu_2^c}^T \nu_2^c +{\nu_3^c}^T \nu_3^c ) + H.c.
\end{array}
\end{equation}
After electroweak symmetry breaking, the $\chi$-fields take VEVs $\langle\chi_1\rangle =\langle\chi_2\rangle $, giving a $\mu - \tau$ invariant neutrino mass matrix (with a maximal atmospheric angle), while the charged lepton mass matrix is diagonal. Here we do not give other details and we refer interested readers to the original paper. But we note that in this model the fermion masses arise from the VEVs of three Higgs doublets, none of which has the same properties as the SM-Higgs field. In the general example that we presented in Eqs.~\eqref{FM1} and~\eqref{FM2}, we have assumed that neutrino masses originate from a dimension-5 Weinberg operator. Here, the neutrino masses instead arise from a type~I seesaw mechanism. But we can see that the use of multiple Higgs fields allows to generate a non-trivial leptonic mixing matrix, which in particular involves a large atmospheric mixing angle and a zero reactor angle. With the measurement of a relatively large reactor angle, such models are ruled out and extensions must be considered. However, it is a good and simple example for the general use of multiple Higgs doublets.\\

The second example that we are going to present is based on~\cite{Morisi:2009sc,Morisi:2011pt}, where all the matter fields, quarks as well as leptons, are assigned to triplet representations of $A_4$. The field content of this supersymmetric model is given in Table~\ref{tab:Multiplet1}. Two pairs of Higgs fields transforming as $A_4$ triplets and with opposite hypercharge (as usual in the MSSM) have been assumed.
\begin{table}[h!]
\begin{center}
  \begin{tabular}{|l||lllll||ll|}
\hline
fields & $\hat{L}$ & $\hat{E}^c$ & $\hat{Q}$& $\hat{U}^c$ & $\hat{D}^c$ & $\hat{H}^u $ & $\hat{H}^{d}$  \\
\hline
$SU(2)_L$ & $\mathbf{2}$ & $\mathbf{1}$ & $\mathbf{2}$ & $\mathbf{1}$ & $\mathbf{1}$ & $\mathbf{2}$ & $\mathbf{2}$ \\
$A_4$ & $\mathbf{3}$ & $\mathbf{3}$ & $\mathbf{3}$ & $\mathbf{3}$ & $\mathbf{3}$ & $\mathbf{3}$ & $\mathbf{3}$ \\
\hline
\end{tabular}
\caption{Scalar and matter asignments of the supersymmetric $A_4$ model.}
\label{tab:Multiplet1}
\end{center}
\end{table}

The most general renormalisable Yukawa superpotential for the charged fermions in the model is
\begin{equation}\label{y}
w_{\text{Yukawa}}
= y^l_{ijk}\hat{L}_i\hat{H}^d_{i}\hat{E}^c_k
+ y^d_{ijk}\hat{Q}_i\hat{H}^d_{i}\hat{D}^c_k
+ y^u_{ijk}\hat{Q}_i\hat{H}^u_{i}\hat{U}^c_k,
\end{equation}
where $y_{ijk}^{u,d,l}$ are $A_4$-tensors. Neutrino masses arise from the dimension-5 operator
\begin{equation}\label{nudim5}
\mathcal{L}_5=
\frac{f_{ijlm}}{\Lambda} \hat{L}_i \hat{L}_j \hat{H}^u_l \hat{H}^u_m + H.c.,
\end{equation}
where $f_{ijlm}$ is a $A_4$-tensor that take into account  all the possible contractions of the product of four $A_4$ triplets.

It is then assumed that the Higgs fields take real VEVs, $\langle H_i^{u,d} \rangle = v_i^{u,d}$. By adding the following $A_4$ soft breaking terms to the scalar potential below the electroweak scale, 
\begin{equation}\label{align0}
\begin{array}{lll}
V_{\rm soft} &=& \sum_{ij}\mu_{ij}H_i^{u*}H_j^u+\sum_{ij}\kappa_{ij}H_i^{d}H_j^u,
\end{array}
\end{equation}
one finds that
\begin{eqnarray}
\label{eq:minima}
\langle{H^u}\rangle=(v^u,\varepsilon_1^u,\varepsilon_2^u),\quad
\langle{H^d}\rangle=(v^d,\varepsilon_2^d,\varepsilon_2^d),
\end{eqnarray}
where $\varepsilon_{1,2}^u\ll v^u$ and $\varepsilon_{1,2}^d\ll v^d$. 
By using the $A_4$ product rules, the charged fermion mass matrix has the following structure,
\begin{equation}
M_f=
\left(
\begin{array}{ccc}
0 & a^f \alpha^f  & b^f  \\
b^f\alpha^f  & 0 & a^f r^f \\
a^f  & b^f r^f & 0
\end{array}
\right),
\label{Mf}
\end{equation}
%
%
where $f$ denotes any charged lepton, or up or down quark, and $a^f=y_1\varepsilon_1^f$, $b^f=y_2\varepsilon_1^f$ [$y_{1,2}$ are the only two couplings arising from the $A_4$-tensors in Eq.~\eqref{y}], $r^f=v^f/\varepsilon_1^f$, and $\alpha^f=\varepsilon_2^f/\varepsilon_1^f$. We have the relations 
\begin{equation}\label{rel}
r^l=r^d,\qquad \alpha^l=\alpha^d,
\end{equation}
which arise form the fact that the same Higgs doublet $H^d$ couples to the lepton and down-quark sectors. The mass matrix in Eq.~\eqref{Mf} is mainly diagonalised on the left by a rotation in the 12-plane, namely
\begin{equation}\label{R12}
\sin\theta^f_{12}\approx\sqrt{\frac{m_1^f}{m_2^f}} \frac{1}{\sqrt{\alpha^f}}.
\end{equation}
It is straightforward to obtain analytical expressions for $a^f$, $b^f$, and $r^f$ from Eq.~\eqref{Mf}. As functions of the charged lepton masses, they can be written as
\begin{equation}
\frac{r^f}{\sqrt{\alpha^f}}\approx \frac{m_{3}^f}{\sqrt{m_{1}^fm_{2}^f}},\quad
a^f\approx\frac{m_{2}^f}{m_{3}^f}\sqrt{\frac{m_{1}^fm_{2}^f}{\alpha^f}},\quad
b^f\approx \sqrt{\frac{m_{1}^fm_{2}^f}{\alpha^f}}.\quad
\end{equation}
Note that $a<b\ll r$. From the relations in Eq.~\eqref{rel}, we have following mass formula relating quark and leptons like in GUT frameworks,
\begin{equation}\label{massrel}
 \frac{m_{\tau}}{\sqrt{m_{e}\, m_{\mu}}}\approx  \frac{m_{b}}{\sqrt{m_{d}\, m_{s}}}\,.
\end{equation}
However in this model only the Cabibbo mixing angle is obtained (the other two mixing angles are too small). Modifications have been recently proposed in order to fit correctly the full CKM mixing matrix~\cite{King:2013hj,Morisi:2013eca}. In particular in~\cite{Morisi:2013eca} meson oscillation phenomenology has been studied. For the phenomenology of multi-Higgs models see, for instance, Refs.~\cite{Toorop:2010ex,Toorop:2010kt}. Vacuum alignment and the Higgs mass spectrum have been discussed in general three Higgs doublet models whose potential is controlled by any discrete symmetry in~\cite{Keus:2013hya}.

\subsection{The seesaw mechanism and sequential dominance}

The seesaw mechanism~\cite{Minkowski:1977sc,Yanagida:1979as,GellMann:1980vs,Glashow:1979nm,Mohapatra:1979ia} sheds light on the smallness of neutrino masses, but it also increases the parameter count considerably due to an undetermined right-handed neutrino Majorana mass matrix. In the diagonal right-handed neutrino and charged lepton basis (the so-called \emph{flavour basis}), there is thus an undetermined neutrino Yukawa matrix. Without the seesaw mechanism, the SM involves three charged fermion Yukawa matrices, but these are non-physical and basis dependent quantities. However, in theories of flavour beyond the SM, the choice of basis may well have physical significance and, in a certain basis defined by the theory, the Yukawa matrices may take simple forms, leading to some predictive power of the model as a result.

There have been many attempts to describe the lepton mixing angles based on the type~I seesaw mechanism combined with sequential dominance (SD)~\cite{King:1998jw,King:1999cm,King:1999mb,King:2002nf}, in which the right-handed neutrinos contribute with sequential strength leading to the prediction of a normal neutrino mass \emph{hierarchy}.
In the flavour basis where the charged lepton mass matrix $M_E$ is diagonal with real positive eigenvalues $m_e, m_{\mu}, m_{\tau}$ and the three right-handed neutrino Majorana mass matrix $M_R$ is also diagonal, with real positive eigenvalues, $M_{\rm atm}, M_{\rm sol}, M_{\rm dec}$,
\begin{equation}
M_{E}=
\left( \begin{array}{ccc}
m_{e} & 0 & 0    \\
0 & m_{ \mu} & 0 \\
0 & 0 & m_{ \tau}
\end{array}
\right), \
M_{R}=
\left( \begin{array}{ccc}
M_{\rm atm} & 0 & 0    \\
0 & M_{\rm sol} & 0 \\
0 & 0 & [M_{\rm dec}]
\end{array}
\right).
\label{seq1}
\end{equation}
We can write the neutrino Dirac mass matrix as
\begin{equation}
m^D=
\left( \begin{array}{ccc}
m^D_{e, \rm atm} & m^D_{e, \rm sol} & [m^D_{e, \rm dec}]   \\
m^D_{\mu, \rm atm} & m^D_{\mu, \rm sol} & [m^D_{\mu, \rm dec}] \\
m^D_{\tau, \rm atm} & m^D_{\tau, \rm sol} & [m^D_{\tau, \rm dec}]
\end{array}
\right)
\equiv \left( \begin{array}{ccc}
m^D_{\rm atm} & m^D_{\rm sol} & [m^D_{\rm dec}]  
\end{array}
\right),
\label{dirac}
\end{equation}
in the convention where the effective Lagrangian after electroweak symmetry breaking, with the Higgs VEV inserted, is given by
\begin{equation}
{\cal L}=-  \overline{E_L} M_{E} E_R  -  \overline{\nu_L} m^{D} N_R 
- \frac{1}{2}\overline{N_R^c} M_{R} N_{R} 
+ H.c.\; ,
\end{equation}
where $\nu_L =(\nu_{e},\nu_{\mu }, \nu_{\tau })$ are the three left-handed neutrino fields which appear together with $E_L=(e_L,\mu_L,\tau_L)$ in the lepton doublets $L=(L_{e},L_{\mu }, L_{\tau })$, and $N_R=(N_{\rm atm},N_{\rm sol}, N_{\rm dec})$ are the three right-handed neutrinos and we have defined the three Dirac column vectors as $m^D_{\rm atm}$, $m^D_{\rm sol}$, $m^D_{\rm dec}$.

The term for the light neutrino masses in the effective Lagrangian (after electroweak symmetry breaking), resulting from integrating out the massive right-handed neutrinos (i.e.\ the seesaw mechanism with the light effective neutrino Majorana mass matrix $m^{\nu}=m^DM_R^{-1}{m^D}^T$) is
\begin{equation}
\mathcal{L}^\nu_{\rm eff} =
 \frac{(\overline{\nu_L} m^D_{\rm atm})({m^D_{\rm atm}}^{T} \nu_L^c)}{M_{\rm atm}}
 +\frac{(\overline{\nu_L} m^D_{\rm sol})({m^D_{\rm sol}}^{T} \nu_L^c)}{M_{\rm sol}}
\ \left[ \ +\frac{(\overline{\nu_L} m^D_{\rm dec})({m^D_{\rm dec}}^{T} \nu_L^c)}{M_{\rm dec}} \ \right] \label{leff}.
\end{equation}

Sequential dominance (SD) then corresponds to the third term being negligible, the second term subdominant, and the first term dominant:
\begin{equation}
\label{SDcond}
\frac{m^D_{\rm atm}{m^D_{\rm atm}}^T}{M_{\rm atm}} \gg
\frac{m^D_{\rm sol}{m^D_{\rm sol}}^T}{M_{\rm sol}} \ \left[ \ \gg
\frac{m^D_{\rm dec}{m^D_{\rm dec}}^T}{M_{\rm dec}}\ \right] \, ,
\end{equation}
which immediately predicts a normal neutrino mass hierarchy (and not only a simple ordering),
\begin{equation}
\label{normal2}
m_3 \gg  m_2 \ \left[\  \gg m_1  \ \right],
\end{equation}
which is the main prediction of SD.

We have labelled the dominant right-handed neutrino and Yukawa couplings mainly responsible for the atmospheric neutrino mass $m_3$ as ``${\rm atm}$'', the subdominant ones mainly responsible for the solar neutrino mass $m_2$ as ``${\rm sol}$'', and the almost decoupled (sub-sub-dominant) ones mainly responsible for $m_1$ as ``${\rm dec}$''. Note that the mass ordering of the right-handed neutrinos is not yet specified. We shall order the right-handed neutrino masses as $M_1<M_2<M_3$, and subsequently identify $M_{\rm atm}, M_{\rm sol}, M_{\rm dec}$ with $M_1,M_2,M_3$ in all possible ways. 

It is clear that, in the limit $m_1\rightarrow 0$, the sub-sub-dominant right-handed neutrino and its associated couplings labelled by ``${\rm dec}$'' decouple completely and the above model reduces to a two right-handed neutrino model. In that limit we simply drop the third terms [in square brackets] in Eqs.~\eqref{seq1}--\eqref{normal2} in anticipation of this.

Constrained sequential dominance (CSD)~\cite{King:2005bj} assumes the SD conditions in Eq.~\eqref{SDcond} and in addition it constrains the the right-handed neutrino mainly responsible for the atmospheric neutrino mass to have couplings to $(\nu_e, \nu_{\mu}, \nu_{\tau})$ [namely $(m^D_{e, \rm atm}, m^D_{\mu, \rm atm} , m^D_{\tau, \rm atm})$] proportional to $(0,1,1)$. Furthermore the right-handed neutrino mainly responsible for the solar neutrino mass has couplings to $(\nu_e, \nu_{\mu}, \nu_{\tau})$ given by $(m^D_{e, \rm sol}, m^D_{\mu, \rm sol}, m^D_{\tau, \rm sol})$ to be proportional to $(1,1,-1)$, leading to TB mixing. CSD2~\cite{Antusch:2011ic} was proposed to give a non-zero reactor angle and is based on the same atmospheric alignment but with the right-handed neutrino mainly responsible for the solar neutrino mass having couplings to $(\nu_e, \nu_{\mu}, \nu_{\tau})$ proportional to $(1,0,-2)$ or $(1,2,0)$. This yields a reactor angle of $\theta_{13}\approx 6^\circ$, which unfortunately is still too small, although the situation can be rescued by invoking charged lepton corrections~\cite{Antusch:2013wn}. The CSD3 model in~\cite{King:2013iva} involves the right-handed neutrino mainly responsible for the solar neutrino mass having couplings to $(\nu_e, \nu_{\mu}, \nu_{\tau})$ proportional to $(1,3,1)$ or $(1,1,3)$ with a relative phase $\mp \pi/3$, yielding a reactor angle of $\theta_{13}\approx 8.5^\circ$ close to the observed value. However, CSD3 predicts approximate TBC mixing with an almost maximal atmospheric mixing angle, which is now disfavoured by the latest global fits and so it may soon be challenged. CSD4~\cite{King:2013xba} involves the same atmospheric neutrino couplings but with the right-handed neutrino mainly responsible for the solar neutrino mass having couplings to $(\nu_e, \nu_{\mu}, \nu_{\tau})$ proportional to $(1,4,2)$ with a relative phase of $\pm 2\pi/5$, i.e.,
\begin{equation}
m^D=
\left( \begin{array}{cc}
m^D_{e, \rm atm} & m^D_{e, \rm sol}    \\
m^D_{\mu, \rm atm} & m^D_{\mu, \rm sol} \\
m^D_{\tau, \rm atm} & m^D_{\tau, \rm sol} 
\end{array}
\right)
=
\left( \begin{array}{cc}
0 & b    \\
a & 4b \\
a & 2b 
\end{array}
\right),
\label{dirac2}
\end{equation}
where $a$ may be taken to be real and $b$ has a phase of $\pm 2\pi/5$. After implementing the seesaw mechanism, this yields, $\theta_{12}\approx 34^{\circ}, \theta_{23}\approx 41^{\circ}, \theta_{13}\approx 9.5^{\circ}$, which closely coincide with the current best-fit values, together with a distinctive prediction for the $CP$ violating oscillation phase: $\delta \approx \pm 106^\circ$.

\subsection{Direct vs.\ indirect approach}

The most predictive models involve family symmetry groups $G$ which admit triplet representations, since it is only in those cases that the solar angle can be predicted. However, the application of family symmetries to model building is not straightforward since the underlying family group $G$ cannot be preserved -- otherwise all three families would be indistinguishable, with degenerate masses and no mixing (think of the consequence of the gauged $SU(3)_C$ QCD colour group where the three colours of quarks are degenerate). Therefore the family symmetry $G$ must be broken either by enlarging the Higgs sector or by introducing ``flavons'' as discussed above. Either way, the question arises of how the family symmetry $G$ is broken, and the simple answer is ``carefully'', since the predictions depend crucially on how $G$ is broken. Although the family symmetry $G$ will be completely broken in the full theory, there will be some ``memory'' or ``relic'' of $G$ in the neutrino sector which differs from that in the charged lepton sector, and it is this difference that is crucial to the predictions (if no relics of $G$ survive in either sector then all predictive power of $G$ is lost).\footnote{Similarly, in more complete theories, different ``relics'' of $G$ may survive in the up- and down-quark sectors, as discussed later.} 

\begin{center}
\begin{figure}[ht]
\begin{tabular}{lcr}
\includegraphics[trim=0cm 5cm 0cm 5cm, width=18cm]{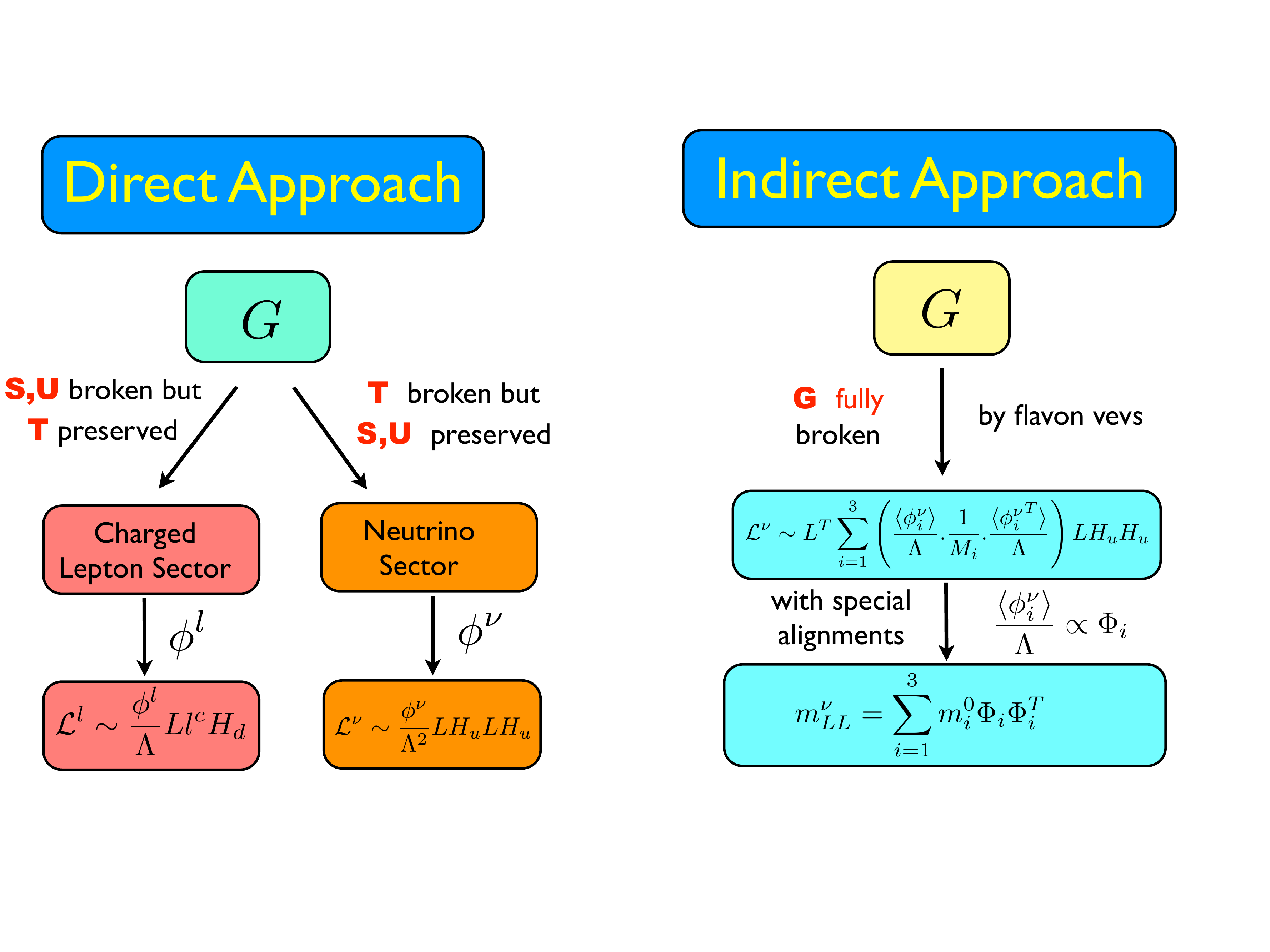} 
\end{tabular}
\caption{\label{directvsindirect} Direct (left panel) vs.\ Indirect (right panel) approaches to model building.}
\end{figure}
\end{center}

There are essentially two ways in which the family symmetry can be broken, with each possibility leading to different ``relics'' of $G$ in the neutrino and charged lepton sectors. Following~\cite{King:2009ap}, we distinguish between the direct and indirect approaches to model building, depending on the way that the family symmetry is broken. In the direct approach different subgroups of the family symmetry survive in the neutrino or charged lepton sectors, while in the indirect approach no subgroup of the family symmetry survives in either sector but the flavons have ``special'' vacuum alignments whose alignment is assisted by the family symmetry, with different flavons appearing in the neutrino and charged lepton sectors. The two approaches are illustrated in Fig.~\ref{directvsindirect}.

In the direct approach, the family symmetry is broken in a very special way such that different subgroups of $G$ are respected in the charged lepton part of the Lagrangian and in the neutrino sector (but the combined theory completely breaks the family symmetry). These two subgroups enforce particular forms of the charged lepton and neutrino mass matrices, such that the lepton mixing matrix matrix is determined purely from symmetry, as we discuss in the example later. The basic idea is that the family symmetry $G$ has three generators $S,T,U$ and is broken by the VEV of a flavon $\phi^l$ down to a subgroup $Z_3$ generated by the order-3 generator $T$, and by the VEV of a flavon $\phi^{\nu}$ down to a subgroup $Z_2\times Z_2$ generated by the order-2 generators $S,U$.

A key feature of the direct approach is that the Klein symmetry of the Majorana neutrino mass matrix, $Z_2\times Z_2$, is identified with a subgroup of the full family symmetry $G$, namely the group generated by $S$ and $U$. While it is always true that any neutrino mass matrix $m^\nu_{LL}$ has some Klein symmetry $Z_2\times Z_2$ (see e.g.~\cite{King:2013eh} for a discussion of this point), it is not always true that the Klein symmetry may be identified with a subgroup of some underlying family symmetry $G$ (which is the case for direct models). In semi-direct models, only one of the two $Z_2$ factors of the Klein symmetry is identified with a subgroup of the family symmetry $G$, for example either the $S$ generator (yielding trimaximal type~II mixing) or the $SU$ combination (yielding trimaximal type~I mixing). 

\begin{center}
\begin{figure}[ht]
\begin{tabular}{lcr}
\includegraphics[width=15cm]{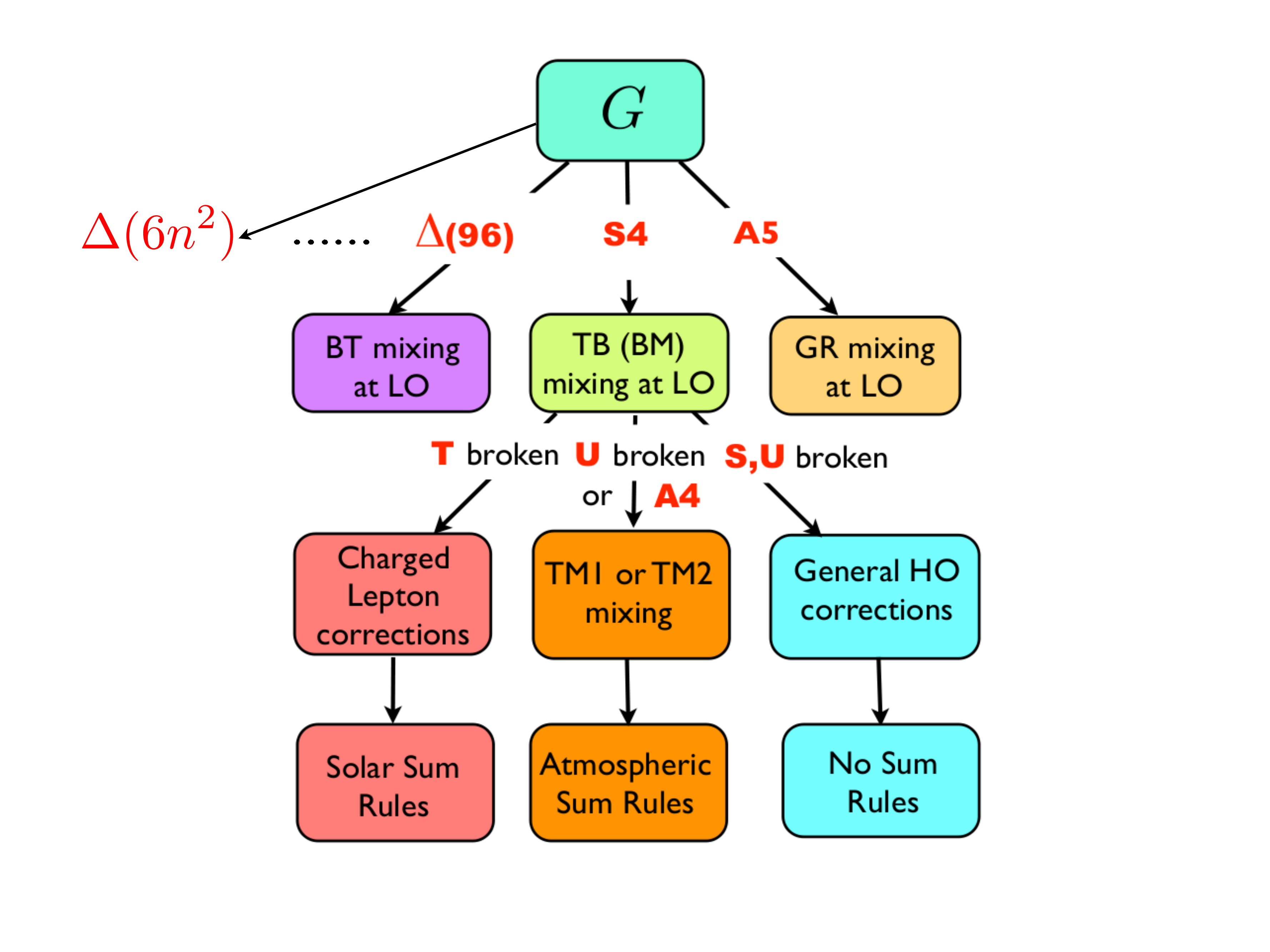} 
\end{tabular}
\caption{\label{direct2}\footnotesize  Direct and semi-direct approaches to model building. The left part of the diagram corresponds to ``larger'' groups such as $\Delta (6n^2)$ which are candidates for direct approaches in which the generators $S,U$ are preserved in the neutrino sector (corresponding to the Klein symmetry) and $T$ is preserved in the charged lepton sector (corresponding to a diagonal charged lepton Yukawa matrix in the $T$ diagonal basis. $\Delta (96)$ gives BT mixing and requires some $T$ violation (i.e.\ charged lepton corrections) to be viable. For the ``smaller'' groups such as $S_4$ which gives TB (or BM) mixing at leading order, one or more of the generators $S,U$ or $T$ must be broken in order to account for the reactor angle. Semi-direct models violate $U$ but preserve either $S$ (giving TM2 mixing) or $SU$ (giving TM1 mixing), distinguished by different types of atmospheric sum rules. The group $A_4$ is necessarily semi-direct since it does not contain $U$.}
\end{figure}
\end{center}

The direct approach typically predicts simple patterns of lepton mixing, such as tri-bimaximal mixing (from for example $S_4$) or golden ratio mixing (from for example $A_5$). However, such simple patterns of lepton mixing predict a zero reactor angle, as a general consequence of the $T$ and $U$ generators being preserved. In the light of a non-zero reactor angle, there are various strategies that have been developed for the direct approach as shown in Fig.~\ref{direct2}. Essentially the choice is to either use larger groups such as $\Delta (6n^2)$~\cite{King:2013vna} (for example $\Delta (96)$~\cite{King:2012in} which gives bi-trimaximal mixing) or to break one or more of the generators $S,T,U$ while keeping a smaller group. If only the $U$ generator is broken then this is known as a semi-direct model. Alternatively one may is keep the direct origin of the Klein symmetry in the neutrino sector based on $S,U$ intact (yielding for example tri-bimaximal neutrino mixing), but breaking the $Z_3^T$ symmetry of the $T$ generator in the charged lepton sector. Then deviations from tri-bimaximal lepton mixing originate entirely from charged lepton corrections.

In fact models can be classified as direct, semi-direct, or indirect, depending on the extent to which a subgroup of the discrete family symmetry can be identified with the Klein symmetry of the neutrino sector~\cite{King:2013eh}. In several of these models quarks are included via $SU(5)$ unification, but typically vacuum alignment does not determine the quark mixing angles. However, in a purely symmetrical approach, the direct approach has been extended to the quark sector, where a subgroup of the discrete family symmetry is used to constrain also the quark mixing angles, in analogy with the procedure followed for the Klein symmetry in the neutrino sector~\cite{Hagedorn:2012pg,Araki:2013rkf,Holthausen:2013vba}, but no realistic model has been proposed. In some such approaches~\cite{Araki:2013rkf,Holthausen:2013vba}, the symmetry groups can be quite large, for example $\Delta (6n^2)$ for large values of $n$~\cite{King:2013vna}. 

In indirect models, the underlying family symmetry $G$ is completely broken by flavon VEVs, such that the Klein symmetry of the neutrino mass matrix, $Z_2\times Z_2$, is \emph{not} identified with any subgroup of $G$. One may ask, then, what is the purpose of the family symmetry? The answer is that the family symmetry $G$ provides flavons $\phi^{\nu}$ with very special vacuum alignments, which however do not collectively respect any subgroup of $G$,\footnote{Even in the case where no special alignments are present (the group is completely broken) there could be some prediction coming from the memory of the flavour group. This is because flavour symmetries constrain the Yukawa couplings reducing the number of free parameters.} although some of them may respect subgroups, while others are determined from orthogonality arguments. In this way the role of the family symmetry $G$ is rather indirect, although still important for vacuum alignment, hence the name indirect models. Typically the indirect approach is used in association with sequential dominance, in which the lepton mixing arises from the neutrino sector as a result of the type~I seesaw mechanism, with one of the right-handed neutrinos being mainly responsible for the atmospheric neutrino mass and a second right-handed neutrino being mainly responsible for the solar neutrino mass while a third right-handed neutrino is approximately decoupled from the seesaw mechanism, leading to a normal neutrino mass hierarchy. 

\subsection{Generalised $\mathbf{CP}$ Symmetry and Geometrical $\mathbf{CP}$ Violation}

Finally we mention the possibility of predicting the $CP$ phases from direct, semi-direct, or indirect models, sometimes referred to as ``Geometrical $CP$ violation''~\cite{deMedeirosVarzielas:2011zw,Varzielas:2012nn,Varzielas:2012pd,Bhattacharyya:2012pi}. The basic starting point is to postulate that $CP$ is conserved in the high energy theory before the family symmetry is broken~\cite{Branco:2011zb}. $CP$ is a discrete symmetry which involves both complex conjugation of fields and inversion of spatial coordinates at the same time,
\begin{equation}
\varphi(x)\stackrel{CP}{\longrightarrow}X_{\mathbf{r}}\,\varphi^{*}(x')\,,
\end{equation}
where $x'=(t,-\mathbf{x})$ and $X_{\mathbf{r}}$ is a matrix of transformations associated with the field
$\varphi(x)$ in the irreducible representation $\mathbf{r}$ of the discrete family symmetry $G$
\cite{Grimus:2003yn}. 

The ``trivial'' $CP$ transformation corresponds to choosing  $X_{\mathbf{r}}$ to be equal to the unit matrix. For a continuous family symmetry, this would be the only choice \cite{Branco:2011zb}. However for a discrete family symmetry, more general $CP$ transformations can be defined, corresponding to various non-trivial choices for the matrix $X_{\mathbf{r}}$ which collectively form the group of unbroken $CP$ transformations $H_{CP}$~\cite{Mohapatra:2012tb}. The $X_{\mathbf{r}}$ matrices must be consistent with the family symmetry transformations. To be precise, if we first perform a $CP$ transformation, then apply a family symmetry transformation, and finally an inverse $CP$ transformation, then the result must be equivalent to some family symmetry transformation \cite{Holthausen:2012dk}:
\begin{equation}
\label{eq:consistency}X_{\mathbf{r}}\rho^{*}_{\mathbf{r}}(g)X^{-1}_{\mathbf{r}}=\rho_{\mathbf{r}}(g'),
\qquad g' \in G\,.
\end{equation}
The full symmetry of the unbroken family symmetry and generalised $CP$ symmetry is written $G \rtimes H_{CP}$ where the semi-direct product symbol reminds us that these two groups do not commute in general for the case of non-Abelian family symmetries. 

The predictions for $CP$ violation depend on how the symmetry group $G \rtimes H_{CP}$, is broken~\cite{Feruglio:2012cw}. First suppose that $H_{CP}$ is spontaneously violated by complex VEVs of flavons which also break the non-Abelian family symmetry $G$. In the case of direct or semi-direct models there will be some subgroup of $G^{\nu} \rtimes H^{\nu}_{CP}$ preserved in the neutrino sector, and some other subgroup $G^e \rtimes H^e_{CP}$ preserved in the charged lepton sector. The surviving symmetries constrain the neutrino mass matrix and charged lepton mass matrix, leading to predictions for $CP$ violating phases as well as constraints on the mixing angles. Typically the oscillation phase is predicted to take simple values such as zero, $\pi$, or $\pm \pi/2$~\cite{Ding:2013hpa,Feruglio:2013hia,Ding:2013bpa,Meroni:2012ty,Li:2013jya}, although other predictions are possible~\cite{Luhn:2013lkn,Ding:2013nsa}.

In the case of indirect models a different possibility has been studied, namely $CP$ can be spontaneously broken by an auxiliary Abelian flavour group $Z_N$ that commutes with the non-Abelian family symmetry $G$~\cite{Antusch:2011sx,Antusch:2013wn,Antusch:2013rla}. This introduces interesting non-trivial phases $2m\pi/N$ with $m=1,\cdots ,N$, where these phases can appear sequentially in the neutrino mass matrix, leading to non-trivial predictions for the oscillation phase, as discussed in the indirect model example below.

\subsection{Example of direct approach}

In the absence of higher order corrections, only very large groups can enable the reactor angle to be obtained from the direct approach where the Klein symmetry is a subgroup of the discrete family group. As shown in the left part of Fig.~\ref{direct2}, examples of such groups are the $\Delta (6n^2)$ series for large $n\geq 4$. For example for $n=4$ we have $\Delta (96)$ which yields bi-trimaximal mixing which requires some charged lepton corrections (corresponding to T generator symmetry violation) to enable it to be viable~\cite{King:2012in}. 

For $\Delta (6n^2)$ with larger values of $n$, the reactor angle can be obtained directly, without the need for charged lepton corrections to be invoked. In such models all the symmetries $S,U$ and $T$ are accurately respected and the model is said to be ``direct'' since all these generators are unbroken subgroups of the underlying $\Delta (6n^2)$ family symmetry. Due to extra auxiliary symmetries, suppose that the flavon $\phi^l$ appears in the charged lepton Yukawa couplings, $\phi^l L l^cH_d$, while the flavon $\phi^{\nu}$ appears in the Weinberg operator responsible for neutrino mass, $\phi^{\nu} L LH_uH_u$ (where we suppress couplings and dimensional scales). Then the charged lepton Yukawa matrix will be invariant under $Z_3$, $Y_eY_e^\dagger=T^\dagger Y_eY_e^\dagger T$, while the neutrino mass matrix $m^\nu_{LL}$ will be invariant under $Z_2\times Z_2$ (known in the trade as the Klein symmetry), namely $S^Tm^\nu_{LL}S=m^\nu_{LL}$ and $U^Tm^\nu_{LL}U=m^\nu_{LL}$. These imply that $[T,Y_eY_e^\dagger]=0$, $[S,m^\nu_{LL}{m^\nu_{LL}}^\dagger]=0$, and $[U,m^\nu_{LL}{m^\nu_{LL}}^\dagger]=0$. As in undergraduate quantum mechanics, where commuting operators can be simultaneously diagonalised, we see that the matrix $V_e$ which diagonalises $Y_eY_e^\dagger$ will also diagonalise  $T$  and the matrix $V_{\nu}$  which diagonalises $m^\nu_{LL}{m^\nu_{LL}}^\dagger$ will also diagonalise  $S$ and $U$. Hence the lepton mixing matrix matrix $U_{\rm }=V_e V_{\nu}^\dagger$ can be obtained purely from group theory by diagonalising $S,T,U$ in some basis (the result being basis invariant). 

Following the above procedure for $\Delta (6n^2)$~\cite{King:2013vna}, the resulting lepton mixing matrix matrix corresponds to TM2 mixing but with a discrete choice of reactor angle,
\begin{eqnarray}
U_{\rm }=\left(
\begin{array}{ccc}
 \sqrt{\frac{2}{3}} \cos (\vartheta ) & \frac{1}{\sqrt{3}} & \sqrt{\frac{2}{3}} \sin (\vartheta ) \\
 -\sqrt{\frac{2}{3}} \sin \left(\frac{\pi }{6}+\vartheta\right) & \frac{1}{\sqrt{3}} & \sqrt{\frac{2}{3}} \cos \left(\frac{\pi
   }{6}+\vartheta\right) \\
 \sqrt{\frac{2}{3}} \sin \left(\frac{\pi }{6}-\vartheta \right) & -\frac{1}{\sqrt{3}} & \sqrt{\frac{2}{3}} \cos \left(\frac{\pi }{6}-\vartheta
   \right) \\
\end{array}
\right)P,
\label{V}
\end{eqnarray}
where $\vartheta =\pi /n, \ldots, \pi/2$ with $n$ dictated by the choice of family symmetry $\Delta (6n^2)$. For example $n=22$ gives a good fit to the reactor angle~\cite{King:2013vna}.

\subsection{Example of indirect approach}

The indirect approach breaks all symmetries $S,U$ and $T$ with the role of the family symmetry being to achieve simple looking vacuum alignments for the neutrino flavons $\phi^{\nu }$,
leading to a simple form for the neutrino mass matrix. In order to understand how indirect models work, it is simplest to consider an example in the basis in which the charged lepton Yukawa matrix is diagonal. At the level of Weinberg operators, there may be two or three Weinberg operators which contribute to the neutrino mass matrix, each being of the special form $(L^T \phi^{\nu })( \phi^{\nu T}  L)H_uH_u$, where $L\sim \mathbf{3}$ and $\phi^{\nu }\sim \overline{\mathbf{3}}$ under the family symmetry (or $\phi^{\nu }\sim \mathbf{3}$ if the group $G$ admits real triplets). However, in most indirect models the Weinberg operators may arise from a type~I seesaw mechanism, including sequential dominance, as we now discuss.

For example, consider the CSD4 couplings such as those in Eq.~\eqref{dirac2}, which were postulated in an {\it ad hoc} way. In realistic models these couplings may arise from flavons, which are triplets $\mathbf{3}$ (or antitriplets $\overline{\mathbf{3}}$ for complex representations) under some family symmetry and have special vacuum alignments, for example~\cite{King:2013iva,King:2013xba},
\begin{equation}
\label{Phi00} 
\langle{\phi}^\nu_{a}\rangle \propto \begin{pmatrix}0 \\ 1 \\ 1\end{pmatrix}, \qquad
\langle{\phi}^\nu_{b}\rangle \propto \begin{pmatrix}1 \\ 4 \\ 2\end{pmatrix},
\end{equation}
where the flavon labelled by $a$ is responsible for the first column and the flavon labelled by $b$ is responsible for the second column in Eq.~\eqref{dirac2}. This is achieved by having couplings in the Lagrangian of the form $L{\phi}^\nu_{a}N^c_{\rm atm}$ and $L{\phi}^\nu_{b}N^c_{\rm sol}$, where the three lepton doublets $L=(L_{e},L_{\mu }, L_{\tau })$ transform as a triplet $\mathbf{3}$ under some family symmetry while $N^c_{\rm atm},N^c_{\rm sol}$ are singlets, as is the Higgs which we have dropped.

Note that $CP$ is assumed to be preserved and it is spontaneously broken when the family and flavour symmetries are broken. In this model, the relevant flavour symmetry is a $Z_5$ auxiliary symmetry that commutes with an $A_4$ family symmetry, leading to the appearance of a discrete phase $e^{4i\pi/5}$~\cite{King:2013iva,King:2013xba}.

After implementing the seesaw mechanism we arrive at the Weinberg operators of the above form. The vacuum alignments in Eq.~\eqref{Phi00} then imply a special form of neutrino mass matrix $m^\nu_{LL} $, generated from summing the two terms $\langle \phi^{\nu }\rangle \langle \phi^{\nu T}\rangle $, yielding~\cite{King:2013iva,King:2013xba},
\begin{equation}
m^\nu_{LL} =
m_a
\begin{pmatrix}
      0&0&0\\0&1&1\\0&1&1 
      \end{pmatrix}
      +
m_be^{4i\pi/5}
\begin{pmatrix}
      1&4&2\\
4&16&8 \\
2&8&4
\end{pmatrix},
\label{mnu}
\end{equation}
which, with $m_a$, $m_b$ real, can reproduce the current best fit lepton mixing matrix mixing parameters~\cite{King:2013iva}, $\theta_{12}\approx 34^{\circ}, \theta_{23}\approx 41^{\circ}, \theta_{13}\approx 9.5^{\circ}$, together with the distinctive prediction for the $CP$ violating oscillation phase $\delta \approx \pm 106^\circ$.

This neutrino mass matrix naturally arises from the seesaw mechanism, since it follows from Eq.~\eqref{Phi00} which corresponds to CSD4. This special form of $m^\nu_{LL} $ has a Klein symmetry $Z_2 \times Z_2$, which cannot be identified with any subgroup of any known family symmetry $G$, even though the vacuum alignments could themselves arise from a group as small as $A_4$~\cite{King:2013iva,King:2013xba}.

\begin{figure}[t]
\centering
\includegraphics[width=0.46\textwidth]{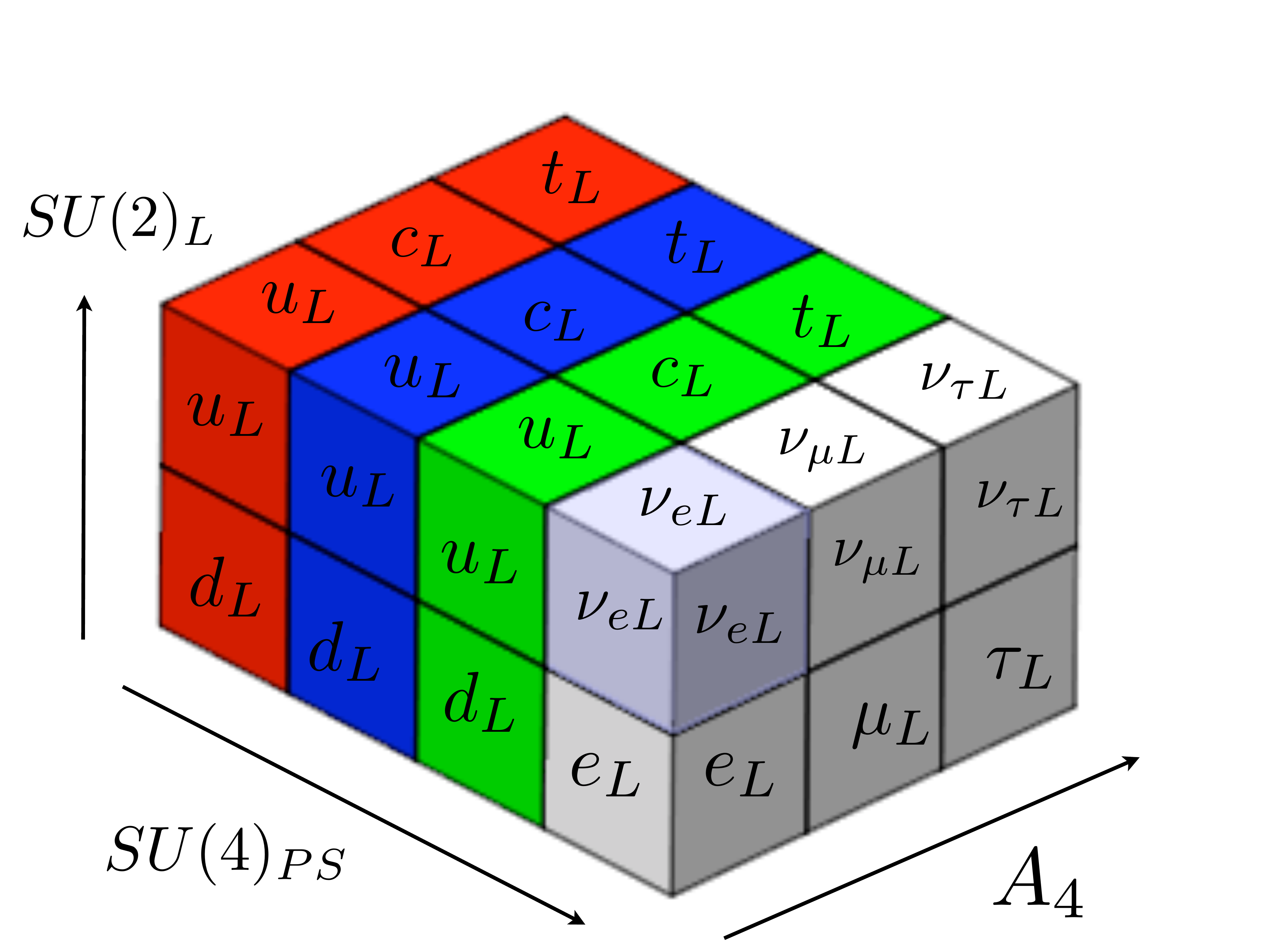}
\includegraphics[width=0.50\textwidth]{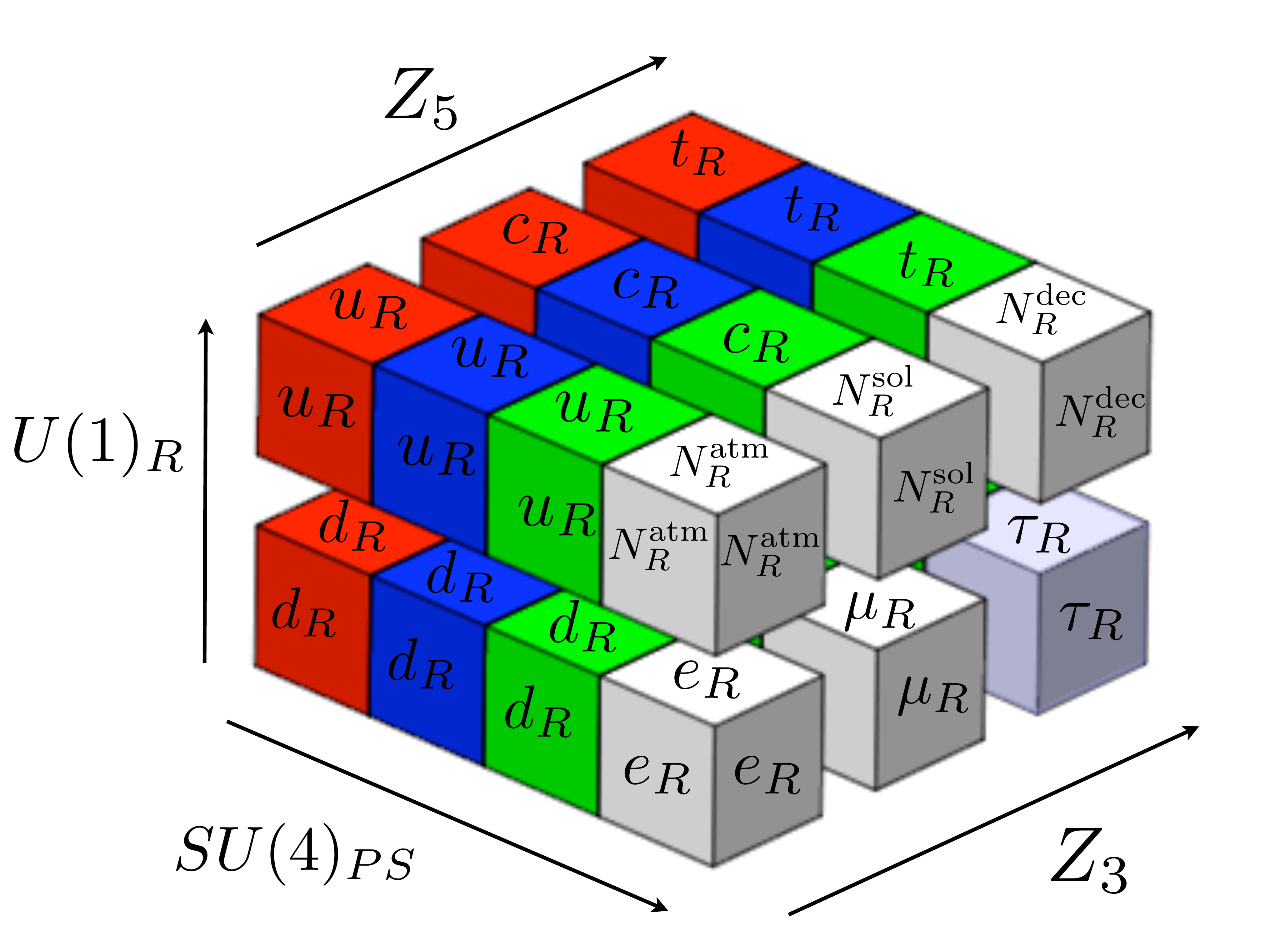}
\vspace*{-4mm}
    \caption{The A4SU421 unification of quarks and leptons in the ``tetra-model''. The left diagram depicts quark-lepton-family unification of the 24 left-handed quarks and leptons denoted collectively as ${\cal Q} $ into a single $(3,4,2,0)$ multiplet of A4SU421. The right diagram shows the 24 right-handed quarks and leptons which form six $A_4$ singlets, ${\cal U}_i$ and ${\cal D}_i$, distinguished by $Z_5$ and $Z_3$,  with quarks and leptons unified in each multiplet.
} \label{421fig}
\vspace*{-2mm}
\end{figure}

Both quark and lepton mixing may be explained by the above indirect model, by extending the gauge group to $SU(4)_{\rm PS}\times SU(2)_L\times U(1)_R$, with the quarks and leptons transforming as in Fig.~\ref{421fig} in the so-called ``tetra-model''~\cite{King:2013hoa}. The structure of the Yukawa matrices depends on the following vacuum alignments, which incorporate the CSD4 vacuum alignments~\cite{King:2013xba} introduced in Eq.~\eqref{Phi00},
\begin{equation}
\langle \phi_{{\cal U}^c_1} \rangle =
\frac{v_{{\cal U}^c_1}}{\sqrt{2}}  \begin{pmatrix}0 \\ 1 \\ 1\end{pmatrix}  \ , \qquad
\langle \phi_{{\cal U}^c_2} \rangle =
\frac{v_{{\cal U}^c_2}}{\sqrt{21}} \begin{pmatrix}1 \\ 4 \\ 2\end{pmatrix} \ , \qquad
\langle \phi_{{\cal U}^c_3} \rangle = v_{{\cal U}^c_3}
\begin{pmatrix} 0 \\ 0 \\1 \end{pmatrix} 
 \ ,\label{phiu}
\end{equation}
and
\begin{equation}
\langle \phi_{{\cal D}^c_1} \rangle =v_{{\cal D}^c_1}
\begin{pmatrix} 1\\0\\0 \end{pmatrix}  \ , \qquad
\langle \phi_{{\cal D}^c_2} \rangle =v_{{\cal D}^c_2}
\begin{pmatrix} 0\\1\\0 \end{pmatrix} \ , \qquad
\langle \phi_{{\cal D}^c_3} \rangle = v_{{\cal D}^c_3}
\begin{pmatrix} 0\\0\\1 \end{pmatrix} ,
 \ \label{phid}
\end{equation}
where $\phi_{{\cal U}^c_i}$ and $\phi_{{\cal D}^c_i}$ are $A_4$ triplets and Pati-Salam singlets. Two Higgs doublets are introduced, $h_u$ and $h_d$, which are $A_4$ singlets. Then the Yukawa couplings arise from the diagonal non-renormalisable operators,
\begin{equation}
h_u(\phi_{{\cal U}^c_i}.{\cal Q}){\cal U}^c_i +h_d(\phi_{{\cal D}^c_i}.{\cal Q}){\cal D}^c_i .
\end{equation}

The up-quark and neutrino Yukawa matrices are obtained from the operators $(\phi_{{\cal U}^c_i}.{\cal Q}){\cal U}^c_i$ by sticking together the three column vectors from Eq.~\eqref{phiu},
\begin{equation}
  Y^{\nu} \sim Y^u \sim  \begin{pmatrix}  0 & b & 0  \\ 
a & 4b & 0\\  a & 2b & c\end{pmatrix},
\end{equation}
where each column is multiplied by a different constant of proportionality. The Yukawa matrices are not expected to be exactly equal due to the decisive Clebsch-Gordan coefficients as discussed in~\cite{King:2013hoa}. Note that the third column is decoupled from the seesaw mechanism according to sequential dominance, so the resulting neutrino mass matrix is as given in Eq.~\eqref{mnu}.

The down-quark and charged lepton Yukawa matrices are similarly obtained from $(\phi_{{\cal D}^c_i}.{\cal Q}){\cal D}^c_i$ by amalgamating the three column vectors in Eq.~\eqref{phid} and are hence diagonal,
\begin{equation}
 Y^d \sim Y^e \sim
  \begin{pmatrix}  y_d & 0  & 0  \\ 0 & y_s & 0\\ 0  & 0 & y_b\end{pmatrix}.
\label{phid1}
\end{equation}
The Yukawa matrices are not expected to be exactly equal due to Clebsch-Gordan coefficients, as discussed in~\cite{King:2013hoa}.

The quark-lepton unification implies that the second column $(1,4,2)^T$ of the neutrino Yukawa matrix is equal to that of the up-quark Yukawa matrix and hence predicts a Cabibbo angle approximately equal to 1/4. The third column (approximately decoupled from the seesaw mechanism) is proportional to $(0,0,1)^T$ at leading order, giving the top-quark Yukawa coupling. Higher order corrections modify the leading order predictions and are responsible for the other quark mixing angles and $CP$ violation as discussed in~\cite{King:2013hoa}.

The above example shows that the indirect approach goes hand in hand with the seesaw mechanism and SD. We emphasise the highly predictive nature of the scheme: the neutrino masses and the entire lepton mixing matrix mixing matrix follow from Eq.~\eqref{mnu}, which involves only two real parameters $m_a$ and $m_b$, with the underlying family symmetry being the minimal choice, $A_4$. The indirect approach is certainly a very attractive possibility in the light of a non-zero reactor angle, but the above model is not a GUT model.

\subsection{\label{sec:GUT}GUTs \& family symmetry}

Grand Unified Theories (GUTs) are the dream of theoretical physics because of their elegance. The SM is made of three gauge groups and the respective gauge couplings are function of the energy scale. There is a tendency of the three gauge couplings to unify at some very high scale.\footnote{To be more precise, the gauge couplings do not meet exactly in the SM and new physics typically is required for the unification like, e.g., SUSY.} This fact has fueled the imagination of theoretical physicist, leading to the idea that the three SM gauge groups could arise from a single gauge group at some very high energy scale. These kinds of theories are very elegant, because they can describe the basic constituents of nature in a very compact and economical way. Moreover, they predict the unification of the three SM gauge couplings into a single one and can explain the quantisation of the electric charge. From a phenomenological point of view, GUTs predict proton decay (its non-observation provides a limit on the GUT scale to be at $10^{15\div 16}$\,GeV or even higher) and in some cases also neutron-antineutron oscillation. These observations would give us strong hints in favour of GUTs. But -- so far -- we can only make theoretical speculations. 

The SM has rank four (namely four diagonal generators), and therefore a grand unified group $G_0$ containing the SM must have at least rank four. Moreover, $G_0$ must have $SU(3)$ as a subgroup (since color is unbroken in the SM) and it must give the correct electric charges. This all implies that the only available rank four group is $SU(5)$. It is also possible to consider groups with a rank bigger then four, and there are many examples. Perhaps the most frequently used ones are the Pati-Salam (PS) group $SU(4)_C\times SU(2)_L\times SU(2)_R$, $SO(10)$ of rank five, and $E_6$ of rank six.  

In $SU(5)$, all SM fields fit into only two irreducible representations, the conjugate five ($F=\overline{\bf 5}$) and the ten ($T={\bf 10}$), which is an antisymmetric $5\times 5$ matrix,
\begin{equation}\label{su5r}
F =
\left(
\begin{array}{c}
d^c_1\\
d^c_2\\
d^c_3\\
e\\
\nu
\end{array}
\right)_L\,,\quad
T=
\left(
\begin{array}{ccccc}
0&u_3^c&-u_2^c&u_1&d_1\\
 & 0   &u_1^c&u_2&d_2\\
 &    &0&u_3&d_3\\
&   &   &  0 & e^c\\
&&&&0
\end{array}
\right)_L\,.\quad
\end{equation}

However, massive neutrinos impose a non-minimality, since in order to generate a Dirac or a Majorana renormalisable mass for neutrino, in $SU(5)$ the right-handed neutrinos must be assigned into the trivial singlet representation $\mathbf{1}$. From this point of view, the PS group and $SO(10)$ are more complete, because right-handed neutrinos does not have to be introduced in a trivial representation. In the PS group, we have two representations $\psi=({\bf 4,2,1})$ and $\psi^c=({\bf \overline{4}, 1,2})$ under $(SU(4)_C,\, SU(2)_L,\, SU(2)_R )$, where
\begin{equation}
\psi=
\left(
\begin{array}{ccccc}
u_1&u_2&u_3&\nu\\
d_1&d_2&d_3&e
\end{array}
\right)_L\,,\quad
\psi^c=
\left(
\begin{array}{ccccc}
u_1^c&u_2^c&u_3^c&\nu^c\\
d_1^c&d_2^c&d_3^c&e^c
\end{array}
\right)_R ,
\end{equation}
so that leptons are basically the fourth component of an extended color $SU(4)_C$. Differently from $SU(5)$, the right-handed neutrinos are not in the trivial representation. In turn, in $SO(10)$ all the SM fields and the right-handed neutrino of one family belong to a single irreducible ${\bf 16}$ representation,
\begin{equation}
{\bf 16} =(u_1,u_2,u_3,d_1,d_2,d_3,u_1^c,u_2^c,u_3^c,e^c,d^c_1,d^c_2,d^c_3,\nu,e,\nu^c).
\end{equation}
$SO(10)$ contains both $SU(5)$ and the PS group as subgroups, and the ${\bf 16}$ irreducible representation decomposes as
\begin{equation}
{\bf 16} = \overline{\bf 5}+{\bf 10}+{\bf 1}\,\quad (SU(5))\quad, \qquad
{\bf 16} = ({\bf 4,2,1}) +({\bf \overline{4}, 1,2})\,\quad ({\rm PS})\quad .
\end{equation}

Our task is to build unified models where the three families transform under some flavour symmetry like, $G_0\times G$. This gives us a constraint on the possible flavour matter assignments. For example, a model where right-handed type quarks transform as singlets under some flavour group, while the left-handed leptonic doublets belong to a non-trivial irreducible representation of $G$, cannot be embedded into $SU(5)$ because such a field must be in the same multiplet $F$. This is an important feature of GUT flavour models, because it greatly restricts the number of valid matter assignments. For this purpose, a very clear example is given by $SO(10)$. Since in this group all the SM fields belong to the same GUT-multiplet, the only possibility that we have is to take three copies of ${\bf 16}_i$ with$i=1,2,3$. Therefore, in $SO(10)$ models the number of right-handed neutrinos must be equal to the number of generations for sequentiality, and we have no freedom at all. Such three families can belong to the, e.g., ${\bf 3}$ or ${\bf 1} \oplus {\bf 2}$ irreducible representations of $G$. Therefore, all the SM fields belong sequentially to  ${\bf 3}$ or ${\bf 1} \oplus {\bf 2}$, and there is indeed no more freedom in the matter assignment.

Many models have been proposed based on the GUT group $SU(5)$ to give TBM lepton mixing, see for instance~\cite{Ma:2006sk,Chen:2007afa,Altarelli:2008bg,Ishimori:2008fi,Burrows:2009pi,Hagedorn:2010th,King:2012in}, and more recently also models yielding a suitable non-zero reactor angle have been proposed, e.g.~\cite{Cooper:2012wf,Hagedorn:2012ut,Antusch:2013rxa}.
Also in the $SO(10)$ framework, models for TBM have been proposed. Models with dominant type~I seesaw have been studied for instance in~\cite{Ross:2002fb,King:2003rf,deMedeirosVarzielas:2005ax,deMedeirosVarzielas:2006fc}, by using  Froggatt-Nielsen diagrams involving Higgs fields in the $\mathbf{45}$ representation of $SO(10)$ with in the non-renormalisable operator $\mathbf{16}\,\mathbf{16}\,\mathbf{10}\,\mathbf{45}$, in~\cite{Morisi:2007ft} by using the operator  $\mathbf{16}\,\mathbf{16}\,\mathbf{10}\,\mathbf{45}\,\mathbf{45}$, and in~\cite{Bazzocchi:2008sp,Blankenburg:2011vw} with the operator $\mathbf{16}\,\mathbf{16}\,\mathbf{120}\,\mathbf{45}$. $SO(10)$ models with dominant type~II and TBM have been studied for instance in~\cite{Hagedorn:2006ug,Grimus:2008tm,King:2009mk,Dutta:2009bj,Altarelli:2010at,BhupalDev:2011gi}.

However TBM is ruled out nowadays and we have to consider models combining GUT with non-Abelian discrete symmetries without TBM, like for instance~\cite{Dermisek:1999vy} which is based on $D_3$ and which has a very economical number of free parameters.

\subsubsection{\label{sec:s4su5}$\boldsymbol{SU(5) \times S_4}$}

Semi-direct models violate $U$ but preserve either $S$ (giving TM2 mixing) or $SU$ (giving TM1 mixing), distinguished by different types of atmospheric sum rules, cf.\ Sec.~\ref{sec:mixing_rules}. The group $A_4$ is necessarily semi-direct, since it does not contain $U$ but only the generators $S$ and $T$. For example, the model in~\cite{King:2011zj} imposes an $A_4$ symmetry, broken spontaneously by a set of flavons, which leads to the second column of the lepton mixing matrix mixing matrix fixed at its trimaximal value (TM2), with the reactor angle undetermined. However, an $S_4$ extension of this idea can explain why the reactor angle is small, the idea being to break the $U$ generator at higher order~\cite{King:2011zj}. This idea can be included in a GUT model~\cite{Hagedorn:2012ut}, as we now discuss.

In this subsection we discuss a supersymmetric $SU(5) \times S_4$ model of~\cite{Hagedorn:2012ut} based on an earlier direct model~\cite{Hagedorn:2010th} of TB mixing. In order to convert this to TM2 mixing, the model of~\cite{Hagedorn:2010th} is augmented with an extra $S_4$ singlet flavon field $\eta$. The three families of $SU(5)$ matter multiplets, $F={\bf \overline{5}}$ and $T={\bf 10}$, transform under $S_4$ as ${\bf 3}$ and ${\bf 2+1}$, respectively. We furthermore introduce three right-handed neutrinos $\nu^c$ which are unified in the ${\bf 3}$ of~$S_4$ and allow for the type~I seesaw mechanism. The Higgs sector consists of $S_4$ singlets and comprises the standard $SU(5)$ Higgses in the ${\bf 5}$ and ${\bf \overline{5}}$, plus an additional Georgi-Jarlskog Higgs in the ${\bf \overline{45}}$. The family symmetry is broken by a set of flavon fields transforming in various representations of $S_4$. In order to control the couplings of the flavon fields to the different matter sectors, we impose a global $U(1)$ shaping symmetry. The complete charge assignments of matter, Higgs, and flavon fields are listed in Table~\ref{tab:S4-assignments}. 

\begin{table}[t]
\begin{center}
$$
\begin{array}{|c|c|c|c|c|c|c|c|c|c|c|c|c|c|c|c||c|}\hline
\!\!\phantom{\Big|}&\multicolumn{4}{c|}{\mathrm{matter~fields}} &
 \multicolumn{3}{c|}{\mathrm{Higgs~fields}}  & 
 \multicolumn{9}{c|}{\mathrm{flavon~fields}} \\ \cline{2-17}
\!\!\phantom{\Big|} & T_3 & T & F & \nu^c & H^{}_{\bf 5} & H_{\bf{\overline{5}}} & H_{\bf{\overline{45}}} &
\phi^u_{\bf 2} & \tilde\phi^u_{\bf 2} & \phi^d_{\bf 3} & \tilde\phi^d_{\bf 3} &
\phi^d_{\bf 2}  & \phi^\nu_{\bf{ 3'}} & \phi^\nu_{\bf 2} & \phi^\nu_{\bf 1} & \eta\\\hline
~\!SU(5)\!\!\!\phantom{\Big|}~ & \bf 10 & \bf 10 & \bf \overline 5 & \bf 1 &\bf  5 &\bf \overline 5 &\bf \overline{45}
&\bf 1&\bf 1&\bf 1&\bf 1&\bf 1&\bf 1&\bf 1&\bf 1& \bf 1\\\hline
S_4\!\!\phantom{\Big|} & \bf 1&\bf 2&\bf 3&\bf 3&\bf 1&\bf 1&\bf 1&\bf 2&\bf 2&\bf 3&\bf 3&\bf
2&\bf 3'&\bf 2&\bf 1 & \bf 1\\ \hline
U(1)\!\!\phantom{\Big|} & 0&5&4&-4&0&0&1&-10&0&-4&-11 &1 &8&8&8& 7 \\\hline
\end{array}
$$
\end{center}
\caption{\label{tab:S4-assignments}Charge assignments of the matter, Higgs, and flavon superfields in the $SU(5) \times S_4$ model of~\cite{Hagedorn:2012ut}. The $U(1)$ shaping symmetry constrains the set of operators allowed in the superpotential.}  
\end{table}

With the model formulated at the effective level, it is straightforward to derive the leading operators of the matter superpotential which are invariant under all symmetries imposed. Assuming a generic messenger mass $M$ of the order the GUT scale and suppressing all dimensionless order one coupling coefficients, we find 
\begin{eqnarray}
w &~\sim~  & T_3T_3H^{}_{\bf 5} + \frac{1}{M} T T  \phi^u_{\bf 2} H^{}_{\bf 5}  +
\frac{1}{M^2} TT \phi^u_{\bf 2} \tilde\phi^u_{\bf 2} H^{}_{\bf 5} 
\nonumber
\\ 
&&+ \frac{1}{M} F T_3 \phi^d_{\bf 3} H_{\bf \overline{5}} + \frac{1}{M^2} (F
\tilde\phi^d_{\bf 3})^{}_{\bf{1}} ( T \phi^d_{\bf 2} )^{}_{\bf{1}} H_{\bf \overline{45}}
+ \frac{1}{M^3} (F \phi^d_{\bf 2} \phi^d_{\bf 2})^{}_{\bf{3}} ( T \tilde\phi^d_{\bf
  3} )^{}_{\bf{3}} H_{\bf \overline{5}} \nonumber
\\
&&+ F \nu^c H^{}_{\bf 5} + \nu^c \nu^c \phi^\nu_{\bf 1} +  \nu^c \nu^c \phi^\nu_{\bf 2} +  \nu^c \nu^c
\phi^\nu_{\bf {3'}}
 {+ \frac{1}{M} \nu^c\nu^c \phi^d_{\bf 2} \eta }
 \ .
\label{eq:s4-nu}
\end{eqnarray} 
It is the last term which provides the source of the higher order correction to the right-handed neutrino mass matrix, which is essential in generating a large reactor angle. In principle, all independent invariant products of the $S_4$ representations have to be considered for each of these terms; in practice, there is often only one possible choice. In our example, the second and the third term of the second line of Eq.~\eqref{eq:s4-nu} would give rise to several independent terms. However, the contractions specified by the subscripts ${\bf 1}$ and ${\bf 3}$ single out a unique choice. Within a given UV completion, the existence and non-existence of certain messenger fields can justify such a construction.

The Yukawa matrices are generated when the flavon fields acquire their VEVs. It has been shown in~\cite{Hagedorn:2010th,Hagedorn:2012ut} that the following alignments can be obtained:
\begin{eqnarray}
\label{VEVup}
&& \langle \phi^u_{\bf 2} \rangle ~ = ~ \varphi^u_{\bf 2} \begin{pmatrix} 0 \\ 1 \end{pmatrix} \; , \;\;
\langle \tilde\phi^u_{\bf 2} \rangle ~=~
\tilde\varphi^u_{\bf 2} \begin{pmatrix}0\\1\end{pmatrix} \; , \;\;
\\ \label{VEVdown}
&& \langle \phi^d_{\bf 3} \rangle ~ = ~ \varphi^d_{\bf 3} \begin{pmatrix}0\\1\\0\end{pmatrix} \; , \;\;
\langle \tilde\phi^d_{\bf 3} \rangle ~ =~
\tilde\varphi^d_{\bf 3} \begin{pmatrix}0\\-1\\1\end{pmatrix} \; , \;\;
\langle \phi^d_{\bf 2} \rangle ~=~ \varphi^d_{\bf 2} \begin{pmatrix}1\\0\end{pmatrix} \; , \;\;
\\ \label{VEVnu}
&& \langle\phi^\nu_{\bf{3'}} \rangle~=~ \varphi^\nu_{\bf{3'}} \begin{pmatrix}1\\1\\1 \end{pmatrix} \; , \;\;
\langle\phi^\nu_{\bf 2}\rangle~=~ \varphi^\nu_{\bf 2} \begin{pmatrix}1\\1 \end{pmatrix} \; , \;\;
\langle \phi^\nu_{\bf 1} \rangle~=~ \varphi^\nu_{\bf 1} \ .
\end{eqnarray}
Inserting these vacuum alignments and the Higgs VEVs $v_u$  and $v_d$ yields a diagonal up-type quark mass matrix as well as non-diagonal down-type quark and charged  lepton mass matrices.

In order to achieve viable GUT scale hierarchies of the quark masses and mixing angles, we assume
\begin{equation}
\varphi_{\bf 2}^u \sim \tilde \varphi_{\bf 2}^u\sim \lambda^4 M \ , \qquad
\varphi_{\bf 3}^d \sim \lambda^2 M \ , \quad
\tilde\varphi_{\bf 3}^d \sim \lambda^3 M \ , \quad 
\varphi_{\bf 2}^d \sim \lambda M \ ,
\end{equation}
where $\lambda$ denotes the Wolfenstein parameter. With these magnitudes, the charged fermion mass matrices are fixed completely,
\begin{equation}
M_u \sim
\begin{pmatrix}
\lambda^8 & 0&0   \\
0 & \lambda^4 & 0 \\
0  & 0 & 1
\end{pmatrix} v_u  \ , ~\quad
M_d \sim
\begin{pmatrix}
0 & \lambda^5 & \lambda^5   \\
\lambda^5 & \lambda^4 &  \lambda^4 \\
0  & 0 & \lambda^2
\end{pmatrix} v_d  \ , ~\quad
M_e \sim
\begin{pmatrix}
0 & \lambda^5 &0     \\
\lambda^5 & 3\lambda^4 &  0 \\
\lambda^5  & 3\lambda^4 & \lambda^2
\end{pmatrix} v_d  \ .
\label{eq:s4masslam}
\end{equation}
The factors of $3$ in $M_e$ originate from the term of Eq.~\eqref{eq:s4-nu} involving the Georgi-Jarlskog (GJ) Higgs field $H_{\bf{\overline{45}}}$. Due to the GJ factor of $(-3)$ and the texture zero in the 11-entry, we obtain viable charged lepton masses. With the vanishing off-diagonals in the third column of $M_e$, there is only a non-trivial 12 mixing in the left-handed charged lepton mixing $V_{e_L}$. This mixing, $\theta^e_{12} \approx \lambda/3$, will contribute to the total lepton mixing matrix mixing as a charged lepton correction. Note that the 12- and 21- entries, which originate from the same superpotential term, have identical absolute values; together with the zero texture in the 11-entry, this allows for a simple realisation of the Gatto-Sartori-Tonin (GST) relation in the $SU(5) \times S_4$ model. 

In the neutrino sector we find the Dirac neutrino mass matrix and the right-handed neutrino mass matrix to be
\begin{equation}
m_{LR} \approx \begin{pmatrix}
                     1 & 0 & 0\\
		     0 & 0 & 1\\
		     0 & 1 & 0
\end{pmatrix} v_u  \, , \quad
M_{RR} \approx \begin{pmatrix}
                     \varphi^\nu_{\bf 1} + 2 \varphi^\nu_{\bf 3'} &
                     \varphi^\nu_{\bf 2} -  \varphi^\nu_{\bf 3'} {+
                     \frac{\varphi^d_{\bf 2}\langle{\eta}\rangle}{M}}
		              & \varphi^\nu_{\bf 2} -\varphi^\nu_{\bf 3'}\\
		       \varphi^\nu_{\bf 2} -  \varphi^\nu_{\bf 3'} {+
                     \frac{\varphi^d_{\bf 2}\langle{\eta}\rangle}{M}}&
                       \varphi^\nu_{\bf 2} + 2\varphi^\nu_{\bf 3'}
		              &  \varphi^\nu_{\bf 1} -  \varphi^\nu_{\bf 3'}\\
		      \varphi^\nu_{\bf 2} - \varphi^\nu_{\bf 3'} &
                      \varphi^\nu_{\bf 1} - \varphi^\nu_{\bf 3'}
		              &  \varphi^\nu_{\bf 2} + 2  \varphi^\nu_{\bf 3'}
                        {+\frac{\varphi^d_{\bf 2}\langle{\eta}\rangle}{M}}
\end{pmatrix}  .\label{matrixneutr}
\end{equation}
The model is formulated in the $T$-diagonal $S_4$ basis as defined in \cite{King:2011zj},
which is related to the basis discussed in subsection Ref.~\ref{subsec:S4} by a basis rotation,
as discussed in Appendix B of Ref.~\cite{Ishimori:2010au}. 
Note that, although the Dirac neutrino Yukawa term does not involve any flavon field, it is not diagonal in this basis. 
As the family symmetry $S_4$ remains unbroken by $m_{LR}$, the mixing pattern of the effective light neutrino mass matrix $m^\nu_{LL}$ (obtained from the type~I seesaw mechanism) is exclusively determined by the structure of $M_{RR}$. Dropping the higher order terms, we note that the leading order structure of $M_{RR}$, and with it $m^\nu_{LL}$,  is of tri-bimaximal form. This can be easily seen by verifying that the flavon alignments of Eq.~\eqref{VEVnu} are left invariant under the $S$ and $U$ transformations. This leading order TBM structure yields light neutrino masses $m^\nu\sim 0.1\,\mathrm{eV}$ if we set $\varphi^\nu_{\bf{1,2,3'}} \sim \lambda^4 M$. As we want to break the TB Klein symmetry by means of the flavon~$\eta$ at higher order, we set $ \langle{\eta}\rangle \sim \lambda^4M$. Then the TB breaking effect is suppressed by one power of $\lambda$ compared to the leading order. The effective flavon  $\phi^d_{\bf 2}\eta$ transforms as an $S_4$ doublet with an alignment proportional to $(1,0)^T$. This alignment breaks the $U$ symmetry but respects $S$. This directly proves that $M_{RR}$ as well as $m^\nu_{LL}$ are both invariant under~$S$, which in turn entails the TM2 neutrino mixing pattern. The physical lepton mixing matrix matrix is obtained from multiplying the TM2 neutrino mixing with the left-handed charged lepton mixing. 

In summary, the measurement of large $\theta_{13}$ has ruled out the original $SU(5) \times S_4$ model~\cite{Hagedorn:2010th}, which predicted accurate tri-bimaximal neutrino mixing plus small charged lepton corrections. However, a modest extension of the particle content~\cite{Hagedorn:2012ut} can induce a breaking of the $U$ symmetry of the TB Klein symmetry at relative order $\lambda$. The required new flavon field allows for large $\theta_{13}$ and does not destroy the successful predictions of the original model, i.e., it does not have any significant effects on the quark or flavon sectors of the model. The resulting $SU(5) \times S_4$ model preserves $S$ but breaks both $U$ and $T$, i.e., it is a semi-direct model with charged lepton corrections.

\section{\label{sec:experiments}{\em Experiment}: Distinguishing models by their predictions}

In this section, we will outline how flavour models could be tested experimentally. The most characteristic signatures of neutrino flavour models are given by so-called \emph{sum rules}, which are concrete relations between different observables. Such sum rules can arise when there are less parameters than observable quantities present in the model sector under consideration, e.g., a relation involving two mixing angles and the cosine of the oscillation phase (mixing sum rules) or if adding the powers of two mass eigenvalues results into the same power of the last eigenvalue (mass sum rules). The origin of the mixing sum rules may be due to the trimaximal structure of the leptonic mixing matrix (leading to atmospheric sum rules) or due to a simple form of that matrix such as bimaximal or tri-bimaximal, corrected by Cabibbo-like charged lepton mixing corrections (leading to solar sum rules). The reason for the mass sum rule typically lies in the flavon couplings, for example if a $3\times 3$ light neutrino mass matrix depends on only two flavon VEVs, the three mass eigenvalues will be related by a sum rule. Such sum rules will be discussed in Secs.~\ref{sec:mixing_rules} and~\ref{sec:mass_rules}.

Furthermore, it could also happen that in a few years from now we might have more information from experiments than anticipated if, e.g., the Dirac $CP$ phase $\delta$ had a value close to maximal $\pm \pi/2$. Such a fortunate situation would in most cases considerably strengthen our ability to experimentally distinguish different models. These possibilities are discussed in Sec.~\ref{sec:luck}.

\subsection{\label{sec:mixing_rules}Mixing sum rules }

\subsubsection{Atmospheric sum rules }

We have already mentioned examples of atmospheric sum rules as being a consequence of TM1 and TM2 mixing, namely 
\begin{equation}
a=r\cos\delta\,,\qquad {\rm TM}_1,
\end{equation}
being a consequence of the preserved $S_4$ generators $T$ and $SU$, and 
\begin{equation}
a=- \frac{1}{2}r\cos\delta\,,\qquad {\rm TM}_2,
\end{equation}
being a consequence of the preserved $A_4$ generators $T$ and $S$. These are semi-direct models since the $U$ generator is not present in each case.

Semi-direct models may be classified in general as arising from finite von Dyck groups~\cite{Hernandez:2012ra}, which contain two preserved generators denoted as $T_{\alpha}$ and $S_i$. In this framework, the above sum rules correspond to particular examples of this general class of semi-direct models for which the general atmospheric sum rule is~\cite{Ballett:2013wya}
\begin{equation} a= a_0 + \lambda r \cos\delta .\label{general-sum rule} \end{equation}
The general class of phenomenologically viable sum rules are given in Table~\ref{all-sumrules-numbers}~\cite{Ballett:2013wya}. In this table, $m$ gives the order of the generator which controls the charge lepton mass matrix, $T^m_\alpha=1$, while $S_i$ is the generator of the von Dyck group that is identified with one of the generators of the Klein symmetry of the neutrino mass matrix (with the other Klein symmetry generator being unrelated to the von  Dyck group, as in semi-direct models). For example, for $A_4$ we identify $S_2\equiv S$, while for $S_4$ we identify $S_2\equiv S$ and $S_1\equiv SU$. Atmospheric sum rules may be tested at next generation neutrino facilities~\cite{Ballett:2013wya}.
The viability of such sum rules has also been discussed recently in \cite{Hall:2013yha}. 

\begin{table}
\centering
\begin{tabular}{c c c c c c}
$G$ & $m$ & $T_\alpha$,$S_i$ & $s$ & $a_0$ & $\lambda$\\
\hline\hline
\multirow{3}{*}{A$_4$} & $3$ & $T_e$,$S_2$ & $0.012$ & $0$ & $-0.5$ \\
& $3$ & $T_\mu$,$S_2$ & $0.012$ & $0$ & $-0.5$ \\
& $3$ & $T_\tau$,$S_2$ & $0.012$ & $0$ & $-0.5$ \\\hline
\multirow{3}{*}{S$_4$} & $3$ & $T_e$,$S_1$ & $-0.024$ & $0$ & $1$ \\
& $4$ & $T_\mu$,$S_2$ &$-0.124$ & $-0.167$ & $-0.408$ \\
& $4$ & $T_\tau$,$S_2$&$-0.124$ & $0.167$ & $-0.408$ \\\hline
\multirow{4}{*}{A$_5$}
& $5$ & $T_e$,$S_1$ &$-0.118$ & $0$ & $1.144$ \\
& $5$ & $T_e$,$S_2$ &$-0.079$ & $0$ & $-0.437$ \\
& $5$ & $T_\mu$,$S_2$ &$0.054$ & $0.067$ & $-0.532$ \\
& $5$ & $T_\tau$,$S_2$ &$0.054$ & $-0.067$ & $-0.532$ \\
\end{tabular}
\caption{\label{all-sumrules-numbers}The phenomenologically viable linearised sum rules of the form $a= a_0 + \lambda r \cos\delta $ (where $a,r$ are the atmospheric and reactor angle deviations from tri-bimaximal mixing and $\delta$ is the $CP$ violating oscillation phase) arising in the Hernandez-Smirnov framework for finite von Dyck groups. }
\end{table}

\subsubsection{\label{sec:patterns-5}Solar mixing sum rules}

The solar mixing sum rules arise from the non-diagonality of the charged lepton Yukawa matrix in the $T$ diagonal basis, corresponding to violation of $T$ generator symmetry. In many models the neutrino mixing matrix has a simple form $U_0$, where $ s_{23}^{\nu}= c_{23}^{\nu}= 1/\sqrt{2}$ and $ s_{13}^{\nu}=0$, while the charged lepton mixing matrix has a CKM-like structure, in the sense that $V_{e_L}$ is dominated by a 12-mixing, i.e.\ its elements $(V_{e_L})_{13}$, $(V_{e_L})_{23}$, $(V_{e_L})_{31}$, and $(V_{e_L})_{32}$ are very small compared to $(V_{e_L})_{12}$ and $(V_{e_L})_{21}$, where in practice we take them to be zero. In this case we are led to a solar sum rule~\cite{King:2005bj,Masina:2005hf,Antusch:2005kw} derived from $U_{\mathrm{}} = V_{e_L} U_0$, which takes the form,
\begin{eqnarray}
U_{\mathrm{ }} &=& \left(\begin{array}{ccc}
\!c^e_{12}& -s^e_{12}e^{-i\delta^e_{12}}&0\!\\
\!s^e_{12}e^{i\delta^e_{12}}&c^e_{12} &0\!\\
\!0&0&1\!
\end{array}
\right)\left( \begin{array}{ccc}
c^{\nu}_{12} & s^{\nu}_{12} & 0\\
 -\frac{s^{\nu}_{12}}{\sqrt{2}}  &  \frac{c^{\nu}_{12}}{\sqrt{2}} & \frac{1}{\sqrt{2}}  \\
\frac{s^{\nu}_{12}}{\sqrt{2}}  &  -\frac{c^{\nu}_{12}}{\sqrt{2}} & \frac{1}{\sqrt{2}} 
\end{array}
\right) 
= \left(\begin{array}{ccc}
\! \cdots & \ \ 
\! \cdots&
\! -\frac{s^e_{12}}{\sqrt{2}}e^{-i\delta^e_{12}} \\
\! \cdots
& \ \
\! \cdots
&
\! \frac{c^e_{12}}{\sqrt{2}}
\!\\
\!\frac{s^{\nu}_{12}}{\sqrt{2}} 
& \ \
\!-\frac{c^{\nu}_{12}}{\sqrt{2}}
&
\! \frac{1}{\sqrt{2}} 
\end{array}
\right) \! . ~~~~~~~
\label{Ucorr}
\end{eqnarray}

We hence obtain the sum rule~\cite{King:2005bj,Masina:2005hf,Antusch:2005kw},
\begin{equation}
\theta_{12}\approx \theta_{12}^{\nu}+ \theta_{13}\cos \delta .
\label{sumrule}
\end{equation}

Given the accurate determination of the reactor angle and the solar angle the sum rule in Eq.~\eqref{sumrule} yields a favoured range of $\cos \delta $ for each of the cases  $\theta_{12}^{\nu}=35.26^\circ, 45^\circ, 31.7^\circ, 36^\circ$ for the cases of TB, BM, GR, GR, namely $\cos \delta \approx -0.2, -1,0.2, -0.2,$ or $\cos \delta \approx -\lambda, -1, \lambda, -\lambda$,  respectively. For example, for TB neutrino mixing, the sum rule in Eq.~\eqref{sumrule} may be written compactly as,
\begin{equation}
s\approx r\,\cos \delta.
\label{sumrule2}
\end{equation}
In order to obtain $s\approx -\lambda^2$ from $r\approx \lambda$, we need to have $\cos \delta \approx -\lambda$.

This approach relies on a Cabibbo-sized charged lepton mixing angle $s^e_{12}\approx \lambda$ in order to account for the observed reactor angle, starting from one of the simple classic patterns of neutrino mixing. This is not straightforward to achieve in realistic models~\cite{King:2012vj}, which would typically prefer smaller charged lepton mixing angles $s^e_{12}\approx \lambda /3$. This suggests that the neutrino mixing angle $\theta_{13}^{\nu}$ is not zero, but has some non-zero value closer to the observed reactor angle. 
Such solar sum rules have been studied and refined recently in \cite{Marzocca:2013cr}.

\subsection{\label{sec:mass_rules}Mass sum rules}

Similar to mixing sum rules, sum rules for the light neutrino mass eigenvalues can drastically increase the testability of neutrino flavour models. Mass sum rules have already been used implicitly (i.e., without writing them down as such) several years ago (see, e.g., Refs.~\cite{Altarelli:2008bg,Hirsch:2008rp,Bazzocchi:2009da}), but the term \emph{mass sum rule} -- in contrast to mixing sum rules -- was only coined later on~\cite{Altarelli:2009kr,Chen:2009um,BarryRodejohann-Classification}. The first systematic study of a few sum rules has been provided in Ref.~\cite{Barry:2010zk}, followed by Ref.~\cite{Dorame:2011eb} in which the four types of sum rules which appear in the literature have been classified. Finally, the most complete study up to now, covering twelve different sum rules derived from more than 50 known models has been provided in 2013~\cite{King:2013psa}.

Indeed, when scanning through the literature available, it turns out that all known sum rules (or, at least, all that we are aware of) derived from concrete models have one of the following structures~\cite{Dorame:2011eb}:
\begin{eqnarray}
&A)&\chi\, \tilde m_2+\xi\, \tilde m_3=\tilde m_1, \nonumber \\
&B)&\frac{\chi }{\tilde m_2}+ \frac{\xi}{\tilde m_3}=\frac{1}{\tilde m_1},\nonumber\\
&C)&\chi\,\sqrt{\tilde m_2}+\xi\, \sqrt{\tilde m_3}=\sqrt{\tilde m_1}, \nonumber\\
&D)&\frac{\chi}{\sqrt{\tilde m_2}}+\frac{\xi }{\sqrt{\tilde m_3}}=\frac{1}{\sqrt{\tilde m_1}}\,,
\label{eq:SR_types}
\end{eqnarray}
where $\chi$ and $\xi$ are model-specific complex parameters and $\tilde m_i$ are the complex neutrino mass eigenvalues. This form is not surprising, when taking into account how sum rules arise. In general, if the light neutrino mass matrix $m^\nu_{LL}$ consists of a product of matrices and is proportional to a power $n$ of some (inverse) mass matrix $M$, where $M$ involves the two decisive flavon couplings and all other matrices in the product only consist of numbers (up to an overall scale), then the power $p$ in the sum rule will be given by $1/n$, $m^\nu_{LL} \propto M^n \Rightarrow p = \frac{1}{n}$~\cite{King:2013psa}. Thus, sum rules of type $A)$ can be expected to appear in models where the neutrino mass matrix $m^\nu_{LL}$ is generated by a dimension-five Weinberg operator~\cite{Weinberg:1979sa} or by a left-handed type~II seesaw matrix $M_L$~\cite{Magg:1980ut,Lazarides:1980nt}. Similarly, type $B)$ sum rules arise in case the right-handed neutrino mass matrix $M_R$ in a type~I seesaw~\cite{Minkowski:1977sc,Ramond:1979py,Yanagida:1979as,GellMann:1980vs,Glashow:1979nm,Mohapatra:1979ia} or scotogenic~\cite{Ma:2006km} framework or the fermion-triplet mass matrix $M_\Sigma$ in a type~III seesaw framework~\cite{Foot:1988aq,Ma:2002pf} are the decisive matrices. Finally, type $C)$ sum rules result from the Dirac neutrino mass matrix $m^D$ in type~I seesaw or from the scotogenic Dirac Yukawa coupling matrix $h_\nu$ and the only known case of a model leading to a type $D)$ sum rule uses the matrix $M_{RS}$ mixing the right-handed neutrinos with the additional singlet neutrinos in a setting based on the inverse seesaw mechanism~\cite{Mohapatra:1986bd,GonzalezGarcia:1988rw}.

This classification has been generalised in Ref.~\cite{King:2013psa}, where it was shown that the most general neutrino mass sum rule of a power $p$ looks like:
\begin{equation}
 A_1 \tilde m_1^p e^{i \chi_1} + A_2 \tilde m_2^p e^{i \chi_2} + A_3 \tilde m_3^p e^{i \chi_3} = 0,
 \label{eq:general_rule}
\end{equation}
with $A_k > 0$, since any possible phase of the complex coefficients is pulled into the factors $e^{i \chi_k}$. Using $\tilde m_k = m_k e^{i \phi_k}$ (where $m_k \geq 0$ and $\phi_k$ are Majorana phases) and defining $B_k \equiv A_k/A_1$ as well as $\Delta \chi_{k1} \equiv \chi_k - \chi_1$ and $\alpha_{k1} \equiv \phi_k - \phi_1$, any sum rule defines a set of parameters,
\begin{equation}
 m_1^p + B_2 \left( m_2 e^{i \alpha_{21}} \right)^p e^{i \Delta \chi_{21}} + B_3 \left( m_3 e^{i \alpha_{31}} \right)^p e^{i \Delta \chi_{31}} = 0\ \ \ \Rightarrow \ \ \ (p, B_2, B_3, \Delta \chi_{21}, \Delta \chi_{31}),
 \label{eq:rule_params}
\end{equation}
which can be used to classify any neutrino mass sum rule. Note that, since Eq.~\eqref{eq:general_rule} is a \emph{complex} equation, any sum rule will lead to \emph{two} conditions constraining the absolute neutrino mass scale and one of the physical Majorana phases $\alpha_{31,21}$. These conditions will then constrain the general effective neutrino mass $m_{ee}$, as measured in experiments on neutrinoless double beta decay ($0\nu\beta\beta$), whose general formula is in the PDG parametrisation given by~\cite{Beringer:1900zz}:
\begin{equation}
 |m_{ee}|_{\rm PDG} = |m_1 c_{12}^2 c_{13}^2 + m_2 s_{12}^2 c_{13}^2 e^{i \alpha_{21}} + m_3 s_{13}^2 e^{i (\alpha_{31} - 2\delta)}|.
 \label{eq:mee_PDG}
\end{equation}

The make things more concrete, it is useful to use the geometrical interpretation of the effective mass~\cite{Lindner:2005kr} and of sum rules~\cite{Barry:2010zk,Dorame:2011eb,King:2013psa}. To be concrete, we illustrate the procedure for two sum rules which appear in realistic models, the first rule being based on an $A_4$~\cite{Barry:2010zk}, $S_4$~\cite{Bazzocchi:2009da,Ding:2010pc}, or $A_5$~\cite{Ding:2011cm,Cooper:2012bd} symmetries and the second one arising from $A_4$~\cite{Barry:2010zk,Ma:2005sha,Ma:2006wm,Honda:2008rs,Brahmachari:2008fn}, $A_5$~\cite{Everett:2008et}, $S_4$~\cite{Bazzocchi:2009pv,Bazzocchi:2009da}, or $\Delta(54)$~\cite{Boucenna:2012qb} groups:
\begin{equation}
 \text{Rule~1: $\frac{1}{\tilde m_1} + \frac{1}{\tilde m_2} = \frac{1}{\tilde m_3}$}\ \ \ , \ \ \ \text{Rule~2: $\tilde m_1 + \tilde m_2 = \tilde m_3$}.
 \label{eq:Example_Rules}
\end{equation}
In terms of our general parameters $(p, B_2, B_3, \Delta \chi_{21}, \Delta \chi_{31})$, cf.\ Eq.~\eqref{eq:rule_params}, these sum rules read:
\begin{equation}
 \text{Rule~1: $(-1, 1, 1, 0, \pi)$}\ \ \ , \ \ \ \text{Rule~2: $(1, 1, 1, 0, \pi)$}.
 \label{eq:Example_Params}
\end{equation}

\begin{center}
\begin{figure}[ht]
\begin{tabular}{lcr}
\includegraphics[width=8cm]{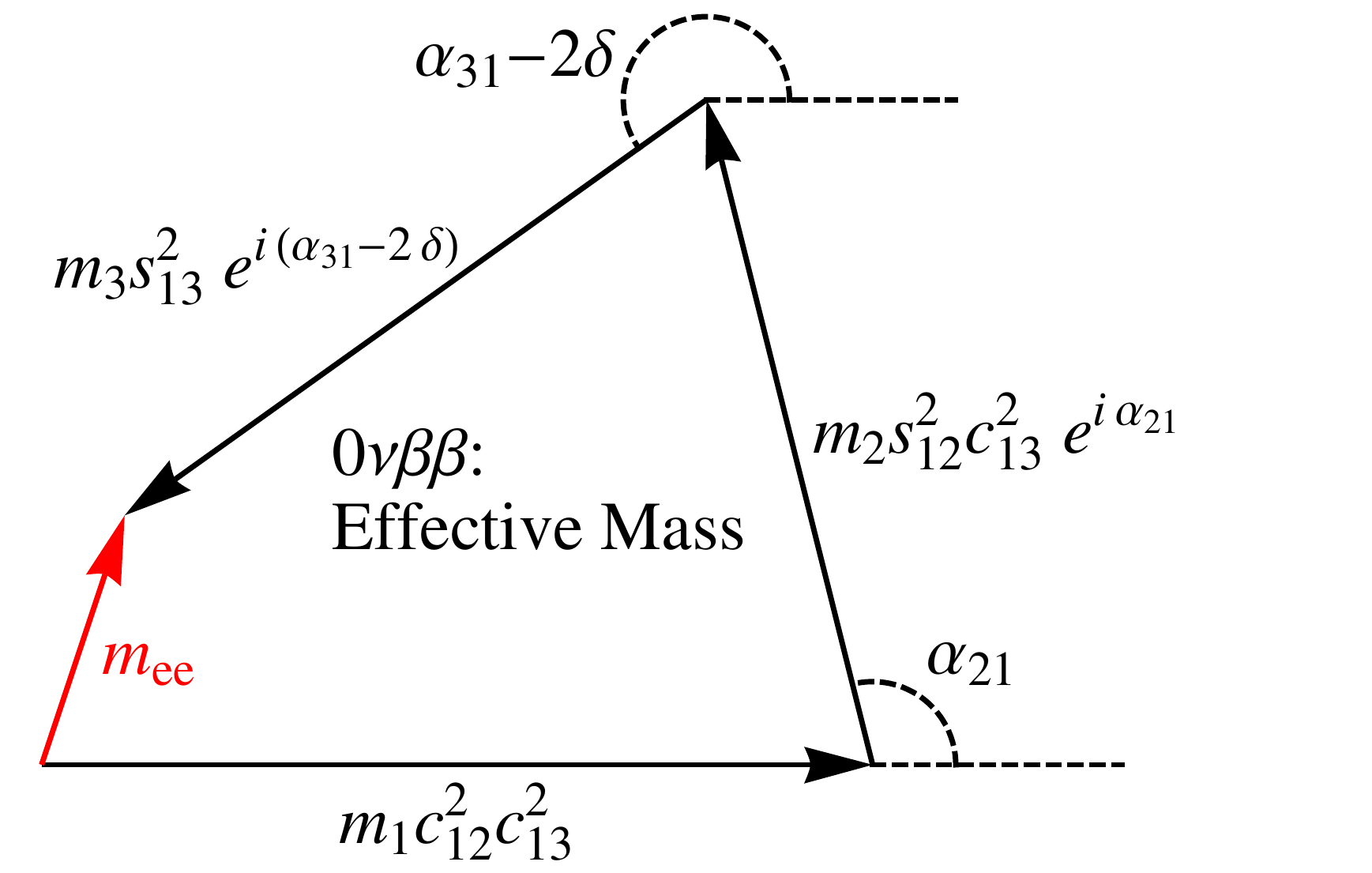}
\end{tabular}
\caption{\label{fig:mee}Geometrical illustration of the effective mass $m_{ee}$ as sum of three 2-dimensional vectors..}
\end{figure}
\end{center}

\begin{center}
\begin{figure}[ht]
\begin{tabular}{lr}
\includegraphics[width=8cm]{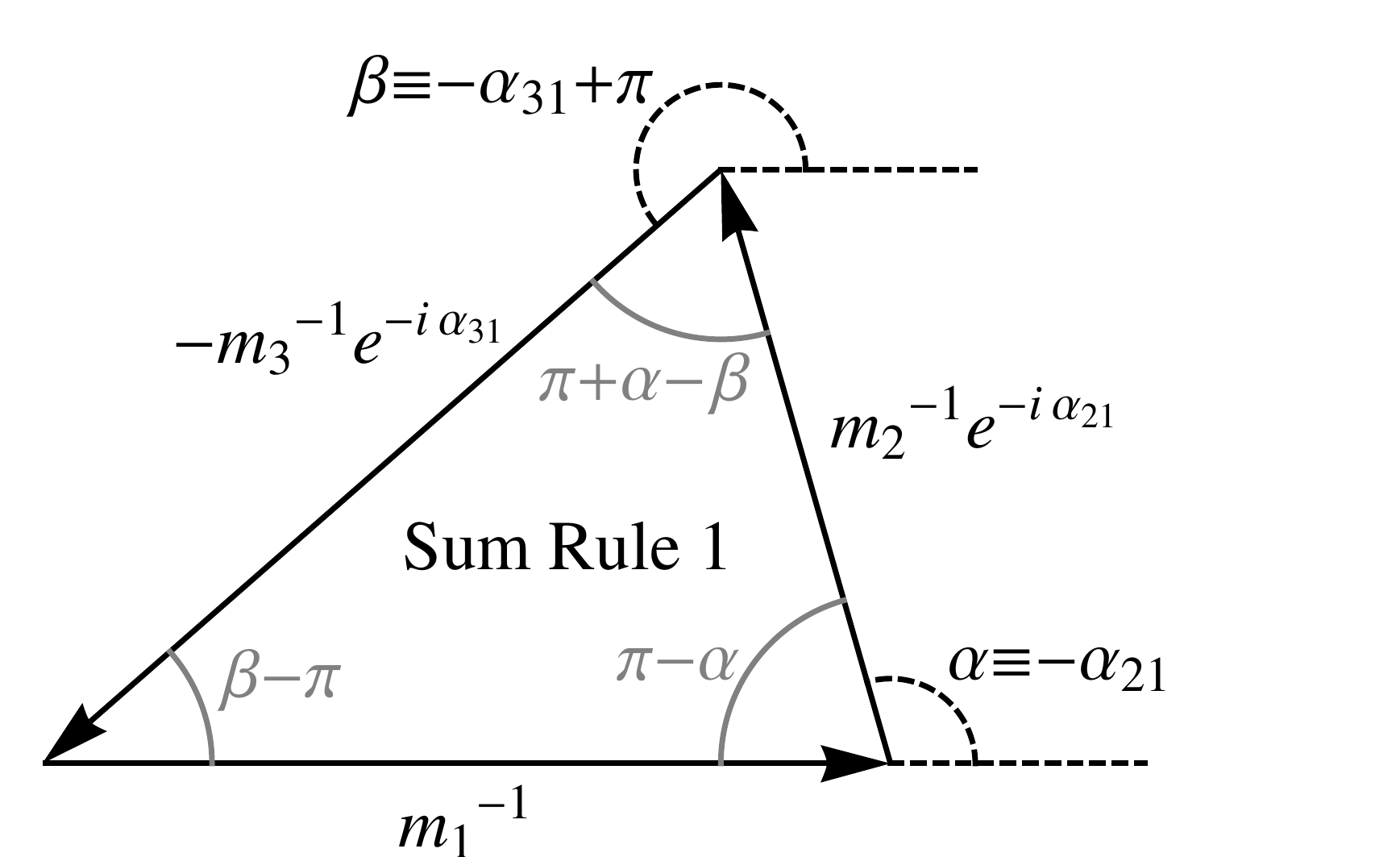} & \includegraphics[width=8cm]{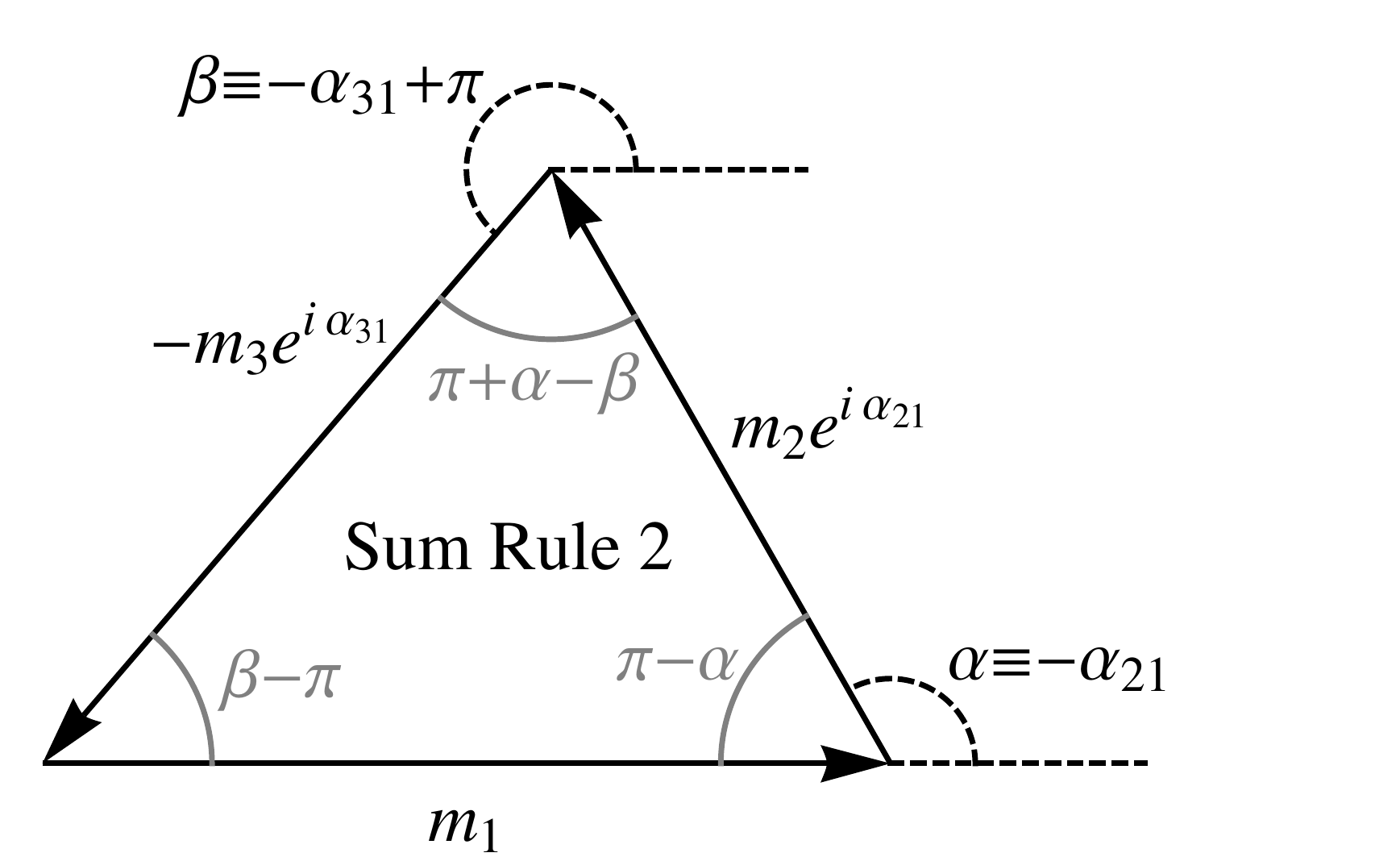}
\end{tabular}
\caption{\label{fig:triangles}Geometrical illustrations of the two example sum rules.}
\end{figure}
\end{center}

The decisive question from a phenomenological point of view is how these sum rules restrict the effective mass $|m_{ee}|$. To do this, it is helpful to recall that the effective mass parameter $|m_{ee}|$, as written explicitly in Eq.~\eqref{eq:mee_PDG}, can be thought of as the absolute value of a vector resulting from the sum of three individual ones, by simply interpreting the complex numbers involved as vectors. This is illustrated in  Fig.~\ref{fig:mee}. Similarly, the two sum rules in Eq.~\eqref{eq:Example_Rules} can be thought of as sums of 2-dimensional vectors, cf.\ Fig.~\ref{fig:triangles}. However, the difference is that the geometrical pictures of sum rules will always be triangles, due to the right-hand sides of Eq.~\eqref{eq:Example_Rules} being zero. This simple geometrical interpretation of sum rules was first proposed in Ref.~\cite{Barry:2010zk}, and then further elaborated on in Refs.~\cite{Dorame:2011eb,King:2013psa}.

\begin{center}
\begin{figure}[ht]
\begin{tabular}{lcr}
\includegraphics[width=7cm]{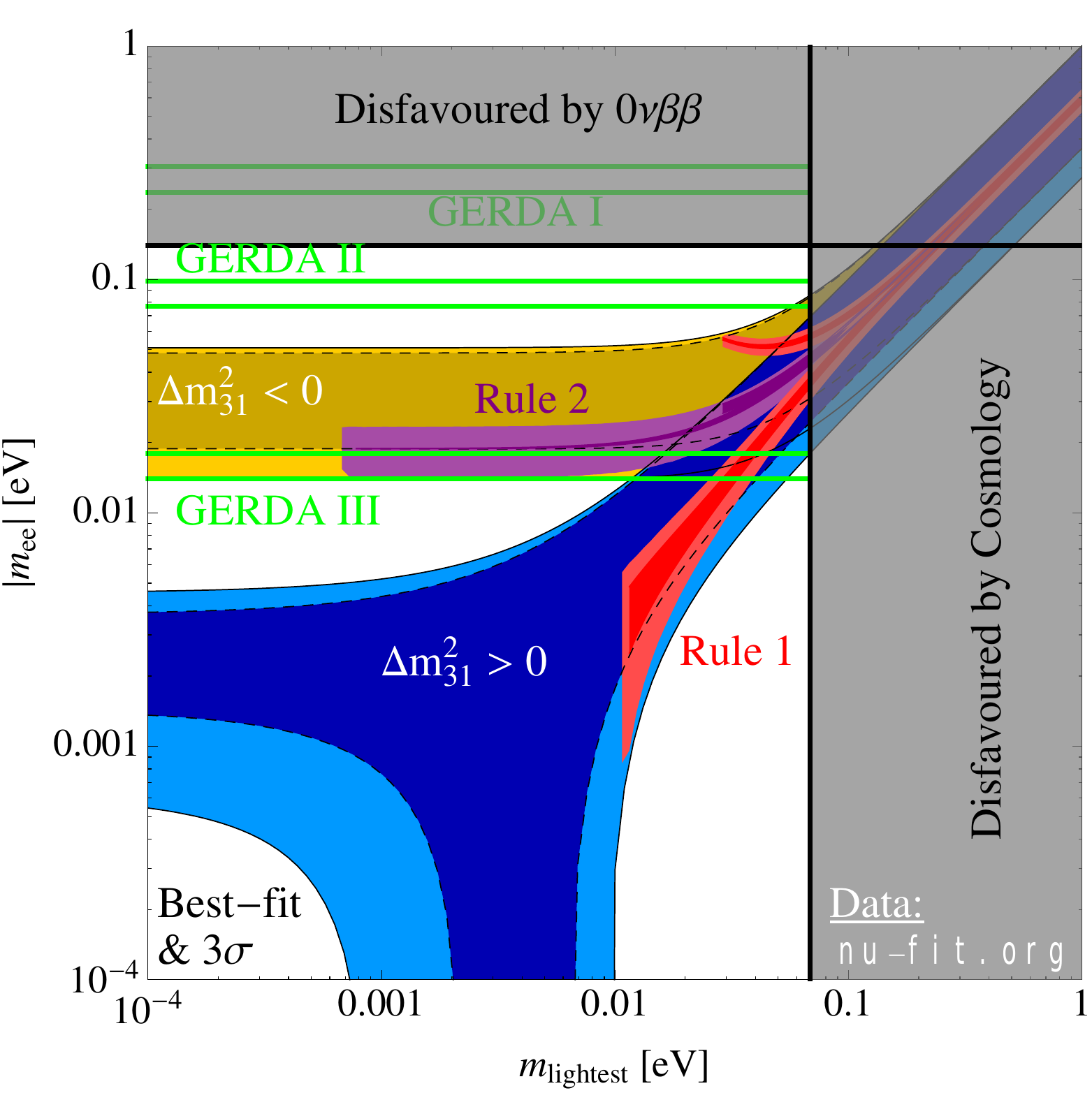} & \includegraphics[width=9cm]{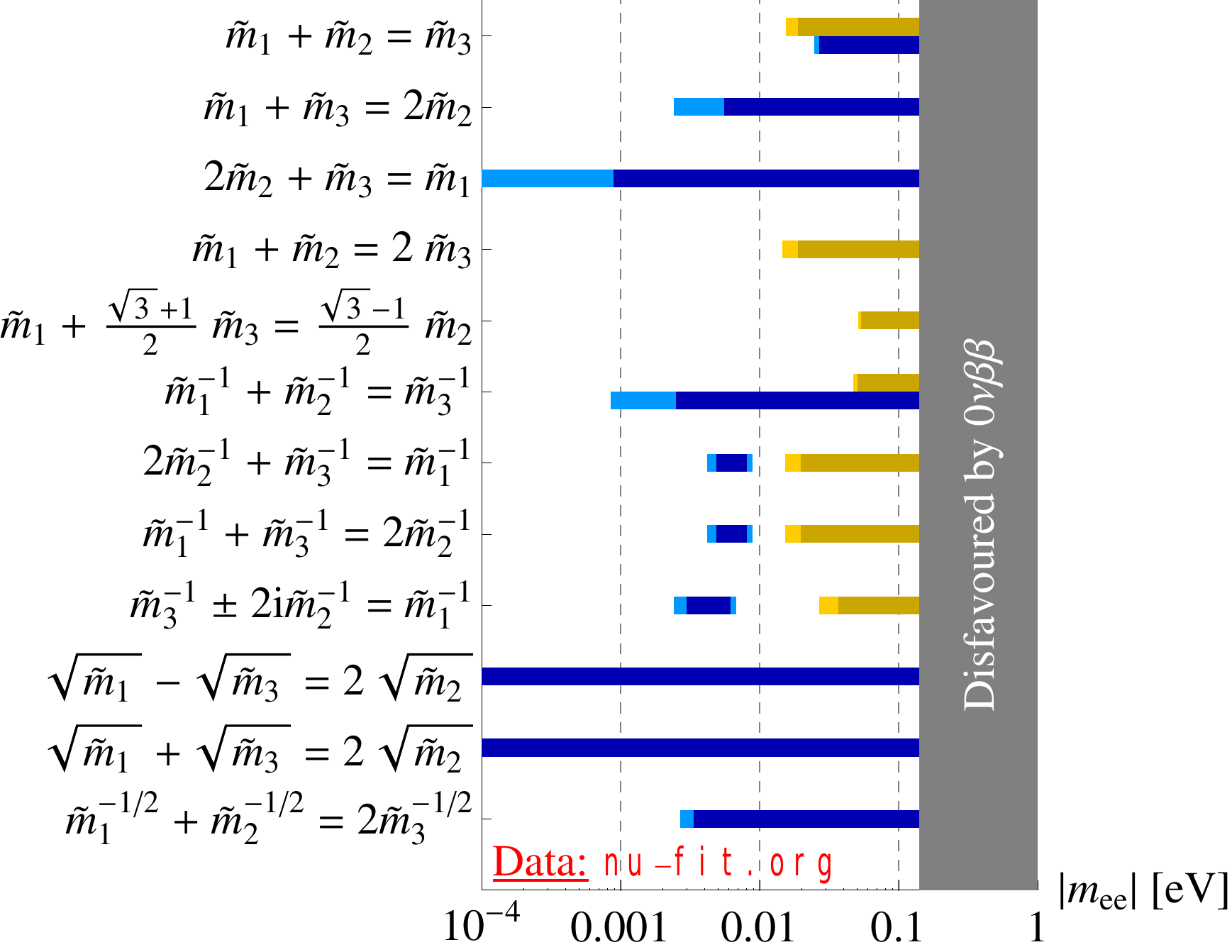}
\end{tabular}
\caption{\label{fig:mee_rules} \emph{Left panel:} Restrictions imposed on the allowed regions of $|m_{ee}|$ by two example sum rules (Rule~1: $\tilde m_1^{-1} + \tilde m_2^{-1} = \tilde m_3^{-1}$, Rule~2: $\tilde m_1 + \tilde m_2 = \tilde m_3$). The GERDA regions are displayed as examples for realistic experimentally accessible ranges. (Plot similar to Fig.~1 in Ref.~\cite{King:2013psa}.) \emph{Right panel:} Derived ranges for the $|m_{ee}|$ from 12~different sum rules, covering more than 50~models in the literature. (Plot similar to Fig.~20 in Ref.~\cite{King:2013psa}.)}
\end{figure}
\end{center}

The restrictions imposed by our example sum rules on $|m_{ee}|$ are displayed in the left panel of Fig.~\ref{fig:mee_rules}. In this plot, we can first of all see the general allowed regions for the effective mass if all known neutrino oscillation parameters are set to their best-fit values (dark blue: NO, dark yellow: IO) or varied within their $3\sigma$ ranges (blue: NO, yellow: IO). In the former case, the thickness of the bands entirely comes from the variation of the Majorana phase $\alpha_{21}$ and the combination $(\alpha_{31} - 2\delta)$ of the second Majorana phase and of the Dirac $CP$ phase, cf.\ Eq.~\eqref{eq:mee_PDG}.\footnote{Note that, typically, the second phase $(\alpha_{31} - 2\delta)$ is re-defined to a new ``Majorana'' phase $\tilde \alpha_{31}$~\cite{Lindner:2005kr,Merle:2006du}. However, as was pointed out in Ref.~\cite{King:2013psa}, this would simultaneously lead to a re-definition of the sum rule under consideration, which makes this step redundant when dealing with sum rules.} When taking into account the sum rules from Eq.~\eqref{eq:Example_Rules}, this general allowed region is restricted further. Explicitly, the conditions imposed by the real and imaginary parts of the two rules read:
\begin{eqnarray}
&\text{Rule~1}:& \frac{1}{m_1} + \frac{\cos \alpha_{21}}{m_2} = \frac{\cos \alpha_{31}}{m_3}\ \ \ \text{\&}\ \ \ \frac{-i \sin \alpha_{21}}{m_2} = \frac{-i \sin \alpha_{31}}{m_3},\nonumber\\
&\text{Rule~2}:& m_1 + m_2 \cos \alpha_{21} = m_3 \cos \alpha_{31}\ \ \ \text{\&}\ \ \ i m_2 \sin \alpha_{21} = i m_3 \sin \alpha_{31}.
 \label{eq:rule_conditions}
\end{eqnarray}
Applying these conditions leads to strong restrictions on $|m_{ee}|$, as can again be seen from the left panel of Fig.~\ref{fig:mee_rules}: the red region results from the restrictions imposed by Rule~1 and the purple region results from Rule~2. As can be seen, both restricted regions stay within the general regions, as they should. Furthermore, both rules allow for both mass orderings. Nevertheless, they impose severe restrictions on the allowed parameter space and, in particular, they restrict the lightest neutrino mass to be bound from below. As an example, the limit (expected sensitivities) for phase~I (phases~II and~III) of the GERDA experiment are indicated by the green lines, where the distance between each pair of horizontal lines reflects the intrinsic nuclear physics uncertainties.

Trying to obtain an analytical understanding, e.g.\ the border of validity Rule~2 for NO (where $m_1 < m_2 < m_3$) can be found using the triangle inequality~\cite{Barry:2010zk}, $|m_1| + |m_2 e^{i \alpha_{21}}| > |m_3 e^{i \alpha_{31}}|$. Equating both sides and using $\Delta m_{\rm sol}^2 \ll \Delta m_{\rm atm}^2$, it is easy to show that $m_{\rm lower}^{\rm NO} \approx \sqrt{\Delta m_{\rm atm}^2/3} \simeq 0.03$~eV, in agreement with the plots (note that the decisive point is the ``edge'' where the purple region exits the blue curve, as this is the cutoff for the NO part of the rule). Inserting the resulting condition, $2 m_{1,2} \simeq m_3 \simeq 2\sqrt{\Delta m_{\rm atm}^2/3}$, into the conditions in Eq.~\eqref{eq:rule_conditions}, they imply that $\alpha_{21,31} \approx 0$. This yields $|m_{ee}| \gtrsim \sqrt{\Delta m_{\rm atm}^2/3} (1 - 3 s_{13}^2)$, which indeed corresponds to the lower bound of the NO region. Many more details and extensive analytical considerations can be found in Ref.~\cite{King:2013psa}.

The most extensive collection available of the restrictions on $|m_{ee}|$ by known neutrino mass sum rules is depicted on the right panel of Fig.~\ref{fig:mee_rules}. This plot covers twelve sum rules which are derived from more than 50~known models from the literature, cf.\ Tab.~\ref{tab:models}, which are based on different flavour symmetries and neutrino mass mechanisms. The figure clearly shows the potential added by sum rules, in particular if external information is added. For example, if by some complementary experiment (different from $0\nu\beta\beta$) we would know that the light neutrino mass ordering was inverted, we could immediately exclude five rules completely (rules 2, 3, 10, 11, and 12 in the right panel of Fig.~\ref{fig:mee_rules}) and five other rules (1, 6 -- 9) would only be narrowed down to their IO (yellow) regions. Then, despite the intrinsic uncertainties imposed by the unknown nuclear matrix elements, all of the remaining rules (1, 4 -- 9) could then be tested with the next (or next-to-next) generation of $0\nu\beta\beta$ experiments, such as GERDA phase~III~\cite{Abt:2004yk,JanicskoCsathy:2009zz,Agostini:2013mzu}.

\begin{table}[t]
\centering
\begin{tabular}{|c|c|c|c|}
\hline
Sum Rule & Group & Seesaw Type& Matrix\\\hline
$\tilde{m}_1+\tilde{m}_2=\tilde{m}_3$ & $A_4$\cite{Barry:2010zk}(\hspace{-0.01cm}\cite{Honda:2008rs,Brahmachari:2008fn,Ding:2010pc,Ma:2005sha,Ma:2006wm}); $S_4$(\hspace{-0.01cm}\cite{Bazzocchi:2009pv}); $A_5$\cite{Everett:2008et}$^*$ &Weinberg& $m^\nu_{LL}$\\
$\tilde{m}_1+\tilde{m}_2=\tilde{m}_3$  &$\Delta(54)$\cite{Boucenna:2012qb}; $S_4$(\hspace{-0.01cm}\cite{Bazzocchi:2009da}) &Type II&$M_L$\\\hline
$\tilde{m}_1+2\tilde{m}_2 = \tilde{m}_3$ & $S_4$\cite{Mohapatra:2012tb}&Type II&$M_L$\\\hline
$2\tilde{m}_2+\tilde{m}_3=\tilde{m}_1$ & $A_4$\cite{Barry:2010zk,Chen:2009um}(\hspace{-0.01cm}\cite{Altarelli:2005yx,Altarelli:2005yp,Altarelli:2006kg,Ma:2006vq,Bazzocchi:2007na,Bazzocchi:2007au,Honda:2008rs,Brahmachari:2008fn,Ma:2005sha,Ma:2006wm,Lin:2008aj,Ma:2009wi,Ciafaloni:2009qs}) &Weinberg&$m^\nu_{LL}$\\
  & $S_4$(\hspace{-0.01cm}\cite{Bazzocchi:2008ej,Feruglio:2013hia})$^{\dagger}$; $T'$\cite{Chen:2009gf,Chen:2009gy}(\hspace{-0.01cm}\cite{Chen:2007afa,Feruglio:2007uu,Ding:2008rj,Merlo:2011hw}); $T_7$(\hspace{-0.01cm}\cite{Luhn:2012bc}) & & \\
$2\tilde{m}_2+\tilde{m}_3=\tilde{m}_1$ & $A_4$(\hspace{-0.01cm}\cite{Fukuyama:2010mz}) & Type II&$M_L$\\\hline
$\tilde{m}_1+\tilde{m}_2=2\tilde{m}_3$ & $S_4$\cite{Ding:2013eca}$^\ddagger$& Dirac$^\ddagger$ &$m^D$\\
$\tilde{m}_1+\tilde{m}_2=2\tilde{m}_3$ & $L_e - L_\mu - L_\tau$(\hspace{-0.01cm}\cite{Lindner:2010wr}) &Type II&$M_L$\\\hline
$\tilde{m}_1 + \frac{\sqrt{3}+1}{2} \tilde{m}_3 = \frac{\sqrt{3}-1}{2} \tilde{m}_2$ & $A_5'$(\hspace{-0.01cm}\cite{Hashimoto:2011tn}) &Weinberg&$m^\nu_{LL}$\\\hline
$\tilde{m}_1^{-1}+\tilde{m}_2^{-1}=\tilde{m}_3^{-1}$ & $A_4$\cite{Barry:2010zk}; $S_4$(\hspace{-0.01cm}\cite{Bazzocchi:2009da,Ding:2010pc}); $A_5$\cite{Ding:2011cm,Cooper:2012bd}&Type I&$M_R$\\
$\tilde{m}_1^{-1}+\tilde{m}_2^{-1}=\tilde{m}_3^{-1}$&$S_4$(\hspace{-0.01cm}\cite{Bazzocchi:2009da})&Type III&$M_{\Sigma}$\\\hline
$2\tilde{m}_2^{-1}+\tilde{m}_3^{-1}=\tilde{m}_1^{-1}$ &$A_4$\cite{Barry:2010zk,Altarelli:2008bg,Altarelli:2009kr,Hagedorn:2009jy,Chen:2009um}(\hspace{-0.01cm}\cite{Altarelli:2005yx,Morisi:2007ft,Adhikary:2008au,Csaki:2008qq,Burrows:2009pi,Ding:2009gh,Mitra:2009jj,Lin:2009bw,delAguila:2010vg,Burrows:2010wz}); $T'$\cite{Chen:2009gy}&Type I&$M_R$\\
$\tilde{m}_1^{-1}+\tilde{m}_3^{-1}=2\tilde{m}_2^{-1}$ & $A_4$(\hspace{-0.01cm}\cite{He:2006dk,Berger:2009tt,Kadosh:2010rm}); $T'$\cite{Lavoura:2012cv}&Type I&$M_R$\\\hline
$\tilde{m}_3^{-1} \pm 2i\tilde{m}_2^{-1}=\tilde{m}_1^{-1}$ & $\Delta(96)$\cite{King:2012in}& Type I&$M_R$\\\hline
$\tilde{m}_1^{1/2}- \tilde{m}_3^{1/2}=2\tilde{m}_2^{1/2}$ & $A_4$(\hspace{-0.01cm}\cite{Hirsch:2008rp})&Type I &$m^D$\\
$\tilde{m}_1^{1/2}+\tilde{m}_3^{1/2}=2\tilde{m}_2^{1/2}$ & $A_4$(\hspace{-0.01cm}\cite{Adulpravitchai:2009gi})&Scotogenic&$h_{\nu}$\\\hline
$\tilde{m}_1^{-1/2}+\tilde{m}_2^{-1/2}=2\tilde{m}_3^{-1/2}$ & $S_4$\cite{Dorame:2012zv}&Inverse&$M_{RS}$\\\hline
\end{tabular}
\caption{\label{tab:models}Sample of sum rules found in the literature, as presented in Ref.~\cite{King:2013psa}. Sum rules grouped together give identical predictions. References in parantheses $(...)$ do not give the sum rules explicitly. $^*$Sum rule only used as a consistency relation. $^\dagger$In Ref.~\cite{Feruglio:2013hia} the Majorana phases were predicted so that the restriction by that concrete model is stronger than the sum rule only. $^\ddagger$Even though this reference predicts a Dirac sum rule.}
\end{table}

This discussion illustrates an important observation: with a bit of luck, we could within the next few years considerably increase our knowledge on the neutrino mass and flavour sectors. This would enable us to probe a big bunch of models and maybe even to select a few prime candidate models which are closest to observations. In the next section, we will shortly discuss these possibilities.

\subsection{\label{sec:luck}With a bit of luck, we could also measure within the coming years...}

We would to conclude this section by pointing out some possibilities for measurements that could realistically be done in the near future -- at least if Nature is kind to us. Currently, it seems that we will sooner or later determine the light neutrino mass ordering, but also a determination of the $\theta_{23}$ octant (in case the true value of $\theta_{23}$ is non-maximal) and even of the Dirac $CP$ phase $\delta$ (at least for a fortunate value) could in principle occur. Thus, with a bit of luck, it may even be that we have a complete picture of neutrino oscillation physics within roughly a decade from now, which would certainly have a big impact on model distinction. Furthermore, we could realistically know within a few years from now whether light (eV scale) sterile neutrinos do exist, or not, and a positive detection would need to be reflected by model building considerations.

\subsubsection{The neutrino mass squared ordering}

The mass ordering
among the different generations is a piece of information which we know for all known fermions except for neutrinos\footnote{For the neutrinos this is often referred to as the ``neutrino mass hierarchy'' or ``hierarchy'', but such a nomenclature
is imprecise and we would advocate the use of the terminology 
``neutrino mass squared ordering'' which is independent of the question
of whether the neutrino masses are hierarchical or quasi-degenerate.}.
Thus, it would be a very valuable handle to probe models which forbid one of the two orderings. The question of whether light neutrino squared masses obey normal ordering 
(NO) or inverted ordering (IO) essentially means that we would like to know the sign of the atmospheric mass square difference, due to $\Delta m_{\rm atm}^2 \simeq \Delta m_{31}^2 \equiv m_3^2 - m_1^2$.

The conventional method to determine the sign is to make use of matter effects in neutrino oscillations for long enough baselines, i.e., ideally more than 1000~km. Indeed, by expanding the oscillation probabilities in the small parameter $\alpha \equiv |\Delta m_{21}^2|/|\Delta m_{31}^2|$, the oscillation probability from muon to electron neutrinos is (for the case of a constant matter potenrial $V$) to $\mathcal{O}(\alpha)$ given by~\cite{Akhmedov:2004ny}:
\begin{eqnarray}
 P(\nu_\mu \to \nu_e) &=& 4 s_{13}^2 s_{23}^2 \frac{\sin^2 [(A-1) \Delta]}{(A-1)^2}\nonumber\\
 && + 2\alpha s_{13} \sin(2\theta_{12}) \sin(2\theta_{23}) \frac{\sin (A \Delta)}{A} \frac{\sin [(A-1) \Delta]}{A-1} \cos \Delta \cos \delta \nonumber\\
 && - 2\alpha s_{13} \sin(2\theta_{12}) \sin(2\theta_{23}) \frac{\sin (A \Delta)}{A} \frac{\sin [(A-1) \Delta]}{A-1} \sin \Delta \sin \delta,
 \label{eq:me-matter}
\end{eqnarray}
where $\Delta \equiv \Delta m^2_{31} L / (4 E)$ and $A \equiv V L / (2 \Delta)$, where $L$ is the baseline of the experiment and $E$ is the energy of the neutrino, such that ${\rm sign}(\Delta) = {\rm sign}(\Delta m_{31}^2)$. When using antineutrinos instead, the corresponding probability can be determined by $P(\bar \nu_\mu \to \bar \nu_e) = P(\nu_\mu \to \nu_e)|_{\delta \to - \delta, V \to -V}$. It is visible that the third term in Eq.~\eqref{eq:me-matter} is the only one sensitive to ${\rm sign}(\Delta)$, but the same difference could be mimicked by a transformation of the $CP$ phase, $\delta \to 2\pi - \delta$. However, when comparing neutrinos ($A>0$) with antineutrinos ($A<0$), then this degeneracy is lifted.

Experimental approaches to the determination of the mass ordering include, e.g., the use of wide band super beam~\cite{Barger:2006vy,Barger:2007jq}. Alternatively, one could try to resolve the atmospheric resonance with liquid argon detectors~\cite{Barger:2012fx} or iron calorimeters~\cite{Samanta:2006sj,Ghosh:2012px}. By studying several oscillation peaks one could also have a chance to determine the mass ordering with shorter baselines~\cite{Li:2013zyd}. Furthermore, several astrophysical probes are possible: one could get some information by studying the CMB polarisation in combination with the 21~cm line~\cite{Oyama:2012tq}, by combining the information from CMB measurements with a limit or a signal from neutrinoless double beta decay~\cite{Maneschg:2008sf}, or by a precision upgrade of the IceCube experiment called PINGU~\cite{Winter:2013ema}.

With many experiments on the way, it is probably a far statement that we will be able to determine the neutrino mass ordering within a few years. However it is remarkable that, if $\theta_{23}$ is in the right octant, PINGU along could probe the whole parameter space by 2020; furthermore, when combining it with complementary data, we can be practically certain to know the mass ordering by then~\cite{Winter:2013ema}.

\subsubsection{The octant of $\theta_{23}$}

Another information which will be valuable for model distinction is about the octant of $\theta_{23}$, i.e., although we know that $\theta_{23} \approx \pi/4$ we could ask whether it is in fact smaller or larger than $\pi/4$. Reparametrising $\theta_{23} = \pi/4 + \epsilon$, it is easy to see that $\sin (2 \theta_{23})$ [and thus the second and third lines in Eq.~\eqref{eq:me-matter}] are sensitive to ${\rm sign}(\epsilon)$ already at $\mathcal{O}(\epsilon)$, while $\sin \theta_{23}$ is not. Interestingly, the decisive terms in Eq.~\eqref{eq:me-matter} also depend on $s_{13}$, and thus they could be larger than previously expected now that $\theta_{13}$ has been measured to be relatively large~\cite{An:2012eh,Ahn:2012nd,Abe:2011fz,Abe:2012tg,Abe:2013nka}.

Two of the main experiments which could be useful in this respect are T2K and NO$\nu$A. T2K~\cite{Abe:2011ks} uses a muon beam line from the J-PARC accelerator facility in Japan, which also contains a near detector complex in a distance of 280~m to the neutrino production target. As far detector, the Super-Kamiokande water \v{C}erenkov detector in a distance of 295~km is used. The far detector is slightly off-axis compared to the neutrino beam, in order to increase the sensitiviy on $\theta_{23}$~\cite{Abe:2013fuq}. NO$\nu$A~\cite{Ayres:2004js}, in turn, uses the NuMI $\nu_\mu$ beam from Fermilab, again off-axis, along with a $300$~t liquid scintillator near detector and a similar far detector at a distance of 810~km~\cite{Bian:2013saa}. Both experiments are running and T2K has already delivered important results (including a 7.3$\sigma$ discovery of a non-zero mixing angle $\theta_{13}$~\cite{Abe:2013nka}). NO$\nu$A, in turn, is expected to yield first results in 2014.

In what regards the octant of $\theta_{23}$, the current T2K results still allow for a maximal angle~\cite{Abe:2013fuq}. Indeed, even for deviations as large as $\epsilon \approx 0.1$, both experiments alone will presumably not be able to resolve the octant alone~\cite{Huber:2009cw}. However, when combined with reactor data, as will be done in practice, we are very likely to determine the sign of $\epsilon$ and by this the true octant with a good precision in the future~\cite{Huber:2009cw,T2K_NExT}.

\subsubsection{The Dirac $\boldsymbol{CP}$ phase $\delta$}

Finally, even the Dirac $CP$ phase $\delta$ could be constrained or even measured in the near future, at least in case it has fortunate values. Indeed, already the newest T2K results~\cite{Abe:2013nka} favour $\delta$ around $3\pi/2 = 270^\circ$ and can exclude part of the parameter space of $\delta$, at least when combined with data from reactor experiments. The regions reported are $[0.19\pi, 0.80\pi]$ ($[0, 1.03\pi]$ and $[1.96\pi,2\pi]$) for NO (IO), excluded at $90\%$~C.L.\ in both cases. However, this exclusion is still relatively mild. Formulated in a positive way, if the $CP$ phase $\delta$ is close to $3\pi/2 = 270^\circ$ then an increased amount of data could actually yield a measurement at some point. After the successful measurement of $\theta_{13}$, the design of T2K which had been initially aimed at measuring an angle $\theta_{13}$ close to zero can be changed to increase the chances for measuring $\delta$~\cite{T2K_NExT}. Furthermore, a boost could be expected if Super-Kamikande was upgraded to Hyper-Kamiokande~\cite{Abe:2011ts}.

\subsubsection{\label{sec:Light_Steriles}Light sterile neutrinos}

Even though we have not yet mentioned them very prominently, light sterile neutrinos have attracted the attention of the field in the last few years because of both, keV sterile neutrinos being candidates for warm Dark Matter~\cite{Dodelson:1993je,Shi:1998km,Bezrukov:2009th,Nemevsek:2012cd,Shaposhnikov:2006xi,Bezrukov:2009yw,Kusenko:2006rh,Petraki:2007gq,Merle:2013wta}\footnote{See Refs.~\cite{Asaka:2005an,Asaka:2005pn} for some more details on the minimal setting, the so-called $\nu$MSM.} and eV sterile neutrinos being indicated by several experiments~\cite{Abazajian:2012ys}. A concise review of the latter topic has been provided recently~\cite{Palazzo:2013me}, which we will partially follow here to briefly outline the current state and possible future developments.

Already several years ago, the LSND accelerator neutrino experiment~\cite{Athanassopoulos:1996jb} observed a clear excess of \emph{appearance} in the $\bar{\nu}_\mu \to \bar{\nu}_e$ channel~\cite{Aguilar:2001ty}. While at that time LSND may have been viewed as outlier~\cite{Slansky:1997bm,Louis:1997bs}, in particular because the (similar) KARMEN experiment could not observe any excess~\cite{Armbruster:2002mp,Church:2002tc}, it later turned out that the dedicated test experiment MiniBooNE unexpectedly confirmed the LSND claim instead of refuting it~\cite{Aguilar-Arevalo:2013pmq}. It even added an additional excess in the $\nu_\mu \to \nu_e$ channel. Due to 3-flavour effects from $\theta_{13} \neq 0$ not playing a very prominent role for high neutrino energies, a big part of the corresponding parameter space could be tested by the ICARUS experiment~\cite{Antonello:2012pq}, but even when putting all the information together, a region around $(\Delta m^2, \sin^2 (2 \theta)) \sim (0.5~{\rm eV}^2, 5\cdot 10^{-3})$ survived~\cite{Palazzo:2013me}.

On the reactor neutrino side, refined calculations of the spectrum of antineutrinos emitted by nuclear power reactors have resulted in an expectation for the flux that is more than $3\%$ higher than what is observed~\cite{Mueller:2011nm,Huber:2011wv}. This suggests event rates which should be about $6\%$ higher than expected and thus, if taken seriously, translates into the observation of a clear deficit ($>3\sigma$) of electron-antineutrinos in experiments with very short baselines (at about 100~m distance from the reactor core or closer)~\cite{Palazzo:2013me}. This observation is by now known as the so-called \emph{reactor anomaly}. If physical, this deficit seems hard to explain in terms of anything else than oscillation into sterile neutrinos, since other possibilities such as decays or decoherence would also significantly affect the spectrum farther away from the reactor (which is not observed), although more exotic alternative interpretations are not excluded. However, the calculations of the spectrum suffer from unknown $\beta$ decay branches which are regarded as free parameters and fitted to the overall spectrum measured by the old ILL-experiment~\cite{VonFeilitzsch:1982jw,Schreckenbach:1985ep,Hahn:1989zr}. Thus, a systematical error in that measurement can, after all, not be fully excluded. However, also in the old SAGE~\cite{Abdurashitov:1998ne} and GALLEX/GNO~\cite{Hampel:1997fc} experiments an unrelated deficit (``\emph{gallium anomaly}'') has been observed~\cite{Abdurashitov:2005tb} at a significance of about $3\sigma$~\cite{Giunti:2012tn}. Putting the reactor data together, one obtains a favoured region around $(\Delta m^2, \sin^2 (2 \theta)) \sim (1~{\rm eV}^2, 0.17)$~\cite{Palazzo:2013me}.

Test experiments~\cite{Agarwalla:2010zu,deGouvea:2011zz,Rubbia:2013ywa,2013arXiv1304.7127K} (in particular NUCIFER~\cite{Porta:2010zz} which will be operated at a distance of only about 10~m from the reactor core) will be able to resolve the situation. However, it should be noted that already now there is a good agreement between the data available if the appearance results by LSND and MiniBooNE, which after all are similar and could thus suffer from the same systematics, are discarded. If the hints persist, then there will be a demand from the model building side not only to explain the active neutrino mixing pattern but also to give an explanation for active-sterile mixing.

\section{\label{sec:outlook}Outlook}

All the solid experimental progress since 1998 supports the three active neutrino paradigm, as described by the lepton mixing matrix matrix involving three measured mixing angles. The outlook for the active neutrino sector lies in the experimental measurements of the $CP$ phases, mass squared ordering,absolute neutrino mass scale and its nature (Dirac or Majorana), together with the quest for ever higher precision of the measured mixing angles. The main motivation for pursuing the answers to these experimental questions lies in the dream of a unified theory which is also provides a robust theory of flavour.

Where do we stand in this quest and what is the outlook? We have reviewed the successes, challenges, and methods used in contemporary flavour model building based on discrete family symmetries. It is noteworthy that there are so many different approaches, and the measurement of the reactor angle has only served to rule out certain cherished models, such as those with tri-bimaximal mixing, without pointing the way to a unique solution to the flavour problem. Authors of such cherished models may be disappointed, even despondent, but others take heart from the knowledge that nature knows best and that a future theory of flavour must be out there, and by knowing the reactor angle we are one step closer to it. 

So, what are the directions for flavour model building following the measurement of the reactor angle? In the direct approach, one is driven to very large groups such as $\Delta(6n^2)$ for large values of $n$ in order to ``predict'' the reactor angle. For some people these groups are getting too large, so many others prefer the semi-direct approach where only a subgroup of the discrete family symmetry is enshrined in the Klein symmetry, leading to TM1 or TM2 mixing for example, where the reactor angle is not predicted, only atmospheric sum rules. Or maybe charged lepton corrections play a role, either with the Klein symmetry preserved leading to solar rules, or in conjunction with semi-direct models?

In keeping with Einstein's remark ``subtle is the Lord'', perhaps discrete symmetries are not realised directly but indirectly? In this case, rather small discrete groups can lead to neutrino mass matrices whose Klein symmetry does not correspond to a subgroup of the family symmetry. This can occur via ``simple'' vacuum alignments, such as $(1,4,2)$, whose origin lies in orthogonality arguments rather than looking for preserved subgroups. By combining vacuum alignments such as $(1,4,2)$ and $(0,1,1) $ using the seesaw mechanism with sequential dominance, we are led to enough complexity to account for all the measured mixing angles so far. Such a model may be extended to include quark mixing, in particular an explanation of the Cabibbo angle as being $\theta_C\approx 1/4$, leading to the tetra-model based on a Pati-Salam type partial unification~\cite{King:2013hoa}.

The outlook in dealing with the flavour problem lies in refining such models, striving for a degree of theoretical elegance and unification that would make such a model a leading candidate for a unified theory of flavour. Perhaps new ideas are required, possibly related to string theory, or perhaps all the ingredients are already lying around and need to be just put together in the right way. The answer to the remaining experimental questions, together with increased precision of the measured leptonic mixing angles, will provide the guiding light to the path towards a unified theory of flavour. However, according to ``anarchy'' a theory of flavour does not exist, so perhaps such a quest will be fruitless? One thing is certain: those that do not seek such a theory have no hope of finding it if it does exist. In this respect the ``anarchists'', or ``anarchy sympathisers'', have already given up and can never succeed. Those driven by dreams of symmetry will continue to believe that finite symmetries may play a part in resolving the flavour problem, in particular aspects related to the ``large'' mixing angles such as the atmospheric, solar, and reactor angles, as well as the Cabibbo angle. The ``small'' quark mixing angles associated with the CKM elements $V_{ub}$ and $V_{cb}$, and quark $CP$ violation, do not seem to be so readily explainable in terms of finite symmetries, and could result from higher order corrections, for example. It is entirely possible that leptonic $CP$ violation, on the other hand, could be a manifestation of an underlying discrete family symmetry, making the experimental pursuit of the neutrino oscillation phase even more important. Perhaps leptonic $CP$ violation is the reason for matter-antimatter asymmetry, which would then link the underlying family symmetry to our very existence. 

We could ask the question: apart from the flavour problem, which possible future directions could exist for the field? Two directions which could be identified are the possible connection between neutrinos and Dark Matter and the field of light sterile neutrinos in general, where by ``light'' we mean with a mass of order eV--keV. However we emphasise that the evidence for light sterile neutrinos is inconclusive at the present time, with several experimental hints all pointing in different directions. Starting with the latter, the challenge for model building would be to explain either tiny (for keV neutrinos, i.e., smaller than about $\theta \sim 10^{-5}$ for a mass of $M_1 = 5$~keV in order to avoid the X-ray bound~\cite{Watson:2006qb,Abazajian:2001vt,Abazajian:2006jc,Boyarsky:2005us,Dolgov:2000ew,Boyarsky:2006fg,RiemerSorensen:2006fh,Abazajian:2006yn,Boyarsky:2006ag,Boyarsky:2007ge,Loewenstein:2008yi,Boyarsky:2006fg,Watson:2011dw,Loewenstein:2012px}) or sizable (for eV neutrinos, i.e., about $\mathcal{O}(0.1)$ in order to match the experimental indications, cf.\ Sec.~\ref{sec:Light_Steriles}) active-sterile mixing, along with a mechanism to motivate the smallness of the sterile neutrino masses, which would certainly add new aspects to the field. First approaches already exist, and a review of many of these ideas has been provided in Ref.~\cite{Merle:2013gea}. Indeed, there are some models which attempt to explain both the mass and mixing patterns of sterile neutrinos purely by flavour symmetries. Known examples involve $L_e - L_\mu - L_\tau$~\cite{Shaposhnikov:2006nn,Lindner:2010wr} symmetry\footnote{While this is typically taken to be a continuous $U(1)$ symmetry, a discretised $Z_4$ version would work equally well.} and $Q_6$~\cite{Araki:2011zg}. Typically, these symmetries are used to predict zero active-sterile mixing with a sterile neutrino generation that is at the same time forced to be massless. As soon as the symmetries are broken, both consequences are alleviated and a small but non-zero mass and mixing are generated. Alternatively, one can separate the mechanism to generate the light neutrino mass scale from the generation of the mixing pattern and use the flavour symmetry exclusively to explain active-sterile mixing. These settings are clearly more versatile and, correspondingly, more proposals have been made. Known examples of this category include models based on $A_4\times Z_3$~\cite{Barry:2011fp,Barry:2011wb} (which use the Froggatt-Nielsen mechanism~\cite{Froggatt:1978nt} to explain the light sterile neutrino mass scale~\cite{Merle:2011yv}), on $A_4$~\cite{Adulpravitchai:2011rq} using the split seesaw mechanism~\cite{Kusenko:2010ik}, and on $A_4$~\cite{Barry:2011wb,Zhang:2011vh} using the extended seesaw mechanism~\cite{Chun:1995js,Zhang:2011vh}. A discussion of many more possibilities to explain a light sterile neutrino mass scale is contained in Ref.~\cite{Merle:2013gea}. Note that mixed settings are also particularly well-suited to generate relatively large, $\mathcal{O}(0.1)$, active-sterile mixing as required for eV sterile neutrinos, due to the generic power of discrete symmetries in this respect.

More generally, we have some theoretical indications that seem to suggest a connection between Dark Matter and neutrinos. Apart from the obvious connection in terms of keV sterile neutrino Dark Matter, there exist several models or mechanisms which generate a light neutrino mass scale and intrinsically involve Dark Matter candidate particles. This statement is particularly true for settings where the light neutrino mass is generated at loop-level. A nice example is provided by the so-called ``scotogenic" model~\cite{Ma:2006km}, where light neutrinos receive a mass only at 1-loop level with the loop involving the Dark Matter particle (either a right-handed TeV-scale neutrino or an inert Higgs field). In fact, the Dark Matter particle even considerably influences the renormalisation group running of the light neutrino parameters~\cite{Bouchand:2012dx}. In the scotogenic model, the fields that are inside the loop are charged under a new parity (an extra $Z_2$ imposed by hand). Hence, the lightest such particle will be stable and thus a potential Dark Matter candidate.\footnote{Note that Dark Matter can be either stable or decaying with a lifetime much longer as the age of the Universe.} Similar ideas have been developed, for example, in variants of the Zee-Babu model~\cite{Lindner:2011it,Schmidt:2012yg}, in the AMEND model~\cite{Farzan:2010mr}, or in the Cocktail model~\cite{Gustafsson:2012vj}. Such setting could even have detectable phenomenology such as, e.g., an enhanced Dark Matter annihilation rate into neutrinos~\cite{Lindner:2010rr}. More generally, the nature of Dark Matter and the origin of its stability are so far unknown. In order to explain the latter, some more ideas have been developed: e.g., in supersymmetric models the Dark Matter is typically stable due to conserved $R$-parity (introduced for a completely different problem, namely to prevent  proton to decay). A ``dark'' parity could also be motivated in GUT frameworks~\cite{Frigerio:2009wf,Kadastik:2009dj}. Recently, the possibility that the $Z_2$ stabilising the Dark Matter could originate from the spontaneous breaking of a flavour symmetry has been investigated~\cite{Hirsch:2010ru,Boucenna:2011tj,Boucenna:2012qb}. For instance in Ref.~\cite{Hirsch:2010ru}, $A_4$ was taken to be the  flavour symmetry group and a scalar $A_4$-triplet $\eta=(\eta_1,\eta_2,\eta_3)$ that obtains a VEV in the $A_4$ direction $\langle \eta \rangle \sim (1,0,0)$ breaks $A_4$ spontaneously to a $Z_2$ symmetry, under which $\eta_1$ is even while $\eta_{2,3}$ are odd and thus the lightest of them is a potential Dark Matter candidate. Another interesting possibility to stabilise the Dark Matter with a flavour symmetry is by an ``accidental" embedding of the flavour group in its double cover, like in~\cite{Lavoura:2012cv} where $T'$, the double cover of $A_4$, has been used. Alternatively, the Dark Matter could be part of a non-trivial representation of the flavour symmetry and thus be stabilised by a more general structure than $Z_2$~\cite{Adulpravitchai:2011ei}. For a review of different stabilisation mechanisms see for instance~\cite{Hambye:2010zb}.
In \cite{Kajiyama:2011gu} the Dark Matter candidate is not stable and decay operators of dimension smaller than 6 are forbidden thanks to the family symmetry.
In this case the Dark Matter candidate lifetime is at least $10^{26}s$ in agreement with cosmological requirements.
\\

As can be seen in these few examples, the field of discrete flavour symmetries could have many more interesting applications than ``just'' the light neutrino flavour sector. In the near future, experiments will hopefully guide us towards the next challenges in this versatile and interesting field.

\section{\label{sec:conc}Summary and Conclusion}

With the measurement of the reactor angle we have entered the era of experimental precision in the lepton mixing, rather like in the quark sector but still of course far behind it. Recall that the quark $CP$ phase $\delta^q$ is known to be about $70^\circ$ to an accuracy of a few degrees, while in the lepton sector the $CP$ phase $\delta^l$ is unknown, although at the time of writing there are some hints that it could be around $270^\circ$ (or $-90^\circ$) -- but it is also consistent with zero at one sigma. It is unfortunate that lepton precision is so far behind quark precision, since the leptonic $CP$ violation may turn out to be more fundamental than that in the quark sector. The reason we say this is that leptonic $CP$ violation is proportional to the reactor angle $\theta_{13}^l$, whereas quark $CP$ violation is proportional to a much smaller angle $\theta_{13}^q$. While we have seen examples of models which can account for the reactor angle as a result of a discrete family symmetry, it is much more difficult to account for very small angles such as $\theta_{13}^q$ from such an approach. Therefore, leptonic $CP$ violation may be related to a discrete family symmetry, while quark $CP$ violation is almost certainly not. On top of that, leptonic $CP$ violation may be entirely responsible for matter-antimatter asymmetry via leptogenesis, while quark $CP$ violation in the SM is much too small to account for the origin of matter for a similar reason (quark mixing angles being too small).
 
Despite the great experimental successes of measuring all the leptonic mixing angles and the magnitudes of neutrino mass squared differences, on the theoretical side we still do not have a final agreed picture for the origin of these parameters. In addition the origin of charged fermion mass hierarchies is unknown, as is quark mixing. Also the origin of light neutrino masses is another big mystery in theoretical physics: we do not even know if neutrinos have Dirac or Majorana nature and what is the mechanism responsible for the light masses of neutrinos. While charged fermion mass hierarchies could arise from an Abelian Froggatt-Nielsen symmetry, or a discrete Abelian group, it seems that fermion mixing can be better understood by means of non-Abelian symmetries. The motivation is that non-Abelian groups admit irreducible representations of dimensions higher than one (for example dimension three) that can lead to large mixing. In particular, first inspired by bimaximal and then tri-bimaximal mixing, discrete non-Abelian groups have been widely used in literature. We have provided a basic introduction to the mathematics of discrete groups and the theory of their representations.

While a maximal atmospheric angle ($45^\circ$) and trimaximal solar angle ($35.26^\circ$) remain viable possibilities, the zero reactor angle is ruled out at about ten sigma, with the measured value around $9^\circ$.
While this destroyed the theoretical consensus was emerging around tri-bimaximal mixing, the reactor angle being large, of the same order as the Cabibbo angle, brings with it fresh opportunities for model building. The present problem is that there are almost too many possibilities, ranging from anarchy to models which directly explain all the lepton mixing matrix parameters, and no consensus has yet emerged comparable to that which surrounded tri-bimaximal mixing prior to the measurement of the reactor angle. 

In the face of all the different theoretical possibilities, we have avoided the temptation to give 
an exhaustive compendium of all the possible models. Instead we found it more useful to discuss the general classification of different types of approaches and models, together with a few concrete examples. We started by considering the use of flavons vs.\ multi-Higgs in family symmetry models. The two classes of models have very different phenomenologies that can eventually give us the possibility to distinguish experimentally between them. Another important argument that can be used to classify different models is by studying the breaking pattern of the family symmetry group. We can have direct or indirect approaches. While in the first case, different subgroups of the family symmetry survive in the neutrino or charged lepton sectors, in the indirect approach no subgroup of the family symmetry survives in either sector but predictions can arise from a ``memory'' of the family symmetry engraved into the vacuum alignments of the flavons. 

It is well appreciated that, with the measurement of the reactor angle, it becomes possible to determine experimentally the Dirac $CP$ violating phase much more easily. This motivates the possibility of predicting the leptonic $CP$ phases in discrete family symmetry models, since such predictions will be tested relatively soon. We have reviewed the recent progress in this direction. We have also discussed the possibility to combine discrete family symmetries with GUT frameworks, and have presented an example of this.

We have reviewed the possibility to distinguish flavor symmetry models in experiments.
For example, we have discussed the atmospheric and solar mixing sum rules, where for instance the atmospheric angle is a function of the Dirac $CP$ phase and the reactor angle. Such sum rules can be tested in future neutrino experiments which are capable of accurately measuring the Dirac $CP$ violating oscillation phase $\delta^l$. Such future experiments may be divided into three categories: Superbeams, Betabeams, and (Low Energy) Neutrino Factory, all at different stages of design and with different possible locations and funding profiles. The most advanced proposal in all respects seems to be Hyper-Kamiokande Superbeam proposal, but time will tell. 
In a similar way, mass sum rules are also possible and we have reviewed both the origin of such sum rules and their prospects for being tested in neutrinoless double beta decay experiments. Such experiments are vital for telling us the scale and nature of neutrino mass, and absolutely essential if we are to understand the origin of neutrino mass.

Perhaps the three active neutrino paradigm is not the whole story. Having introduced ``heavy'' sterile neutrinos for the seesaw mechanism, the possibility remains that some of these sterile neutrinos might be very light, of order eV--keV mass for example, in which case they may be referred to as ``light'' sterile neutrinos, leading to observable effects in short-baseline neutrino oscillation experiments. As a matter of terminology, ``heavy'' sterile neutrinos as used in the seesaw mechanism are often called simply ``right-handed neutrinos'', while ``light'' sterile neutrinos are often called simply ``sterile neutrinos'' for short. While there is no conclusive evidence for such ``sterile neutrinos'', there remains fragments of evidence from different experiments (albeit pointing in different directions), and a future experimental programme is on the table to settle the issue. Neutrino physics has never failed to surprise theorists, so keeping one eye on the possibility of sterile neutrinos seems a prudent strategy.

For the general reader, we remark that, while neutrinos do not seem important in our everyday lives (although in the future they could find applications, e.g.\ in geophysics or reactor physics) they are highly important cosmologically, with possible relevance for Dark Matter, the matter-antimatter asymmetry, inflation, and even Dark Energy. Although we have not focussed on such issues in this review, it is worth mentioning that the lightest fermions may be the most important ones cosmologically.

In this review we have focussed on the role of discrete symmetries which may play a role in understanding neutrino mass and especially leptonic mixing. We have seen that the existence of large mixing angles motivates theories based on discrete non-Abelian family symmetries. Ironically, large mixing angles simultaneously motivate the idea of anarchy, where there is no underlying theory. The ghost of Einstein's question of quantum mechanics returns to haunt us in the present day as: does God play dice with the lepton mixing angles? If so, then why not also with the quark mixing angles? If the reactor angle were very small it could have been explained by a discrete symmetry, whereas it would not have fitted very well with anarchy. Anarchists have claimed that the discovery of a large reactor angle of similar magnitude to the Cabibbo angle could be circumstantial evidence for anarchy. Anarchists further claim that they expected a large reactor angle before it was measured.
On the other hand, model builders counter that large angles can also be explained by discrete symmetries.
Model builders also point out that some approaches such as sequential dominance have long predicted a reactor angle of order $m_2/m_3$ \cite{King:2002nf}.
However such predictions were neglected in the premature euphoria that surrounded tri-bimaximal mixing. 

In the light of the reactor angle measurement, such approaches have been revisited and refined
(for example CSD4), and many new symmetry approaches have been developed, to account for the measured reactor angle and also at the same time to predict the so far unmeasured $CP$ violating phase. It remains an exciting prospect that the large lepton mixing angles could arise from an underlying theory of flavour based on discrete symmetry, where such a theory could be extended to the quark sector, perhaps in the framework of a GUT. In such frameworks, large mixing angles, including also the reactor angle and the Cabibbo angle, are the key that could unlock the whole theory of flavour. This point of view is the polar opposite of the view of ``anarchists'' who argue that large mixing angles indicate that we should give up on the problem of flavour.
What are we to make of such diametrically opposing claims and what is the way forwards?

In order to resolve this question, in our view we must continue to 
construct theories of flavour which are capable of giving accurate predictions of lepton mixing matrix parameters that can be confronted with ever more precise experimental data, in the hope that a leading candidate theory of flavour that explains everything will eventually emerge. In any case, it is clear that flavour model building will remain an active area of research in the future, where such an activity both complements and motivates the high precision neutrino experimental programme. We hope that this review article, which spans the spectrum of mathematics, model building and experiment, may serve to inspire the next generation of younger researchers in the quest for a robust theory of flavour and unification.

\section*{Acknowledgements}
SFK acknowledges partial support from the STFC Consolidated ST/J000396/1 and the EU ITN grant UNILHC 237920. AM acknowledges financial support by a Marie Curie Intra-European Fellowship within the 7th European Community Framework Programme FP7-PEOPLE-2011-IEF, contract PIEF-GA-2011-297557. SFK and AM both acknowledge partial support from the European Union FP7 ITN-INVISIBLES (Marie Curie Actions, PITN-GA-2011-289442). SM was funded by the German Research Foundation (DFG) and the University of W\"urzburg in the funding programme Open Access Publishing and DFG grant WI 2639/4-1. MT acknowledges partial support from JSPS Grand-in-Aid for Scientific Research, 21340055 and 24654062.



\bibliographystyle{apsrev}
\bibliography{minirevstefano,minirevsteve}

\begin{thebibliography}{100}
\expandafter\ifx\csname bibnamefont\endcsname\relax
  \def\bibnamefont#1{#1}\fi
\expandafter\ifx\csname bibfnamefont\endcsname\relax
  \def\bibfnamefont#1{#1}\fi
\expandafter\ifx\csname url\endcsname\relax
  \def\url#1{\texttt{#1}}\fi
\expandafter\ifx\csname urlprefix\endcsname\relax\def\urlprefix{URL }\fi
\providecommand{\bibinfo}[2]{#2}
\providecommand{\eprint}[2][]{\url{#2}}

\bibitem{Beringer:1900zz}
\bibinfo{author}{\bibfnamefont{J.}~\bibnamefont{Beringer}} \emph{et~al.}
  (\bibinfo{collaboration}{Particle Data Group}), \bibinfo{journal}{Phys. Rev.}
  \textbf{\bibinfo{volume}{D86}}, \bibinfo{pages}{010001}
  (\bibinfo{year}{2012}).

\bibitem{Wolfenstein:1981kw}
\bibinfo{author}{\bibfnamefont{L.}~\bibnamefont{Wolfenstein}},
  \bibinfo{journal}{Nucl. Phys.} \textbf{\bibinfo{volume}{B186}},
  \bibinfo{pages}{147} (\bibinfo{year}{1981}).

\bibitem{Valle:1982yw}
\bibinfo{author}{\bibfnamefont{J.~W.~F.} \bibnamefont{Valle}},
  \bibinfo{journal}{Phys. Rev.} \textbf{\bibinfo{volume}{D27}},
  \bibinfo{pages}{1672} (\bibinfo{year}{1983}).

\bibitem{Allahverdi:2010us}
\bibinfo{author}{\bibfnamefont{R.}~\bibnamefont{Allahverdi}},
  \bibinfo{author}{\bibfnamefont{B.}~\bibnamefont{Dutta}}, \bibnamefont{and}
  \bibinfo{author}{\bibfnamefont{R.~N.} \bibnamefont{Mohapatra}},
  \bibinfo{journal}{Phys. Lett.} \textbf{\bibinfo{volume}{B695}},
  \bibinfo{pages}{181} (\bibinfo{year}{2011}), \eprint{1008.1232}.

\bibitem{Schechter:1980gr}
\bibinfo{author}{\bibfnamefont{J.}~\bibnamefont{Schechter}} \bibnamefont{and}
  \bibinfo{author}{\bibfnamefont{J.~W.~F.} \bibnamefont{Valle}},
  \bibinfo{journal}{Phys. Rev.} \textbf{\bibinfo{volume}{D22}},
  \bibinfo{pages}{2227} (\bibinfo{year}{1980}).

\bibitem{Xing:2007uq}
\bibinfo{author}{\bibfnamefont{Z.-z.} \bibnamefont{Xing}},
  \bibinfo{journal}{Chin. Phys.} \textbf{\bibinfo{volume}{C32}},
  \bibinfo{pages}{96} (\bibinfo{year}{2008}), \eprint{0706.0052}.

\bibitem{Weinberg:1979sa}
\bibinfo{author}{\bibfnamefont{S.}~\bibnamefont{Weinberg}},
  \bibinfo{journal}{Phys. Rev. Lett.} \textbf{\bibinfo{volume}{43}},
  \bibinfo{pages}{1566} (\bibinfo{year}{1979}).

\bibitem{Minkowski:1977sc}
\bibinfo{author}{\bibfnamefont{P.}~\bibnamefont{Minkowski}},
  \bibinfo{journal}{Phys. Lett.} \textbf{\bibinfo{volume}{B67}},
  \bibinfo{pages}{421} (\bibinfo{year}{1977}).

\bibitem{Yanagida:1979as}
\bibinfo{author}{\bibfnamefont{T.}~\bibnamefont{Yanagida}},
  \bibinfo{journal}{Conf. Proc.} \textbf{\bibinfo{volume}{C7902131}},
  \bibinfo{pages}{95} (\bibinfo{year}{1979}).

\bibitem{GellMann:1980vs}
\bibinfo{author}{\bibfnamefont{M.}~\bibnamefont{Gell-Mann}},
  \bibinfo{author}{\bibfnamefont{P.}~\bibnamefont{Ramond}}, \bibnamefont{and}
  \bibinfo{author}{\bibfnamefont{R.}~\bibnamefont{Slansky}},
  \bibinfo{journal}{Conf. Proc.} \textbf{\bibinfo{volume}{C790927}},
  \bibinfo{pages}{315} (\bibinfo{year}{1979}), \bibinfo{note}{to be published
  in Supergravity, P. van Nieuwenhuizen \& D. Z. Freedman (eds.), North Holland
  Publ. Co., 1979}.

\bibitem{Mohapatra:1979ia}
\bibinfo{author}{\bibfnamefont{R.~N.} \bibnamefont{Mohapatra}}
  \bibnamefont{and}
  \bibinfo{author}{\bibfnamefont{G.}~\bibnamefont{Senjanovic}},
  \bibinfo{journal}{Phys. Rev. Lett.} \textbf{\bibinfo{volume}{44}},
  \bibinfo{pages}{912} (\bibinfo{year}{1980}).

\bibitem{Magg:1980ut}
\bibinfo{author}{\bibfnamefont{M.}~\bibnamefont{Magg}} \bibnamefont{and}
  \bibinfo{author}{\bibfnamefont{C.}~\bibnamefont{Wetterich}},
  \bibinfo{journal}{Phys. Lett.} \textbf{\bibinfo{volume}{B94}},
  \bibinfo{pages}{61} (\bibinfo{year}{1980}).

\bibitem{Wetterich:1981bx}
\bibinfo{author}{\bibfnamefont{C.}~\bibnamefont{Wetterich}},
  \bibinfo{journal}{Nucl. Phys.} \textbf{\bibinfo{volume}{B187}},
  \bibinfo{pages}{343} (\bibinfo{year}{1981}).

\bibitem{Mohapatra:1980yp}
\bibinfo{author}{\bibfnamefont{R.~N.} \bibnamefont{Mohapatra}}
  \bibnamefont{and}
  \bibinfo{author}{\bibfnamefont{G.}~\bibnamefont{Senjanovic}},
  \bibinfo{journal}{Phys. Rev.} \textbf{\bibinfo{volume}{D23}},
  \bibinfo{pages}{165} (\bibinfo{year}{1981}).

\bibitem{Cheng:1980qt}
\bibinfo{author}{\bibfnamefont{T.~P.} \bibnamefont{Cheng}} \bibnamefont{and}
  \bibinfo{author}{\bibfnamefont{L.-F.} \bibnamefont{Li}},
  \bibinfo{journal}{Phys. Rev.} \textbf{\bibinfo{volume}{D22}},
  \bibinfo{pages}{2860} (\bibinfo{year}{1980}).

\bibitem{Foot:1988aq}
\bibinfo{author}{\bibfnamefont{R.}~\bibnamefont{Foot}},
  \bibinfo{author}{\bibfnamefont{H.}~\bibnamefont{Lew}},
  \bibinfo{author}{\bibfnamefont{X.~G.} \bibnamefont{He}}, \bibnamefont{and}
  \bibinfo{author}{\bibfnamefont{G.~C.} \bibnamefont{Joshi}},
  \bibinfo{journal}{Z. Phys.} \textbf{\bibinfo{volume}{C44}},
  \bibinfo{pages}{441} (\bibinfo{year}{1989}).

\bibitem{Ma:1998dn}
\bibinfo{author}{\bibfnamefont{E.}~\bibnamefont{Ma}},
  \bibinfo{journal}{Phys.Rev.Lett.} \textbf{\bibinfo{volume}{81}},
  \bibinfo{pages}{1171} (\bibinfo{year}{1998}), \eprint{hep-ph/9805219}.

\bibitem{Mohapatra:1986bd}
\bibinfo{author}{\bibfnamefont{R.~N.} \bibnamefont{Mohapatra}}
  \bibnamefont{and} \bibinfo{author}{\bibfnamefont{J.~W.~F.}
  \bibnamefont{Valle}}, \bibinfo{journal}{Phys. Rev.}
  \textbf{\bibinfo{volume}{D34}}, \bibinfo{pages}{1642} (\bibinfo{year}{1986}).

\bibitem{Akhmedov:1995vm}
\bibinfo{author}{\bibfnamefont{E.~K.} \bibnamefont{Akhmedov}},
  \bibinfo{author}{\bibfnamefont{M.}~\bibnamefont{Lindner}},
  \bibinfo{author}{\bibfnamefont{E.}~\bibnamefont{Schnapka}}, \bibnamefont{and}
  \bibinfo{author}{\bibfnamefont{J.~W.~F.} \bibnamefont{Valle}},
  \bibinfo{journal}{Phys. Rev.} \textbf{\bibinfo{volume}{D53}},
  \bibinfo{pages}{2752} (\bibinfo{year}{1996}), \eprint{hep-ph/9509255}.

\bibitem{Akhmedov:1995ip}
\bibinfo{author}{\bibfnamefont{E.~K.} \bibnamefont{Akhmedov}},
  \bibinfo{author}{\bibfnamefont{M.}~\bibnamefont{Lindner}},
  \bibinfo{author}{\bibfnamefont{E.}~\bibnamefont{Schnapka}}, \bibnamefont{and}
  \bibinfo{author}{\bibfnamefont{J.~W.~F.} \bibnamefont{Valle}},
  \bibinfo{journal}{Phys. Lett.} \textbf{\bibinfo{volume}{B368}},
  \bibinfo{pages}{270} (\bibinfo{year}{1996}), \eprint{hep-ph/9507275}.

\bibitem{Malinsky:2005bi}
\bibinfo{author}{\bibfnamefont{M.}~\bibnamefont{Malinsky}},
  \bibinfo{author}{\bibfnamefont{J.~C.} \bibnamefont{Romao}}, \bibnamefont{and}
  \bibinfo{author}{\bibfnamefont{J.~W.~F.} \bibnamefont{Valle}},
  \bibinfo{journal}{Phys. Rev. Lett.} \textbf{\bibinfo{volume}{95}},
  \bibinfo{pages}{161801} (\bibinfo{year}{2005}), \eprint{hep-ph/0506296}.

\bibitem{Ma:2006km}
\bibinfo{author}{\bibfnamefont{E.}~\bibnamefont{Ma}}, \bibinfo{journal}{Phys.
  Rev.} \textbf{\bibinfo{volume}{D73}}, \bibinfo{pages}{077301}
  (\bibinfo{year}{2006}), \eprint{hep-ph/0601225}.

\bibitem{Barabash:2011mf}
\bibinfo{author}{\bibfnamefont{A.~S.} \bibnamefont{Barabash}},
  \bibinfo{journal}{Phys. Atom. Nucl.} \textbf{\bibinfo{volume}{74}},
  \bibinfo{pages}{603} (\bibinfo{year}{2011}), \eprint{1104.2714}.

\bibitem{Rodejohann:2011mu}
\bibinfo{author}{\bibfnamefont{W.}~\bibnamefont{Rodejohann}},
  \bibinfo{journal}{Int. J. Mod. Phys.} \textbf{\bibinfo{volume}{E20}},
  \bibinfo{pages}{1833} (\bibinfo{year}{2011}), \eprint{1106.1334}.

\bibitem{Vergados:2012xy}
\bibinfo{author}{\bibfnamefont{J.~D.} \bibnamefont{Vergados}},
  \bibinfo{author}{\bibfnamefont{H.}~\bibnamefont{Ejiri}}, \bibnamefont{and}
  \bibinfo{author}{\bibfnamefont{F.}~\bibnamefont{Simkovic}},
  \bibinfo{journal}{Rept. Prog. Phys.} \textbf{\bibinfo{volume}{75}},
  \bibinfo{pages}{106301} (\bibinfo{year}{2012}), \eprint{1205.0649}.

\bibitem{Schechter:1981bd}
\bibinfo{author}{\bibfnamefont{J.}~\bibnamefont{Schechter}} \bibnamefont{and}
  \bibinfo{author}{\bibfnamefont{J.~W.~F.} \bibnamefont{Valle}},
  \bibinfo{journal}{Phys. Rev.} \textbf{\bibinfo{volume}{D25}},
  \bibinfo{pages}{2951} (\bibinfo{year}{1982}).

\bibitem{Duerr:2011zd}
\bibinfo{author}{\bibfnamefont{M.}~\bibnamefont{Duerr}},
  \bibinfo{author}{\bibfnamefont{M.}~\bibnamefont{Lindner}}, \bibnamefont{and}
  \bibinfo{author}{\bibfnamefont{A.}~\bibnamefont{Merle}},
  \bibinfo{journal}{JHEP} \textbf{\bibinfo{volume}{1106}}, \bibinfo{pages}{091}
  (\bibinfo{year}{2011}), \eprint{1105.0901}.

\bibitem{Luhn:2011ip}
\bibinfo{author}{\bibfnamefont{C.}~\bibnamefont{Luhn}}, \bibinfo{journal}{JHEP}
  \textbf{\bibinfo{volume}{1103}}, \bibinfo{pages}{108} (\bibinfo{year}{2011}),
  \eprint{1101.2417}.

\bibitem{Merle:2011vy}
\bibinfo{author}{\bibfnamefont{A.}~\bibnamefont{Merle}} \bibnamefont{and}
  \bibinfo{author}{\bibfnamefont{R.}~\bibnamefont{Zwicky}},
  \bibinfo{journal}{JHEP} \textbf{\bibinfo{volume}{1202}}, \bibinfo{pages}{128}
  (\bibinfo{year}{2012}), \eprint{1110.4891}.

\bibitem{Merle:2012xr}
\bibinfo{author}{\bibfnamefont{A.}~\bibnamefont{Merle}} \bibnamefont{and}
  \bibinfo{author}{\bibfnamefont{R.}~\bibnamefont{Zwicky}} pp.
  \bibinfo{pages}{191--198} (\bibinfo{year}{2012}), \eprint{1210.6239}.

\bibitem{Harrison:2002er}
\bibinfo{author}{\bibfnamefont{P.~F.} \bibnamefont{Harrison}},
  \bibinfo{author}{\bibfnamefont{D.~H.} \bibnamefont{Perkins}},
  \bibnamefont{and} \bibinfo{author}{\bibfnamefont{W.~G.} \bibnamefont{Scott}},
  \bibinfo{journal}{Phys. Lett.} \textbf{\bibinfo{volume}{B530}},
  \bibinfo{pages}{167} (\bibinfo{year}{2002}), \eprint{hep-ph/0202074}.

\bibitem{GonzalezGarcia:2012sz}
\bibinfo{author}{\bibfnamefont{M.~C.} \bibnamefont{Gonzalez-Garcia}},
  \bibinfo{author}{\bibfnamefont{M.}~\bibnamefont{Maltoni}},
  \bibinfo{author}{\bibfnamefont{J.}~\bibnamefont{Salvado}}, \bibnamefont{and}
  \bibinfo{author}{\bibfnamefont{T.}~\bibnamefont{Schwetz}},
  \bibinfo{journal}{JHEP} \textbf{\bibinfo{volume}{1212}}, \bibinfo{pages}{123}
  (\bibinfo{year}{2012}), \eprint{1209.3023}.

\bibitem{Capozzi:2013csa}
\bibinfo{author}{\bibfnamefont{F.}~\bibnamefont{Capozzi}},
  \bibinfo{author}{\bibfnamefont{G.~L.} \bibnamefont{Fogli}},
  \bibinfo{author}{\bibfnamefont{E.}~\bibnamefont{Lisi}},
  \bibinfo{author}{\bibfnamefont{A.}~\bibnamefont{Marrone}},
  \bibinfo{author}{\bibfnamefont{D.}~\bibnamefont{Montanino}}, \emph{et~al.}
  (\bibinfo{year}{2013}), \eprint{1312.2878}.

\bibitem{Tortola:2012te}
\bibinfo{author}{\bibfnamefont{D.~V.} \bibnamefont{Forero}},
  \bibinfo{author}{\bibfnamefont{M.}~\bibnamefont{Tortola}}, \bibnamefont{and}
  \bibinfo{author}{\bibfnamefont{J.~W.~F.} \bibnamefont{Valle}},
  \bibinfo{journal}{Phys. Rev.} \textbf{\bibinfo{volume}{D86}},
  \bibinfo{pages}{073012} (\bibinfo{year}{2012}), \eprint{1205.4018}.

\bibitem{Ma:2004zv}
\bibinfo{author}{\bibfnamefont{E.}~\bibnamefont{Ma}},
  \bibinfo{journal}{Phys.Rev.} \textbf{\bibinfo{volume}{D70}},
  \bibinfo{pages}{031901} (\bibinfo{year}{2004}), \eprint{hep-ph/0404199}.

\bibitem{Altarelli:2005yp}
\bibinfo{author}{\bibfnamefont{G.}~\bibnamefont{Altarelli}} \bibnamefont{and}
  \bibinfo{author}{\bibfnamefont{F.}~\bibnamefont{Feruglio}},
  \bibinfo{journal}{Nucl. Phys.} \textbf{\bibinfo{volume}{B720}},
  \bibinfo{pages}{64} (\bibinfo{year}{2005}), \eprint{hep-ph/0504165}.

\bibitem{Altarelli:2005yx}
\bibinfo{author}{\bibfnamefont{G.}~\bibnamefont{Altarelli}} \bibnamefont{and}
  \bibinfo{author}{\bibfnamefont{F.}~\bibnamefont{Feruglio}},
  \bibinfo{journal}{Nucl. Phys.} \textbf{\bibinfo{volume}{B741}},
  \bibinfo{pages}{215} (\bibinfo{year}{2006}), \eprint{hep-ph/0512103}.

\bibitem{Babu:2005se}
\bibinfo{author}{\bibfnamefont{K.~S.} \bibnamefont{Babu}} \bibnamefont{and}
  \bibinfo{author}{\bibfnamefont{X.-G.} \bibnamefont{He}}
  (\bibinfo{year}{2005}), \eprint{hep-ph/0507217}.

\bibitem{deMedeirosVarzielas:2005qg}
\bibinfo{author}{\bibfnamefont{I.}~\bibnamefont{de~Medeiros~Varzielas}},
  \bibinfo{author}{\bibfnamefont{S.~F.} \bibnamefont{King}}, \bibnamefont{and}
  \bibinfo{author}{\bibfnamefont{G.~G.} \bibnamefont{Ross}},
  \bibinfo{journal}{Phys. Lett.} \textbf{\bibinfo{volume}{B644}},
  \bibinfo{pages}{153} (\bibinfo{year}{2007}), \eprint{hep-ph/0512313}.

\bibitem{Ma:2001dn}
\bibinfo{author}{\bibfnamefont{E.}~\bibnamefont{Ma}} \bibnamefont{and}
  \bibinfo{author}{\bibfnamefont{G.}~\bibnamefont{Rajasekaran}},
  \bibinfo{journal}{Phys. Rev.} \textbf{\bibinfo{volume}{D64}},
  \bibinfo{pages}{113012} (\bibinfo{year}{2001}), \eprint{hep-ph/0106291}.

\bibitem{Babu:2002dz}
\bibinfo{author}{\bibfnamefont{K.~S.} \bibnamefont{Babu}},
  \bibinfo{author}{\bibfnamefont{E.}~\bibnamefont{Ma}}, \bibnamefont{and}
  \bibinfo{author}{\bibfnamefont{J.~W.~F.} \bibnamefont{Valle}},
  \bibinfo{journal}{Phys. Lett.} \textbf{\bibinfo{volume}{B552}},
  \bibinfo{pages}{207} (\bibinfo{year}{2003}), \eprint{hep-ph/0206292}.

\bibitem{Grimus:2008vg}
\bibinfo{author}{\bibfnamefont{W.}~\bibnamefont{Grimus}} \bibnamefont{and}
  \bibinfo{author}{\bibfnamefont{L.}~\bibnamefont{Lavoura}},
  \bibinfo{journal}{JHEP} \textbf{\bibinfo{volume}{0904}}, \bibinfo{pages}{013}
  (\bibinfo{year}{2009}), \eprint{0811.4766}.

\bibitem{Mohapatra:2006pu}
\bibinfo{author}{\bibfnamefont{R.~N.} \bibnamefont{Mohapatra}},
  \bibinfo{author}{\bibfnamefont{S.}~\bibnamefont{Nasri}}, \bibnamefont{and}
  \bibinfo{author}{\bibfnamefont{H.-B.} \bibnamefont{Yu}},
  \bibinfo{journal}{Phys. Lett.} \textbf{\bibinfo{volume}{B639}},
  \bibinfo{pages}{318} (\bibinfo{year}{2006}), \eprint{hep-ph/0605020}.

\bibitem{Lam:2008rs}
\bibinfo{author}{\bibfnamefont{C.~S.} \bibnamefont{Lam}},
  \bibinfo{journal}{Phys. Rev. Lett.} \textbf{\bibinfo{volume}{101}},
  \bibinfo{pages}{121602} (\bibinfo{year}{2008}), \eprint{0804.2622}.

\bibitem{Bazzocchi:2008ej}
\bibinfo{author}{\bibfnamefont{F.}~\bibnamefont{Bazzocchi}} \bibnamefont{and}
  \bibinfo{author}{\bibfnamefont{S.}~\bibnamefont{Morisi}},
  \bibinfo{journal}{Phys. Rev.} \textbf{\bibinfo{volume}{D80}},
  \bibinfo{pages}{096005} (\bibinfo{year}{2009}), \eprint{0811.0345}.

\bibitem{Feruglio:2007uu}
\bibinfo{author}{\bibfnamefont{F.}~\bibnamefont{Feruglio}},
  \bibinfo{author}{\bibfnamefont{C.}~\bibnamefont{Hagedorn}},
  \bibinfo{author}{\bibfnamefont{Y.}~\bibnamefont{Lin}}, \bibnamefont{and}
  \bibinfo{author}{\bibfnamefont{L.}~\bibnamefont{Merlo}},
  \bibinfo{journal}{Nucl. Phys.} \textbf{\bibinfo{volume}{B775}},
  \bibinfo{pages}{120} (\bibinfo{year}{2007}), \eprint{hep-ph/0702194}.

\bibitem{Carr:2007qw}
\bibinfo{author}{\bibfnamefont{P.~D.} \bibnamefont{Carr}} \bibnamefont{and}
  \bibinfo{author}{\bibfnamefont{P.~H.} \bibnamefont{Frampton}}
  (\bibinfo{year}{2007}), \eprint{hep-ph/0701034}.

\bibitem{deMedeirosVarzielas:2006fc}
\bibinfo{author}{\bibfnamefont{I.}~\bibnamefont{de~Medeiros~Varzielas}},
  \bibinfo{author}{\bibfnamefont{S.~F.} \bibnamefont{King}}, \bibnamefont{and}
  \bibinfo{author}{\bibfnamefont{G.~G.} \bibnamefont{Ross}},
  \bibinfo{journal}{Phys. Lett.} \textbf{\bibinfo{volume}{B648}},
  \bibinfo{pages}{201} (\bibinfo{year}{2007}), \eprint{hep-ph/0607045}.

\bibitem{Ishimori:2008uc}
\bibinfo{author}{\bibfnamefont{H.}~\bibnamefont{Ishimori}},
  \bibinfo{author}{\bibfnamefont{T.}~\bibnamefont{Kobayashi}},
  \bibinfo{author}{\bibfnamefont{H.}~\bibnamefont{Okada}},
  \bibinfo{author}{\bibfnamefont{Y.}~\bibnamefont{Shimizu}}, \bibnamefont{and}
  \bibinfo{author}{\bibfnamefont{M.}~\bibnamefont{Tanimoto}},
  \bibinfo{journal}{JHEP} \textbf{\bibinfo{volume}{0904}}, \bibinfo{pages}{011}
  (\bibinfo{year}{2009}), \eprint{0811.4683}.

\bibitem{Abe:2013nka}
\bibinfo{author}{\bibfnamefont{K.}~\bibnamefont{Abe}} \emph{et~al.}
  (\bibinfo{collaboration}{collaboration for the T2K})  (\bibinfo{year}{2013}),
  \eprint{1311.4750}.

\bibitem{An:2013zwz}
\bibinfo{author}{\bibfnamefont{F.~P.} \bibnamefont{An}} \emph{et~al.}
  (\bibinfo{collaboration}{Daya Bay Collaboration})  (\bibinfo{year}{2013}),
  \eprint{1310.6732}.

\bibitem{Adamson:2013whj}
\bibinfo{author}{\bibfnamefont{P.}~\bibnamefont{Adamson}} \emph{et~al.}
  (\bibinfo{collaboration}{MINOS Collaboration}), \bibinfo{journal}{Phys. Rev.
  Lett.} \textbf{\bibinfo{volume}{110}}, \bibinfo{pages}{251801}
  (\bibinfo{year}{2013}), \eprint{1304.6335}.

\bibitem{Ahn:2012nd}
\bibinfo{author}{\bibfnamefont{J.~K.} \bibnamefont{Ahn}} \emph{et~al.}
  (\bibinfo{collaboration}{RENO collaboration}), \bibinfo{journal}{Phys. Rev.
  Lett.} \textbf{\bibinfo{volume}{108}}, \bibinfo{pages}{191802}
  (\bibinfo{year}{2012}), \eprint{1204.0626}.

\bibitem{Ma:2011yi}
\bibinfo{author}{\bibfnamefont{E.}~\bibnamefont{Ma}} \bibnamefont{and}
  \bibinfo{author}{\bibfnamefont{D.}~\bibnamefont{Wegman}},
  \bibinfo{journal}{Phys. Rev. Lett.} \textbf{\bibinfo{volume}{107}},
  \bibinfo{pages}{061803} (\bibinfo{year}{2011}), \eprint{1106.4269}.

\bibitem{Antusch:2011qg}
\bibinfo{author}{\bibfnamefont{S.}~\bibnamefont{Antusch}} \bibnamefont{and}
  \bibinfo{author}{\bibfnamefont{V.}~\bibnamefont{Maurer}},
  \bibinfo{journal}{Phys. Rev.} \textbf{\bibinfo{volume}{D84}},
  \bibinfo{pages}{117301} (\bibinfo{year}{2011}), \eprint{1107.3728}.

\bibitem{Acosta:2012qf}
\bibinfo{author}{\bibfnamefont{J.~A.} \bibnamefont{Acosta}},
  \bibinfo{author}{\bibfnamefont{A.}~\bibnamefont{Aranda}},
  \bibinfo{author}{\bibfnamefont{M.~A.} \bibnamefont{Buen-Abad}},
  \bibnamefont{and} \bibinfo{author}{\bibfnamefont{A.~D.} \bibnamefont{Rojas}},
  \bibinfo{journal}{Phys. Lett.} \textbf{\bibinfo{volume}{B718}},
  \bibinfo{pages}{1413} (\bibinfo{year}{2013}), \eprint{1207.6093}.

\bibitem{King:2007pr}
\bibinfo{author}{\bibfnamefont{S.~F.} \bibnamefont{King}},
  \bibinfo{journal}{Phys. Lett.} \textbf{\bibinfo{volume}{B659}},
  \bibinfo{pages}{244} (\bibinfo{year}{2008}), \eprint{0710.0530}.

\bibitem{Grimus:2008tt}
\bibinfo{author}{\bibfnamefont{W.}~\bibnamefont{Grimus}} \bibnamefont{and}
  \bibinfo{author}{\bibfnamefont{L.}~\bibnamefont{Lavoura}},
  \bibinfo{journal}{JHEP} \textbf{\bibinfo{volume}{0809}}, \bibinfo{pages}{106}
  (\bibinfo{year}{2008}), \eprint{0809.0226}.

\bibitem{Vissani:1997pa}
\bibinfo{author}{\bibfnamefont{F.}~\bibnamefont{Vissani}}
  (\bibinfo{year}{1997}), \eprint{hep-ph/9708483}.

\bibitem{Altarelli:2009gn}
\bibinfo{author}{\bibfnamefont{G.}~\bibnamefont{Altarelli}},
  \bibinfo{author}{\bibfnamefont{F.}~\bibnamefont{Feruglio}}, \bibnamefont{and}
  \bibinfo{author}{\bibfnamefont{L.}~\bibnamefont{Merlo}},
  \bibinfo{journal}{JHEP} \textbf{\bibinfo{volume}{0905}}, \bibinfo{pages}{020}
  (\bibinfo{year}{2009}), \eprint{0903.1940}.

\bibitem{King:2012vj}
\bibinfo{author}{\bibfnamefont{S.~F.} \bibnamefont{King}},
  \bibinfo{journal}{Phys. Lett.} \textbf{\bibinfo{volume}{B718}},
  \bibinfo{pages}{136} (\bibinfo{year}{2012}), \eprint{1205.0506}.

\bibitem{King:2011ab}
\bibinfo{author}{\bibfnamefont{S.~F.} \bibnamefont{King}} \bibnamefont{and}
  \bibinfo{author}{\bibfnamefont{C.}~\bibnamefont{Luhn}},
  \bibinfo{journal}{JHEP} \textbf{\bibinfo{volume}{1203}}, \bibinfo{pages}{036}
  (\bibinfo{year}{2012}), \eprint{1112.1959}.

\bibitem{Morisi:2011pm}
\bibinfo{author}{\bibfnamefont{S.}~\bibnamefont{Morisi}},
  \bibinfo{author}{\bibfnamefont{K.~M.} \bibnamefont{Patel}}, \bibnamefont{and}
  \bibinfo{author}{\bibfnamefont{E.}~\bibnamefont{Peinado}},
  \bibinfo{journal}{Phys. Rev.} \textbf{\bibinfo{volume}{D84}},
  \bibinfo{pages}{053002} (\bibinfo{year}{2011}), \eprint{1107.0696}.

\bibitem{Boucenna:2012xb}
\bibinfo{author}{\bibfnamefont{S.~M.} \bibnamefont{Boucenna}},
  \bibinfo{author}{\bibfnamefont{S.}~\bibnamefont{Morisi}},
  \bibinfo{author}{\bibfnamefont{M.}~\bibnamefont{Tortola}}, \bibnamefont{and}
  \bibinfo{author}{\bibfnamefont{J.~W.~F.} \bibnamefont{Valle}},
  \bibinfo{journal}{Phys. Rev.} \textbf{\bibinfo{volume}{D86}},
  \bibinfo{pages}{051301} (\bibinfo{year}{2012}), \eprint{1206.2555}.

\bibitem{Ding:2012wh}
\bibinfo{author}{\bibfnamefont{G.-J.} \bibnamefont{Ding}},
  \bibinfo{author}{\bibfnamefont{S.}~\bibnamefont{Morisi}}, \bibnamefont{and}
  \bibinfo{author}{\bibfnamefont{J.~W.~F.} \bibnamefont{Valle}},
  \bibinfo{journal}{Phys. Rev.} \textbf{\bibinfo{volume}{D87}},
  \bibinfo{pages}{053013} (\bibinfo{year}{2013}), \eprint{1211.6506}.

\bibitem{King:2012in}
\bibinfo{author}{\bibfnamefont{S.~F.} \bibnamefont{King}},
  \bibinfo{author}{\bibfnamefont{C.}~\bibnamefont{Luhn}}, \bibnamefont{and}
  \bibinfo{author}{\bibfnamefont{A.~J.} \bibnamefont{Stuart}},
  \bibinfo{journal}{Nucl. Phys.} \textbf{\bibinfo{volume}{B867}},
  \bibinfo{pages}{203} (\bibinfo{year}{2013}), \eprint{1207.5741}.

\bibitem{Toorop:2011jn}
\bibinfo{author}{\bibfnamefont{R.~d.~A.} \bibnamefont{Toorop}},
  \bibinfo{author}{\bibfnamefont{F.}~\bibnamefont{Feruglio}}, \bibnamefont{and}
  \bibinfo{author}{\bibfnamefont{C.}~\bibnamefont{Hagedorn}},
  \bibinfo{journal}{Phys. Lett.} \textbf{\bibinfo{volume}{B703}},
  \bibinfo{pages}{447} (\bibinfo{year}{2011}), \eprint{1107.3486}.

\bibitem{Datta:2003qg}
\bibinfo{author}{\bibfnamefont{A.}~\bibnamefont{Datta}},
  \bibinfo{author}{\bibfnamefont{F.-S.} \bibnamefont{Ling}}, \bibnamefont{and}
  \bibinfo{author}{\bibfnamefont{P.}~\bibnamefont{Ramond}},
  \bibinfo{journal}{Nucl.Phys.} \textbf{\bibinfo{volume}{B671}},
  \bibinfo{pages}{383} (\bibinfo{year}{2003}), \eprint{hep-ph/0306002}.

\bibitem{Everett:2008et}
\bibinfo{author}{\bibfnamefont{L.~L.} \bibnamefont{Everett}} \bibnamefont{and}
  \bibinfo{author}{\bibfnamefont{A.~J.} \bibnamefont{Stuart}},
  \bibinfo{journal}{Phys.Rev.} \textbf{\bibinfo{volume}{D79}},
  \bibinfo{pages}{085005} (\bibinfo{year}{2009}), \eprint{0812.1057}.

\bibitem{Feruglio:2011qq}
\bibinfo{author}{\bibfnamefont{F.}~\bibnamefont{Feruglio}} \bibnamefont{and}
  \bibinfo{author}{\bibfnamefont{A.}~\bibnamefont{Paris}},
  \bibinfo{journal}{JHEP} \textbf{\bibinfo{volume}{1103}}, \bibinfo{pages}{101}
  (\bibinfo{year}{2011}), \eprint{1101.0393}.

\bibitem{Rodejohann:2008ir}
\bibinfo{author}{\bibfnamefont{W.}~\bibnamefont{Rodejohann}},
  \bibinfo{journal}{Phys.Lett.} \textbf{\bibinfo{volume}{B671}},
  \bibinfo{pages}{267} (\bibinfo{year}{2009}), \eprint{0810.5239}.

\bibitem{Adulpravitchai:2009bg}
\bibinfo{author}{\bibfnamefont{A.}~\bibnamefont{Adulpravitchai}},
  \bibinfo{author}{\bibfnamefont{A.}~\bibnamefont{Blum}}, \bibnamefont{and}
  \bibinfo{author}{\bibfnamefont{W.}~\bibnamefont{Rodejohann}},
  \bibinfo{journal}{New J.Phys.} \textbf{\bibinfo{volume}{11}},
  \bibinfo{pages}{063026} (\bibinfo{year}{2009}), \eprint{0903.0531}.

\bibitem{Hall:1999sn}
\bibinfo{author}{\bibfnamefont{L.~J.} \bibnamefont{Hall}},
  \bibinfo{author}{\bibfnamefont{H.}~\bibnamefont{Murayama}}, \bibnamefont{and}
  \bibinfo{author}{\bibfnamefont{N.}~\bibnamefont{Weiner}},
  \bibinfo{journal}{Phys. Rev. Lett.} \textbf{\bibinfo{volume}{84}},
  \bibinfo{pages}{2572} (\bibinfo{year}{2000}), \eprint{hep-ph/9911341}.

\bibitem{deGouvea:2012ac}
\bibinfo{author}{\bibfnamefont{A.}~\bibnamefont{de~Gouvea}} \bibnamefont{and}
  \bibinfo{author}{\bibfnamefont{H.}~\bibnamefont{Murayama}}
  (\bibinfo{year}{2012}), \eprint{1204.1249}.

\bibitem{Altarelli:2012ia}
\bibinfo{author}{\bibfnamefont{G.}~\bibnamefont{Altarelli}},
  \bibinfo{author}{\bibfnamefont{F.}~\bibnamefont{Feruglio}},
  \bibinfo{author}{\bibfnamefont{I.}~\bibnamefont{Masina}}, \bibnamefont{and}
  \bibinfo{author}{\bibfnamefont{L.}~\bibnamefont{Merlo}},
  \bibinfo{journal}{JHEP} \textbf{\bibinfo{volume}{1211}}, \bibinfo{pages}{139}
  (\bibinfo{year}{2012}), \eprint{1207.0587}.

\bibitem{Altarelli:2010gt}
\bibinfo{author}{\bibfnamefont{G.}~\bibnamefont{Altarelli}} \bibnamefont{and}
  \bibinfo{author}{\bibfnamefont{F.}~\bibnamefont{Feruglio}},
  \bibinfo{journal}{Rev. Mod. Phys.} \textbf{\bibinfo{volume}{82}},
  \bibinfo{pages}{2701} (\bibinfo{year}{2010}), \eprint{1002.0211}.

\bibitem{Ishimori:2010au}
\bibinfo{author}{\bibfnamefont{H.}~\bibnamefont{Ishimori}},
  \bibinfo{author}{\bibfnamefont{T.}~\bibnamefont{Kobayashi}},
  \bibinfo{author}{\bibfnamefont{H.}~\bibnamefont{Ohki}},
  \bibinfo{author}{\bibfnamefont{Y.}~\bibnamefont{Shimizu}},
  \bibinfo{author}{\bibfnamefont{H.}~\bibnamefont{Okada}}, \emph{et~al.},
  \bibinfo{journal}{Prog. Theor. Phys. Suppl.} \textbf{\bibinfo{volume}{183}},
  \bibinfo{pages}{1} (\bibinfo{year}{2010}), \eprint{1003.3552}.

\bibitem{King:2013eh}
\bibinfo{author}{\bibfnamefont{S.~F.} \bibnamefont{King}} \bibnamefont{and}
  \bibinfo{author}{\bibfnamefont{C.}~\bibnamefont{Luhn}},
  \bibinfo{journal}{Rept. Prog. Phys.} \textbf{\bibinfo{volume}{76}},
  \bibinfo{pages}{056201} (\bibinfo{year}{2013}), \eprint{1301.1340}.

\bibitem{Ishimori:2012zz}
\bibinfo{author}{\bibfnamefont{H.}~\bibnamefont{Ishimori}},
  \bibinfo{author}{\bibfnamefont{T.}~\bibnamefont{Kobayashi}},
  \bibinfo{author}{\bibfnamefont{H.}~\bibnamefont{Ohki}},
  \bibinfo{author}{\bibfnamefont{H.}~\bibnamefont{Okada}},
  \bibinfo{author}{\bibfnamefont{Y.}~\bibnamefont{Shimizu}}, \emph{et~al.},
  \bibinfo{journal}{Lect. Notes Phys.} \textbf{\bibinfo{volume}{858}},
  \bibinfo{pages}{pp.1} (\bibinfo{year}{2012}).

\bibitem{Ishimori:2013woa}
\bibinfo{author}{\bibfnamefont{H.}~\bibnamefont{Ishimori}},
  \bibinfo{author}{\bibfnamefont{T.}~\bibnamefont{Kobayashi}},
  \bibinfo{author}{\bibfnamefont{Y.}~\bibnamefont{Shimizu}},
  \bibinfo{author}{\bibfnamefont{H.}~\bibnamefont{Ohki}},
  \bibinfo{author}{\bibfnamefont{H.}~\bibnamefont{Okada}}, \emph{et~al.},
  \bibinfo{journal}{Fortsch. Phys.} \textbf{\bibinfo{volume}{61}},
  \bibinfo{pages}{441} (\bibinfo{year}{2013}).

\bibitem{Georgi:1982jb}
\bibinfo{author}{\bibfnamefont{H.}~\bibnamefont{Georgi}},
  \bibinfo{journal}{Front. Phys.} \textbf{\bibinfo{volume}{54}},
  \bibinfo{pages}{1} (\bibinfo{year}{1982}).

\bibitem{Frampton:1994rk}
\bibinfo{author}{\bibfnamefont{P.~H.} \bibnamefont{Frampton}} \bibnamefont{and}
  \bibinfo{author}{\bibfnamefont{T.~W.} \bibnamefont{Kephart}},
  \bibinfo{journal}{Int. J. Mod. Phys.} \textbf{\bibinfo{volume}{A10}},
  \bibinfo{pages}{4689} (\bibinfo{year}{1995}), \eprint{hep-ph/9409330}.

\bibitem{Ludl:2009ft}
\bibinfo{author}{\bibfnamefont{P.~O.} \bibnamefont{Ludl}}
  (\bibinfo{year}{2009}), \eprint{0907.5587}.

\bibitem{Grimus:2005mu}
\bibinfo{author}{\bibfnamefont{W.}~\bibnamefont{Grimus}} \bibnamefont{and}
  \bibinfo{author}{\bibfnamefont{L.}~\bibnamefont{Lavoura}},
  \bibinfo{journal}{JHEP} \textbf{\bibinfo{volume}{0508}}, \bibinfo{pages}{013}
  (\bibinfo{year}{2005}), \eprint{hep-ph/0504153}.

\bibitem{ramond}
\bibinfo{author}{\bibfnamefont{P.}~\bibnamefont{Ramond}},
  \emph{\bibinfo{title}{Group Theory: A Physicist's Survey}}
  (\bibinfo{publisher}{Cambridge University Press}, \bibinfo{year}{2010}).

\bibitem{miller}
\bibinfo{author}{\bibfnamefont{G.~A.} \bibnamefont{Miller}},
  \bibinfo{author}{\bibfnamefont{H.~F.} \bibnamefont{Dickson}},
  \bibnamefont{and} \bibinfo{author}{\bibfnamefont{L.~E.}
  \bibnamefont{Blichfeldt}}, \emph{\bibinfo{title}{Theory and Applications of
  Finite Groups}} (\bibinfo{publisher}{John Wiley and Sons. New York},
  \bibinfo{year}{1916}).

\bibitem{Hamermesh}
\bibinfo{author}{\bibfnamefont{M.}~\bibnamefont{Hamermesh}},
  \emph{\bibinfo{title}{Group Theory and its Application to Physical Problems}}
  (\bibinfo{publisher}{Addison-Wesley, Reading, Mass.}, \bibinfo{year}{1962}).

\bibitem{Fairbairn}
\bibinfo{author}{\bibfnamefont{W.~M.} \bibnamefont{Fairbairn}},
  \bibinfo{author}{\bibfnamefont{T.}~\bibnamefont{Fulton}}, \bibnamefont{and}
  \bibinfo{author}{\bibfnamefont{W.~H.} \bibnamefont{Klink}},
  \bibinfo{journal}{J. Math. Phys. 5, 1038}  (\bibinfo{year}{1964}).

\bibitem{Grimus:2003kq}
\bibinfo{author}{\bibfnamefont{W.}~\bibnamefont{Grimus}} \bibnamefont{and}
  \bibinfo{author}{\bibfnamefont{L.}~\bibnamefont{Lavoura}},
  \bibinfo{journal}{Phys. Lett.} \textbf{\bibinfo{volume}{B572}},
  \bibinfo{pages}{189} (\bibinfo{year}{2003}), \eprint{hep-ph/0305046}.

\bibitem{Kubo:2003iw}
\bibinfo{author}{\bibfnamefont{J.}~\bibnamefont{Kubo}},
  \bibinfo{author}{\bibfnamefont{A.}~\bibnamefont{Mondragon}},
  \bibinfo{author}{\bibfnamefont{M.}~\bibnamefont{Mondragon}},
  \bibnamefont{and}
  \bibinfo{author}{\bibfnamefont{E.}~\bibnamefont{Rodriguez-Jauregui}},
  \bibinfo{journal}{Prog. Theor. Phys.} \textbf{\bibinfo{volume}{109}},
  \bibinfo{pages}{795} (\bibinfo{year}{2003}), \eprint{hep-ph/0302196}.

\bibitem{Mondragon:2007af}
\bibinfo{author}{\bibfnamefont{A.}~\bibnamefont{Mondragon}},
  \bibinfo{author}{\bibfnamefont{M.}~\bibnamefont{Mondragon}},
  \bibnamefont{and} \bibinfo{author}{\bibfnamefont{E.}~\bibnamefont{Peinado}},
  \bibinfo{journal}{Phys. Rev.} \textbf{\bibinfo{volume}{D76}},
  \bibinfo{pages}{076003} (\bibinfo{year}{2007}), \eprint{0706.0354}.

\bibitem{Morisi:2009sc}
\bibinfo{author}{\bibfnamefont{S.}~\bibnamefont{Morisi}} \bibnamefont{and}
  \bibinfo{author}{\bibfnamefont{E.}~\bibnamefont{Peinado}},
  \bibinfo{journal}{Phys. Rev.} \textbf{\bibinfo{volume}{D80}},
  \bibinfo{pages}{113011} (\bibinfo{year}{2009}), \eprint{0910.4389}.

\bibitem{Morisi:2011pt}
\bibinfo{author}{\bibfnamefont{S.}~\bibnamefont{Morisi}},
  \bibinfo{author}{\bibfnamefont{E.}~\bibnamefont{Peinado}},
  \bibinfo{author}{\bibfnamefont{Y.}~\bibnamefont{Shimizu}}, \bibnamefont{and}
  \bibinfo{author}{\bibfnamefont{J.~W.~F.} \bibnamefont{Valle}},
  \bibinfo{journal}{Phys. Rev.} \textbf{\bibinfo{volume}{D84}},
  \bibinfo{pages}{036003} (\bibinfo{year}{2011}), \eprint{1104.1633}.

\bibitem{King:2013hj}
\bibinfo{author}{\bibfnamefont{S.~F.} \bibnamefont{King}},
  \bibinfo{author}{\bibfnamefont{S.}~\bibnamefont{Morisi}},
  \bibinfo{author}{\bibfnamefont{E.}~\bibnamefont{Peinado}}, \bibnamefont{and}
  \bibinfo{author}{\bibfnamefont{J.~W.~F.} \bibnamefont{Valle}},
  \bibinfo{journal}{Phys. Lett.} \textbf{\bibinfo{volume}{B724}},
  \bibinfo{pages}{68} (\bibinfo{year}{2013}), \eprint{1301.7065}.

\bibitem{Morisi:2013eca}
\bibinfo{author}{\bibfnamefont{S.}~\bibnamefont{Morisi}},
  \bibinfo{author}{\bibfnamefont{M.}~\bibnamefont{Nebot}},
  \bibinfo{author}{\bibfnamefont{K.~M.} \bibnamefont{Patel}},
  \bibinfo{author}{\bibfnamefont{E.}~\bibnamefont{Peinado}}, \bibnamefont{and}
  \bibinfo{author}{\bibfnamefont{J.~W.~F.} \bibnamefont{Valle}},
  \bibinfo{journal}{Phys. Rev.} \textbf{\bibinfo{volume}{D88}},
  \bibinfo{pages}{036001} (\bibinfo{year}{2013}), \eprint{1303.4394}.

\bibitem{Toorop:2010ex}
\bibinfo{author}{\bibfnamefont{R.}~\bibnamefont{de~Adelhart~Toorop}},
  \bibinfo{author}{\bibfnamefont{F.}~\bibnamefont{Bazzocchi}},
  \bibinfo{author}{\bibfnamefont{L.}~\bibnamefont{Merlo}}, \bibnamefont{and}
  \bibinfo{author}{\bibfnamefont{A.}~\bibnamefont{Paris}},
  \bibinfo{journal}{JHEP} \textbf{\bibinfo{volume}{1103}}, \bibinfo{pages}{035}
  (\bibinfo{year}{2011}), \eprint{1012.1791}.

\bibitem{Toorop:2010kt}
\bibinfo{author}{\bibfnamefont{R.}~\bibnamefont{de~Adelhart~Toorop}},
  \bibinfo{author}{\bibfnamefont{F.}~\bibnamefont{Bazzocchi}},
  \bibinfo{author}{\bibfnamefont{L.}~\bibnamefont{Merlo}}, \bibnamefont{and}
  \bibinfo{author}{\bibfnamefont{A.}~\bibnamefont{Paris}},
  \bibinfo{journal}{JHEP} \textbf{\bibinfo{volume}{1103}}, \bibinfo{pages}{040}
  (\bibinfo{year}{2011}), \eprint{1012.2091}.

\bibitem{Keus:2013hya}
\bibinfo{author}{\bibfnamefont{V.}~\bibnamefont{Keus}},
  \bibinfo{author}{\bibfnamefont{S.~F.} \bibnamefont{King}}, \bibnamefont{and}
  \bibinfo{author}{\bibfnamefont{S.}~\bibnamefont{Moretti}}
  (\bibinfo{year}{2013}), \eprint{1310.8253}.

\bibitem{Glashow:1979nm}
\bibinfo{author}{\bibfnamefont{S.~L.} \bibnamefont{Glashow}},
  \bibinfo{journal}{NATO Adv. Study Inst. Ser. B Phys.}
  \textbf{\bibinfo{volume}{59}}, \bibinfo{pages}{687} (\bibinfo{year}{1980}),
  \bibinfo{note}{preliminary version given at Colloquium in Honor of A.
  Visconti, Marseille-Luminy Univ., Jul 1979}.

\bibitem{King:1998jw}
\bibinfo{author}{\bibfnamefont{S.~F.} \bibnamefont{King}},
  \bibinfo{journal}{Phys. Lett.} \textbf{\bibinfo{volume}{B439}},
  \bibinfo{pages}{350} (\bibinfo{year}{1998}), \eprint{hep-ph/9806440}.

\bibitem{King:1999cm}
\bibinfo{author}{\bibfnamefont{S.~F.} \bibnamefont{King}},
  \bibinfo{journal}{Nucl. Phys.} \textbf{\bibinfo{volume}{B562}},
  \bibinfo{pages}{57} (\bibinfo{year}{1999}), \eprint{hep-ph/9904210}.

\bibitem{King:1999mb}
\bibinfo{author}{\bibfnamefont{S.~F.} \bibnamefont{King}},
  \bibinfo{journal}{Nucl. Phys.} \textbf{\bibinfo{volume}{B576}},
  \bibinfo{pages}{85} (\bibinfo{year}{2000}), \eprint{hep-ph/9912492}.

\bibitem{King:2002nf}
\bibinfo{author}{\bibfnamefont{S.~F.} \bibnamefont{King}},
  \bibinfo{journal}{JHEP} \textbf{\bibinfo{volume}{0209}}, \bibinfo{pages}{011}
  (\bibinfo{year}{2002}), \eprint{hep-ph/0204360}.

\bibitem{King:2005bj}
\bibinfo{author}{\bibfnamefont{S.~F.} \bibnamefont{King}},
  \bibinfo{journal}{JHEP} \textbf{\bibinfo{volume}{0508}}, \bibinfo{pages}{105}
  (\bibinfo{year}{2005}), \eprint{hep-ph/0506297}.

\bibitem{Antusch:2011ic}
\bibinfo{author}{\bibfnamefont{S.}~\bibnamefont{Antusch}},
  \bibinfo{author}{\bibfnamefont{S.~F.} \bibnamefont{King}},
  \bibinfo{author}{\bibfnamefont{C.}~\bibnamefont{Luhn}}, \bibnamefont{and}
  \bibinfo{author}{\bibfnamefont{M.}~\bibnamefont{Spinrath}},
  \bibinfo{journal}{Nucl. Phys.} \textbf{\bibinfo{volume}{B856}},
  \bibinfo{pages}{328} (\bibinfo{year}{2012}), \eprint{1108.4278}.

\bibitem{Antusch:2013wn}
\bibinfo{author}{\bibfnamefont{S.}~\bibnamefont{Antusch}},
  \bibinfo{author}{\bibfnamefont{S.~F.} \bibnamefont{King}}, \bibnamefont{and}
  \bibinfo{author}{\bibfnamefont{M.}~\bibnamefont{Spinrath}},
  \bibinfo{journal}{Phys. Rev.} \textbf{\bibinfo{volume}{D87}},
  \bibinfo{pages}{096018} (\bibinfo{year}{2013}), \eprint{1301.6764}.

\bibitem{King:2013iva}
\bibinfo{author}{\bibfnamefont{S.~F.} \bibnamefont{King}},
  \bibinfo{journal}{JHEP} \textbf{\bibinfo{volume}{1307}}, \bibinfo{pages}{137}
  (\bibinfo{year}{2013}), \eprint{1304.6264}.

\bibitem{King:2013xba}
\bibinfo{author}{\bibfnamefont{S.~F.} \bibnamefont{King}}
  (\bibinfo{year}{2013}), \eprint{1305.4846}.

\bibitem{King:2009ap}
\bibinfo{author}{\bibfnamefont{S.~F.} \bibnamefont{King}} \bibnamefont{and}
  \bibinfo{author}{\bibfnamefont{C.}~\bibnamefont{Luhn}},
  \bibinfo{journal}{JHEP} \textbf{\bibinfo{volume}{0910}}, \bibinfo{pages}{093}
  (\bibinfo{year}{2009}), \eprint{0908.1897}.

\bibitem{King:2013vna}
\bibinfo{author}{\bibfnamefont{S.~F.} \bibnamefont{King}},
  \bibinfo{author}{\bibfnamefont{T.}~\bibnamefont{Neder}}, \bibnamefont{and}
  \bibinfo{author}{\bibfnamefont{A.~J.} \bibnamefont{Stuart}},
  \bibinfo{journal}{Phys. Lett.} \textbf{\bibinfo{volume}{B726}},
  \bibinfo{pages}{312} (\bibinfo{year}{2013}), \eprint{1305.3200}.

\bibitem{Hagedorn:2012pg}
\bibinfo{author}{\bibfnamefont{C.}~\bibnamefont{Hagedorn}} \bibnamefont{and}
  \bibinfo{author}{\bibfnamefont{D.}~\bibnamefont{Meloni}},
  \bibinfo{journal}{Nucl. Phys.} \textbf{\bibinfo{volume}{B862}},
  \bibinfo{pages}{691} (\bibinfo{year}{2012}), \eprint{1204.0715}.

\bibitem{Araki:2013rkf}
\bibinfo{author}{\bibfnamefont{T.}~\bibnamefont{Araki}},
  \bibinfo{author}{\bibfnamefont{H.}~\bibnamefont{Ishida}},
  \bibinfo{author}{\bibfnamefont{H.}~\bibnamefont{Ishimori}},
  \bibinfo{author}{\bibfnamefont{T.}~\bibnamefont{Kobayashi}},
  \bibnamefont{and}
  \bibinfo{author}{\bibfnamefont{A.}~\bibnamefont{Ogasahara}},
  \bibinfo{journal}{Phys. Rev.} \textbf{\bibinfo{volume}{D88}},
  \bibinfo{pages}{096002} (\bibinfo{year}{2013}), \eprint{1309.4217}.

\bibitem{Holthausen:2013vba}
\bibinfo{author}{\bibfnamefont{M.}~\bibnamefont{Holthausen}} \bibnamefont{and}
  \bibinfo{author}{\bibfnamefont{K.~S.} \bibnamefont{Lim}},
  \bibinfo{journal}{Phys. Rev.} \textbf{\bibinfo{volume}{D88}},
  \bibinfo{pages}{033018} (\bibinfo{year}{2013}), \eprint{1306.4356}.

\bibitem{deMedeirosVarzielas:2011zw}
\bibinfo{author}{\bibfnamefont{I.}~\bibnamefont{de~Medeiros~Varzielas}}
  \bibnamefont{and}
  \bibinfo{author}{\bibfnamefont{D.}~\bibnamefont{Emmanuel-Costa}},
  \bibinfo{journal}{Phys. Rev.} \textbf{\bibinfo{volume}{D84}},
  \bibinfo{pages}{117901} (\bibinfo{year}{2011}), \eprint{1106.5477}.

\bibitem{Varzielas:2012nn}
\bibinfo{author}{\bibfnamefont{I.}~\bibnamefont{de~Medeiros~Varzielas}},
  \bibinfo{author}{\bibfnamefont{D.}~\bibnamefont{Emmanuel-Costa}},
  \bibnamefont{and} \bibinfo{author}{\bibfnamefont{P.}~\bibnamefont{Leser}},
  \bibinfo{journal}{Phys. Lett.} \textbf{\bibinfo{volume}{B716}},
  \bibinfo{pages}{193} (\bibinfo{year}{2012}), \eprint{1204.3633}.

\bibitem{Varzielas:2012pd}
\bibinfo{author}{\bibfnamefont{I.}~\bibnamefont{de~Medeiros~Varzielas}},
  \bibinfo{journal}{JHEP} \textbf{\bibinfo{volume}{1208}}, \bibinfo{pages}{055}
  (\bibinfo{year}{2012}), \eprint{1205.3780}.

\bibitem{Bhattacharyya:2012pi}
\bibinfo{author}{\bibfnamefont{G.}~\bibnamefont{Bhattacharyya}},
  \bibinfo{author}{\bibfnamefont{I.}~\bibnamefont{de~Medeiros~Varzielas}},
  \bibnamefont{and} \bibinfo{author}{\bibfnamefont{P.}~\bibnamefont{Leser}},
  \bibinfo{journal}{Phys. Rev. Lett.} \textbf{\bibinfo{volume}{109}},
  \bibinfo{pages}{241603} (\bibinfo{year}{2012}), \eprint{1210.0545}.

\bibitem{Branco:2011zb}
\bibinfo{author}{\bibfnamefont{G.~C.} \bibnamefont{Branco}},
  \bibinfo{author}{\bibfnamefont{R.~G.} \bibnamefont{Felipe}},
  \bibnamefont{and} \bibinfo{author}{\bibfnamefont{F.~R.}
  \bibnamefont{Joaquim}}, \bibinfo{journal}{Rev. Mod. Phys.}
  \textbf{\bibinfo{volume}{84}}, \bibinfo{pages}{515} (\bibinfo{year}{2012}),
  \eprint{1111.5332}.

\bibitem{Grimus:2003yn}
\bibinfo{author}{\bibfnamefont{W.}~\bibnamefont{Grimus}} \bibnamefont{and}
  \bibinfo{author}{\bibfnamefont{L.}~\bibnamefont{Lavoura}},
  \bibinfo{journal}{Phys. Lett.} \textbf{\bibinfo{volume}{B579}},
  \bibinfo{pages}{113} (\bibinfo{year}{2004}), \eprint{hep-ph/0305309}.

\bibitem{Mohapatra:2012tb}
\bibinfo{author}{\bibfnamefont{R.~N.} \bibnamefont{Mohapatra}}
  \bibnamefont{and} \bibinfo{author}{\bibfnamefont{C.~C.} \bibnamefont{Nishi}},
  \bibinfo{journal}{Phys. Rev.} \textbf{\bibinfo{volume}{D86}},
  \bibinfo{pages}{073007} (\bibinfo{year}{2012}), \eprint{1208.2875}.

\bibitem{Holthausen:2012dk}
\bibinfo{author}{\bibfnamefont{M.}~\bibnamefont{Holthausen}},
  \bibinfo{author}{\bibfnamefont{M.}~\bibnamefont{Lindner}}, \bibnamefont{and}
  \bibinfo{author}{\bibfnamefont{M.~A.} \bibnamefont{Schmidt}},
  \bibinfo{journal}{JHEP} \textbf{\bibinfo{volume}{1304}}, \bibinfo{pages}{122}
  (\bibinfo{year}{2013}), \eprint{1211.6953}.

\bibitem{Feruglio:2012cw}
\bibinfo{author}{\bibfnamefont{F.}~\bibnamefont{Feruglio}},
  \bibinfo{author}{\bibfnamefont{C.}~\bibnamefont{Hagedorn}}, \bibnamefont{and}
  \bibinfo{author}{\bibfnamefont{R.}~\bibnamefont{Ziegler}},
  \bibinfo{journal}{JHEP} \textbf{\bibinfo{volume}{1307}}, \bibinfo{pages}{027}
  (\bibinfo{year}{2013}), \eprint{1211.5560}.

\bibitem{Ding:2013hpa}
\bibinfo{author}{\bibfnamefont{G.-J.} \bibnamefont{Ding}},
  \bibinfo{author}{\bibfnamefont{S.~F.} \bibnamefont{King}},
  \bibinfo{author}{\bibfnamefont{C.}~\bibnamefont{Luhn}}, \bibnamefont{and}
  \bibinfo{author}{\bibfnamefont{A.~J.} \bibnamefont{Stuart}},
  \bibinfo{journal}{JHEP} \textbf{\bibinfo{volume}{1305}}, \bibinfo{pages}{084}
  (\bibinfo{year}{2013}), \eprint{1303.6180}.

\bibitem{Feruglio:2013hia}
\bibinfo{author}{\bibfnamefont{F.}~\bibnamefont{Feruglio}},
  \bibinfo{author}{\bibfnamefont{C.}~\bibnamefont{Hagedorn}}, \bibnamefont{and}
  \bibinfo{author}{\bibfnamefont{R.}~\bibnamefont{Ziegler}}
  (\bibinfo{year}{2013}), \eprint{1303.7178}.

\bibitem{Ding:2013bpa}
\bibinfo{author}{\bibfnamefont{G.-J.} \bibnamefont{Ding}},
  \bibinfo{author}{\bibfnamefont{S.~F.} \bibnamefont{King}}, \bibnamefont{and}
  \bibinfo{author}{\bibfnamefont{A.~J.} \bibnamefont{Stuart}},
  \bibinfo{journal}{JHEP} \textbf{\bibinfo{volume}{1312}}, \bibinfo{pages}{006}
  (\bibinfo{year}{2013}), \eprint{1307.4212}.

\bibitem{Meroni:2012ty}
\bibinfo{author}{\bibfnamefont{A.}~\bibnamefont{Meroni}},
  \bibinfo{author}{\bibfnamefont{S.~T.} \bibnamefont{Petcov}},
  \bibnamefont{and} \bibinfo{author}{\bibfnamefont{M.}~\bibnamefont{Spinrath}},
  \bibinfo{journal}{Phys. Rev.} \textbf{\bibinfo{volume}{D86}},
  \bibinfo{pages}{113003} (\bibinfo{year}{2012}), \eprint{1205.5241}.

\bibitem{Li:2013jya}
\bibinfo{author}{\bibfnamefont{C.-C.} \bibnamefont{Li}} \bibnamefont{and}
  \bibinfo{author}{\bibfnamefont{G.-J.} \bibnamefont{Ding}}
  (\bibinfo{year}{2013}), \eprint{1312.4401}.

\bibitem{Luhn:2013lkn}
\bibinfo{author}{\bibfnamefont{C.}~\bibnamefont{Luhn}}, \bibinfo{journal}{Nucl.
  Phys.} \textbf{\bibinfo{volume}{B875}}, \bibinfo{pages}{80}
  (\bibinfo{year}{2013}), \eprint{1306.2358}.

\bibitem{Ding:2013nsa}
\bibinfo{author}{\bibfnamefont{G.-J.} \bibnamefont{Ding}} \bibnamefont{and}
  \bibinfo{author}{\bibfnamefont{Y.-L.} \bibnamefont{Zhou}}
  (\bibinfo{year}{2013}), \eprint{1312.5222}.

\bibitem{Antusch:2011sx}
\bibinfo{author}{\bibfnamefont{S.}~\bibnamefont{Antusch}},
  \bibinfo{author}{\bibfnamefont{S.~F.} \bibnamefont{King}},
  \bibinfo{author}{\bibfnamefont{C.}~\bibnamefont{Luhn}}, \bibnamefont{and}
  \bibinfo{author}{\bibfnamefont{M.}~\bibnamefont{Spinrath}},
  \bibinfo{journal}{Nucl. Phys.} \textbf{\bibinfo{volume}{B850}},
  \bibinfo{pages}{477} (\bibinfo{year}{2011}), \eprint{1103.5930}.

\bibitem{Antusch:2013rla}
\bibinfo{author}{\bibfnamefont{S.}~\bibnamefont{Antusch}},
  \bibinfo{author}{\bibfnamefont{M.}~\bibnamefont{Holthausen}},
  \bibinfo{author}{\bibfnamefont{M.~A.} \bibnamefont{Schmidt}},
  \bibnamefont{and} \bibinfo{author}{\bibfnamefont{M.}~\bibnamefont{Spinrath}},
  \bibinfo{journal}{Nucl. Phys.} \textbf{\bibinfo{volume}{B877}},
  \bibinfo{pages}{752} (\bibinfo{year}{2013}), \eprint{1307.0710}.

\bibitem{King:2013hoa}
\bibinfo{author}{\bibfnamefont{S.~F.} \bibnamefont{King}}
  (\bibinfo{year}{2013}), \eprint{1311.3295}.

\bibitem{Ma:2006sk}
\bibinfo{author}{\bibfnamefont{E.}~\bibnamefont{Ma}},
  \bibinfo{author}{\bibfnamefont{H.}~\bibnamefont{Sawanaka}}, \bibnamefont{and}
  \bibinfo{author}{\bibfnamefont{M.}~\bibnamefont{Tanimoto}},
  \bibinfo{journal}{Phys. Lett.} \textbf{\bibinfo{volume}{B641}},
  \bibinfo{pages}{301} (\bibinfo{year}{2006}), \eprint{hep-ph/0606103}.

\bibitem{Chen:2007afa}
\bibinfo{author}{\bibfnamefont{M.-C.} \bibnamefont{Chen}} \bibnamefont{and}
  \bibinfo{author}{\bibfnamefont{K.}~\bibnamefont{Mahanthappa}},
  \bibinfo{journal}{Phys. Lett.} \textbf{\bibinfo{volume}{B652}},
  \bibinfo{pages}{34} (\bibinfo{year}{2007}), \eprint{0705.0714}.

\bibitem{Altarelli:2008bg}
\bibinfo{author}{\bibfnamefont{G.}~\bibnamefont{Altarelli}},
  \bibinfo{author}{\bibfnamefont{F.}~\bibnamefont{Feruglio}}, \bibnamefont{and}
  \bibinfo{author}{\bibfnamefont{C.}~\bibnamefont{Hagedorn}},
  \bibinfo{journal}{JHEP} \textbf{\bibinfo{volume}{0803}}, \bibinfo{pages}{052}
  (\bibinfo{year}{2008}), \eprint{0802.0090}.

\bibitem{Ishimori:2008fi}
\bibinfo{author}{\bibfnamefont{H.}~\bibnamefont{Ishimori}},
  \bibinfo{author}{\bibfnamefont{Y.}~\bibnamefont{Shimizu}}, \bibnamefont{and}
  \bibinfo{author}{\bibfnamefont{M.}~\bibnamefont{Tanimoto}},
  \bibinfo{journal}{Prog. Theor. Phys.} \textbf{\bibinfo{volume}{121}},
  \bibinfo{pages}{769} (\bibinfo{year}{2009}), \eprint{0812.5031}.

\bibitem{Burrows:2009pi}
\bibinfo{author}{\bibfnamefont{T.~J.} \bibnamefont{Burrows}} \bibnamefont{and}
  \bibinfo{author}{\bibfnamefont{S.~F.} \bibnamefont{King}},
  \bibinfo{journal}{Nucl. Phys.} \textbf{\bibinfo{volume}{B835}},
  \bibinfo{pages}{174} (\bibinfo{year}{2010}), \eprint{0909.1433}.

\bibitem{Hagedorn:2010th}
\bibinfo{author}{\bibfnamefont{C.}~\bibnamefont{Hagedorn}},
  \bibinfo{author}{\bibfnamefont{S.~F.} \bibnamefont{King}}, \bibnamefont{and}
  \bibinfo{author}{\bibfnamefont{C.}~\bibnamefont{Luhn}},
  \bibinfo{journal}{JHEP} \textbf{\bibinfo{volume}{1006}}, \bibinfo{pages}{048}
  (\bibinfo{year}{2010}), \eprint{1003.4249}.

\bibitem{Cooper:2012wf}
\bibinfo{author}{\bibfnamefont{I.~K.} \bibnamefont{Cooper}},
  \bibinfo{author}{\bibfnamefont{S.~F.} \bibnamefont{King}}, \bibnamefont{and}
  \bibinfo{author}{\bibfnamefont{C.}~\bibnamefont{Luhn}},
  \bibinfo{journal}{JHEP} \textbf{\bibinfo{volume}{1206}}, \bibinfo{pages}{130}
  (\bibinfo{year}{2012}), \eprint{1203.1324}.

\bibitem{Hagedorn:2012ut}
\bibinfo{author}{\bibfnamefont{C.}~\bibnamefont{Hagedorn}},
  \bibinfo{author}{\bibfnamefont{S.~F.} \bibnamefont{King}}, \bibnamefont{and}
  \bibinfo{author}{\bibfnamefont{C.}~\bibnamefont{Luhn}},
  \bibinfo{journal}{Phys. Lett.} \textbf{\bibinfo{volume}{B717}},
  \bibinfo{pages}{207} (\bibinfo{year}{2012}), \eprint{1205.3114}.

\bibitem{Antusch:2013rxa}
\bibinfo{author}{\bibfnamefont{S.}~\bibnamefont{Antusch}},
  \bibinfo{author}{\bibfnamefont{S.~F.} \bibnamefont{King}}, \bibnamefont{and}
  \bibinfo{author}{\bibfnamefont{M.}~\bibnamefont{Spinrath}}
  (\bibinfo{year}{2013}), \eprint{1311.0877}.

\bibitem{Ross:2002fb}
\bibinfo{author}{\bibfnamefont{G.~G.} \bibnamefont{Ross}} \bibnamefont{and}
  \bibinfo{author}{\bibfnamefont{L.}~\bibnamefont{Velasco-Sevilla}},
  \bibinfo{journal}{Nucl. Phys.} \textbf{\bibinfo{volume}{B653}},
  \bibinfo{pages}{3} (\bibinfo{year}{2003}), \eprint{hep-ph/0208218}.

\bibitem{King:2003rf}
\bibinfo{author}{\bibfnamefont{S.~F.} \bibnamefont{King}} \bibnamefont{and}
  \bibinfo{author}{\bibfnamefont{G.~G.} \bibnamefont{Ross}},
  \bibinfo{journal}{Phys. Lett.} \textbf{\bibinfo{volume}{B574}},
  \bibinfo{pages}{239} (\bibinfo{year}{2003}), \eprint{hep-ph/0307190}.

\bibitem{deMedeirosVarzielas:2005ax}
\bibinfo{author}{\bibfnamefont{I.}~\bibnamefont{de~Medeiros~Varzielas}}
  \bibnamefont{and} \bibinfo{author}{\bibfnamefont{G.~G.} \bibnamefont{Ross}},
  \bibinfo{journal}{Nucl. Phys.} \textbf{\bibinfo{volume}{B733}},
  \bibinfo{pages}{31} (\bibinfo{year}{2006}), \eprint{hep-ph/0507176}.

\bibitem{Morisi:2007ft}
\bibinfo{author}{\bibfnamefont{S.}~\bibnamefont{Morisi}},
  \bibinfo{author}{\bibfnamefont{M.}~\bibnamefont{Picariello}},
  \bibnamefont{and}
  \bibinfo{author}{\bibfnamefont{E.}~\bibnamefont{Torrente-Lujan}},
  \bibinfo{journal}{Phys. Rev.} \textbf{\bibinfo{volume}{D75}},
  \bibinfo{pages}{075015} (\bibinfo{year}{2007}), \eprint{hep-ph/0702034}.

\bibitem{Bazzocchi:2008sp}
\bibinfo{author}{\bibfnamefont{F.}~\bibnamefont{Bazzocchi}},
  \bibinfo{author}{\bibfnamefont{M.}~\bibnamefont{Frigerio}}, \bibnamefont{and}
  \bibinfo{author}{\bibfnamefont{S.}~\bibnamefont{Morisi}},
  \bibinfo{journal}{Phys. Rev.} \textbf{\bibinfo{volume}{D78}},
  \bibinfo{pages}{116018} (\bibinfo{year}{2008}), \eprint{0809.3573}.

\bibitem{Blankenburg:2011vw}
\bibinfo{author}{\bibfnamefont{G.}~\bibnamefont{Blankenburg}} \bibnamefont{and}
  \bibinfo{author}{\bibfnamefont{S.}~\bibnamefont{Morisi}},
  \bibinfo{journal}{JHEP} \textbf{\bibinfo{volume}{1201}}, \bibinfo{pages}{016}
  (\bibinfo{year}{2012}), \eprint{1109.3396}.

\bibitem{Hagedorn:2006ug}
\bibinfo{author}{\bibfnamefont{C.}~\bibnamefont{Hagedorn}},
  \bibinfo{author}{\bibfnamefont{M.}~\bibnamefont{Lindner}}, \bibnamefont{and}
  \bibinfo{author}{\bibfnamefont{R.~N.} \bibnamefont{Mohapatra}},
  \bibinfo{journal}{JHEP} \textbf{\bibinfo{volume}{0606}}, \bibinfo{pages}{042}
  (\bibinfo{year}{2006}), \eprint{hep-ph/0602244}.

\bibitem{Grimus:2008tm}
\bibinfo{author}{\bibfnamefont{W.}~\bibnamefont{Grimus}} \bibnamefont{and}
  \bibinfo{author}{\bibfnamefont{H.}~\bibnamefont{Kuhbock}},
  \bibinfo{journal}{Phys. Rev.} \textbf{\bibinfo{volume}{D77}},
  \bibinfo{pages}{055008} (\bibinfo{year}{2008}), \eprint{0710.1585}.

\bibitem{King:2009mk}
\bibinfo{author}{\bibfnamefont{S.~F.} \bibnamefont{King}} \bibnamefont{and}
  \bibinfo{author}{\bibfnamefont{C.}~\bibnamefont{Luhn}},
  \bibinfo{journal}{Nucl. Phys.} \textbf{\bibinfo{volume}{B820}},
  \bibinfo{pages}{269} (\bibinfo{year}{2009}), \eprint{0905.1686}.

\bibitem{Dutta:2009bj}
\bibinfo{author}{\bibfnamefont{B.}~\bibnamefont{Dutta}},
  \bibinfo{author}{\bibfnamefont{Y.}~\bibnamefont{Mimura}}, \bibnamefont{and}
  \bibinfo{author}{\bibfnamefont{R.~N.} \bibnamefont{Mohapatra}},
  \bibinfo{journal}{JHEP} \textbf{\bibinfo{volume}{1005}}, \bibinfo{pages}{034}
  (\bibinfo{year}{2010}), \eprint{0911.2242}.

\bibitem{Altarelli:2010at}
\bibinfo{author}{\bibfnamefont{G.}~\bibnamefont{Altarelli}} \bibnamefont{and}
  \bibinfo{author}{\bibfnamefont{G.}~\bibnamefont{Blankenburg}},
  \bibinfo{journal}{JHEP} \textbf{\bibinfo{volume}{1103}}, \bibinfo{pages}{133}
  (\bibinfo{year}{2011}), \eprint{1012.2697}.

\bibitem{BhupalDev:2011gi}
\bibinfo{author}{\bibfnamefont{P.~S.} \bibnamefont{Bhupal~Dev}},
  \bibinfo{author}{\bibfnamefont{R.~N.} \bibnamefont{Mohapatra}},
  \bibnamefont{and} \bibinfo{author}{\bibfnamefont{M.}~\bibnamefont{Severson}},
  \bibinfo{journal}{Phys. Rev.} \textbf{\bibinfo{volume}{D84}},
  \bibinfo{pages}{053005} (\bibinfo{year}{2011}), \eprint{1107.2378}.

\bibitem{Dermisek:1999vy}
\bibinfo{author}{\bibfnamefont{R.}~\bibnamefont{Dermisek}} \bibnamefont{and}
  \bibinfo{author}{\bibfnamefont{S.}~\bibnamefont{Raby}},
  \bibinfo{journal}{Phys. Rev.} \textbf{\bibinfo{volume}{D62}},
  \bibinfo{pages}{015007} (\bibinfo{year}{2000}), \eprint{hep-ph/9911275}.

\bibitem{King:2011zj}
\bibinfo{author}{\bibfnamefont{S.~F.} \bibnamefont{King}} \bibnamefont{and}
  \bibinfo{author}{\bibfnamefont{C.}~\bibnamefont{Luhn}},
  \bibinfo{journal}{JHEP} \textbf{\bibinfo{volume}{1109}}, \bibinfo{pages}{042}
  (\bibinfo{year}{2011}), \eprint{1107.5332}.

\bibitem{Hernandez:2012ra}
\bibinfo{author}{\bibfnamefont{D.}~\bibnamefont{Hernandez}} \bibnamefont{and}
  \bibinfo{author}{\bibfnamefont{A.~Y.} \bibnamefont{Smirnov}},
  \bibinfo{journal}{Phys. Rev.} \textbf{\bibinfo{volume}{D86}},
  \bibinfo{pages}{053014} (\bibinfo{year}{2012}), \eprint{1204.0445}.

\bibitem{Ballett:2013wya}
\bibinfo{author}{\bibfnamefont{P.}~\bibnamefont{Ballett}},
  \bibinfo{author}{\bibfnamefont{S.~F.} \bibnamefont{King}},
  \bibinfo{author}{\bibfnamefont{C.}~\bibnamefont{Luhn}},
  \bibinfo{author}{\bibfnamefont{S.}~\bibnamefont{Pascoli}}, \bibnamefont{and}
  \bibinfo{author}{\bibfnamefont{M.~A.} \bibnamefont{Schmidt}}
  (\bibinfo{year}{2013}), \eprint{1308.4314}.

\bibitem{Hall:2013yha}
\bibinfo{author}{\bibfnamefont{L.}~\bibnamefont{Hall}} \bibnamefont{and}
  \bibinfo{author}{\bibfnamefont{G.}~\bibnamefont{Ross}},
  \bibinfo{journal}{JHEP} \textbf{\bibinfo{volume}{1311}}, \bibinfo{pages}{091}
  (\bibinfo{year}{2013}), \eprint{1303.6962}.

\bibitem{Masina:2005hf}
\bibinfo{author}{\bibfnamefont{I.}~\bibnamefont{Masina}},
  \bibinfo{journal}{Phys. Lett.} \textbf{\bibinfo{volume}{B633}},
  \bibinfo{pages}{134} (\bibinfo{year}{2006}), \eprint{hep-ph/0508031}.

\bibitem{Antusch:2005kw}
\bibinfo{author}{\bibfnamefont{S.}~\bibnamefont{Antusch}} \bibnamefont{and}
  \bibinfo{author}{\bibfnamefont{S.~F.} \bibnamefont{King}},
  \bibinfo{journal}{Phys. Lett.} \textbf{\bibinfo{volume}{B631}},
  \bibinfo{pages}{42} (\bibinfo{year}{2005}), \eprint{hep-ph/0508044}.

\bibitem{Marzocca:2013cr}
\bibinfo{author}{\bibfnamefont{D.}~\bibnamefont{Marzocca}},
  \bibinfo{author}{\bibfnamefont{S.}~\bibnamefont{Petcov}},
  \bibinfo{author}{\bibfnamefont{A.}~\bibnamefont{Romanino}}, \bibnamefont{and}
  \bibinfo{author}{\bibfnamefont{M.}~\bibnamefont{Sevilla}},
  \bibinfo{journal}{JHEP} \textbf{\bibinfo{volume}{1305}}, \bibinfo{pages}{073}
  (\bibinfo{year}{2013}), \eprint{1302.0423}.

\bibitem{Hirsch:2008rp}
\bibinfo{author}{\bibfnamefont{M.}~\bibnamefont{Hirsch}},
  \bibinfo{author}{\bibfnamefont{S.}~\bibnamefont{Morisi}}, \bibnamefont{and}
  \bibinfo{author}{\bibfnamefont{J.~W.~F.} \bibnamefont{Valle}},
  \bibinfo{journal}{Phys. Rev.} \textbf{\bibinfo{volume}{D78}},
  \bibinfo{pages}{093007} (\bibinfo{year}{2008}), \eprint{0804.1521}.

\bibitem{Bazzocchi:2009da}
\bibinfo{author}{\bibfnamefont{F.}~\bibnamefont{Bazzocchi}},
  \bibinfo{author}{\bibfnamefont{L.}~\bibnamefont{Merlo}}, \bibnamefont{and}
  \bibinfo{author}{\bibfnamefont{S.}~\bibnamefont{Morisi}},
  \bibinfo{journal}{Phys. Rev.} \textbf{\bibinfo{volume}{D80}},
  \bibinfo{pages}{053003} (\bibinfo{year}{2009}), \eprint{0902.2849}.

\bibitem{Altarelli:2009kr}
\bibinfo{author}{\bibfnamefont{G.}~\bibnamefont{Altarelli}} \bibnamefont{and}
  \bibinfo{author}{\bibfnamefont{D.}~\bibnamefont{Meloni}},
  \bibinfo{journal}{J. Phys.} \textbf{\bibinfo{volume}{G36}},
  \bibinfo{pages}{085005} (\bibinfo{year}{2009}), \eprint{0905.0620}.

\bibitem{Chen:2009um}
\bibinfo{author}{\bibfnamefont{M.-C.} \bibnamefont{Chen}} \bibnamefont{and}
  \bibinfo{author}{\bibfnamefont{S.~F.} \bibnamefont{King}},
  \bibinfo{journal}{JHEP} \textbf{\bibinfo{volume}{0906}}, \bibinfo{pages}{072}
  (\bibinfo{year}{2009}), \eprint{0903.0125}.

\bibitem{BarryRodejohann-Classification}
\bibinfo{author}{\bibfnamefont{J.}~\bibnamefont{Barry}} \bibnamefont{and}
  \bibinfo{author}{\bibfnamefont{W.}~\bibnamefont{Rodejohann}},
  \bibinfo{journal}{Phys. Rev.} \textbf{\bibinfo{volume}{D81}},
  \bibinfo{pages}{093002} (\bibinfo{year}{2010}), \eprint{1003.2385}.

\bibitem{Barry:2010zk}
\bibinfo{author}{\bibfnamefont{J.}~\bibnamefont{Barry}} \bibnamefont{and}
  \bibinfo{author}{\bibfnamefont{W.}~\bibnamefont{Rodejohann}},
  \bibinfo{journal}{Nucl. Phys.} \textbf{\bibinfo{volume}{B842}},
  \bibinfo{pages}{33} (\bibinfo{year}{2011}), \eprint{1007.5217}.

\bibitem{Dorame:2011eb}
\bibinfo{author}{\bibfnamefont{L.}~\bibnamefont{Dorame}},
  \bibinfo{author}{\bibfnamefont{D.}~\bibnamefont{Meloni}},
  \bibinfo{author}{\bibfnamefont{S.}~\bibnamefont{Morisi}},
  \bibinfo{author}{\bibfnamefont{E.}~\bibnamefont{Peinado}}, \bibnamefont{and}
  \bibinfo{author}{\bibfnamefont{J.~W.~F.} \bibnamefont{Valle}},
  \bibinfo{journal}{Nucl. Phys.} \textbf{\bibinfo{volume}{B861}},
  \bibinfo{pages}{259} (\bibinfo{year}{2012}), \eprint{1111.5614}.

\bibitem{King:2013psa}
\bibinfo{author}{\bibfnamefont{S.~F.} \bibnamefont{King}},
  \bibinfo{author}{\bibfnamefont{A.}~\bibnamefont{Merle}}, \bibnamefont{and}
  \bibinfo{author}{\bibfnamefont{A.~J.} \bibnamefont{Stuart}},
  \bibinfo{journal}{JHEP} \textbf{\bibinfo{volume}{1312}}, \bibinfo{pages}{005}
  (\bibinfo{year}{2013}), \eprint{1307.2901}.

\bibitem{Lazarides:1980nt}
\bibinfo{author}{\bibfnamefont{G.}~\bibnamefont{Lazarides}},
  \bibinfo{author}{\bibfnamefont{Q.}~\bibnamefont{Shafi}}, \bibnamefont{and}
  \bibinfo{author}{\bibfnamefont{C.}~\bibnamefont{Wetterich}},
  \bibinfo{journal}{Nucl. Phys.} \textbf{\bibinfo{volume}{B181}},
  \bibinfo{pages}{287} (\bibinfo{year}{1981}).

\bibitem{Ramond:1979py}
\bibinfo{author}{\bibfnamefont{P.}~\bibnamefont{Ramond}},
  \emph{\bibinfo{title}{{The Family Group in Grand Unified Theories}}}
  (\bibinfo{year}{1979}), \eprint{hep-ph/9809459}.

\bibitem{Ma:2002pf}
\bibinfo{author}{\bibfnamefont{E.}~\bibnamefont{Ma}} \bibnamefont{and}
  \bibinfo{author}{\bibfnamefont{D.~P.} \bibnamefont{Roy}},
  \bibinfo{journal}{Nucl. Phys.} \textbf{\bibinfo{volume}{B644}},
  \bibinfo{pages}{290} (\bibinfo{year}{2002}), \eprint{hep-ph/0206150}.

\bibitem{GonzalezGarcia:1988rw}
\bibinfo{author}{\bibfnamefont{M.~C.} \bibnamefont{Gonzalez-Garcia}}
  \bibnamefont{and} \bibinfo{author}{\bibfnamefont{J.~W.~F.}
  \bibnamefont{Valle}}, \bibinfo{journal}{Phys. Lett.}
  \textbf{\bibinfo{volume}{B216}}, \bibinfo{pages}{360} (\bibinfo{year}{1989}).

\bibitem{Lindner:2005kr}
\bibinfo{author}{\bibfnamefont{M.}~\bibnamefont{Lindner}},
  \bibinfo{author}{\bibfnamefont{A.}~\bibnamefont{Merle}}, \bibnamefont{and}
  \bibinfo{author}{\bibfnamefont{W.}~\bibnamefont{Rodejohann}},
  \bibinfo{journal}{Phys. Rev.} \textbf{\bibinfo{volume}{D73}},
  \bibinfo{pages}{053005} (\bibinfo{year}{2006}), \eprint{hep-ph/0512143}.

\bibitem{Ding:2010pc}
\bibinfo{author}{\bibfnamefont{G.-J.} \bibnamefont{Ding}},
  \bibinfo{journal}{Nucl. Phys.} \textbf{\bibinfo{volume}{B846}},
  \bibinfo{pages}{394} (\bibinfo{year}{2011}), \eprint{1006.4800}.

\bibitem{Ding:2011cm}
\bibinfo{author}{\bibfnamefont{G.-J.} \bibnamefont{Ding}},
  \bibinfo{author}{\bibfnamefont{L.~L.} \bibnamefont{Everett}},
  \bibnamefont{and} \bibinfo{author}{\bibfnamefont{A.~J.}
  \bibnamefont{Stuart}}, \bibinfo{journal}{Nucl. Phys.}
  \textbf{\bibinfo{volume}{B857}}, \bibinfo{pages}{219} (\bibinfo{year}{2012}),
  \eprint{1110.1688}.

\bibitem{Cooper:2012bd}
\bibinfo{author}{\bibfnamefont{I.~K.} \bibnamefont{Cooper}},
  \bibinfo{author}{\bibfnamefont{S.~F.} \bibnamefont{King}}, \bibnamefont{and}
  \bibinfo{author}{\bibfnamefont{A.~J.} \bibnamefont{Stuart}},
  \bibinfo{journal}{Nucl. Phys.} \textbf{\bibinfo{volume}{B875}},
  \bibinfo{pages}{650} (\bibinfo{year}{2013}), \eprint{1212.1066}.

\bibitem{Ma:2005sha}
\bibinfo{author}{\bibfnamefont{E.}~\bibnamefont{Ma}}, \bibinfo{journal}{Phys.
  Rev.} \textbf{\bibinfo{volume}{D72}}, \bibinfo{pages}{037301}
  (\bibinfo{year}{2005}), \eprint{hep-ph/0505209}.

\bibitem{Ma:2006wm}
\bibinfo{author}{\bibfnamefont{E.}~\bibnamefont{Ma}}, \bibinfo{journal}{Mod.
  Phys. Lett.} \textbf{\bibinfo{volume}{A21}}, \bibinfo{pages}{2931}
  (\bibinfo{year}{2006}), \eprint{hep-ph/0607190}.

\bibitem{Honda:2008rs}
\bibinfo{author}{\bibfnamefont{M.}~\bibnamefont{Honda}} \bibnamefont{and}
  \bibinfo{author}{\bibfnamefont{M.}~\bibnamefont{Tanimoto}},
  \bibinfo{journal}{Prog. Theor. Phys.} \textbf{\bibinfo{volume}{119}},
  \bibinfo{pages}{583} (\bibinfo{year}{2008}), \eprint{0801.0181}.

\bibitem{Brahmachari:2008fn}
\bibinfo{author}{\bibfnamefont{B.}~\bibnamefont{Brahmachari}},
  \bibinfo{author}{\bibfnamefont{S.}~\bibnamefont{Choubey}}, \bibnamefont{and}
  \bibinfo{author}{\bibfnamefont{M.}~\bibnamefont{Mitra}},
  \bibinfo{journal}{Phys. Rev.} \textbf{\bibinfo{volume}{D77}},
  \bibinfo{pages}{073008} (\bibinfo{year}{2008}), \eprint{0801.3554}.

\bibitem{Bazzocchi:2009pv}
\bibinfo{author}{\bibfnamefont{F.}~\bibnamefont{Bazzocchi}},
  \bibinfo{author}{\bibfnamefont{L.}~\bibnamefont{Merlo}}, \bibnamefont{and}
  \bibinfo{author}{\bibfnamefont{S.}~\bibnamefont{Morisi}},
  \bibinfo{journal}{Nucl. Phys.} \textbf{\bibinfo{volume}{B816}},
  \bibinfo{pages}{204} (\bibinfo{year}{2009}), \eprint{0901.2086}.

\bibitem{Boucenna:2012qb}
\bibinfo{author}{\bibfnamefont{M.~S.} \bibnamefont{Boucenna}},
  \bibinfo{author}{\bibfnamefont{S.}~\bibnamefont{Morisi}},
  \bibinfo{author}{\bibfnamefont{E.}~\bibnamefont{Peinado}},
  \bibinfo{author}{\bibfnamefont{Y.}~\bibnamefont{Shimizu}}, \bibnamefont{and}
  \bibinfo{author}{\bibfnamefont{J.~W.~F.} \bibnamefont{Valle}},
  \bibinfo{journal}{Phys. Rev.} \textbf{\bibinfo{volume}{D86}},
  \bibinfo{pages}{073008} (\bibinfo{year}{2012}), \eprint{1204.4733}.

\bibitem{Merle:2006du}
\bibinfo{author}{\bibfnamefont{A.}~\bibnamefont{Merle}} \bibnamefont{and}
  \bibinfo{author}{\bibfnamefont{W.}~\bibnamefont{Rodejohann}},
  \bibinfo{journal}{Phys.Rev.} \textbf{\bibinfo{volume}{D73}},
  \bibinfo{pages}{073012} (\bibinfo{year}{2006}), \eprint{hep-ph/0603111}.

\bibitem{Abt:2004yk}
\bibinfo{author}{\bibfnamefont{I.}~\bibnamefont{Abt}},
  \bibinfo{author}{\bibfnamefont{M.~F.} \bibnamefont{Altmann}},
  \bibinfo{author}{\bibfnamefont{A.}~\bibnamefont{Bakalyarov}},
  \bibinfo{author}{\bibfnamefont{I.}~\bibnamefont{Barabanov}},
  \bibinfo{author}{\bibfnamefont{C.}~\bibnamefont{Bauer}}, \emph{et~al.}
  (\bibinfo{year}{2004}), \eprint{hep-ex/0404039}.

\bibitem{JanicskoCsathy:2009zz}
\bibinfo{author}{\bibfnamefont{J.}~\bibnamefont{Janicsko-Csathy}}
  (\bibinfo{collaboration}{GERDA Collaboration}), \bibinfo{journal}{Nucl. Phys.
  Proc. Suppl.} \textbf{\bibinfo{volume}{188}}, \bibinfo{pages}{68}
  (\bibinfo{year}{2009}).

\bibitem{Agostini:2013mzu}
\bibinfo{author}{\bibfnamefont{M.}~\bibnamefont{Agostini}} \emph{et~al.}
  (\bibinfo{collaboration}{GERDA Collaboration}), \bibinfo{journal}{Phys. Rev.
  Lett.} \textbf{\bibinfo{volume}{111}}, \bibinfo{pages}{122503}
  (\bibinfo{year}{2013}), \eprint{1307.4720}.

\bibitem{Altarelli:2006kg}
\bibinfo{author}{\bibfnamefont{G.}~\bibnamefont{Altarelli}},
  \bibinfo{author}{\bibfnamefont{F.}~\bibnamefont{Feruglio}}, \bibnamefont{and}
  \bibinfo{author}{\bibfnamefont{Y.}~\bibnamefont{Lin}},
  \bibinfo{journal}{Nucl. Phys.} \textbf{\bibinfo{volume}{B775}},
  \bibinfo{pages}{31} (\bibinfo{year}{2007}), \eprint{hep-ph/0610165}.

\bibitem{Ma:2006vq}
\bibinfo{author}{\bibfnamefont{E.}~\bibnamefont{Ma}}, \bibinfo{journal}{Mod.
  Phys. Lett.} \textbf{\bibinfo{volume}{A22}}, \bibinfo{pages}{101}
  (\bibinfo{year}{2007}), \eprint{hep-ph/0610342}.

\bibitem{Bazzocchi:2007na}
\bibinfo{author}{\bibfnamefont{F.}~\bibnamefont{Bazzocchi}},
  \bibinfo{author}{\bibfnamefont{S.}~\bibnamefont{Kaneko}}, \bibnamefont{and}
  \bibinfo{author}{\bibfnamefont{S.}~\bibnamefont{Morisi}},
  \bibinfo{journal}{JHEP} \textbf{\bibinfo{volume}{0803}}, \bibinfo{pages}{063}
  (\bibinfo{year}{2008}), \eprint{0707.3032}.

\bibitem{Bazzocchi:2007au}
\bibinfo{author}{\bibfnamefont{F.}~\bibnamefont{Bazzocchi}},
  \bibinfo{author}{\bibfnamefont{S.}~\bibnamefont{Morisi}}, \bibnamefont{and}
  \bibinfo{author}{\bibfnamefont{M.}~\bibnamefont{Picariello}},
  \bibinfo{journal}{Phys. Lett.} \textbf{\bibinfo{volume}{B659}},
  \bibinfo{pages}{628} (\bibinfo{year}{2008}), \eprint{0710.2928}.

\bibitem{Lin:2008aj}
\bibinfo{author}{\bibfnamefont{Y.}~\bibnamefont{Lin}}, \bibinfo{journal}{Nucl.
  Phys.} \textbf{\bibinfo{volume}{B813}}, \bibinfo{pages}{91}
  (\bibinfo{year}{2009}), \eprint{0804.2867}.

\bibitem{Ma:2009wi}
\bibinfo{author}{\bibfnamefont{E.}~\bibnamefont{Ma}}, \bibinfo{journal}{Mod.
  Phys. Lett.} \textbf{\bibinfo{volume}{A25}}, \bibinfo{pages}{2215}
  (\bibinfo{year}{2010}), \eprint{0908.3165}.

\bibitem{Ciafaloni:2009qs}
\bibinfo{author}{\bibfnamefont{P.}~\bibnamefont{Ciafaloni}},
  \bibinfo{author}{\bibfnamefont{M.}~\bibnamefont{Picariello}},
  \bibinfo{author}{\bibfnamefont{A.}~\bibnamefont{Urbano}}, \bibnamefont{and}
  \bibinfo{author}{\bibfnamefont{E.}~\bibnamefont{Torrente-Lujan}},
  \bibinfo{journal}{Phys. Rev.} \textbf{\bibinfo{volume}{D81}},
  \bibinfo{pages}{016004} (\bibinfo{year}{2010}), \eprint{0909.2553}.

\bibitem{Chen:2009gf}
\bibinfo{author}{\bibfnamefont{M.-C.} \bibnamefont{Chen}} \bibnamefont{and}
  \bibinfo{author}{\bibfnamefont{K.~T.} \bibnamefont{Mahanthappa}},
  \bibinfo{journal}{Phys. Lett.} \textbf{\bibinfo{volume}{B681}},
  \bibinfo{pages}{444} (\bibinfo{year}{2009}), \eprint{0904.1721}.

\bibitem{Chen:2009gy}
\bibinfo{author}{\bibfnamefont{M.-C.} \bibnamefont{Chen}},
  \bibinfo{author}{\bibfnamefont{K.}~\bibnamefont{Mahanthappa}},
  \bibnamefont{and} \bibinfo{author}{\bibfnamefont{F.}~\bibnamefont{Yu}},
  \bibinfo{journal}{Phys. Rev.} \textbf{\bibinfo{volume}{D81}},
  \bibinfo{pages}{036004} (\bibinfo{year}{2010}), \eprint{0907.3963}.

\bibitem{Ding:2008rj}
\bibinfo{author}{\bibfnamefont{G.-J.} \bibnamefont{Ding}},
  \bibinfo{journal}{Phys. Rev.} \textbf{\bibinfo{volume}{D78}},
  \bibinfo{pages}{036011} (\bibinfo{year}{2008}), \eprint{0803.2278}.

\bibitem{Merlo:2011hw}
\bibinfo{author}{\bibfnamefont{L.}~\bibnamefont{Merlo}},
  \bibinfo{author}{\bibfnamefont{S.}~\bibnamefont{Rigolin}}, \bibnamefont{and}
  \bibinfo{author}{\bibfnamefont{B.}~\bibnamefont{Zaldivar}},
  \bibinfo{journal}{JHEP} \textbf{\bibinfo{volume}{1111}}, \bibinfo{pages}{047}
  (\bibinfo{year}{2011}), \eprint{1108.1795}.

\bibitem{Luhn:2012bc}
\bibinfo{author}{\bibfnamefont{C.}~\bibnamefont{Luhn}},
  \bibinfo{author}{\bibfnamefont{K.~M.} \bibnamefont{Parattu}},
  \bibnamefont{and}
  \bibinfo{author}{\bibfnamefont{A.}~\bibnamefont{Wingerter}},
  \bibinfo{journal}{JHEP} \textbf{\bibinfo{volume}{1212}}, \bibinfo{pages}{096}
  (\bibinfo{year}{2012}), \eprint{1210.1197}.

\bibitem{Fukuyama:2010mz}
\bibinfo{author}{\bibfnamefont{T.}~\bibnamefont{Fukuyama}},
  \bibinfo{author}{\bibfnamefont{H.}~\bibnamefont{Sugiyama}}, \bibnamefont{and}
  \bibinfo{author}{\bibfnamefont{K.}~\bibnamefont{Tsumura}},
  \bibinfo{journal}{Phys. Rev.} \textbf{\bibinfo{volume}{D82}},
  \bibinfo{pages}{036004} (\bibinfo{year}{2010}), \eprint{1005.5338}.

\bibitem{Ding:2013eca}
\bibinfo{author}{\bibfnamefont{G.-J.} \bibnamefont{Ding}} \bibnamefont{and}
  \bibinfo{author}{\bibfnamefont{Y.-L.} \bibnamefont{Zhou}},
  \bibinfo{journal}{Nucl. Phys.} \textbf{\bibinfo{volume}{B876}},
  \bibinfo{pages}{418} (\bibinfo{year}{2013}), \eprint{1304.2645}.

\bibitem{Lindner:2010wr}
\bibinfo{author}{\bibfnamefont{M.}~\bibnamefont{Lindner}},
  \bibinfo{author}{\bibfnamefont{A.}~\bibnamefont{Merle}}, \bibnamefont{and}
  \bibinfo{author}{\bibfnamefont{V.}~\bibnamefont{Niro}},
  \bibinfo{journal}{JCAP} \textbf{\bibinfo{volume}{1101}}, \bibinfo{pages}{034}
  (\bibinfo{year}{2011}), \eprint{1011.4950}.

\bibitem{Hashimoto:2011tn}
\bibinfo{author}{\bibfnamefont{K.}~\bibnamefont{Hashimoto}} \bibnamefont{and}
  \bibinfo{author}{\bibfnamefont{H.}~\bibnamefont{Okada}}
  (\bibinfo{year}{2011}), \eprint{1110.3640}.

\bibitem{Hagedorn:2009jy}
\bibinfo{author}{\bibfnamefont{C.}~\bibnamefont{Hagedorn}},
  \bibinfo{author}{\bibfnamefont{E.}~\bibnamefont{Molinaro}}, \bibnamefont{and}
  \bibinfo{author}{\bibfnamefont{S.}~\bibnamefont{Petcov}},
  \bibinfo{journal}{JHEP} \textbf{\bibinfo{volume}{0909}}, \bibinfo{pages}{115}
  (\bibinfo{year}{2009}), \eprint{0908.0240}.

\bibitem{Adhikary:2008au}
\bibinfo{author}{\bibfnamefont{B.}~\bibnamefont{Adhikary}} \bibnamefont{and}
  \bibinfo{author}{\bibfnamefont{A.}~\bibnamefont{Ghosal}},
  \bibinfo{journal}{Phys. Rev.} \textbf{\bibinfo{volume}{D78}},
  \bibinfo{pages}{073007} (\bibinfo{year}{2008}), \eprint{0803.3582}.

\bibitem{Csaki:2008qq}
\bibinfo{author}{\bibfnamefont{C.}~\bibnamefont{Csaki}},
  \bibinfo{author}{\bibfnamefont{C.}~\bibnamefont{Delaunay}},
  \bibinfo{author}{\bibfnamefont{C.}~\bibnamefont{Grojean}}, \bibnamefont{and}
  \bibinfo{author}{\bibfnamefont{Y.}~\bibnamefont{Grossman}},
  \bibinfo{journal}{JHEP} \textbf{\bibinfo{volume}{0810}}, \bibinfo{pages}{055}
  (\bibinfo{year}{2008}), \eprint{0806.0356}.

\bibitem{Ding:2009gh}
\bibinfo{author}{\bibfnamefont{G.-J.} \bibnamefont{Ding}} \bibnamefont{and}
  \bibinfo{author}{\bibfnamefont{J.-F.} \bibnamefont{Liu}},
  \bibinfo{journal}{JHEP} \textbf{\bibinfo{volume}{1005}}, \bibinfo{pages}{029}
  (\bibinfo{year}{2010}), \eprint{0911.4799}.

\bibitem{Mitra:2009jj}
\bibinfo{author}{\bibfnamefont{M.}~\bibnamefont{Mitra}},
  \bibinfo{journal}{JHEP} \textbf{\bibinfo{volume}{1011}}, \bibinfo{pages}{026}
  (\bibinfo{year}{2010}), \eprint{0912.5291}.

\bibitem{Lin:2009bw}
\bibinfo{author}{\bibfnamefont{Y.}~\bibnamefont{Lin}}, \bibinfo{journal}{Nucl.
  Phys.} \textbf{\bibinfo{volume}{B824}}, \bibinfo{pages}{95}
  (\bibinfo{year}{2010}), \eprint{0905.3534}.

\bibitem{delAguila:2010vg}
\bibinfo{author}{\bibfnamefont{F.}~\bibnamefont{del Aguila}},
  \bibinfo{author}{\bibfnamefont{A.}~\bibnamefont{Carmona}}, \bibnamefont{and}
  \bibinfo{author}{\bibfnamefont{J.}~\bibnamefont{Santiago}},
  \bibinfo{journal}{JHEP} \textbf{\bibinfo{volume}{1008}}, \bibinfo{pages}{127}
  (\bibinfo{year}{2010}), \eprint{1001.5151}.

\bibitem{Burrows:2010wz}
\bibinfo{author}{\bibfnamefont{T.~J.} \bibnamefont{Burrows}} \bibnamefont{and}
  \bibinfo{author}{\bibfnamefont{S.~F.} \bibnamefont{King}},
  \bibinfo{journal}{Nucl. Phys.} \textbf{\bibinfo{volume}{B842}},
  \bibinfo{pages}{107} (\bibinfo{year}{2011}), \eprint{1007.2310}.

\bibitem{He:2006dk}
\bibinfo{author}{\bibfnamefont{X.-G.} \bibnamefont{He}},
  \bibinfo{author}{\bibfnamefont{Y.-Y.} \bibnamefont{Keum}}, \bibnamefont{and}
  \bibinfo{author}{\bibfnamefont{R.~R.} \bibnamefont{Volkas}},
  \bibinfo{journal}{JHEP} \textbf{\bibinfo{volume}{0604}}, \bibinfo{pages}{039}
  (\bibinfo{year}{2006}), \eprint{hep-ph/0601001}.

\bibitem{Berger:2009tt}
\bibinfo{author}{\bibfnamefont{J.}~\bibnamefont{Berger}} \bibnamefont{and}
  \bibinfo{author}{\bibfnamefont{Y.}~\bibnamefont{Grossman}},
  \bibinfo{journal}{JHEP} \textbf{\bibinfo{volume}{1002}}, \bibinfo{pages}{071}
  (\bibinfo{year}{2010}), \eprint{0910.4392}.

\bibitem{Kadosh:2010rm}
\bibinfo{author}{\bibfnamefont{A.}~\bibnamefont{Kadosh}} \bibnamefont{and}
  \bibinfo{author}{\bibfnamefont{E.}~\bibnamefont{Pallante}},
  \bibinfo{journal}{JHEP} \textbf{\bibinfo{volume}{1008}}, \bibinfo{pages}{115}
  (\bibinfo{year}{2010}), \eprint{1004.0321}.

\bibitem{Lavoura:2012cv}
\bibinfo{author}{\bibfnamefont{L.}~\bibnamefont{Lavoura}},
  \bibinfo{author}{\bibfnamefont{S.}~\bibnamefont{Morisi}}, \bibnamefont{and}
  \bibinfo{author}{\bibfnamefont{J.~W.~F.} \bibnamefont{Valle}},
  \bibinfo{journal}{JHEP} \textbf{\bibinfo{volume}{1302}}, \bibinfo{pages}{118}
  (\bibinfo{year}{2013}), \eprint{1205.3442}.

\bibitem{Adulpravitchai:2009gi}
\bibinfo{author}{\bibfnamefont{A.}~\bibnamefont{Adulpravitchai}},
  \bibinfo{author}{\bibfnamefont{M.}~\bibnamefont{Lindner}}, \bibnamefont{and}
  \bibinfo{author}{\bibfnamefont{A.}~\bibnamefont{Merle}},
  \bibinfo{journal}{Phys. Rev.} \textbf{\bibinfo{volume}{D80}},
  \bibinfo{pages}{055031} (\bibinfo{year}{2009}), \eprint{0907.2147}.

\bibitem{Dorame:2012zv}
\bibinfo{author}{\bibfnamefont{L.}~\bibnamefont{Dorame}},
  \bibinfo{author}{\bibfnamefont{S.}~\bibnamefont{Morisi}},
  \bibinfo{author}{\bibfnamefont{E.}~\bibnamefont{Peinado}},
  \bibinfo{author}{\bibfnamefont{J.~W.~F.} \bibnamefont{Valle}},
  \bibnamefont{and} \bibinfo{author}{\bibfnamefont{A.~D.} \bibnamefont{Rojas}},
  \bibinfo{journal}{Phys. Rev.} \textbf{\bibinfo{volume}{D86}},
  \bibinfo{pages}{056001} (\bibinfo{year}{2012}), \eprint{1203.0155}.

\bibitem{Akhmedov:2004ny}
\bibinfo{author}{\bibfnamefont{E.~K.} \bibnamefont{Akhmedov}},
  \bibinfo{author}{\bibfnamefont{R.}~\bibnamefont{Johansson}},
  \bibinfo{author}{\bibfnamefont{M.}~\bibnamefont{Lindner}},
  \bibinfo{author}{\bibfnamefont{T.}~\bibnamefont{Ohlsson}}, \bibnamefont{and}
  \bibinfo{author}{\bibfnamefont{T.}~\bibnamefont{Schwetz}},
  \bibinfo{journal}{JHEP} \textbf{\bibinfo{volume}{0404}}, \bibinfo{pages}{078}
  (\bibinfo{year}{2004}), \eprint{hep-ph/0402175}.

\bibitem{Barger:2006vy}
\bibinfo{author}{\bibfnamefont{V.}~\bibnamefont{Barger}},
  \bibinfo{author}{\bibfnamefont{M.}~\bibnamefont{Dierckxsens}},
  \bibinfo{author}{\bibfnamefont{M.}~\bibnamefont{Diwan}},
  \bibinfo{author}{\bibfnamefont{P.}~\bibnamefont{Huber}},
  \bibinfo{author}{\bibfnamefont{C.}~\bibnamefont{Lewis}}, \emph{et~al.},
  \bibinfo{journal}{Phys. Rev.} \textbf{\bibinfo{volume}{D74}},
  \bibinfo{pages}{073004} (\bibinfo{year}{2006}), \eprint{hep-ph/0607177}.

\bibitem{Barger:2007jq}
\bibinfo{author}{\bibfnamefont{V.}~\bibnamefont{Barger}},
  \bibinfo{author}{\bibfnamefont{P.}~\bibnamefont{Huber}},
  \bibinfo{author}{\bibfnamefont{D.}~\bibnamefont{Marfatia}}, \bibnamefont{and}
  \bibinfo{author}{\bibfnamefont{W.}~\bibnamefont{Winter}},
  \bibinfo{journal}{Phys. Rev.} \textbf{\bibinfo{volume}{D76}},
  \bibinfo{pages}{053005} (\bibinfo{year}{2007}), \eprint{hep-ph/0703029}.

\bibitem{Barger:2012fx}
\bibinfo{author}{\bibfnamefont{V.}~\bibnamefont{Barger}},
  \bibinfo{author}{\bibfnamefont{R.}~\bibnamefont{Gandhi}},
  \bibinfo{author}{\bibfnamefont{P.}~\bibnamefont{Ghoshal}},
  \bibinfo{author}{\bibfnamefont{S.}~\bibnamefont{Goswami}},
  \bibinfo{author}{\bibfnamefont{D.}~\bibnamefont{Marfatia}}, \emph{et~al.},
  \bibinfo{journal}{Phys. Rev. Lett.} \textbf{\bibinfo{volume}{109}},
  \bibinfo{pages}{091801} (\bibinfo{year}{2012}), \eprint{1203.6012}.

\bibitem{Samanta:2006sj}
\bibinfo{author}{\bibfnamefont{A.}~\bibnamefont{Samanta}},
  \bibinfo{journal}{Phys. Lett.} \textbf{\bibinfo{volume}{B673}},
  \bibinfo{pages}{37} (\bibinfo{year}{2009}), \eprint{hep-ph/0610196}.

\bibitem{Ghosh:2012px}
\bibinfo{author}{\bibfnamefont{A.}~\bibnamefont{Ghosh}},
  \bibinfo{author}{\bibfnamefont{T.}~\bibnamefont{Thakore}}, \bibnamefont{and}
  \bibinfo{author}{\bibfnamefont{S.}~\bibnamefont{Choubey}},
  \bibinfo{journal}{JHEP} \textbf{\bibinfo{volume}{1304}}, \bibinfo{pages}{009}
  (\bibinfo{year}{2013}), \eprint{1212.1305}.

\bibitem{Li:2013zyd}
\bibinfo{author}{\bibfnamefont{Y.-F.} \bibnamefont{Li}},
  \bibinfo{author}{\bibfnamefont{J.}~\bibnamefont{Cao}},
  \bibinfo{author}{\bibfnamefont{Y.}~\bibnamefont{Wang}}, \bibnamefont{and}
  \bibinfo{author}{\bibfnamefont{L.}~\bibnamefont{Zhan}},
  \bibinfo{journal}{Phys. Rev.} \textbf{\bibinfo{volume}{D88}},
  \bibinfo{pages}{013008} (\bibinfo{year}{2013}), \eprint{1303.6733}.

\bibitem{Oyama:2012tq}
\bibinfo{author}{\bibfnamefont{Y.}~\bibnamefont{Oyama}},
  \bibinfo{author}{\bibfnamefont{A.}~\bibnamefont{Shimizu}}, \bibnamefont{and}
  \bibinfo{author}{\bibfnamefont{K.}~\bibnamefont{Kohri}},
  \bibinfo{journal}{Phys. Lett.} \textbf{\bibinfo{volume}{B718}},
  \bibinfo{pages}{1186} (\bibinfo{year}{2013}), \eprint{1205.5223}.

\bibitem{Maneschg:2008sf}
\bibinfo{author}{\bibfnamefont{W.}~\bibnamefont{Maneschg}},
  \bibinfo{author}{\bibfnamefont{A.}~\bibnamefont{Merle}}, \bibnamefont{and}
  \bibinfo{author}{\bibfnamefont{W.}~\bibnamefont{Rodejohann}},
  \bibinfo{journal}{Europhys. Lett.} \textbf{\bibinfo{volume}{85}},
  \bibinfo{pages}{51002} (\bibinfo{year}{2009}), \eprint{0812.0479}.

\bibitem{Winter:2013ema}
\bibinfo{author}{\bibfnamefont{W.}~\bibnamefont{Winter}},
  \bibinfo{journal}{Phys. Rev.} \textbf{\bibinfo{volume}{D88}},
  \bibinfo{pages}{013013} (\bibinfo{year}{2013}), \eprint{1305.5539}.

\bibitem{An:2012eh}
\bibinfo{author}{\bibfnamefont{F.~P.} \bibnamefont{An}} \emph{et~al.}
  (\bibinfo{collaboration}{DAYA-BAY Collaboration}), \bibinfo{journal}{Phys.
  Rev. Lett.} \textbf{\bibinfo{volume}{108}}, \bibinfo{pages}{171803}
  (\bibinfo{year}{2012}), \eprint{1203.1669}.

\bibitem{Abe:2011fz}
\bibinfo{author}{\bibfnamefont{Y.}~\bibnamefont{Abe}} \emph{et~al.}
  (\bibinfo{collaboration}{DOUBLE-CHOOZ Collaboration}),
  \bibinfo{journal}{Phys. Rev. Lett.} \textbf{\bibinfo{volume}{108}},
  \bibinfo{pages}{131801} (\bibinfo{year}{2012}), \eprint{1112.6353}.

\bibitem{Abe:2012tg}
\bibinfo{author}{\bibfnamefont{Y.}~\bibnamefont{Abe}} \emph{et~al.}
  (\bibinfo{collaboration}{Double Chooz Collaboration}),
  \bibinfo{journal}{Phys. Rev.} \textbf{\bibinfo{volume}{D86}},
  \bibinfo{pages}{052008} (\bibinfo{year}{2012}), \eprint{1207.6632}.

\bibitem{Abe:2011ks}
\bibinfo{author}{\bibfnamefont{K.}~\bibnamefont{Abe}} \emph{et~al.}
  (\bibinfo{collaboration}{T2K Collaboration}), \bibinfo{journal}{Nucl.
  Instrum. Meth.} \textbf{\bibinfo{volume}{A659}}, \bibinfo{pages}{106}
  (\bibinfo{year}{2011}), \eprint{1106.1238}.

\bibitem{Abe:2013fuq}
\bibinfo{author}{\bibfnamefont{K.}~\bibnamefont{Abe}} \emph{et~al.}
  (\bibinfo{collaboration}{T2K Collaboration}), \bibinfo{journal}{Phys. Rev.
  Lett. 111,} \textbf{\bibinfo{volume}{211803}} (\bibinfo{year}{2013}),
  \eprint{1308.0465}.

\bibitem{Ayres:2004js}
\bibinfo{author}{\bibfnamefont{D.}~\bibnamefont{Ayres}} \emph{et~al.}
  (\bibinfo{collaboration}{NOvA Collaboration})  (\bibinfo{year}{2004}),
  \eprint{hep-ex/0503053}.

\bibitem{Bian:2013saa}
\bibinfo{author}{\bibfnamefont{J.}~\bibnamefont{Bian}}  (\bibinfo{year}{2013}),
  \eprint{1309.7898}.

\bibitem{Huber:2009cw}
\bibinfo{author}{\bibfnamefont{P.}~\bibnamefont{Huber}},
  \bibinfo{author}{\bibfnamefont{M.}~\bibnamefont{Lindner}},
  \bibinfo{author}{\bibfnamefont{T.}~\bibnamefont{Schwetz}}, \bibnamefont{and}
  \bibinfo{author}{\bibfnamefont{W.}~\bibnamefont{Winter}},
  \bibinfo{journal}{JHEP} \textbf{\bibinfo{volume}{0911}}, \bibinfo{pages}{044}
  (\bibinfo{year}{2009}), \eprint{0907.1896}.

\bibitem{T2K_NExT}
\bibinfo{author}{\bibfnamefont{M.}~\bibnamefont{Malek}},
  \emph{\bibinfo{title}{{Life After $\nu_e$ Appearance: What NEXT for T2K?}}}
  (\bibinfo{year}{2013}), \bibinfo{note}{talk given at the \emph{What NExT?
  NExT meeting on BSM physics in light of LHC, Planck results and $\theta_{13}$
  discovery}, Southampton, UK, 27 November 2013; available online at
  \url{https://indico.nbi.ku.dk/getFile.py/access?contribId=16&resId=0&materialId=slides&confId=616}.}

\bibitem{Abe:2011ts}
\bibinfo{author}{\bibfnamefont{K.}~\bibnamefont{Abe}},
  \bibinfo{author}{\bibfnamefont{T.}~\bibnamefont{Abe}},
  \bibinfo{author}{\bibfnamefont{H.}~\bibnamefont{Aihara}},
  \bibinfo{author}{\bibfnamefont{Y.}~\bibnamefont{Fukuda}},
  \bibinfo{author}{\bibfnamefont{Y.}~\bibnamefont{Hayato}}, \emph{et~al.}
  (\bibinfo{year}{2011}), \eprint{1109.3262}.

\bibitem{Dodelson:1993je}
\bibinfo{author}{\bibfnamefont{S.}~\bibnamefont{Dodelson}} \bibnamefont{and}
  \bibinfo{author}{\bibfnamefont{L.~M.} \bibnamefont{Widrow}},
  \bibinfo{journal}{Phys. Rev. Lett.} \textbf{\bibinfo{volume}{72}},
  \bibinfo{pages}{17} (\bibinfo{year}{1994}), \eprint{hep-ph/9303287}.

\bibitem{Shi:1998km}
\bibinfo{author}{\bibfnamefont{X.-D.} \bibnamefont{Shi}} \bibnamefont{and}
  \bibinfo{author}{\bibfnamefont{G.~M.} \bibnamefont{Fuller}},
  \bibinfo{journal}{Phys. Rev. Lett.} \textbf{\bibinfo{volume}{82}},
  \bibinfo{pages}{2832} (\bibinfo{year}{1999}), \eprint{astro-ph/9810076}.

\bibitem{Bezrukov:2009th}
\bibinfo{author}{\bibfnamefont{F.}~\bibnamefont{Bezrukov}},
  \bibinfo{author}{\bibfnamefont{H.}~\bibnamefont{Hettmansperger}},
  \bibnamefont{and} \bibinfo{author}{\bibfnamefont{M.}~\bibnamefont{Lindner}},
  \bibinfo{journal}{Phys. Rev.} \textbf{\bibinfo{volume}{D81}},
  \bibinfo{pages}{085032} (\bibinfo{year}{2010}), \eprint{0912.4415}.

\bibitem{Nemevsek:2012cd}
\bibinfo{author}{\bibfnamefont{M.}~\bibnamefont{Nemevsek}},
  \bibinfo{author}{\bibfnamefont{G.}~\bibnamefont{Senjanovic}},
  \bibnamefont{and} \bibinfo{author}{\bibfnamefont{Y.}~\bibnamefont{Zhang}},
  \bibinfo{journal}{JCAP} \textbf{\bibinfo{volume}{1207}}, \bibinfo{pages}{006}
  (\bibinfo{year}{2012}), \eprint{1205.0844}.

\bibitem{Shaposhnikov:2006xi}
\bibinfo{author}{\bibfnamefont{M.}~\bibnamefont{Shaposhnikov}}
  \bibnamefont{and} \bibinfo{author}{\bibfnamefont{I.}~\bibnamefont{Tkachev}},
  \bibinfo{journal}{Phys. Lett.} \textbf{\bibinfo{volume}{B639}},
  \bibinfo{pages}{414} (\bibinfo{year}{2006}), \eprint{hep-ph/0604236}.

\bibitem{Bezrukov:2009yw}
\bibinfo{author}{\bibfnamefont{F.}~\bibnamefont{Bezrukov}} \bibnamefont{and}
  \bibinfo{author}{\bibfnamefont{D.}~\bibnamefont{Gorbunov}},
  \bibinfo{journal}{JHEP} \textbf{\bibinfo{volume}{1005}}, \bibinfo{pages}{010}
  (\bibinfo{year}{2010}), \eprint{0912.0390}.

\bibitem{Kusenko:2006rh}
\bibinfo{author}{\bibfnamefont{A.}~\bibnamefont{Kusenko}},
  \bibinfo{journal}{Phys. Rev. Lett.} \textbf{\bibinfo{volume}{97}},
  \bibinfo{pages}{241301} (\bibinfo{year}{2006}), \eprint{hep-ph/0609081}.

\bibitem{Petraki:2007gq}
\bibinfo{author}{\bibfnamefont{K.}~\bibnamefont{Petraki}} \bibnamefont{and}
  \bibinfo{author}{\bibfnamefont{A.}~\bibnamefont{Kusenko}},
  \bibinfo{journal}{Phys. Rev.} \textbf{\bibinfo{volume}{D77}},
  \bibinfo{pages}{065014} (\bibinfo{year}{2008}), \eprint{0711.4646}.

\bibitem{Merle:2013wta}
\bibinfo{author}{\bibfnamefont{A.}~\bibnamefont{Merle}},
  \bibinfo{author}{\bibfnamefont{V.}~\bibnamefont{Niro}}, \bibnamefont{and}
  \bibinfo{author}{\bibfnamefont{D.}~\bibnamefont{Schmidt}}
  (\bibinfo{year}{2013}), \eprint{1306.3996}.

\bibitem{Asaka:2005an}
\bibinfo{author}{\bibfnamefont{T.}~\bibnamefont{Asaka}},
  \bibinfo{author}{\bibfnamefont{S.}~\bibnamefont{Blanchet}}, \bibnamefont{and}
  \bibinfo{author}{\bibfnamefont{M.}~\bibnamefont{Shaposhnikov}},
  \bibinfo{journal}{Phys. Lett.} \textbf{\bibinfo{volume}{B631}},
  \bibinfo{pages}{151} (\bibinfo{year}{2005}), \eprint{hep-ph/0503065}.

\bibitem{Asaka:2005pn}
\bibinfo{author}{\bibfnamefont{T.}~\bibnamefont{Asaka}} \bibnamefont{and}
  \bibinfo{author}{\bibfnamefont{M.}~\bibnamefont{Shaposhnikov}},
  \bibinfo{journal}{Phys. Lett.} \textbf{\bibinfo{volume}{B620}},
  \bibinfo{pages}{17} (\bibinfo{year}{2005}), \eprint{hep-ph/0505013}.

\bibitem{Abazajian:2012ys}
\bibinfo{author}{\bibfnamefont{K.~N.} \bibnamefont{Abazajian}},
  \bibinfo{author}{\bibfnamefont{M.~A.} \bibnamefont{Acero}},
  \bibinfo{author}{\bibfnamefont{S.~K.} \bibnamefont{Agarwalla}},
  \bibinfo{author}{\bibfnamefont{A.~A.} \bibnamefont{Aguilar-Arevalo}},
  \bibinfo{author}{\bibfnamefont{C.~H.} \bibnamefont{Albright}}, \emph{et~al.}
  (\bibinfo{year}{2012}), \eprint{1204.5379}.

\bibitem{Palazzo:2013me}
\bibinfo{author}{\bibfnamefont{A.}~\bibnamefont{Palazzo}},
  \bibinfo{journal}{Mod. Phys. Lett.} \textbf{\bibinfo{volume}{A28}},
  \bibinfo{pages}{1330004} (\bibinfo{year}{2013}), \eprint{1302.1102}.

\bibitem{Athanassopoulos:1996jb}
\bibinfo{author}{\bibfnamefont{C.}~\bibnamefont{Athanassopoulos}} \emph{et~al.}
  (\bibinfo{collaboration}{LSND Collaboration}), \bibinfo{journal}{Phys. Rev.
  Lett.} \textbf{\bibinfo{volume}{77}}, \bibinfo{pages}{3082}
  (\bibinfo{year}{1996}), \eprint{nucl-ex/9605003}.

\bibitem{Aguilar:2001ty}
\bibinfo{author}{\bibfnamefont{A.}~\bibnamefont{Aguilar-Arevalo}} \emph{et~al.}
  (\bibinfo{collaboration}{LSND Collaboration}), \bibinfo{journal}{Phys. Rev.}
  \textbf{\bibinfo{volume}{D64}}, \bibinfo{pages}{112007}
  (\bibinfo{year}{2001}), \eprint{hep-ex/0104049}.

\bibitem{Slansky:1997bm}
\bibinfo{author}{\bibfnamefont{R.}~\bibnamefont{Slansky}},
  \bibinfo{author}{\bibfnamefont{S.}~\bibnamefont{Raby}},
  \bibinfo{author}{\bibfnamefont{J.~T.} \bibnamefont{Goldman}},
  \bibnamefont{and} \bibinfo{author}{\bibfnamefont{G.}~\bibnamefont{Garvey}},
  \bibinfo{journal}{Los Alamos Sci.} \textbf{\bibinfo{volume}{25}},
  \bibinfo{pages}{28} (\bibinfo{year}{1997}).

\bibitem{Louis:1997bs}
\bibinfo{author}{\bibfnamefont{B.}~\bibnamefont{Louis}},
  \bibinfo{author}{\bibfnamefont{V.}~\bibnamefont{Sandberg}}, \bibnamefont{and}
  \bibinfo{author}{\bibfnamefont{H.}~\bibnamefont{White}},
  \bibinfo{journal}{Los Alamos Sci.} \textbf{\bibinfo{volume}{25}},
  \bibinfo{pages}{92} (\bibinfo{year}{1997}).

\bibitem{Armbruster:2002mp}
\bibinfo{author}{\bibfnamefont{B.}~\bibnamefont{Armbruster}} \emph{et~al.}
  (\bibinfo{collaboration}{KARMEN Collaboration}), \bibinfo{journal}{Phys.
  Rev.} \textbf{\bibinfo{volume}{D65}}, \bibinfo{pages}{112001}
  (\bibinfo{year}{2002}), \eprint{hep-ex/0203021}.

\bibitem{Church:2002tc}
\bibinfo{author}{\bibfnamefont{E.~D.} \bibnamefont{Church}},
  \bibinfo{author}{\bibfnamefont{K.}~\bibnamefont{Eitel}},
  \bibinfo{author}{\bibfnamefont{G.~B.} \bibnamefont{Mills}}, \bibnamefont{and}
  \bibinfo{author}{\bibfnamefont{M.}~\bibnamefont{Steidl}},
  \bibinfo{journal}{Phys. Rev.} \textbf{\bibinfo{volume}{D66}},
  \bibinfo{pages}{013001} (\bibinfo{year}{2002}), \eprint{hep-ex/0203023}.

\bibitem{Aguilar-Arevalo:2013pmq}
\bibinfo{author}{\bibfnamefont{A.~A.} \bibnamefont{Aguilar-Arevalo}}
  \emph{et~al.} (\bibinfo{collaboration}{MiniBooNE Collaboration}),
  \bibinfo{journal}{Phys. Rev. Lett.} \textbf{\bibinfo{volume}{110}},
  \bibinfo{pages}{161801} (\bibinfo{year}{2013}), \eprint{1207.4809}.

\bibitem{Antonello:2012pq}
\bibinfo{author}{\bibfnamefont{M.}~\bibnamefont{Antonello}},
  \bibinfo{author}{\bibfnamefont{B.}~\bibnamefont{Baibussinov}},
  \bibinfo{author}{\bibfnamefont{P.}~\bibnamefont{Benetti}},
  \bibinfo{author}{\bibfnamefont{E.}~\bibnamefont{Calligarich}},
  \bibinfo{author}{\bibfnamefont{N.}~\bibnamefont{Canci}}, \emph{et~al.},
  \bibinfo{journal}{Eur. Phys. J.} \textbf{\bibinfo{volume}{C73}},
  \bibinfo{pages}{2345} (\bibinfo{year}{2013}), \eprint{1209.0122}.

\bibitem{Mueller:2011nm}
\bibinfo{author}{\bibfnamefont{T.~A.} \bibnamefont{Mueller}},
  \bibinfo{author}{\bibfnamefont{D.}~\bibnamefont{Lhuillier}},
  \bibinfo{author}{\bibfnamefont{M.}~\bibnamefont{Fallot}},
  \bibinfo{author}{\bibfnamefont{A.}~\bibnamefont{Letourneau}},
  \bibinfo{author}{\bibfnamefont{S.}~\bibnamefont{Cormon}}, \emph{et~al.},
  \bibinfo{journal}{Phys. Rev.} \textbf{\bibinfo{volume}{C83}},
  \bibinfo{pages}{054615} (\bibinfo{year}{2011}), \eprint{1101.2663}.

\bibitem{Huber:2011wv}
\bibinfo{author}{\bibfnamefont{P.}~\bibnamefont{Huber}},
  \bibinfo{journal}{Phys. Rev.} \textbf{\bibinfo{volume}{C84}},
  \bibinfo{pages}{024617} (\bibinfo{year}{2011}), \eprint{1106.0687}.

\bibitem{VonFeilitzsch:1982jw}
\bibinfo{author}{\bibfnamefont{F.}~\bibnamefont{Von~Feilitzsch}},
  \bibinfo{author}{\bibfnamefont{A.~A.} \bibnamefont{Hahn}}, \bibnamefont{and}
  \bibinfo{author}{\bibfnamefont{K.}~\bibnamefont{Schreckenbach}},
  \bibinfo{journal}{Phys. Lett.} \textbf{\bibinfo{volume}{B118}},
  \bibinfo{pages}{162} (\bibinfo{year}{1982}).

\bibitem{Schreckenbach:1985ep}
\bibinfo{author}{\bibfnamefont{K.}~\bibnamefont{Schreckenbach}},
  \bibinfo{author}{\bibfnamefont{G.}~\bibnamefont{Colvin}},
  \bibinfo{author}{\bibfnamefont{W.}~\bibnamefont{Gelletly}}, \bibnamefont{and}
  \bibinfo{author}{\bibfnamefont{F.}~\bibnamefont{Von~Feilitzsch}},
  \bibinfo{journal}{Phys. Lett.} \textbf{\bibinfo{volume}{B160}},
  \bibinfo{pages}{325} (\bibinfo{year}{1985}).

\bibitem{Hahn:1989zr}
\bibinfo{author}{\bibfnamefont{A.~A.} \bibnamefont{Hahn}},
  \bibinfo{author}{\bibfnamefont{K.}~\bibnamefont{Schreckenbach}},
  \bibinfo{author}{\bibfnamefont{G.}~\bibnamefont{Colvin}},
  \bibinfo{author}{\bibfnamefont{B.}~\bibnamefont{Krusche}},
  \bibinfo{author}{\bibfnamefont{W.}~\bibnamefont{Gelletly}}, \emph{et~al.},
  \bibinfo{journal}{Phys. Lett.} \textbf{\bibinfo{volume}{B218}},
  \bibinfo{pages}{365} (\bibinfo{year}{1989}).

\bibitem{Abdurashitov:1998ne}
\bibinfo{author}{\bibfnamefont{J.~N.} \bibnamefont{Abdurashitov}} \emph{et~al.}
  (\bibinfo{collaboration}{SAGE Collaboration}), \bibinfo{journal}{Phys. Rev.}
  \textbf{\bibinfo{volume}{C59}}, \bibinfo{pages}{2246} (\bibinfo{year}{1999}),
  \eprint{hep-ph/9803418}.

\bibitem{Hampel:1997fc}
\bibinfo{author}{\bibfnamefont{W.}~\bibnamefont{Hampel}} \emph{et~al.}
  (\bibinfo{collaboration}{GALLEX Collaboration}), \bibinfo{journal}{Phys.
  Lett.} \textbf{\bibinfo{volume}{B420}}, \bibinfo{pages}{114}
  (\bibinfo{year}{1998}).

\bibitem{Abdurashitov:2005tb}
\bibinfo{author}{\bibfnamefont{J.~N.} \bibnamefont{Abdurashitov}},
  \bibinfo{author}{\bibfnamefont{V.~N.} \bibnamefont{Gavrin}},
  \bibinfo{author}{\bibfnamefont{S.~V.} \bibnamefont{Girin}},
  \bibinfo{author}{\bibfnamefont{V.~V.} \bibnamefont{Gorbachev}},
  \bibinfo{author}{\bibfnamefont{P.~P.} \bibnamefont{Gurkina}}, \emph{et~al.},
  \bibinfo{journal}{Phys. Rev.} \textbf{\bibinfo{volume}{C73}},
  \bibinfo{pages}{045805} (\bibinfo{year}{2006}), \eprint{nucl-ex/0512041}.

\bibitem{Giunti:2012tn}
\bibinfo{author}{\bibfnamefont{C.}~\bibnamefont{Giunti}},
  \bibinfo{author}{\bibfnamefont{M.}~\bibnamefont{Laveder}},
  \bibinfo{author}{\bibfnamefont{Y.~F.} \bibnamefont{Li}},
  \bibinfo{author}{\bibfnamefont{Q.~Y.} \bibnamefont{Liu}}, \bibnamefont{and}
  \bibinfo{author}{\bibfnamefont{H.~W.} \bibnamefont{Long}},
  \bibinfo{journal}{Phys. Rev.} \textbf{\bibinfo{volume}{D86}},
  \bibinfo{pages}{113014} (\bibinfo{year}{2012}), \eprint{1210.5715}.

\bibitem{Agarwalla:2010zu}
\bibinfo{author}{\bibfnamefont{S.~K.} \bibnamefont{Agarwalla}}
  \bibnamefont{and} \bibinfo{author}{\bibfnamefont{P.}~\bibnamefont{Huber}},
  \bibinfo{journal}{Phys. Lett.} \textbf{\bibinfo{volume}{B696}},
  \bibinfo{pages}{359} (\bibinfo{year}{2011}), \eprint{1007.3228}.

\bibitem{deGouvea:2011zz}
\bibinfo{author}{\bibfnamefont{A.}~\bibnamefont{de~Gouvea}} \bibnamefont{and}
  \bibinfo{author}{\bibfnamefont{W.-C.} \bibnamefont{Huang}},
  \bibinfo{journal}{Phys. Rev.} \textbf{\bibinfo{volume}{D85}},
  \bibinfo{pages}{053006} (\bibinfo{year}{2012}), \eprint{1110.6122}.

\bibitem{Rubbia:2013ywa}
\bibinfo{author}{\bibfnamefont{C.}~\bibnamefont{Rubbia}},
  \bibinfo{author}{\bibfnamefont{A.}~\bibnamefont{Guglielmi}},
  \bibinfo{author}{\bibfnamefont{F.}~\bibnamefont{Pietropaolo}},
  \bibnamefont{and} \bibinfo{author}{\bibfnamefont{P.}~\bibnamefont{Sala}}
  (\bibinfo{year}{2013}), \eprint{1304.2047}.

\bibitem{2013arXiv1304.7127K}
\bibinfo{author}{\bibfnamefont{U.}~\bibnamefont{Kose}}  (\bibinfo{year}{2013}),
  \eprint{1304.7127}.

\bibitem{Porta:2010zz}
\bibinfo{author}{\bibfnamefont{A.}~\bibnamefont{Porta}}
  (\bibinfo{collaboration}{Nucifer Collaboration}), \bibinfo{journal}{J. Phys.
  Conf. Ser.} \textbf{\bibinfo{volume}{203}}, \bibinfo{pages}{012092}
  (\bibinfo{year}{2010}).

\bibitem{Watson:2006qb}
\bibinfo{author}{\bibfnamefont{C.~R.} \bibnamefont{Watson}},
  \bibinfo{author}{\bibfnamefont{J.~F.} \bibnamefont{Beacom}},
  \bibinfo{author}{\bibfnamefont{H.}~\bibnamefont{Yuksel}}, \bibnamefont{and}
  \bibinfo{author}{\bibfnamefont{T.~P.} \bibnamefont{Walker}},
  \bibinfo{journal}{Phys. Rev.} \textbf{\bibinfo{volume}{D74}},
  \bibinfo{pages}{033009} (\bibinfo{year}{2006}), \eprint{astro-ph/0605424}.

\bibitem{Abazajian:2001vt}
\bibinfo{author}{\bibfnamefont{K.}~\bibnamefont{Abazajian}},
  \bibinfo{author}{\bibfnamefont{G.~M.} \bibnamefont{Fuller}},
  \bibnamefont{and} \bibinfo{author}{\bibfnamefont{W.~H.}
  \bibnamefont{Tucker}}, \bibinfo{journal}{Astrophys. J.}
  \textbf{\bibinfo{volume}{562}}, \bibinfo{pages}{593} (\bibinfo{year}{2001}),
  \eprint{astro-ph/0106002}.

\bibitem{Abazajian:2006jc}
\bibinfo{author}{\bibfnamefont{K.~N.} \bibnamefont{Abazajian}},
  \bibinfo{author}{\bibfnamefont{M.}~\bibnamefont{Markevitch}},
  \bibinfo{author}{\bibfnamefont{S.~M.} \bibnamefont{Koushiappas}},
  \bibnamefont{and} \bibinfo{author}{\bibfnamefont{R.~C.}
  \bibnamefont{Hickox}}, \bibinfo{journal}{Phys. Rev.}
  \textbf{\bibinfo{volume}{D75}}, \bibinfo{pages}{063511}
  (\bibinfo{year}{2007}), \eprint{astro-ph/0611144}.

\bibitem{Boyarsky:2005us}
\bibinfo{author}{\bibfnamefont{A.}~\bibnamefont{Boyarsky}},
  \bibinfo{author}{\bibfnamefont{A.}~\bibnamefont{Neronov}},
  \bibinfo{author}{\bibfnamefont{O.}~\bibnamefont{Ruchayskiy}},
  \bibnamefont{and}
  \bibinfo{author}{\bibfnamefont{M.}~\bibnamefont{Shaposhnikov}},
  \bibinfo{journal}{Mon. Not. Roy. Astron. Soc.}
  \textbf{\bibinfo{volume}{370}}, \bibinfo{pages}{213} (\bibinfo{year}{2006}),
  \eprint{astro-ph/0512509}.

\bibitem{Dolgov:2000ew}
\bibinfo{author}{\bibfnamefont{A.~D.} \bibnamefont{Dolgov}} \bibnamefont{and}
  \bibinfo{author}{\bibfnamefont{S.~H.} \bibnamefont{Hansen}},
  \bibinfo{journal}{Astropart. Phys.} \textbf{\bibinfo{volume}{16}},
  \bibinfo{pages}{339} (\bibinfo{year}{2002}), \eprint{hep-ph/0009083}.

\bibitem{Boyarsky:2006fg}
\bibinfo{author}{\bibfnamefont{A.}~\bibnamefont{Boyarsky}},
  \bibinfo{author}{\bibfnamefont{A.}~\bibnamefont{Neronov}},
  \bibinfo{author}{\bibfnamefont{O.}~\bibnamefont{Ruchayskiy}},
  \bibinfo{author}{\bibfnamefont{M.}~\bibnamefont{Shaposhnikov}},
  \bibnamefont{and} \bibinfo{author}{\bibfnamefont{I.}~\bibnamefont{Tkachev}},
  \bibinfo{journal}{Phys. Rev. Lett.} \textbf{\bibinfo{volume}{97}},
  \bibinfo{pages}{261302} (\bibinfo{year}{2006}), \eprint{astro-ph/0603660}.

\bibitem{RiemerSorensen:2006fh}
\bibinfo{author}{\bibfnamefont{S.}~\bibnamefont{Riemer-Sorensen}},
  \bibinfo{author}{\bibfnamefont{S.~H.} \bibnamefont{Hansen}},
  \bibnamefont{and} \bibinfo{author}{\bibfnamefont{K.}~\bibnamefont{Pedersen}},
  \bibinfo{journal}{Astrophys. J.} \textbf{\bibinfo{volume}{644}},
  \bibinfo{pages}{L33} (\bibinfo{year}{2006}), \eprint{astro-ph/0603661}.

\bibitem{Abazajian:2006yn}
\bibinfo{author}{\bibfnamefont{K.}~\bibnamefont{Abazajian}} \bibnamefont{and}
  \bibinfo{author}{\bibfnamefont{S.~M.} \bibnamefont{Koushiappas}},
  \bibinfo{journal}{Phys. Rev.} \textbf{\bibinfo{volume}{D74}},
  \bibinfo{pages}{023527} (\bibinfo{year}{2006}), \eprint{astro-ph/0605271}.

\bibitem{Boyarsky:2006ag}
\bibinfo{author}{\bibfnamefont{A.}~\bibnamefont{Boyarsky}},
  \bibinfo{author}{\bibfnamefont{J.}~\bibnamefont{Nevalainen}},
  \bibnamefont{and}
  \bibinfo{author}{\bibfnamefont{O.}~\bibnamefont{Ruchayskiy}},
  \bibinfo{journal}{Astron. Astrophys.} \textbf{\bibinfo{volume}{471}},
  \bibinfo{pages}{51} (\bibinfo{year}{2007}), \eprint{astro-ph/0610961}.

\bibitem{Boyarsky:2007ge}
\bibinfo{author}{\bibfnamefont{A.}~\bibnamefont{Boyarsky}},
  \bibinfo{author}{\bibfnamefont{D.}~\bibnamefont{Malyshev}},
  \bibinfo{author}{\bibfnamefont{A.}~\bibnamefont{Neronov}}, \bibnamefont{and}
  \bibinfo{author}{\bibfnamefont{O.}~\bibnamefont{Ruchayskiy}},
  \bibinfo{journal}{Mon. Not. Roy. Astron. Soc.}
  \textbf{\bibinfo{volume}{387}}, \bibinfo{pages}{1345} (\bibinfo{year}{2008}),
  \eprint{0710.4922}.

\bibitem{Loewenstein:2008yi}
\bibinfo{author}{\bibfnamefont{M.}~\bibnamefont{Loewenstein}},
  \bibinfo{author}{\bibfnamefont{A.}~\bibnamefont{Kusenko}}, \bibnamefont{and}
  \bibinfo{author}{\bibfnamefont{P.~L.} \bibnamefont{Biermann}},
  \bibinfo{journal}{Astrophys. J.} \textbf{\bibinfo{volume}{700}},
  \bibinfo{pages}{426} (\bibinfo{year}{2009}), \eprint{0812.2710}.

\bibitem{Watson:2011dw}
\bibinfo{author}{\bibfnamefont{C.~R.} \bibnamefont{Watson}},
  \bibinfo{author}{\bibfnamefont{Z.-Y.} \bibnamefont{Li}}, \bibnamefont{and}
  \bibinfo{author}{\bibfnamefont{N.~K.} \bibnamefont{Polley}},
  \bibinfo{journal}{JCAP} \textbf{\bibinfo{volume}{1203}}, \bibinfo{pages}{018}
  (\bibinfo{year}{2012}), \eprint{1111.4217}.

\bibitem{Loewenstein:2012px}
\bibinfo{author}{\bibfnamefont{M.}~\bibnamefont{Loewenstein}} \bibnamefont{and}
  \bibinfo{author}{\bibfnamefont{A.}~\bibnamefont{Kusenko}},
  \bibinfo{journal}{Astrophys. J.} \textbf{\bibinfo{volume}{751}},
  \bibinfo{pages}{82} (\bibinfo{year}{2012}), \eprint{1203.5229}.

\bibitem{Merle:2013gea}
\bibinfo{author}{\bibfnamefont{A.}~\bibnamefont{Merle}}, \bibinfo{journal}{Int.
  J. Mod. Phys.} \textbf{\bibinfo{volume}{D22}}, \bibinfo{pages}{1330020}
  (\bibinfo{year}{2013}), \eprint{1302.2625}.

\bibitem{Shaposhnikov:2006nn}
\bibinfo{author}{\bibfnamefont{M.}~\bibnamefont{Shaposhnikov}},
  \bibinfo{journal}{Nucl. Phys.} \textbf{\bibinfo{volume}{B763}},
  \bibinfo{pages}{49} (\bibinfo{year}{2007}), \eprint{hep-ph/0605047}.

\bibitem{Araki:2011zg}
\bibinfo{author}{\bibfnamefont{T.}~\bibnamefont{Araki}} \bibnamefont{and}
  \bibinfo{author}{\bibfnamefont{Y.~F.} \bibnamefont{Li}},
  \bibinfo{journal}{Phys. Rev.} \textbf{\bibinfo{volume}{D85}},
  \bibinfo{pages}{065016} (\bibinfo{year}{2012}), \eprint{1112.5819}.

\bibitem{Barry:2011fp}
\bibinfo{author}{\bibfnamefont{J.}~\bibnamefont{Barry}},
  \bibinfo{author}{\bibfnamefont{W.}~\bibnamefont{Rodejohann}},
  \bibnamefont{and} \bibinfo{author}{\bibfnamefont{H.}~\bibnamefont{Zhang}},
  \bibinfo{journal}{JCAP} \textbf{\bibinfo{volume}{1201}}, \bibinfo{pages}{052}
  (\bibinfo{year}{2012}), \eprint{1110.6382}.

\bibitem{Barry:2011wb}
\bibinfo{author}{\bibfnamefont{J.}~\bibnamefont{Barry}},
  \bibinfo{author}{\bibfnamefont{W.}~\bibnamefont{Rodejohann}},
  \bibnamefont{and} \bibinfo{author}{\bibfnamefont{H.}~\bibnamefont{Zhang}},
  \bibinfo{journal}{JHEP} \textbf{\bibinfo{volume}{1107}}, \bibinfo{pages}{091}
  (\bibinfo{year}{2011}), \eprint{1105.3911}.

\bibitem{Froggatt:1978nt}
\bibinfo{author}{\bibfnamefont{C.~D.} \bibnamefont{Froggatt}} \bibnamefont{and}
  \bibinfo{author}{\bibfnamefont{H.~B.} \bibnamefont{Nielsen}},
  \bibinfo{journal}{Nucl. Phys.} \textbf{\bibinfo{volume}{B147}},
  \bibinfo{pages}{277} (\bibinfo{year}{1979}).

\bibitem{Merle:2011yv}
\bibinfo{author}{\bibfnamefont{A.}~\bibnamefont{Merle}} \bibnamefont{and}
  \bibinfo{author}{\bibfnamefont{V.}~\bibnamefont{Niro}},
  \bibinfo{journal}{JCAP} \textbf{\bibinfo{volume}{1107}}, \bibinfo{pages}{023}
  (\bibinfo{year}{2011}), \eprint{1105.5136}.

\bibitem{Adulpravitchai:2011rq}
\bibinfo{author}{\bibfnamefont{A.}~\bibnamefont{Adulpravitchai}}
  \bibnamefont{and}
  \bibinfo{author}{\bibfnamefont{R.}~\bibnamefont{Takahashi}},
  \bibinfo{journal}{JHEP} \textbf{\bibinfo{volume}{1109}}, \bibinfo{pages}{127}
  (\bibinfo{year}{2011}), \eprint{1107.3829}.

\bibitem{Kusenko:2010ik}
\bibinfo{author}{\bibfnamefont{A.}~\bibnamefont{Kusenko}},
  \bibinfo{author}{\bibfnamefont{F.}~\bibnamefont{Takahashi}},
  \bibnamefont{and} \bibinfo{author}{\bibfnamefont{T.~T.}
  \bibnamefont{Yanagida}}, \bibinfo{journal}{Phys. Lett.}
  \textbf{\bibinfo{volume}{B693}}, \bibinfo{pages}{144} (\bibinfo{year}{2010}),
  \eprint{1006.1731}.

\bibitem{Zhang:2011vh}
\bibinfo{author}{\bibfnamefont{H.}~\bibnamefont{Zhang}},
  \bibinfo{journal}{Phys. Lett.} \textbf{\bibinfo{volume}{B714}},
  \bibinfo{pages}{262} (\bibinfo{year}{2012}), \eprint{1110.6838}.

\bibitem{Chun:1995js}
\bibinfo{author}{\bibfnamefont{E.~J.} \bibnamefont{Chun}},
  \bibinfo{author}{\bibfnamefont{A.~S.} \bibnamefont{Joshipura}},
  \bibnamefont{and} \bibinfo{author}{\bibfnamefont{A.~Y.}
  \bibnamefont{Smirnov}}, \bibinfo{journal}{Phys. Lett.}
  \textbf{\bibinfo{volume}{B357}}, \bibinfo{pages}{608} (\bibinfo{year}{1995}),
  \eprint{hep-ph/9505275}.

\bibitem{Bouchand:2012dx}
\bibinfo{author}{\bibfnamefont{R.}~\bibnamefont{Bouchand}} \bibnamefont{and}
  \bibinfo{author}{\bibfnamefont{A.}~\bibnamefont{Merle}},
  \bibinfo{journal}{JHEP} \textbf{\bibinfo{volume}{1207}}, \bibinfo{pages}{084}
  (\bibinfo{year}{2012}), \eprint{1205.0008}.

\bibitem{Lindner:2011it}
\bibinfo{author}{\bibfnamefont{M.}~\bibnamefont{Lindner}},
  \bibinfo{author}{\bibfnamefont{D.}~\bibnamefont{Schmidt}}, \bibnamefont{and}
  \bibinfo{author}{\bibfnamefont{T.}~\bibnamefont{Schwetz}},
  \bibinfo{journal}{Phys. Lett.} \textbf{\bibinfo{volume}{B705}},
  \bibinfo{pages}{324} (\bibinfo{year}{2011}), \eprint{1105.4626}.

\bibitem{Schmidt:2012yg}
\bibinfo{author}{\bibfnamefont{D.}~\bibnamefont{Schmidt}},
  \bibinfo{author}{\bibfnamefont{T.}~\bibnamefont{Schwetz}}, \bibnamefont{and}
  \bibinfo{author}{\bibfnamefont{T.}~\bibnamefont{Toma}},
  \bibinfo{journal}{Phys. Rev.} \textbf{\bibinfo{volume}{D85}},
  \bibinfo{pages}{073009} (\bibinfo{year}{2012}), \eprint{1201.0906}.

\bibitem{Farzan:2010mr}
\bibinfo{author}{\bibfnamefont{Y.}~\bibnamefont{Farzan}},
  \bibinfo{author}{\bibfnamefont{S.}~\bibnamefont{Pascoli}}, \bibnamefont{and}
  \bibinfo{author}{\bibfnamefont{M.~A.} \bibnamefont{Schmidt}},
  \bibinfo{journal}{JHEP} \textbf{\bibinfo{volume}{1010}}, \bibinfo{pages}{111}
  (\bibinfo{year}{2010}), \eprint{1005.5323}.

\bibitem{Gustafsson:2012vj}
\bibinfo{author}{\bibfnamefont{M.}~\bibnamefont{Gustafsson}},
  \bibinfo{author}{\bibfnamefont{J.~M.} \bibnamefont{No}}, \bibnamefont{and}
  \bibinfo{author}{\bibfnamefont{M.~A.} \bibnamefont{Rivera}},
  \bibinfo{journal}{Phys. Rev. Lett.} \textbf{\bibinfo{volume}{110}},
  \bibinfo{pages}{211802} (\bibinfo{year}{2013}), \eprint{1212.4806}.

\bibitem{Lindner:2010rr}
\bibinfo{author}{\bibfnamefont{M.}~\bibnamefont{Lindner}},
  \bibinfo{author}{\bibfnamefont{A.}~\bibnamefont{Merle}}, \bibnamefont{and}
  \bibinfo{author}{\bibfnamefont{V.}~\bibnamefont{Niro}},
  \bibinfo{journal}{Phys. Rev.} \textbf{\bibinfo{volume}{D82}},
  \bibinfo{pages}{123529} (\bibinfo{year}{2010}), \eprint{1005.3116}.

\bibitem{Frigerio:2009wf}
\bibinfo{author}{\bibfnamefont{M.}~\bibnamefont{Frigerio}} \bibnamefont{and}
  \bibinfo{author}{\bibfnamefont{T.}~\bibnamefont{Hambye}},
  \bibinfo{journal}{Phys. Rev.} \textbf{\bibinfo{volume}{D81}},
  \bibinfo{pages}{075002} (\bibinfo{year}{2010}), \eprint{0912.1545}.

\bibitem{Kadastik:2009dj}
\bibinfo{author}{\bibfnamefont{M.}~\bibnamefont{Kadastik}},
  \bibinfo{author}{\bibfnamefont{K.}~\bibnamefont{Kannike}}, \bibnamefont{and}
  \bibinfo{author}{\bibfnamefont{M.}~\bibnamefont{Raidal}},
  \bibinfo{journal}{Phys. Rev.} \textbf{\bibinfo{volume}{D81}},
  \bibinfo{pages}{015002} (\bibinfo{year}{2010}), \eprint{0903.2475}.

\bibitem{Hirsch:2010ru}
\bibinfo{author}{\bibfnamefont{M.}~\bibnamefont{Hirsch}},
  \bibinfo{author}{\bibfnamefont{S.}~\bibnamefont{Morisi}},
  \bibinfo{author}{\bibfnamefont{E.}~\bibnamefont{Peinado}}, \bibnamefont{and}
  \bibinfo{author}{\bibfnamefont{J.~W.~F.} \bibnamefont{Valle}},
  \bibinfo{journal}{Phys. Rev.} \textbf{\bibinfo{volume}{D82}},
  \bibinfo{pages}{116003} (\bibinfo{year}{2010}), \eprint{1007.0871}.

\bibitem{Boucenna:2011tj}
\bibinfo{author}{\bibfnamefont{M.~S.} \bibnamefont{Boucenna}},
  \bibinfo{author}{\bibfnamefont{M.}~\bibnamefont{Hirsch}},
  \bibinfo{author}{\bibfnamefont{S.}~\bibnamefont{Morisi}},
  \bibinfo{author}{\bibfnamefont{E.}~\bibnamefont{Peinado}},
  \bibinfo{author}{\bibfnamefont{M.}~\bibnamefont{Taoso}}, \emph{et~al.},
  \bibinfo{journal}{JHEP} \textbf{\bibinfo{volume}{1105}}, \bibinfo{pages}{037}
  (\bibinfo{year}{2011}), \eprint{1101.2874}.

\bibitem{Adulpravitchai:2011ei}
\bibinfo{author}{\bibfnamefont{A.}~\bibnamefont{Adulpravitchai}},
  \bibinfo{author}{\bibfnamefont{B.}~\bibnamefont{Batell}}, \bibnamefont{and}
  \bibinfo{author}{\bibfnamefont{J.}~\bibnamefont{Pradler}},
  \bibinfo{journal}{Phys. Lett.} \textbf{\bibinfo{volume}{B700}},
  \bibinfo{pages}{207} (\bibinfo{year}{2011}), \eprint{1103.3053}.

\bibitem{Hambye:2010zb}
\bibinfo{author}{\bibfnamefont{T.}~\bibnamefont{Hambye}},
  \bibinfo{journal}{PoS} \textbf{\bibinfo{volume}{IDM2010}},
  \bibinfo{pages}{098} (\bibinfo{year}{2011}), \eprint{1012.4587}.

\bibitem{Kajiyama:2011gu}
\bibinfo{author}{\bibfnamefont{Y.}~\bibnamefont{Kajiyama}},
  \bibinfo{author}{\bibfnamefont{K.}~\bibnamefont{Kannike}}, \bibnamefont{and}
  \bibinfo{author}{\bibfnamefont{M.}~\bibnamefont{Raidal}},
  \bibinfo{journal}{Phys.Rev.} \textbf{\bibinfo{volume}{D85}},
  \bibinfo{pages}{033008} (\bibinfo{year}{2012}), \eprint{1111.1270}.

\end{thebibliography}

\end{document}